\documentclass[a4paper,12pt]{article}
\usepackage{epsf}
\newcommand{\tr}{\mbox{tr}\,}
\newcommand{\re}{\mbox{Re}\,}
\newcommand{\im}{\mbox{Im}\,}
\newcommand{\Tr}{\mbox{Tr}}

\makeatletter
\@addtoreset{equation}{section}
\@addtoreset{table}{section}
\@addtoreset{figure}{section}
\makeatother

\begin{document}
\pagestyle{plain}
\title{\rightline{\normalsize HUB-EP-99/67}\vskip 1truecm
QCD forces and heavy quark bound states}
\author {Gunnar S.\ Bali\thanks{E-mail: bali@physik.hu-berlin.de}\\
\normalsize\em Humboldt-Universit\"at zu Berlin, Institut f\"ur Physik,\\
\normalsize\em Invalidenstr.~110, 10115 Berlin, Germany}
\maketitle

\begin{abstract}
\baselineskip=20pt
The present knowledge of QCD confining forces between
static test charges is summarised, with an emphasis on lattice results.
Recent developments in
relating QCD potentials to quarkonium properties by use of
effective field theory methods are presented.
The validity of non-relativistic QCD and the adiabatic
approximation with respect to heavy quark bound states is explored.
Besides the static potential and relativistic correction terms,
the spectra of glueballs and gluinoballs,
hybrid excitations of the
QCD flux tube between fundamental colour sources,
potentials between charges in various representations of the
$SU(3)$ gauge group, and multi-particle interactions are discussed.
Some implications for quarkonia systems and quark-gluon
hybrid mesons are drawn.
\end{abstract}

\thispagestyle{empty}

\newpage \pagenumbering{arabic}
\tableofcontents
\newpage

\section{Motivation}
The phenomenology of strong interactions contains three fundamental
ingredients: the confinement of colour charges, chiral symmetry breaking
and asymptotic freedom.
The latter requirement culminated in the invention of quantum chromodynamics
(QCD)
some 25 years ago. Predicting low energy
properties of strongly interacting matter still represents a
serious theoretical challenge. This is particularly disappointing since
non-perturbative techniques are not only
important in QCD but also for an understanding of
physics beyond the standard model or perturbation theory.
For instance
a rigorous proof is still lacking that shows QCD as the microscopic theory of
strong interactions to give rise to the macroscopic properties of
chiral symmetry breaking and quark confinement.

So far Lattice Gauge Theory~\cite{Wilson:1974sk} constitutes the only
known entirely non-perturbative regularisation scheme.
By numerically simulating gauge theories on a lattice, one can in principle
predict properties of interacting
QCD matter without any non-QCD input (except for the
quark masses).
Such simulations have provided
convincing evidence not only for quark confinement~\cite{Creutz:1980zw}
but also
for chiral symmetry
breaking.
Moreover,
at finite temperature,
pure gauge theories are found to undergo a confinement-deconfinement
phase transition~\cite{Kuti:1981gh,McLerran:1981pk,McLerran:1981pb}
while chiral symmetry is restored at high
temperature~\cite{Gupta:1986vt,Gottlieb:1987eg},
in QCD with sea quarks.
The accuracy of these results
has been tremendously improved during the past decade
with the availability of more
powerful computers and advanced numerical techniques.

Unfortunately, the speed and memory of present day computers still
allows only for
``solving'' relatively simple QCD problems to a satisfactory precision.
One particular weakness that the standard lattice methodology
shares with, for instance, the QCD sum rule
approach~\cite{Shifman:1979bx,Novikov:1978cn}
is the difficulty in calculating
properties of radially excited hadrons. In simple potential models, however,
the spectrum of such excitations can easily be computed.
Such models have been successfully applied  in quarkonium
physics since the discovery of the $J/\psi$ resonance more than two decades ago~\cite{Kang:1975cq,Appelquist:1975ya,Eichten:1975af,Eichten:1977jk,Gromes:1977np,Richardson:1979bt,Quigg:1979vr,Eichten:1980ms,Martin:1980jx,Martin:1981rm,Buchmuller:1981su,Quigg:1981bj}.

A Hamiltonian representation
in terms of functions of simple dynamical variables such
as distance, angular momentum, relative momentum and spin
allows for an understanding of the underlying system that is rather
transparent and intuitive.
One would like to clarify what component of the
success of this simple picture
results from the freedom of choice in constructing a phenomenological
Hamiltonian
and what part indeed reflects fundamental properties of the underlying bound
state dynamics. Not long ago, a semi-relativistic
Hamiltonian that governs heavy quarkonia bound states has been
directly derived from
QCD~\cite{Eichten:1979pu,Eichten:1981mw,Peskin:1983up,Gromes:1984pm,Barchielli:1988zs,Barchielli:1990zp,Chen:1995dg,Bali:1997am,Bali:1998pi,Brambilla:1999ja,Brambilla:2000gk}.
Starting from a non-relativistic expansion of the
QCD Lagrangian (NRQCD)~\cite{Caswell:1986ui,Thacker:1991bm,Lepage:1992tx},
the gluonic degrees of freedom have been separated
from the heavy quark dynamics into functions of the canonical
coordinates (the potentials) and integrated out
by means of lattice simulations~\cite{Bali:1997am}.
The resulting Hamiltonian incorporates many properties
of the previously proposed purely phenomenological or QCD inspired models.

Heavy quarks closely resemble static test charges
which can be used to probe microscopic properties of the QCD vacuum,
in particular the anatomy of the confinement mechanism. Indeed, from
charmonium
spectroscopy and even more so from bottomonia
states, a lot has been learned about the nature and properties of QCD
confining forces.
Either motivated by experimental input or by QCD itself, many effective
models of
low energy aspects have been proposed, in
particular bag
models~\cite{Chodos:1974je,DeGrand:1975cf,Hasenfratz:1978dt,Hasenfratz:1980jv,DeTar:1983rw},
strong coupling and flux tube
models~\cite{Kogut:1976zr,Buchmuller:1982fr,Isgur:1983wj,Isgur:1985bm},
bosonic string models~\cite{Goddard:1973qh,Chodos:1974gt,Luscher:1981iy},
the
stochastic vacuum model~\cite{Dosch:1987sk,Dosch:1988ha,Simonov:1988rn},
dual QCD~\cite{Baker:1983bt,Baker:1986mh,Baker:1995nq,Baker:1996mk}
and the Abelian Higgs model~\cite{Maedan:1988yi},
instanton based
models~\cite{Shuryak:1982ff,Diakonov:1986eg,Schafer:1998wv} and
relativistic quark
models~\cite{Godfrey:1985xj}.
Many of these models are either expected to apply best to
a non-relativistic setting or can most easily be solved
in the situation of slowly moving colour charges.

In view of the fact that many problems like
properties of complex nuclei are unlikely ever to be solved
from {\em first principles} alone, to some extent modelling
and approximations will always be required.
Recently, using the stochastic vacuum model
as well as dual QCD and
the minimal area law, that is common to the strong coupling
limit and
string pictures,
the potentials within the quarkonium bound state Hamiltonian
have been computed~\cite{Baker:1996mu,Brambilla:1997aq}, and compared to
lattice results
to test the underlying assumptions
in the non-relativistic setting~\cite{Baker:1997bg,Brambilla:1998bt}.
It is a challenge for lattice
simulations to realise simple QCD situations in which low energy models
can be thoroughly checked.

Predictions of low energy quantities
like hadron masses and form factors
are the obvious phenomenological application of lattice QCD methods.
In view of the new $b$ physics experiments Babar, Belle, HERA-B and LHCb,
precise non-perturbative QCD contributions to weak decay constants
are required to relate experimental input to the least well determined CKM
matrix elements. Heavy-light systems are also thought to be
sensitive towards CP violations. In view of the proposed
linear electron colliders NLC and TESLA
a calculation
of the top production rate, $e^+e^-\rightarrow t\bar{t}$, near threshold
is required to precisely determine the top quark mass and even in this
high energy regime non-perturbative effects might turn out to play
an substantial r\^ole.
Therefore, developing heavy quark methods and verifying their accurateness
against precision experimental data from quarkonium systems
is of utmost interest.
Even quarkonia themselves contain valuable information.
For instance, one would expect cleaner discriminatory signals
for heavy quark-gluon hybrid states, that should exist as a consequence of QCD,
than for their light hybrid counterparts. Moreover,
the first $B_c$ mesons have recently been discovered and it is a challenge to
predict their spectrum. Last but not least, quarkonia systems contain
information on the $c$ and $b$ quark masses that are fundamental parameters
of the Standard Model.

This report is organised as follows: in Section~\ref{spect},
phenomenological evidence for linear confinement from the spectrum
of light mesons and quarkonia is presented.
In Section~\ref{field}, a brief introduction to the lattice methodology
is provided before the
present knowledge on the static
QCD potential will be reviewed in Section~\ref{potential}. In view of
latest results from lattice simulations including sea quarks, particular
emphasis is put on the ``breaking'' of the hadronic string in full QCD.
Subsequently, in Section~\ref{potential2}
static forces
in more complicated situations, in particular
hybrid potentials, bound states involving static gluinos,
potentials between charges in
higher representations of the $SU(N)$ colour group, and multi-body
forces are discussed.
In Section~\ref{relativistic}, attention is paid to
relativistic corrections to the static potential and the applicability
of the adiabatic approximation.
The results are then applied to quarkonium systems
in Section~\ref{quarkonia}.

\section{The hadron spectrum}
\label{spect}
The discovery of asymptotically free constituents of hadronic matter
in deep inelastic scattering experiments gave birth to
QCD as the generally accepted theory of strong interactions.
However, the most precise experimental data to-date, the
hadron spectrum, have been obtained
in the low energy region and not at the high energies
necessary to resolve the quark-gluon sub-structure of hadrons.
While perturbative QCD (pQCD) should be applicable to
high energy scattering problems to some extent,
solving QCD in the low energy region
poses a serious problem to theorists: 
not only does one have to deal with a strongly coupled system
but also with
a relativistic many-body bound state problem.
Moreover, unlike in the prototype gauge theory, QED, even on the
classical level the QCD vacuum structure is non-trivial, giving rise
to instanton induced effects for example.

It is instructive to consider the historical developments that
culminated in the discovery of QCD, in particular since the pre-QCD
era was dominated by concepts that were almost exclusively inspired
by non-perturbative phenomenology, such as the resonance spectrum.
General $S$-matrix properties
and dispersive relations~\cite{Regge:1959mz,Chew:1961ev} formed the
formal basis of such pre-QCD developments.
A serious conceptual problem of the $S$-matrix approach (also known as the
bootstrap) is the fact that the
unitarity of tree level scattering amplitudes is
broken as soon as one allows for virtual
point-like quanta of spin larger than one
to be exchanged between external particles.
This observation was one of the motivations for
Veneziano's duality
conjecture~\cite{Veneziano:1968yb} and the dual resonance model
of the late 60s which finally culminated in the invention of string
theories~\cite{Neveu:1971rx,Goto:1971ce,Goddard:1973qh,Nambu:1974zg}. 

While the $S$-matrix framework
addressed dynamical issues of strong
interactions, the na\"{\i}ve $SU_F(3)$ quark
model~\cite{Gell-Mann:1964nj,Zweig:1964jf}
served well in classifying all known hadronic states, in particular
after it had been extended by the colour $SU(3)$ degrees of
freedom~\cite{Greenberg:1964pe,Han:1965pf}.
However,
the quark model alone did not relate to any dynamical questions
of the underlying interaction. For instance,
no explanation was provided for the alignment of particles of mass
$m$ and spin $J$
along almost linear Regge trajectories in the $m^2-J$
plane~\cite{Chew:1962eu,Regge:1959mz}.
Bosonic string theories finally did not only resolve the unitarity puzzle
of the $S$-matrix theory
but also offered an explanation for the linearity of Regge
trajectories~\cite{Goto:1971ce,Goddard:1973qh,Nambu:1974zg}. 
However, string theories encountered internal inconsistencies when
formulated in four space-time dimensions~\cite{Brink:1973ja}
and were also incompatible
with the Bj{\o}rken scaling observed
in $e^-p$ collisions~\cite{Bjorken:1969dy}.
An explanation for the latter
was provided by the invention of
partons~\cite{Feynman:1969ej,Bjorken:1969ja,Gross:1969jf}
and asymptotic
freedom.

With the advent of QCD dynamics~\cite{Fritzsch:1973pi,Weinberg:1973un},
these partons were identified
as the quarks of the eightfold way and became the accepted
elementary constituents
of hadronic matter: the string theory
of strong interactions
that had been developed in parallel
survived only
as a possible low energy effective theory, in four space-time dimensions.
While QCD --- unlike
all preceeding suggestions ---
certainly
explains asymptotic freedom, it is still unproven that it
indeed results in collective phenomena such as the confinement of
quarks and gluons or chiral symmetry breaking. However,
lattice simulations provide convincing evidence.

It is legitimate to
speculate whether QCD really contains all low energy information:
is the set of
fundamental parameters that describes the hadron spectrum compatible
with the parameters needed to explain high energy scattering experiments or
is there place for new physics? For example
a (hypothetical) gluino with mass of a few GeV would
affect the running of the QCD coupling between $m_Z$
and typical hadronic scales that are smaller by
two orders of magnitude.
Is QCD the right theory at all? If so, quark-gluon
hybrids and glueballs should show up in the particle spectrum.
Although these general questions are not central to this article they
motivate continued phenomenological interest in QCD itself from a general
perspective.

The discovery of states composed of heavy quarks,
namely
charmonia in 1974 and bottomonia in 1977, enabled aspects
of strong interaction dynamics to be probed
in a non-relativistic setting.
By means of simple potential models a wealth of data on energy levels
and decay rates
could be explained. The question arises: if these models yield the
right particle spectrum, can they eventually be derived from QCD?
What do such models tell us about QCD and what does QCD tell us about such
models?

Before addressing these questions in later Sections, here
some aspects of hadron spectroscopy that relate to flux tube and potential
models are summarised.

\subsection{Regge trajectories}
\label{sec:regtra}
\begin{table}
\caption{Light meson masses.}
\label{mesons}

\begin{center}
\begin{tabular}{c|c|c}
state&$J^{P(C)}$&$m/\mbox{MeV}$\\\hline
$\pi$           &$0^{-+}$&138\\
$b_1(1235)$     &$1^{+-}$&1229(3)\\
$\pi_2(1670)$   &$2^{-+}$&1670(20)\\\hline
$\rho(770)$     &$1^{--}$&770(1)\\
$a_2(1320)$     &$2^{++}$&1318(1)\\
$\rho_3(1690)$  &$3^{--}$&1691(5)\\
$a_4(2040)$     &$4^{++}$&2020(16)\\\hline
$\omega(782)$   &$1^{--}$&782\\
$f_2(1270)$     &$2^{++}$&1275(1)\\
$\omega_3(1670)$&$3^{--}$&1667(4)\\
$f_4(2050)$    &$4^{++}$&2044(11)\\\hline
$\phi(1020)$    &$1^{--}$&1019\\
$f_2'(1525)$    &$2^{++}$&1525(5)\\
$\phi_3(1850)$  &$3^{--}$&1854(7)\\\hline
$\eta$          &$0^{-+}$&547\\
$h_1(1170)$     &$1^{+-}$&1170(20)\\\hline
$K$             &$0^-   $&495\\
$K_1(1270)$     &$1^+   $&1273(7)\\
$K_2(1770)$     &$2^-   $&1773(8)\\\hline
$K^*(892)$      &$1^-   $&893\\
$K_2^*(1430)$   &$2^+   $&1428(2)\\
$K_3^*(1780)$   &$3^-   $&1776(7)\\
$K_4^*(2045)$   &$4^+   $&2045(9)
\end{tabular}
\end{center}
\end{table}
Since the early sixties it has been noticed that mesons as well as
baryons of mass $m$ and spin $J$ group themselves into almost linear,
so-called Regge
trajectories~\cite{Regge:1959mz,Chew:1961ev,Chew:1962eu}
in the $m^2-J$ plane up to spins as high as
$J=11/2$. In Table~\ref{mesons} the light meson spectrum is summarised.
Only resonances that are confirmed in the Review of Particle
Properties~\cite{Caso:1998tx}
have been included.
The $\pi$, $K^*$, $K_2^*$ and $K$ triplets
have been replaced by their weighted mass averages.
The second column of the Table represents the $J^{PC}$ assignment.
Each increase of the orbital angular momentum
by one unit results
in a switch of both, parity and charge assignments.

\begin{figure}[thb]
\centerline{\epsfxsize=10truecm\epsffile{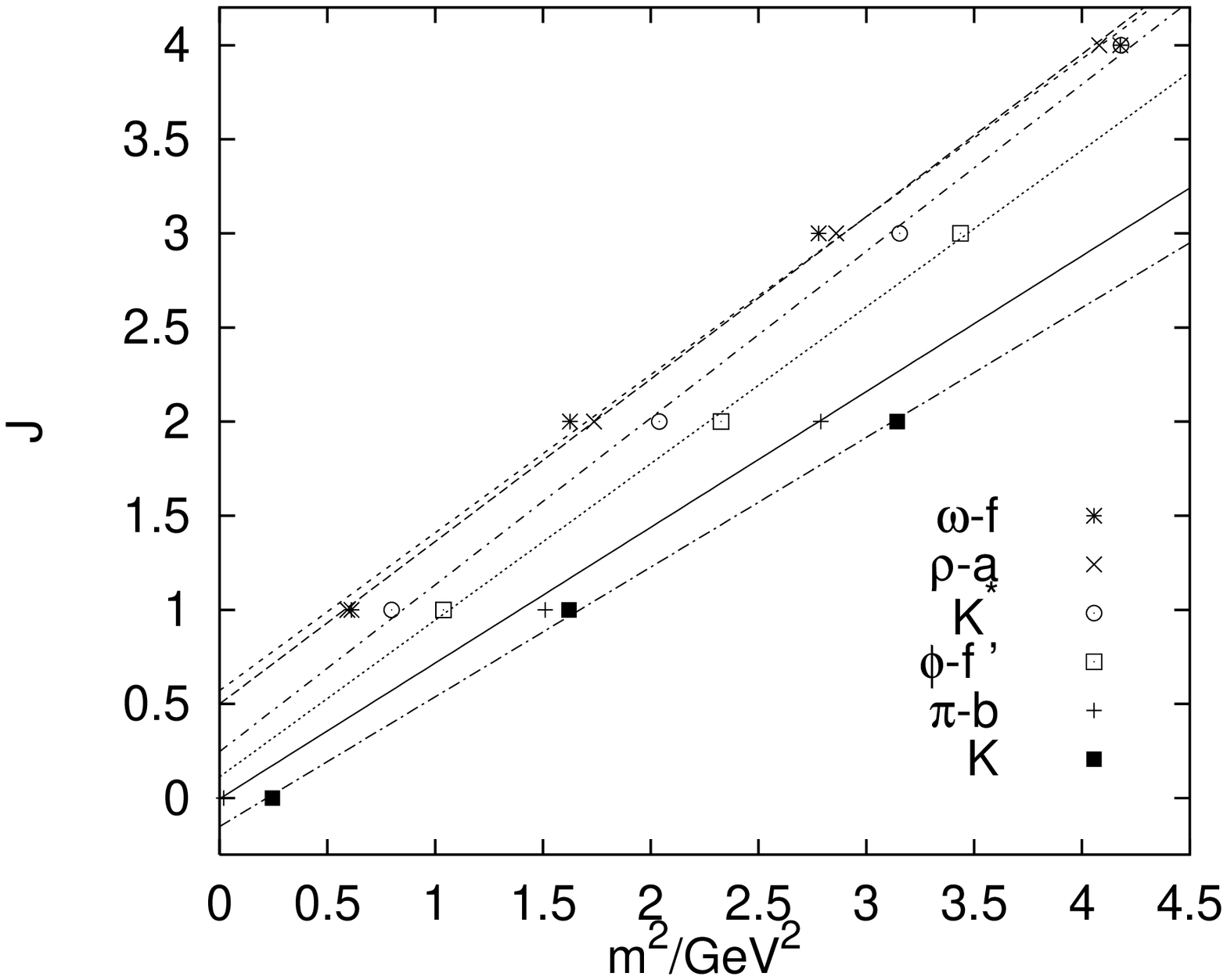}}
\caption{Regge trajectories.}
\label{figregge}
\end{figure}

\begin{table}
\caption{String tensions from Regge trajectories.}
\label{regge}

\begin{center}
\begin{tabular}{c|c|c}
trajectory&$\sqrt{\sigma}/\mbox{MeV}$&$\Delta J$\\\hline
$\pi,b_1,\ldots$&469(6)&0.06\\
$\rho,a_2,\ldots$&429(2)&0.03\\
$\omega,f_2,\ldots$&436(8)&0.12\\
$\phi,f_2',\ldots$&437(5)&0.06\\
$K,K_1,\ldots$&480(4)&0.04\\
$K^*,K^*_2,\ldots$&424(5)&0.07
\end{tabular}
\end{center}
\end{table}

The data of Table~\ref{mesons} is displayed in Figure~\ref{figregge},
together with linear fits of the form,
\begin{equation}
J(m)=\alpha(0)+\alpha'm^2.
\end{equation}
Similar plots can be made for the baryon spectrum.
$\alpha(0)$ is known as the Regge intersect and,
\begin{equation}
\label{eq:regges}
\alpha'=\frac{1}{2\pi\sigma},
\end{equation}
as the Regge slope. The resulting values for the
``string tension'', $\sigma$, are displayed
in Table~\ref{regge}. While statistical errors on the data points
increase with $J$, the applicability of the relativistic string model
that, as we shall see below,
predicts the linear dependence is expected to improve with $J$.
Therefore, in the fits we have decided to ignore the experimental
errors and give all points equal weight.
$\Delta J$ denotes the root mean square
deviation between fitted angular momenta and data points, normalised by the
root of the degrees of freedom (i.e.\ the number of data points minus two)
and reflects the overall quality of a fit.

\begin{figure}[thb]
\centerline{\epsfxsize=8truecm\epsffile{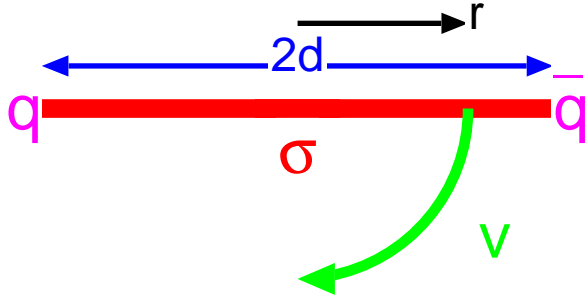}}
\caption{Rotating quarks, connected by a string of constant energy density.}
\label{figrot}
\end{figure}

A simple explanation of the linear behaviour is provided by the relativistic
string model~\cite{Goto:1971ce,Goddard:1973qh}: imagine a rotating string
of length $2d$ with a constant energy density per unit length,
$\sigma$ (Figure~\ref{figrot}).
If this string spans between (approximately)
massless quarks, we might expect
those quarks to move at (almost)
the speed of light, $c=1$, with respect to the centre of mass.
The velocity as a function of the distance from the centre of the string, $r$,
in this set-up is given by, $v(r)=r/d$. From this, we calculate the energy
stored in the rotating string,
\begin{equation}
m=2\int_0^d\frac{dr\,\sigma}{\sqrt{1-v^2(r)}}=\pi d\sigma,
\end{equation}
and angular momentum,
\begin{equation}
J=2\int_0^d\frac{dr\,\sigma\,rv(r)}{\sqrt{1-v^2(r)}}=\frac{\pi^2d^2\sigma}{2}
=\frac{1}{2\pi\sigma}m^2,
\end{equation}
which results in the relation of Eq.~(\ref{eq:regges}) between
Regge slope, $\alpha'$, and string tension, $\sigma$.
This crude approximation can of course be improved.
For example, one can allow for a rest mass of the quarks.
Velocities smaller than $c$ will
result in a slight increase
of the Regge slope. The assumption that the string energy entirely consists
of a longitudinal electric component in the
co-rotating frame yields predictions for
spin-orbit splittings~\cite{Buchmuller:1982fr} etc..

For the two Regge trajectories starting with a pseudo-scalar ($\pi$ and $K$),
one finds values, $470\,\mbox{MeV}<\sqrt{\sigma}< 480$~MeV, while
all other numbers scatter between
424 and 437 MeV. The value extracted from the $\rho,a_2,\ldots$ trajectory,
which is the most linear one, is $\sqrt{\sigma}=(429\pm 2)$~MeV.

\subsection{Quarkonia}
\label{sec:quark}
Soon after the discovery of the $J/\psi$ meson in $e^+e^-$ annihilation,
the possibility of a non-relativistic treatment of such states, in analogy to
the positronium of electrodynamics, was
suggested~\cite{Appelquist:1975ya}.
Quarkonia, i.e.\ mesonic states that contain two heavy constituent quarks,
either charm or bottom\footnote{Due to the large weak decay
rate, $t\rightarrow b W^+$,
the top quark does not appear as a constituent in bound states~(see e.g.\
Ref.~\cite{Quigg:1997fy}).},
owe their name to this analogy. Within the quark model,
the quark anti-quark system can be
characterised
by its total spin, ${\mathbf S}={\mathbf S_1}+{\mathbf S_2}$
($s=0$ or $s=1$), the relative orbital angular momentum,
${\mathbf L}$, and the total spin,
${\mathbf J}={\mathbf L}+{\mathbf S}$.
Within the standard
spectroscopic notation, $n^{2s+1}l_J$,
$n$ denotes the radial excitation while
$l=0$ is labelled by the letter
$S$, $l=1$ by $P$, $l=2$ by $D$ etc.. 
The parity of a quark anti-quark state is given by,
$P=(-1)^{l+1}$, while the charge conjugation operator (if quark and
anti-quark share the same flavour) has eigenvalue,
$C=(-1)^{l+s}$.

In making the above $J^{PC}$ assignments, we ignore
the possibility of the gluonic
degrees of freedom contributing to the quantum numbers.
This simplification results in certain combinations to be quark model
forbidden (or spin-exotic),
namely, $J^{PC}=0^{+-},0^{--},1^{-+},2^{+-},3^{-+},\ldots$.
Another aspect is that some $J^{PC}$ assignments can be generated in
various ways. For instance, ${}^3S_1$ and ${}^3D_1$ states both
result in $J^{PC}=1^{--}$. As soon as gluons are introduced,
the relative angular momentum, ${\mathbf L}$, is not conserved
anymore and physical vector particles will in general be superpositions
of excitations from these two channels: strictly speaking, only
the number of nodes of the wave function, $n$, the spin $J$, parity $P$,
charge $C$ (in the case of flavour singlet mesons),
and the constituent quark content
(neglecting annihilation processes and weak decays)
represent ``good" quantum numbers.

\begin{table}
\caption{Classification of charmonium and bottomonium states.}
\label{quarkon}

\begin{center}
\begin{tabular}{c|c|c|c}
$n^{2s+1}l_J$&$J^{PC}$&$c\bar{c}$&$b\bar{b}$\\\hline
$1^1S_0$&$0^{-+}$&$\eta_c$&$\eta_b$\\
$1^3S_1$&$1^{--}$&$J/\psi$&$\Upsilon$\\
$2^3S_1$&$1^{--}$&$\psi(2S)$&$\Upsilon(2S)$\\
$1^1P_1$&$1^{+-}$&$h_c$&$h_b$\\
$1^3P_0$&$0^{++}$&$\chi_{c0}$&$\chi_{b0}$\\
$1^3P_1$&$1^{++}$&$\chi_{c1}$&$\chi_{b1}$\\
$1^3P_2$&$2^{++}$&$\chi_{c2}$&$\chi_{b2}$
\end{tabular}
\end{center}
\end{table}

In Table~\ref{quarkon}, we have compiled quantum numbers
and names for some members of the $J/\psi$ and $\Upsilon$ families.
Little is known experimentally about $B_c$  mesons, which are
bound states of
a $\bar{b}$ and a $c$ quark. For these particles an additional
peculiarity has to be considered:
charge and total spin are no longer ``good" quark model
quantum numbers.
For $l\geq 1$ this results in mixing between the $J=l$
would-be singlet and would-be
triplet states.

\begin{figure}[thb]
\centerline{\epsfxsize=10truecm\epsffile{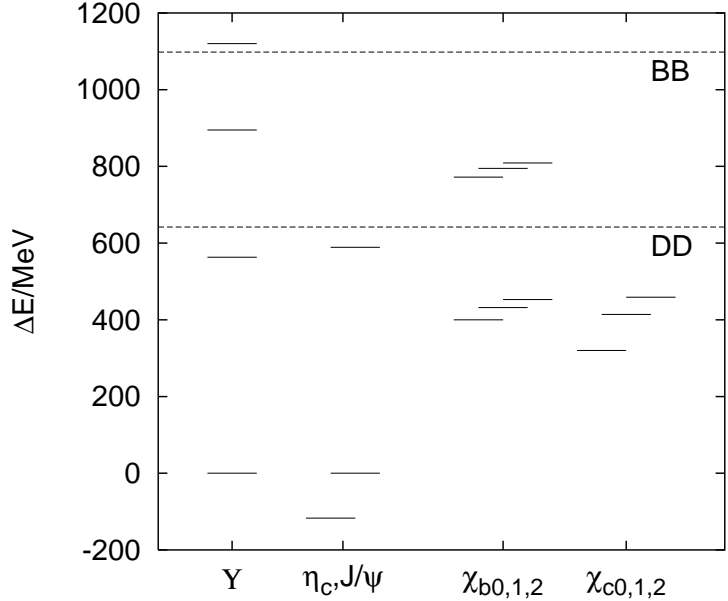}}
\caption{Energy splittings, $\Delta E$, for charmonia and bottomonia
with respect to the
$\Upsilon$ and $J/\psi$ triplet $S$ wave states.}
\label{figspect}
\end{figure}

In Figure~\ref{figspect}, all experimentally determined
splittings with respect to the $1^3S_1$ state for 
the $\Upsilon$ and $J/\psi$ families are depicted.
We have restricted ourselves to states,
listed in the Review of Particle
Properties~\cite{Caso:1998tx}, that are below the
$D\overline{D}$ and $B\overline{B}$ thresholds (dashed horizontal lines)
for charmonia and bottomonia,
respectively, with the
exception of the $\Upsilon(4S)$.
While the mass of the $J/\psi$ (3.097~GeV) considerably
differs from that of the $\Upsilon$ (9.46~GeV), indicating
a substantial difference in the quark masses,
$m_b\approx 3m_c$, both $2^3S_1-1^3S_1$ splittings
agree within 5~\% (589 and 563~MeV).
We define the spin averaged $\chi$ mass by,
\begin{equation}
m_{\overline{{}^3\!P}}=\frac{1}{9}\left(m_{{}^3\!P_0}+3m_{{}^3\!P_1}+
5m_{{}^3\!P_2}\right)\approx m_{{}^1\!P_1}.
\end{equation}
Again, within a few per cent, the $1\overline{{}^3P}-1^3S_1$ splittings
agree (429~MeV vs.\ 440~MeV).
Unfortunately, while the $\eta_c$ has been discovered, no 
pseudo-scalar $b\bar{b}$ ${}^1S_0$ meson has yet been seen, such that
a consistent comparison with respect to spin averaged $S$ state masses,
\begin{equation}
m_{\overline{S}}=\frac{1}{4}\left(m_{{}^1\!S_0}+3m_{{}^3\!S_1}\right),
\end{equation}
is not possible.

While the $2S-1S$ and $1P-1S$ splittings seem
to agree within a few per cent,
the fine structure splittings between the $P$ states
come out to be almost three times as large in the charm case,
compared to that for the
bottom,
\begin{equation}
\frac{m_{\chi_{b2}}-m_{\chi_{b0}}}{m_{\chi_{c2}}-m_{\chi_{c0}}}
=0.38(1)\approx
\frac{53\,\mbox{MeV}}{141\,\mbox{MeV}}.
\end{equation}
This is consistent with the expectation that in the limit of infinite
quark mass, fine structure splittings will eventually completely disappear,
in analogy to hydrogen-like
systems. However, for the 
ratio between the respective
$m_{\chi_2}-m_{\chi_1}$ splittings one finds a different
numerical value, 0.47(2), indicating a more complicated dependence on
the inverse quark mass than mere proportionality.

For sufficiently heavy quarks, one might hope that
the characteristic time scale associated with the relative movement of the
constituent quarks is much larger than that associated with
the gluonic (or sea quark)
degrees of freedom~\cite{Appelquist:1975ya}.
In this case the
adiabatic (or Born-Oppenheimer) approximation applies and the effect
of gluons and sea quarks can be represented by an averaged
instantaneous interaction potential between the heavy quark sources.
Moreover, the bound state problem will essentially become non-relativistic and
the dynamics will, to first approximation, be controlled by
the Schr\"odinger equation,
\begin{equation}
\label{eq:schroe}
\left[\frac{{\mathbf p}^2}{2\mu_R}+V(r)\right]\psi_{nll_3}({\mathbf r})
=E_{nl}\psi_{nll_3}({\mathbf r}),
\end{equation}
with a potential, $V(r)$ ($r=|{\mathbf r}|$), or, if
spin effects are taken into account, semi-relativistic Pauli-Thomas-like
extensions. In the adiabatic approximation quarkonia are
the positronium of QCD. However, unlike in QED where the interaction potential
can be calculated perturbatively and the spectrum predicted, we are faced with
the inverse problem of determining or guessing the interaction potential
and the reduced quark mass, $\mu_R=m/2$, from
the observed spectrum, $E_{nl}$, and decay rates. The latter can be
related to properties of the wave function at the origin~\cite{Quigg:1979vr}.
If the adiabatic approximation is justified we would expect, to leading order
in a semi-relativistic expansion, the same
potential to explain $c\bar{c}$ as well as $b\bar{b}$ spectra
since QCD interactions are flavour blind.
On the other hand it is clear that the adiabatic approximation
will at least fail for unstable excitations like the $\Psi(3S)$ or
$\Upsilon(4S)$ since decays cannot be accounted for by a one channel
Hamiltonian with a real potential.

\begin{table}
\caption{The scaling of level splittings, $\Delta E$, inverse length scales,
$\langle r^{-1}\rangle$, and the squared velocity, $\langle v^2\rangle$, with
reduced mass and coupling.}
\label{tab:potsca}

\begin{center}
\begin{tabular}{c|c|c|c}
potential&$\Delta E$&$\langle r^{-1}\rangle$&$\langle v^2\rangle$\\\hline
Coulomb, Eq.~(\ref{eq:coul2})&$e^2\mu_R$&$e\mu_R$&$e^2$\\
logarithmic, Eq.~(\ref{eq:log2})&$C$&$C^{-1/2}\mu_R^{1/2}r_0^{-1}$&$C\mu_R^{-1}$\\
linear, Eq.~(\ref{eq:lin2})&$\sigma^{2/3}\mu_R^{-1/3}$&$\sigma^{1/3}\mu_R^{1/3}$&
$\sigma^{2/3}\mu_R^{-4/3}$
\end{tabular}
\end{center}
\end{table}

In Appendix~\ref{app:schr}, we
derive general properties of the spectrum for power law and logarithmic
potentials. The main results for a Coulomb potential,
\begin{equation}
\label{eq:coul2}
V(r)=-\frac{e}{r},
\end{equation}
a logarithmic potential,
\begin{equation}
\label{eq:log2}
V(r)=C\ln\left(\frac{r}{r_0}\right),
\end{equation}
and a linear potential,
\begin{equation}
\label{eq:lin2}
V(r)=\sigma\,r,
\end{equation}
are displayed in Table~\ref{tab:potsca}.

From the spin-averaged
quarkonia spectra it is evident that the underlying potential
cannot be purely Coulomb type. Otherwise, the $2S-1S$ splitting
would be approximately degenerate with the lowest lying
$nP-1S$ splitting and, moreover, $\Upsilon$ splittings would be
enhanced with respect to  $J/\psi$ splittings by the ratio of the
quark masses, $m_b/m_c\approx 3$.
However, a logarithmic potential that would explain the approximate mass
independence of spin-averaged splittings
is incompatible with tree level perturbation theory, i.e.\
Eq.~(\ref{eq:coul2}), with $e=(4/3)\alpha_s$.

The Cornell potential~\cite{Eichten:1975af},
\begin{equation}
V(r)=-\frac{e}{r}+\sigma r,
\end{equation}
contains the perturbative expectation plus
an additional linear term.
The parameters $e$ and $\sigma$
can be adjusted such that within the range of charm and bottom
quark masses, the linear dependence of the Rydberg energy on
$\mu_R$ is compensated by the $1/\mu_R^{1/3}$ behaviour expected from the
large distance linear term: within the distance scales relevant for the
quarkonium
bound state problem, the Cornell potential looks effectively logarithmic.
At quark masses larger than $m_b$,
the Coulomb term will eventually dominate and splittings will diverge
in proportion with $\mu_R$. Note that the Cornell potential predicts
the average velocity,
$\langle v^2\rangle\propto \Delta E/\mu_R$, to saturate at the
value $\langle v^2\rangle = e^2$ for large quark mass while from
the approximate
equality of bottomonia and charmonia level splittings one would
expect $\langle v^2_b\rangle/\langle v^2_c\rangle\approx m_c/m_b$.
$\langle v^2\rangle$ quantifies the quality of the
non-relativistic approximation while the applicability of the
adiabatic approximation is more complicated
to establish from a QCD perspective.

Before the discovery of the $\Upsilon(2S)$, fits to the spin averaged
quarkonia spectra resulted in parameter values~\cite{Eichten:1977jk},
$e\approx 0.25$ and $\sqrt{\sigma}\approx 455$~MeV. After
inclusion of the
$\Upsilon$ states, that probe the potential at smaller distances,
values like~\cite{Quigg:1979vr},
$e\approx 0.51$ and $\sqrt{\sigma}\approx 412$~MeV,
and~\cite{Eichten:1980ms},
$e\approx 0.52$ and $\sqrt{\sigma}\approx 427$~MeV, emerged.
However, within the 
region, $0.2\,\mbox{fm}<r<1$~fm, which is effectively probed
by spin-averaged quarkonia splittings, the $e\approx 0.25$
parametrisation only marginally
differs from the $e\approx  0.5$ parametrisations; the higher value
of the Coulomb coefficient is compensated for by a smaller slope, $\sigma$.
Interestingly, the slope of the Cornell potential is in qualitative agreement
with $\sqrt{\sigma}\approx 430$~MeV, the
estimate of the string tension from Regge trajectories
of light mesons, discussed in
Section~\ref{sec:regtra}.

While the spin averaged spectrum probes the potential at distances
$r>0.2$~fm, fine structure splittings are sensitive
towards the Lorentz and spin structure of the
interacting force as well as to the functional form of the
potential at short distances.
We shall discuss this in detail in Section~\ref{quarkonia}.

\section{Lattice methods}
\label{field}
Lattice QCD was invented by Wilson~\cite{Wilson:1974sk}
shortly after QCD emerged as
the prime candidate for a consistent theory of strong interactions.
The main intention was to define an entirely non-perturbative
regularisation scheme for QCD, based on the principle of local
gauge invariance. Besides regulating the theory, the lattice lends
itself to strong coupling expansion techniques
in terms of the inverse QCD coupling,
\begin{equation}
\label{eq:betadef}
\beta=\frac{2N}{g^2}=\frac{2N}{4\pi\alpha_s}.
\end{equation}
Such techniques complement the conventional perturbative
weak coupling expansion and have, in particular in the Hamiltonian
formulation of lattice QCD~\cite{Kogut:1975ag}, stimulated the
flux tube model of Ref.~\cite{Isgur:1985bm}.
However, so far nobody has managed to analytically relate
the strong coupling limit of QCD to weak coupling results.
For instance, in $U(1)$ as well as in $SU(N)$ gauge theories
one obtains an area law for Wilson loops [Eq.~(\ref{eq:wl})],
i.e.\ confinement, in the
strong coupling limit. While in $(3+1)$-dimensional $U(1)$ lattice
gauge theory the strong coupling regime
is separated from a non-confining weak coupling region by a phase transition,
in $SU(N)$ one would hope that no such phase transition at finite $\beta$
exists and confinement survives at weak coupling.

Besides offering new analytical insight and techniques, the lattice
approach to QCD lends itself to treatment on a
computer~\cite{Creutz:1979dw,Wilson:1979wp,Creutz:1980zw}.
To allow for a numerical evaluation of expectation values
it is convenient to work in Euclidean space-time in which a
path integral measure can be defined.
Moreover, the time evolution operator becomes
anti-Hermitian which results in $n$-point correlation
functions decaying exponentially, rather than exhibiting
oscillatory behaviour.
Most results that are obtained in Euclidean space can
be related to the space-like region of the Minkowski world
and can in principle be analytically continued into the
time-like region of interest.
With results that have been obtained on a discrete set
of points with finite precision, however, such a continuation
is anything but straight forward.
Fortunately, unless one is interested in real time processes
like particle scattering, this is in general not required.
In particular the mass spectrum remains unaffected by
the rotation to imaginary time as long as reflection
positivity
holds~\cite{Osterwalder:1975tc},
which is the case at least for the lattice actions discussed
in this article.

In what follows,
the aspects of lattice simulations that are
relevant for our discussion are
summarised. For
a more detailed introduction to Lattice Gauge Theories
the reader may consult several
books~\cite{Creutz:1984mg,Itzykson:1989sx,Rothe:1992nt,Montvay:1994cy}
and review
articles~\cite{Drouffe:1978dn,Wilson:1979wp,Kogut:1983ds,Hasenfratz:1985pd,Gupta:1997nd,Wittig:1999kb}.
The conventions that are adapted throughout
the article are detailed in Appendix~\ref{app:conv}.

\subsection{What can the lattice do?}
The lattice allows for a {\em first principles}
numerical evaluation of expectation values
of a given quantum field
theory that is defined by an action $S$ in Euclidean space-time.
However, the accessible lattice volumes
and resolutions are limited by the available (finite)
computer performance and memory. 

The obvious strength of lattice methods are hadron mass predictions.
Only recently computers have become powerful enough to allow
for a determination of the infinite volume
light hadron spectrum in the continuum limit
in the quenched approximation\footnote{In the quenched approximation,
vacuum polarisation effects due to sea quarks are neglected by replacing
the fermionic part of the action by a constant. In the language of
perturbative QCD this amounts to neglecting quark loops.} to QCD within
uncertainties of a few per cent~\cite{Aoki:1999yr}.
To this accuracy the quenched spectrum has been found to
differ from experiment.
Some collaborations have started to systematically explore 
QCD with two flavours of light
sea quarks and the first precision results
indeed indicate deviations from the quenched
approximation in the direction of the experimental
values~\cite{Kuramashi:1999ij}.
Even if one is unimpressed by post-dictions of hadron masses
that have been known with high precision for decades
such simulations allow fundamental standard model
parameters to be fixed from
low energy input data, like quark
masses~\cite{Davies:1994pz,Eicker:1997ws,Garden:1999fg,AliKhan:1999zp}
and the QCD running
coupling~\cite{Capitani:1998mq}.
Of course, as we shall see, a wealth
of other applications of phenomenological
importance exists.

Unfortunately, only the lowest
radial excitations of a hadronic state are
accessible in practice. Lattice predictions
are restricted to rather simple systems too.
Even the deuteron is beyond the reach of present day super-computers.
Therefore, it is desirable to supplement lattice simulations by
analytical methods. The computer alone acts as a {\em black box}.
In order to understand
and interprete the output values and to predict their dependence on
the input parameters, some modelling is required.
Vice versa, the lattice itself is a strong tool to validate models and
approximations. Unlike in the ``real'' world, one can vary
the quark masses, $m_i$,
the number of colours, $N$,
the number of flavours, $n_f$, the temperature, the spatial volume,
the space-time dimension and even the
boundary conditions in order to expose models to thorough
tests in many situations.

\subsection{The method}
\begin{figure}[thb]
\centerline{\epsfxsize=13truecm\epsffile{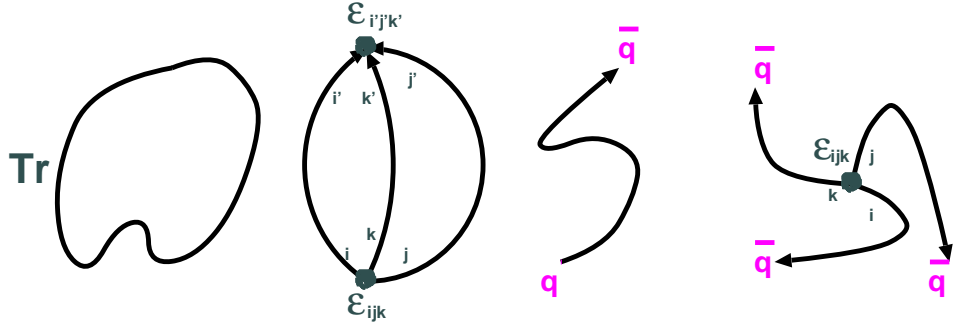}}
\caption{Examples of gauge invariant objects. Lines correspond to
gauge transporters.}
\label{fig:gaugeinv}
\end{figure}

In a lattice simulation, Euclidean space-time is
discretised on a torus\footnote{For fermions anti-periodic boundary
conditions are chosen in the temporal direction.}
with $L_{\sigma}^3L_{\tau}$ lattice points or sites,
$x=na, n_i=0,1,\ldots, L_{\sigma}-1$, $n_4=0,1,\ldots,L_{\tau}-1$, separated
by the lattice spacing\footnote{For simplicity, we assume
$a_1=a_2=a_3=a_4=a$, $L_1=L_2=L_3=L_{\sigma}$.}, $a$, that provides
an ultra-violet cut-off on the gluon momenta, $q\leq \pi/a$, and regulates the
theory. Two adjacent points are connected by an
oriented bond or link, $(x,\mu)$. While Dirac quark fields, $q^i_x$,
are represented by $4\times N$ tuples\footnote{The superscript, $i=1,\ldots
n_f$, runs over the flavours. The factor ``4'' is due to the Dirac components.}
at lattice sites, $x$, gauge fields,
\begin{equation}
U_{x,\mu}={\mathcal P}\left[\exp\left(i\int_x^{x+a\hat{\mu}}\!dx'_{\mu}\,
A_{\mu}(x')\right)\right]\in SU(N),
\end{equation}
are ``link variables''. ${\mathcal P}$ denotes path ordering of the argument
and $\hat{\mu}$ is a unit vector pointing into $\mu$ direction.
We further define, $U_{x,-\mu}=U^{\dagger}_{x-\hat{\mu},\mu}$.
The transformation property of a lattice fermion field under
gauge transformations, $\Omega_x\in SU(N)$, is [Eq.~(\ref{eq:transphi})],
\begin{equation}
q(x)\rightarrow \Omega(x)q(x),\quad \bar{q}(x)\rightarrow
\bar{q}(x)\Omega^{\dagger}(x).
\end{equation}
From Eq.~(\ref{eq:transa}),
\begin{equation}
\label{eq:transa2}
A_{\mu}\rightarrow A_{\mu}^{\Omega}=\Omega[A_{\mu}-i\partial_{\mu}]
\Omega^{\dagger}=-i\Omega D_{\mu}\Omega^{\dagger},
\end{equation}
one can derive the the transformation property
of links,
\begin{equation}
U_{x,\mu}\rightarrow U_{x,\mu}^{\Omega}=\Omega_x U_{x,\mu}
\Omega_{x+a\hat{\mu}}^{\dagger}.
\end{equation}
It is easy to see that the trace of a product of links along a closed loop
is gauge invariant. Other gauge invariant objects are
$N$ gauge transporters whose colour indices are
contracted by completely antisymmetric tensors of rank $N$
at a common start and a common end point,
a quark and an anti-quark field
that are connected by a gauge transporter or a state of $N$
quarks whose colours are transported to a common point,
where they are anti-symmetrically contracted.
The situation is depicted in Figure~\ref{fig:gaugeinv} for $N=3$.

\begin{figure}[thb]
\centerline{\epsfxsize=8truecm\epsffile{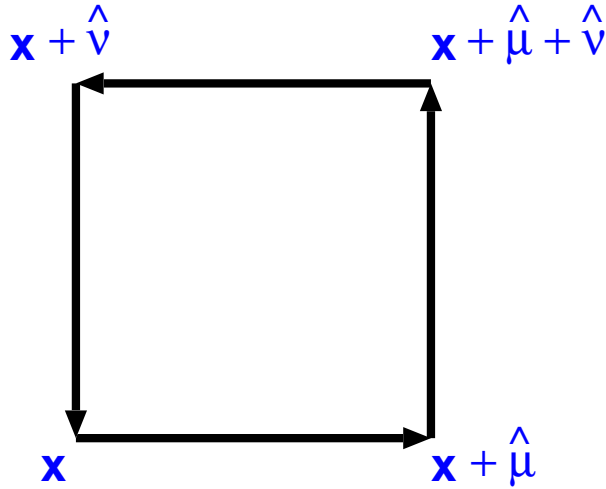}}
\caption{The plaquette, $U_{x,\mu\nu}$.}
\label{fig:plaq}
\end{figure}

The simplest non-trivial gauge invariant object that can be constructed
is the product of four links, enclosing an elementary square,
\begin{equation}
\label{eq:plaq}
U_{x,\mu\nu}=U_{x,\mu}U_{x+a\hat{\mu},\nu}U^{\dagger}_{x+a\hat{\nu},\mu}
U^{\dagger}_{x,\nu},
\end{equation}
the ``plaquette'' (Figure~\ref{fig:plaq}).
The plaquette determines the local curvature of the gauge fields within
the group manifold, i.e.\ it is related to the field strength tensor,
\begin{eqnarray}
\label{eq:fieldstr}
U_{x,\mu\nu}&=&\exp\left(ia^2{\mathcal F}_{x,\mu\nu}\right),\\
\label{eq:fields2}
F^{\alpha\beta}_{\mu\nu}
\left(x+\frac{a}{2}\hat{\mu}+\frac{a}{2}\hat{\nu}\right)
&=&\left({\mathcal F}^{\alpha\beta}_{x,\mu\nu}
-\delta^{\alpha\beta}\Tr\,{\mathcal F}_{x,\mu\nu}\right)
\left[1+{\mathcal O}(a^2)\right],
\end{eqnarray}
where
we denote the normalised trace of
an element in a $D$-dimensional representation of the gauge group
by $\mbox{Tr}_D$ or $\Tr$,
\begin{equation}
\label{eq:trno}
\Tr\,{\mathbf 1}_D=\mbox{Tr}_D {\mathbf 1}_D = \frac{1}{D}\tr{\mathbf 1}_D
=\frac{1}{D}\sum_{i=1}^D\delta_{ii}=1.
\end{equation}
For the fundamental representation above, we have $D=N$.
$\alpha,\beta=1,\ldots,N$ label the colours and $F_{\mu\nu}=F_{\mu\nu}^aT^a$,
where the $N\times N$ matrices $T^a$ denote the gauge group
generators in the fundamental representation.
Note that
${\mathcal F}_{\mu\mu}=0$ and that
${\mathcal F}_{\mu\nu}=-{\mathcal F}_{\nu\mu}^\dagger$ is anti-hermitian,
as a consequence of
$U_{\mu\nu}=U^{-1}_{\nu\mu}=U^{\dagger}_{\nu\mu}$.

Discretised lattice actions are formulated in a manifestly gauge-invariant
way and should approach the continuum action in the limit,
$a\rightarrow 0$. Since the action depends on couplings rather than
directly on the lattice spacing, it is not {\em a priori} clear
if this limit can be realised.
We shall discuss the approach to the continuum limit
below. For the moment, we remark that
from the asymptotic freedom of
perturbative QCD we expect $a$ to approach zero
as $g\rightarrow 0$, i.e.\
$\beta\rightarrow\infty$.

The simplest gluonic action is the so-called Wilson action,
\begin{equation}
\label{eq:wilglue}
S_W[U]=\beta\sum_{x,\mu>\nu}
\left[1-\re\Tr\left(U_{x,\mu\nu}\right)\right],
\end{equation}
where $\Tr$ denotes the normalised trace of
Eq.~(\ref{eq:trno}). From Eqs.~(\ref{eq:yangmills})
and (\ref{eq:fieldstr}) it is easy to see that
$S_W=S_{YM}[1+{\mathcal O}(a^2)]$. The constant term
in the action is irrelevant
as it cancels from expectation values.
The choice of the action is far from unique. For instance an
alternative form,
suggested by Manton~\cite{Manton:1980ts},
has been used in the glueball studies of
Refs.~\cite{Berg:1984yj,Michael:1988nd}.
The action can in principle be systematically improved
to approximate the continuum action to a higher order in
$a$~\cite{Symanzik:1983dc,Wilson:1993dy}.
This Symanzik improvement programme has first been
applied to Yang-Mills lattice gauge theory by L\"uscher and
Weisz~\cite{Weisz:1983zw,Luscher:1985xn,Luscher:1985zq}.
In a classical theory, all the coefficients of higher dimensional
operators that are added to the plaquette of the Wilson action can easily
be determined. However, in the quantum field theory case of interest,
the coefficients are subject to radiative corrections, and have to be
determined non-perturbatively to fully eliminate the ${\mathcal O}(a^2)$
lattice artefacts of the Wilson action. Although this has not been achieved
yet, impressive
results on static
potentials~\cite{Alford:1995hw,Pennanen:1997ni,Morningstar:1998da}, the
glueball
spectrum~\cite{Morningstar:1997ff,Morningstar:1999rf,Shakespeare:1998uu}
and thermodynamics~\cite{Beinlich:1997ia}
have recently
been obtained with Symanzik improved gluonic actions
with coefficients, approximated by a mean field (``tadpole'')
estimate~\cite{Parisi:1980pe,Lepage:1993xa}.
An alternative improved gluonic action that has been used in recent
lattice studies~\cite{Iwasaki:1997sn}
is the renormalisation group improved Iwasaki
action~\cite{Iwasaki:1983ck,Iwasaki:1984cj}.
The renormalisation group approach towards an improved
continuum limit behaviour
has been systematised in the work of
Hasenfratz and Niedermayer~\cite{Hasenfratz:1994sp} on ``perfect'' lattice
actions. Approximately perfect actions have been constructed
for example in
Refs.~\cite{DeGrand:1995ji,Niedermayer:1997eb,Bietenholz:1996cy}.

A na\"{\i}ve discretisation of the Dirac fermionic action of
Eq.~(\ref{eq:Dirac})
suffers under the fermion doubling problem
(cf.\ Refs.~\cite{Creutz:1984mg,Montvay:1994cy}). 
The simplest way to remove the
un-wanted modes is to give them extra mass by adding an irrelevant term,
$-a\bar{q}D_{\mu}D_{\mu}q$, to the action. This results
in Wilson fermions~\cite{Wilson:1975id},
\begin{equation}
\label{eq:wilsonf}
S_f[U,q,\bar{q}]=\sum_{x,y}\bar{q}_xM_{xy}(U)q_y,
\end{equation}
where
\begin{equation}
M_{xy}=\delta_{xy}-
\kappa\sum_{\mu}
\left[\left(1-\gamma_{\mu}\right)U_{x,\mu}\delta_{x+\hat{\mu},y}
+\left(1+\gamma_{\mu}\right)U^{\dagger}_{x-{\hat{\mu}},\mu}
\delta_{x-\hat{\mu},y}\right].
\end{equation}
One of the disadvantages of this solution is
that continuum fermions are only approximated
up to ${\mathcal O}(a)$
lattice artefacts. Remember that the gauge action was
correct up to
${\mathcal O}(a^2)$ errors.
The parameter, $\kappa$, is related to the inverse
bare quark mass,
\begin{equation}
ma=\frac{1}{2}\left(\frac{1}{\kappa}-\frac{1}{\kappa_c}\right),
\end{equation}
where $\kappa_c\geq 1/8$ approaches the free field
($U_{x,\mu}={\mathbf 1}$)
limit, $\kappa_c=1/8$, as $\beta\rightarrow\infty$.
Note that the quark fields in Eq.~(\ref{eq:wilsonf}) have been rescaled,
\begin{equation}
q\rightarrow\sqrt{\frac{a^3}{2\kappa}}q.
\end{equation}
Another popular alternative is the Kogut-Susskind
action~\cite{Susskind:1977jm} which is correct up to
${\mathcal O}(a^2)$ lattice artefacts. However, it requires
four mass degenerate quark flavours.
The Sheikoleshlami-Wohlert action~\cite{Sheikholeslami:1985ij}
is an ${\mathcal O}(a)$ Symanzik improved variant
of the Wilson fermionic action. The coefficient of the additional term
is known non-perturbatively~\cite{Luscher:1997ug}.
Other suggestions of Symanzik improved fermionic actions
have been put forward for instance by Naik~\cite{Naik:1989bn}
and Eguchi~\cite{Eguchi:1984xr}.
Domain wall fermions have been
suggested~\cite{Kaplan:1992bt,Shamir:1993zy},
in order to realise (approximate) chiral symmetry in the lattice theory.
These fermions have received renewed attention since they have
been found to fulfil the
Ginsparg-Wilson relation~\cite{Ginsparg:1982bj,Luscher:1998du}.
They share this feature with other
fermionic actions like the ``perfect'' action
of Ref.~\cite{Hasenfratz:1998ri}
and the action derived by use of the overlap
formalism~\cite{Narayanan:1993wx,Narayanan:1995gw} in
Ref.~\cite{Neuberger:1997fp}. However, we are
interested in quite the opposite
of massless fermions, namely
heavy quarks, such that these exciting new developments are of limited
interest in the present context.

Expectation values of operators, $O$, are
determined by the computation of the path integral,
\begin{equation}
\label{eq:vacexp}
\langle O\rangle=\frac{1}{Z}\int[dU][dq][d\bar{q}]\,
O[U]e^{-S[U,q,\bar{q}]}.
\end{equation}
The normalisation factor, or partition function, $Z$,
is such that $\langle {\mathbf 1}\rangle =1$.
The shorthand notation, $q$, represents $\{q^i_x\}$ and
$U$ stands for all gauge fields, $\{U_{x,\mu}\}$.
The high-dimensional integral is evaluated by means of a 
(stochastic) Monte-Carlo method
as an average over an ensemble of $n$ {\em representative} gauge
configurations\footnote{The basic numerical techniques employed
to generate these configurations are e.g.\ explained
in Ref.~\cite{Montvay:1994cy} and references therein.},
${\mathcal C}_i=\{U_{x,\mu}^{(i)}\}, i=1,\ldots,n$:
\begin{equation}
\langle O\rangle=\frac{1}{n}\sum_{i=1}^n O[{\mathcal C}_i]+\Delta O.
\end{equation}
Therefore, the result on the
expectation value is subject to a statistical error, $\Delta O$,
that will decrease like $1/\sqrt{n}$:
the more {\em measurements} are taken, the more precise the prediction becomes.
For this reason one might speak of {\em lattice measurements} and
{\em lattice experiments}, in analogy to {\em ``real'' experiments}.
The method represents an {\em exact} approach in the sense that
the statistical errors can in principle be made arbitrarily small
by increasing the sample size, $n$.

\subsection{Getting the physics right}
In general, the action
that is simulated depends on $n_f$ quark masses, $m_i$,
as well as on a bare 
QCD coupling, $g$. By varying $g$
and $m_i$ the lattice spacing, $a(g,m_i)$, is changed. 
Lattice QCD is a {\em first principles} approach in that
no additional parameters are introduced, apart from those that are inherent
to QCD, mentioned above. In order to fit
these $n_f+1$ parameters, $n_f+1$ low energy quantities  are matched
to their experimental values: the lattice spacing, $a(g,m_i)$,
can be obtained for instance by fixing
$m_{\rho}$ as determined on the lattice
to the experimental value.
The lattice parameters that correspond to physical
$m_u\approx m_d$ can then be obtained by adjusting $m_{\pi}/m_{\rho}$;
the right $m_s$ can be reproduced
by adjusting $m_K/m_{\rho}$ or $m_\phi/m_\rho$
to experiment etc..

If the right theory is being simulated 
all experimental mass ratios should be reproduced in the continuum limit,
$a\rightarrow 0$, which will be reached as $g\rightarrow 0$,
such that it becomes irrelevant what set of
experimental input quantities has been chosen initially.
In practice, the available computer speed and memory are
finite and simulations are often performed within the quenched approximation,
neglecting sea quark effects, or at un-physically heavy quark masses.
Therefore, unless
controlled extrapolations to the right number of flavours, $n_f$,
and masses
of sea quarks, $m_i$, are performed,
residual scale uncertainties that depend on the
choice of experimental input parameters will survive in the continuum limit.
Once the scale and quark masses
have been set, everything else becomes a prediction.

Lattice results in general need to be extrapolated to the
(continuum) limit, $a\rightarrow 0$, at fixed physical volume.
The functional form of this extrapolation is theoretically well
understood and under control. This claim is substantiated by the fact
that simulations with different
lattice discretisations of the continuum QCD action yield compatible
results after the continuum extrapolation has been performed.
For high energies, an overlap between certain quenched lattice computations
and perturbative QCD has been confirmed
too~\cite{Capitani:1998mq,Luscher:1994gh}, excluding the
possibility of fixed points of the $\beta$-function at finite
values of the coupling, other than $g=0$.
After taking the continuum limit, an infinite volume
extrapolation should be performed. In most cases, results on hadron
masses from quenched evaluations
on lattices with spatial extent, $L_{\sigma}a>2$~fm,
are virtually indistinguishable from the infinite volume limit
within typical statistical errors down to pion masses, $m_{\pi}\approx
m_{\rho}/3$.
However, for QCD with sea quarks
the available information is not yet sufficient for definite conclusions,
in particular as one might expect a substantial dependence of
the on-set of finite size effects on the sea quark mass(es).
The typical lattice spacings used in light hadron spectroscopy
cover the region $0.05\,\mbox{fm}<a< 0.2$~fm.

The effective infinite volume limit
of realistically light pions cannot be realised
at a reasonable computational cost, neither in quenched
nor in full QCD.
Therefore, in practice another extrapolation is
required. This extrapolation to the physical light quark mass
is theoretically less well under control than those to
the continuum and infinite volume limits.
The parametrisations used are in general
motivated by chiral perturbation theory
and the related theoretical uncertainties
are the dominant source of error in latest state-of-the-art spectrum
calculations~\cite{Aoki:1999yr}.
Ideally, the Monte Carlo sample size $n$ is chosen
such that the statistical precision is
smaller or similar in size than the systematic uncertainty due to the
extrapolations involved.

\subsection{Mass determinations}
\label{sec:mass}
In order to extract the ground state mass of a state with quantum
numbers $\alpha$, one starts from a connected gauge invariant correlation
function,
\begin{equation}
\label{eq:correl}
C_{\alpha}(t)=\langle 0|\Psi_{\alpha}^{\dagger}(t) \Psi_{\alpha}(0)
|0\rangle-\left|\langle 0|\Psi_{\alpha}|0\rangle\right|^2,
\end{equation}
where $|0\rangle$ denotes the vacuum state\footnote{In 
Eq.~(\ref{eq:vacexp}) we have employed the short-hand notation,
$\langle O\rangle=\langle 0|O|0\rangle$, for the vacuum expectation
value of the operator $O$.}.
$\alpha$ contains the momentum and the $J^{PC}$ quantum numbers
of the state of interest as well
as the constituent quark content, i.e.\ isospin, strangeness etc..
In most cases, one is interested in the rest mass. Therefore, $\Psi_{\alpha}$
usually involves a summation over all spatial positions, ${\mathbf x}$, within
a time slice to project onto spatial momentum,
${\mathbf p}={\mathbf 0}$. Any other
lattice momentum can be singled out by taking the corresponding discrete
Fourier transform.
Due to the translational invariance on the lattice, it is sufficient
to project only either source or sink onto the desired momentum state.

In what follows, we will for simplicity assume, $L_{\tau}a\rightarrow\infty$.
At finite $L_{\tau}a$ additional contributions arise from the propagation
into the negative time
direction around the periodically closed temporal boundary.
Such effects
can easily be taken into account whenever they turn out to be numerically
relevant. By inserting a complete set of eigenstates of the Hamiltonian,
$|\Phi_{\alpha,n}\rangle$,
into Eq.~(\ref{eq:correl}),
one obtains,
\begin{equation}
\label{eq:mani}
C_{\alpha}(t)=\sum_n |c_n(\alpha)|^2 e^{-E_{n}(\alpha)t}.
\end{equation}
with
\begin{equation}
c_n(\alpha)=\langle\Phi_{\alpha,n} |\Psi_{\alpha}(0)|0\rangle.
\end{equation}
$E_n(\alpha)$ is the energy eigenvalue of the state
$|\Phi_{\alpha,n}\rangle$, $e^{-Ht}|\Phi_{\alpha,n}\rangle=
e^{-E_n(\alpha)t}|\Phi_{\alpha,n}\rangle$, and,
$\Psi^{\dagger}_{\alpha}(t)=e^{Ht}\Psi_{\alpha}^{\dagger}(0)e^{-Ht}$.
In the limit, $t\rightarrow\infty$, the ground state mass,
\begin{equation}
\label{eq:limit}
E_0(\alpha)=-\lim_{t\rightarrow\infty}\frac{d}{dt}\ln C_{\alpha}(t),
\end{equation}
can be extracted. The above formula converges exponentially fast and
is, therefore, suitable for numerical studies.
In general, $\Psi_{\alpha}$ can be any linear combination
of $\Phi_{\alpha,n}$ and its choice is not unique. This observation
is exploited in iterative
smearing or fuzzing
techniques~\cite{Teper:1987wt,Albanese:1987ds,Gusken:1989ad,Perantonis:1989uz,Bali:1992ab,Gupta:1993vp,Lacock:1995qx} that seek
to prepare an initial state with optimised overlap to the
level of interest. This will then allow
the infinite time
limit of Eq.~(\ref{eq:limit}) to be effectively realised
at moderate temporal separations, $t$.
In principle, not only a single correlation function but
a whole cross-correlation matrix between differently optimised
$\Psi$'s can be measured.
In doing so, there is the chance that
by diagonalising the matrix and employing sophisticated multi-exponential
fitting techniques not only the ground state energy can be extracted
but also
those of the lowest one or two
radial excitations~\cite{Davies:1994mp,Philipsen:1999wf,Morningstar:1999rf,Eicker:1998vx,Spitz:1999tu}.

In Eq.~(\ref{eq:mani}) we have adapted the normalisation convention,
$\langle \Phi_{\alpha,m}|\Phi_{\alpha,n}\rangle=\delta_{mn}$,
$\sum_n |\Phi_{\alpha,n}\rangle\langle \Phi_{\alpha,n}|={\mathbf 1}$.
This results in $0\leq |c_n(\alpha)|^2\leq 1$ and
$\sum_n |c_n(\alpha)|^2 = 1$. The deviation of
$|c_0(\alpha)|^2$, the ground state overlap, from the optimal value,
$|c_0(\alpha)|^2=1$, determines the quality of the smeared operator,
$\Psi_{\alpha}$. It should be noted that
if $\Psi_{\alpha}$ contains Dirac spinors,
e.g.\ if it is a pion creation operator,
the standard normalisation condition would be,
$\langle \Phi_{\pi,m}|\Phi_{\pi,n}\rangle=2m_{\pi,m}\delta_{mn}$,
instead.
As a consequence, Eq.~(\ref{eq:mani}) is replaced by,
\begin{equation}
\label{eq:mani2}
C_{\pi}(t)=\sum_n \frac{|c_{\pi,n}|^2}{2m_{\pi,n}} e^{-m_{\pi,n}t}.
\end{equation}

For the manipulations yielding Eq.~(\ref{eq:mani}) we have assumed the
existence of a positive definite self-adjoint Hamiltonian.
L\"uscher~\cite{Luscher:1977ms}
has shown that the Wilson gluonic and fermionic lattice actions
fulfil both,
reflection positivity~\cite{Osterwalder:1973dx,Osterwalder:1975tc} with
respect to hyperplanes going through lattice sites and through the centre
of temporal lattice links (see also Ref.~\cite{Osterwalder:1978pc}).
This feature implies the existence of a positive
transfer matrix and the possibility of analytical
continuation to Minkowski space-time.
Another important consequence of reflection positivity is
that the coefficients of the series in Eq.~(\ref{eq:mani}),
are non-negative and that, therefore, the limit of Eq.~(\ref{eq:limit})
is approached monotonically from above. General properties of the transfer
matrix for continuum limit improved actions are discussed
in Ref.~\cite{Luscher:1984is}.

\subsection{The continuum limit}
A continuum limit of the lattice theory can be defined at fixed points
associated to phase transitions of second or higher order in the
space spanned by the bare couplings of the action. 
In the vicinity
of such a phase transition any correlation length, $\xi/a$, diverges
which implies, $a\rightarrow 0$, if we associate $\xi$ to a physical
distance or mass, $\xi=1/m$. Moreover, universality sets
in, i.e.\ the behaviour of
different correlation lengths is governed by one and the same critical
exponent.
This results in ratios between two correlation lengths, or masses,
to saturate at constant values: the system forgets the lattice
spacing, $a$. One refers to this behaviour as ``scaling''. In the case
of the Wilson gluonic action, the leading order violations of scaling are
expected to be proportional to $a^2$ while for the Wilson fermionic action,
they are only linear in $a$.

The Callan-Symanzik $\beta$-function,
\begin{equation}
\label{rge}
\beta(\alpha_s)=\frac{d\alpha_s}{d\ln \mu^2}=
-\beta_0\alpha_s^2-\beta_1\alpha_s^3-\beta_2\alpha_s^4
-\ldots,
\end{equation}
parameterises the variation of the QCD coupling, $\alpha_s=g^2/(4\pi)$,
with a scale $\mu$. Perturbative QCD tells us, $\beta_0>0$ and $\beta_1>0$,
which implies asymptotic freedom: the limit $\alpha_s=0$ is reached with
$\mu\rightarrow\infty$, i.e.\ the continuum limit of lattice QCD,
$a\rightarrow 0$,
corresponds to\footnote{Here, $\beta$ represents the inverse
lattice coupling of Eq.~(\ref{eq:betadef}) and not
the $\beta$ function.} $\beta\rightarrow\infty$.
Far away from the phase transition,
no unique $\beta$-function can be defined; due to the occurrence of
power corrections, different masses will in general run
differently
as a function of the bare coupling. Lattice results seem to imply
that in zero temperature $SU(N)$ gauge theory
no fixed point other than $\alpha_s=0$ exists.

\begin{figure}[thb]
\centerline{\epsfxsize=10truecm\epsffile{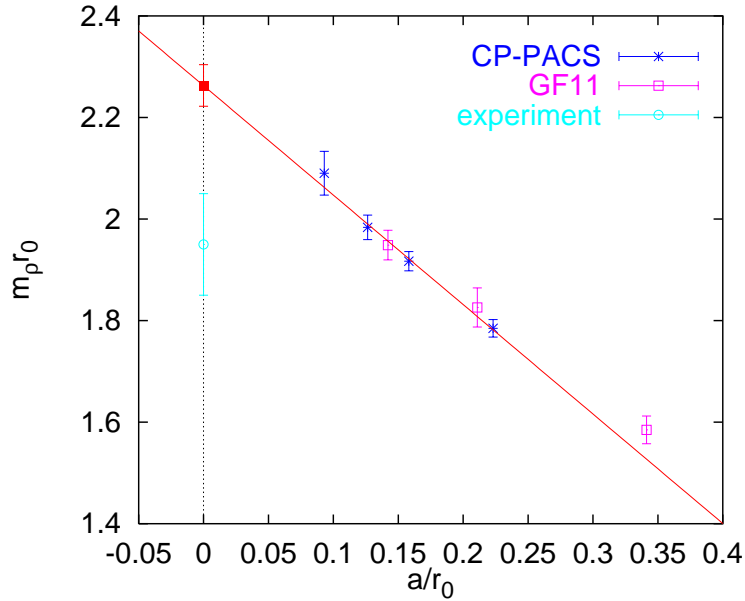}}
\caption{The ratio, $m_{\rho}r_0$, in quenched QCD,
extrapolated to the continuum limit.
The $\rho$ masses in lattice units are taken
from Refs.~\cite{Aoki:1999yr} (CP-PACS) and
\cite{Butler:1994em} (GF11).}
\label{figrho}
\end{figure}

While the coefficients $\beta_0$ and $\beta_1$ within Eq.~(\ref{rge})
are universal, higher order 
coefficients depend on the renormalisation scheme.
Integrating Eq.~(\ref{rge}) yields,
\begin{equation}
\label{rundown}
\mu=\Lambda\exp\left(\int_{\alpha(\Lambda)}^{\alpha(\mu)}
\frac{d\alpha}{2\beta(\alpha)}\right),
\end{equation}
where we define the integration constant, the so-called
QCD $\Lambda$-parameter,
via the two loop relation,
\begin{equation}
\label{eq:weakc}
\Lambda = \lim_{\mu\rightarrow\infty} \mu\, \exp\left(
-\frac{1}{2\beta_0\alpha(\mu)}\right)
\left[\beta_0\alpha(\mu)\right]^{-\frac{\beta_1}{2\beta_0}}.
\end{equation}
In Appendix~\ref{app:run}, we display results on the coefficients $\beta_i$
of Eq.~(\ref{rge}) for reference and detail how to translate between
different schemes.

In QCD with sea quarks, the lattice cut-off, $a$, will not only depend
on the coupling but also on the bare quark masses of the Lagrangian.
This dependence can be parameterised into quark mass anomalous dimension 
functions. The continuum limit of a theory with $n_f$ different quark masses 
will be taken along a trajectory on which
$n_f$ physical mass ratios are kept fixed. In the approximation to QCD with two
degenerate light quark masses for instance the physical curve
$m_{\pi}/m_{\rho}\approx 2/11$ would serve this purpose.

\begin{figure}[thb]
\centerline{\epsfxsize=10truecm\epsffile{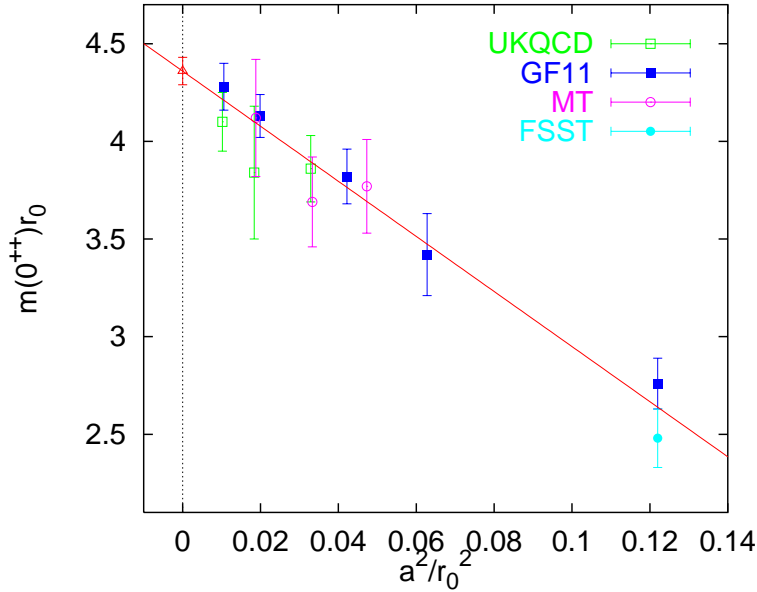}}
\caption{The scalar glueball mass in units of $r_0$ as a function of $a^2$.
The glueball masses in lattice units are taken
from Refs.~\cite{Bali:1993fb} (UKQCD),
\cite{Sexton:1995kd} (GF11), \cite{Michael:1988wf,Michael:1989jr} (MT)
and \cite{deForcrand:1985rs} (FSST).}
\label{figglue}
\end{figure}

In Figure~\ref{figrho}, we show a continuum limit extrapolation
of the quantity $m_{\rho}r_0$,
where $r_0$ is a length scale implicitly defined through the static
potential~\cite{Sommer:1993ce}, $V(r)$,
\begin{equation}
\left.\frac{dV(r)}{dr}\right|_{r=r_0}=1.65.
\end{equation}
From bottomonium phenomenology~\cite{Sommer:1993ce,Bali:1997am,Bali:1998pi},
we can assign the experimental 
value, $r_0^{-1}=(394\pm 20)$~MeV,
while $m_{\rho}\approx 770$~MeV.
The data on $m_{\rho}$ has been obtained in the quenched approximation to QCD,
by use of the Wilson fermionic and
gluonic action by the GF11 and CP-PACS
collaborations~\cite{Butler:1994em,Aoki:1999yr}. The corresponding $r_0$
values have been obtained from the interpolating
formula of the ALPHA collaboration~\cite{Guagnelli:1998ud}
for $5.7\leq\beta\leq 6.57$,
\begin{equation}
\label{eq:interp}
a/r_0=\exp\left\{-\left[d_0+d_1(\beta-6)+d_2(\beta-6)^2
+d_3(\beta-6)^3\right]\right\},
\end{equation}
with $d_0=1.6805, d_1=1.7139, d_2=-0.8155, d_3=0.6667$.

The leading order scaling violations of $m_{\rho}r_0$
are expected to be proportional
to the lattice spacing, $a$.
The data points cover the range, $5.7\leq\beta\leq 6.47$,
or, $0.17\,\mbox{fm}\geq a\geq 0.047$~fm. Only the
CP-PACS results have been used in the linear fit.
In the continuum limit the ratio $m_{\rho}r_0$ deviates from the
phenomenological estimate by about 15~\%, indicating the limitations
of the quenched
approximation. In Ref.~\cite{Aoki:1999yr} deviations of some
quenched ratios between masses of light hadrons from experiment
of up to 10~\% have been observed.

Due to the substantial slope of the extrapolation,
the result obtained on the finest lattice with a resolution of about
4~GeV still
deviates by almost 10~\% from the continuum limit extrapolated value.
This is different from the situation regarding the glueball spectrum where
leading order lattice artefacts are proportional to $a^2$. In
Figure~\ref{figglue}, we display the continuum limit extrapolation for the
lightest quenched glueball mass that has scalar
quantum numbers, $J^{PC}=0^{++}$.
The $\beta$ range covered in the Figure, $5.7\leq\beta\leq
6.4$, is about the same as that of Figure~\ref{figrho}. However,
within statistical errors, the $\beta=6.4$ results are compatible
with the continuum limit and this despite the fact that the
scalar glueball behaves rather pathologically~\cite{Morningstar:1999rf}
in the sense that the slope of this
extrapolation is much larger than in any other of the glueball channels.
The continuum limit extrapolated mass comes out to be
$m(0^{++})=1.485(35)$~GeV or $m(0^{++})=1.720(50)$~GeV, depending on whether
the scale is set from the $\rho$-mass or $r_0$, respectively; clearly,
the dominant source of uncertainty is quenching.

\begin{figure}[thb]
\centerline{\epsfxsize=10truecm\epsffile{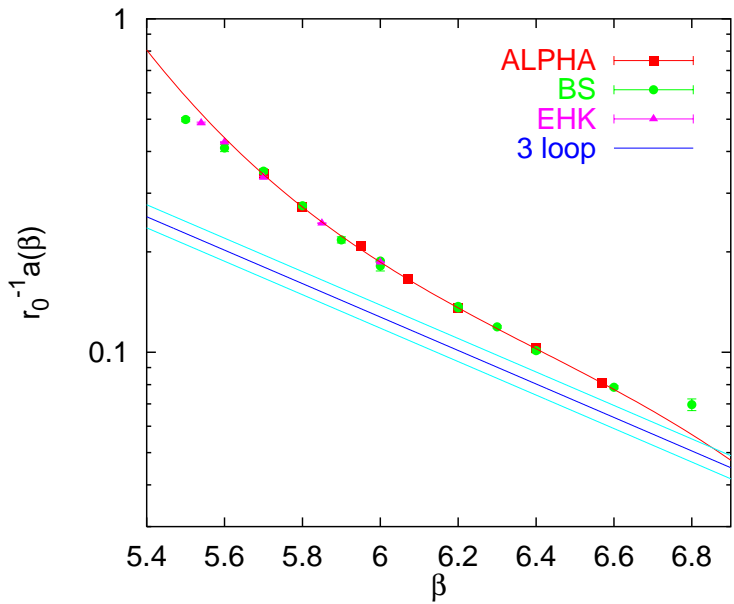}}
\caption{The scale, $r_0$, in lattice units against the coupling
$\beta$. The data are from Refs.~\cite{Guagnelli:1998ud} (ALPHA),
\cite{Schilling:1993bk,Bali:1997bj} (BS) and \cite{Edwards:1997xf} (EHK).}
\label{figasym}
\end{figure}

\begin{table}[thb]
\caption{The scale $r_0$ in lattice units, obtained from
$SU(3)$ gauge theory simulations with Wilson action.}
\label{r0tab}

\begin{center}
\begin{tabular}{c|c|c|c|c}
$\beta$&\multicolumn{4}{|c}{$r_0/a$}\\\cline{2-5}
&\cite{Schilling:1993bk,Bali:1997bj} (BS)&\cite{Edwards:1997xf} (EHK)&
\cite{Guagnelli:1998ud} (ALPHA)&Eq.~(\ref{eq:interp})\\\hline
5.5&2.005(29)&&\\
5.54&&2.054(13)&\\
5.6&2.439(62)&2.344(8)&\\
5.7&2.863(47)&2.990(24)&2.922(9)&2.930\\
5.8&3.636(46)&&3.673(5)&3.668\\
5.85&&4.103(12)&&4.067\\
5.9&4.601(97)&&&4.483\\
5.95&&&4.808(12)&4.917\\
6.0&5.328(31)&5.369(9)&&5.368\\
6.07&&&6.033(17)&6.030\\
6.2&7.290(34)&&7.380(26)&7.360\\
6.3&8.391(72)&&&8.493\\
6.4&9.89(16)&&9.74(5)&9.760\\
6.57&&&12.38(7)&12.38\\
6.6&12.73(14)&&&12.93\\
6.8&14.36(8)&&&
\end{tabular}
\end{center}
\end{table}

In Figure~\ref{figasym}, we plot $r_0^{-1}a$ obtained from quenched Wilson action
simulations~\cite{Guagnelli:1998ud,Schilling:1993bk,Bali:1997bj,Edwards:1997xf}
versus the bare coupling, $\beta$. The results are also displayed in
Table~\ref{r0tab}.
Within the range, $5.5\leq\beta\leq 6.8$,
the lattice spacing varies by a factor of about $7$. The interpolating
curve for $5.7\leq\beta\leq 6.57$, Eq.~(\ref{eq:interp}),
is included into the
plot as well as an estimate obtained by converting the
result~\cite{Capitani:1998mq}, $\Lambda_{SF}^{(0)}r_0=0.294(24)$, into
the bare lattice scheme~\cite{Bode:1998hd} at high energy ($1000\,r_0^{-1}$)
and running the coupling down to lower scales via Eq.~(\ref{rundown}),
using the three loop approximation of the $\beta$-function,
Eqs.~(\ref{rge}), (\ref{beta0}), (\ref{beta1}) and (\ref{beta2l}).
Taking into account the logarithmic scale, deviations from asymptotic scaling
are quite substantial, at least for $\beta\leq 6.4$.
One of the reasons for this failure of perturbation theory at
energy scales of several GeV are large renormalisations
of the lattice action~\cite{Parisi:1980pe}, due to contributions from
tadpole diagrams~\cite{Lepage:1993xa}. One might hope to partially cancel
such contributions by defining an effective
coupling~\cite{Parisi:1980pe,Makeenko:1982bb,Samuel:1985ub,Fingberg:1993ju,Bali:1993ru,Lepage:1993xa}
from the average
plaquette value, measured on the lattice and, indeed, such a procedure
somewhat
reduces the amount of violations of asymptotic
scaling~\cite{Fingberg:1993ju,Bali:1993ru}.

\section{The static QCD potential}
\label{potential}
We shall introduce the Wegner-Wilson loop and derive its relation to
the static potential. Subsequently, expectations on this potential
from exact considerations, strong coupling and string arguments
as well as perturbation theory and quarkonia phenomenology
are presented. Lattice results are then
reviewed. Finally, the behaviour of the potential at short distances,
the breaking of the hadronic string and aspects of the
confinement mechanism are discussed.

\subsection{Wilson loops}
\label{sec:will}
\begin{figure}[thb]
\centerline{\epsfxsize=10truecm\epsffile{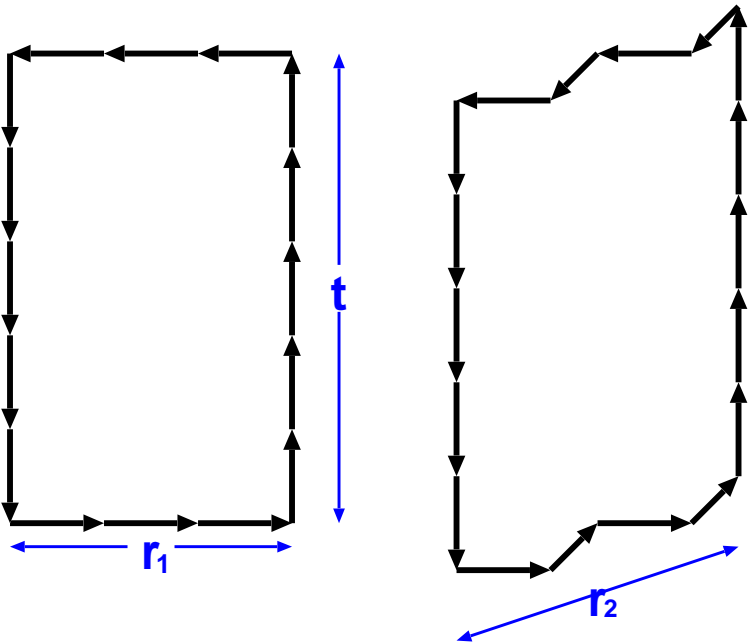}}
\caption{Examples of rectangular on- and off-axis Wilson loops with temporal
extent, $t=5a$, and spatial extents, $r_1=3a$, and,
$r_2=2\sqrt{2}\,a$, respectively.}
\label{figwilson}
\end{figure}

The Wegner-Wilson loop has originally been introduced by
Wegner~\cite{Wegner:1984qt}
as an order parameter in $Z_2$ gauge theory.
It is defined as the trace of the product of gauge variables along
a closed oriented contour, $\delta C$, enclosing an area, $C$,
\begin{equation}
\label{eq:wl}
W(C)=\Tr\left\{
{\mathcal P}\left[\exp\left(
i\int_{\delta C}\!dx_{\mu}\,A_{\mu}(x)\right)\right]\right\}=
\Tr\left(\prod_{(x,\mu)\in\delta C}U_{x,\mu}\right).
\end{equation}
While the loop, determined on a gauge configuration, $\{U_{x,\mu}\}$,
is in general complex, its expectation value is real, due to
charge invariance: in Euclidean space we have,
$\langle W(C)\rangle=\langle W^*(C)\rangle=\langle W(C)\rangle^*=0$.
It is straight forward to generalise the above Wilson loop
to any non-fundamental representation, $D$, of the gauge field,
just by replacing
the variables, $U_{x,\mu}$, with the corresponding links, $U_{x,\mu}^D$.
The arguments below, relating the Wilson loop to the potential
energy of static sources go through, independent of the representation
according to which the sources transform under local
gauge transformations. In what follows, we will denote a Wilson loop,
enclosing a rectangular contour with one purely spatial distance,
${\mathbf r}$, and one temporal separation, $t$, by $W({\mathbf r},t)$.
Examples of Wilson loops on a lattice for two different choices
of contours, $\delta C$, are displayed in Figure~\ref{figwilson}.

In Wilson's original work~\cite{Wilson:1974sk},
the Wilson loop has been related to
the potential energy of a pair of static colour sources, by use of
transfer matrix arguments. However,
it took a few years until
Brown and Weisberger attempted to
derive the connection between the Wilson loop
and the effective
potential between heavy, not necessarily static, quarks in a mesonic
bound state~\cite{Brown:1979ya}.
Later on mass dependent
corrections to the static potential have been
derived along similar lines~\cite{Eichten:1979pu,Eichten:1981mw}.
In Section~\ref{sec:gluon}, 
we will discuss these developments in detail.
Here, we
derive the connection between a Wilson loop
and the static potential between colour sources
which highlights similarities with
the situation in classical electrodynamics and which is close to
Wilson's spirit.

For this purpose we start from the Euclidean Yang-Mills action,
Eq.~(\ref{eq:yangmills}),
\begin{equation}
S=\frac{1}{4g^2}\int\!d^4x\,F_{\mu\nu}^aF_{\mu\nu}^a.
\end{equation}
The canonically conjugated momentum to the field, $A_{i}^a$, is given by
the functional derivative,
\begin{equation}
\label{eq:can1}
\pi^a_{i}=\frac{\delta S}{\delta(\partial_4A_{i}^a)}=\frac{1}{g^2}
F_{4i}^a=-\frac{1}{g}E_i^a.
\end{equation}
The anti-symmetry of the field strength tensor implies, $\pi^a_4=0$.
In order to obtain a Hamiltonian formulation of the
gauge theory, we fix the temporal gauge, $A_4^a=0$. In infinite volume,
such gauges can always be found. On a toroidal lattice
this is possible up to one time slice $t'$, which we demand to be outside of
the Wilson loop contour, $t'>t$.

The canonically conjugated momentum,
\begin{equation}
\label{eq:can2}
\pi_{\mu}^a=-i\frac{\delta}{\delta A_{\mu}^a},
\end{equation}
now fulfils the usual commutation relations,
\begin{equation}
[A_{j}^a,\pi_{\mu}^b]=i\delta_{j \mu}\delta^{ab},
\end{equation}
and we can construct the Hamiltonian,
\begin{equation}
H=\int\!d^3x\,\left(\pi_{\mu}^a\partial_4A_{\mu}^a-\frac{1}{4g^2}
F_{\mu\nu}^aF_{\mu\nu}^a\right)
=\frac{1}{2}\int\!d^3x\,\left(E_i^aE_i^a-B_i^aB_i^a\right),
\end{equation}
that acts onto states, $\Psi[A_{\mu}]$. In Euclidean metric,
the magnetic contribution to the total energy is negative.

A gauge transformation,
$\Omega$,
can for instance be represented as a bundle
of $SU(N)$ matrices in some representation $D$,
$\Omega_D({\mathbf x})=e^{i\omega^a({\mathbf x})T_D^a}$.
We wish to derive the
operator representation
of the group generators, $T_R^a$, that acts on the Hilbert space
of wave functionals.
For this purpose
we start from,
\begin{equation}
R(\Omega)\Psi=
\left[1+i\int\! d^3x\, \omega^a({\mathbf x})
T_R^a({\mathbf x})
+\cdots\right]\Psi=\Psi+\delta \Psi.
\end{equation}
From Eq.~(\ref{eq:transa2}) one easily sees that,
$\delta A_i=A_i^{\Omega}-A_i=-(\partial_i\omega+i[A_i,\omega])$.
We obtain,
\begin{equation}
\delta\Psi=
\int\!d^3\!x\,\frac{\delta\Psi}{\delta A_i({\mathbf x})}
\delta A_i({\mathbf x})
=\int\!d^3\!x\,\omega({\mathbf x})
D_i\frac{\delta\Psi}{\delta A_i({\mathbf x})}
=-\frac{i}{g}\int\!d^3\!x\,\omega^a({\mathbf x})(D_iE_i)^a({\mathbf x})
\Psi,
\end{equation}
where we have performed a partial integration and have
made use of the equivalence,
\begin{equation}
\frac{\delta}{\delta A_i}=-\frac{i}{g}E_i,
\end{equation}
of Eqs.~(\ref{eq:can1}) and (\ref{eq:can2}).
Hence we obtain the representation,
\begin{equation}
T_R^a=-\frac{1}{g}(D_iE_i)^a:
\end{equation}
the covariant divergence of the electric field operator is the
generator of gauge transformations!

Let us assume that the wave functional is a singlet
under gauge transformations,
$
R(\Omega)\Psi[A_{\mu}]=\Psi[A_{\mu}].
$
This implies,
\begin{equation}
(D_iE_i)^a\Psi=0,
\end{equation}
which is Gau\ss{}' law in the absence of sources:
$\Psi$ lies in the eigenspace of $D_iE_i$
that corresponds to the eigenvalue zero.
Let us next place an external source in fundamental representation of the
colour group at position ${\mathbf r}$.
In this case, the associated wave functional,
$\Psi_{\alpha},\alpha = 1,\ldots,N$, transforms in a non-trivial way,
\begin{equation}
[R(\Omega)\Psi]_{\alpha}
=\Omega_{\alpha\beta}\Psi_{\beta},
\end{equation}
This implies,
\begin{equation}
\label{eq:gau}
(D_iE_i)^a\Psi=-g\delta^3({\mathbf r})T^a\Psi,
\end{equation}
which again resembles Gau\ss{}' law, this time
for a point-like colour charge at position\footnote{Of course,
on a torus, such a state cannot be constructed. Note also that in our Euclidean
space-time conventions Gau\ss{}' law reads, ${\mathbf D}_i{\mathbf E}_i
({\mathbf x})=-\rho({\mathbf x})$, where $rho$ denotes the charge density.}
${\mathbf r}$.
For non-fundamental representations, $D$,
Eq.~(\ref{eq:gau}) remains valid under the replacement,
$T^a\rightarrow T^a_D$.

Let us now place a fundamental
source at position ${\mathbf 0}$ and an anti-source
at position ${\mathbf r}$. The wave functional, $\Psi_{\mathbf r}$,
which is an $N\times N$ matrix in colour space
will transform according to,
\begin{equation}
\Psi^{\Omega}_{\mathbf r,\alpha\beta}
=\Omega_{\alpha\gamma}({\mathbf 0})\Omega^{*}_{\beta\delta}({\mathbf r})
\Psi_{\mathbf r,\gamma\delta}.
\end{equation}
One object with the correct transformation property
is a gauge transporter (Schwinger line) from ${\mathbf 0}$ to
${\mathbf r}$,
\begin{equation}
\Psi_{\mathbf r}
=\frac{1}{\sqrt{N}}
U^{\dagger}({\mathbf r},t)=\frac{1}{\sqrt{N}}
{\mathcal P}\left[\exp\left(
i\int_{\mathbf 0}^{\mathbf r} d{\mathbf x}\,{\mathbf A}({\mathbf x},t)\right)
\right],
\end{equation}
which on the lattice corresponds to the ordered product of link variables
along a connection between the two points.
Since we are in temporal gauge, $A_4(x)=0$, the correlation function
between two such lines at time-like separation, $t$, 
is the Wilson loop,
\begin{equation}
\langle W({\mathbf r},t)\rangle=\frac{1}{N}
\langle U_{\alpha\beta}({\mathbf r},t)
U^{\dagger}_{\beta\alpha}({\mathbf r},0)\rangle,
\end{equation}
which, being a gauge invariant object, will give the same result in any gauge.
Other choices of $\Psi_{\mathbf r}$, e.g.\ linear combinations of
spatial gauge transporters, connecting ${\mathbf 0}$ with ${\mathbf r}$,
define generalised (or smeared) Wilson loops, $W_{\Psi}({\mathbf r},t)$.

Following the discussion of Section~\ref{sec:mass}, we insert a complete
set of transfer matrix eigenstates, $|\Phi_{{\mathbf r},n}\rangle$,
within the sector of the Hilbert space
that corresponds to a charge and anti-charge in fundamental
representation at distance ${\mathbf r}$, and expect the Wilson loop
in the limit, $L_{\tau}a\gg t$,
to behave like,
\begin{equation}
\label{eq:specw}
\langle W_{\Psi}({\mathbf r},t)\rangle
=\sum_n
\left|\left\langle \Phi_{\mathbf r,n}\right|\Psi_{{\mathbf r}}\left|
0
\right\rangle
\right|^2
e^{-E_n({\mathbf r})t},
\end{equation}
where the normalisation convention
is such that, $\langle\Phi_n|\Phi_n
\rangle=\langle\Psi^{\dagger}\Psi\rangle=1$, and the completeness of
eigenstates implies, $\sum_n |\langle \Phi_n|\Psi|0\rangle|^2=1$.
Note that no disconnected part has to be subtracted from
the correlation function since
$\Psi_{\mathbf r}$ is distinguished from
the vacuum state by its colour indices. 
$E_n({\mathbf r})$
denote the energy levels. The ground state
contribution, $E_0({\mathbf r})$, that will
dominate in the limit of large $t$ can
be identified as the static potential.

The gauge transformation properties of the colour state discussed above,
which determine the colour group representation of the static sources
and their separation, ${\mathbf r}$, do not yet completely
determine the state in question: the sources will be connected by
an elongated chromo-electric
flux tube.
This vortex can for instance be in a rotational state
with spin $\Lambda\neq 0$ about the inter-source
axis. Moreover, under interchange of
the ends
the state can transform evenly (g) or oddly (u).
Finally, in the case of $\Lambda = 0$, it can transform symmetrically
or anti-symmetrically
under reflections with respect to a plane containing the sources.
It is possible to
single out sectors within a given irreducible representation of the
relevant cylindrical symmetry group~\cite{Landau:1987qm}, $D_{\infty h}$,
with an adequate choice of $\Psi$. 
A straight line connection between the sources
corresponds to the $D_{\infty h}$ quantum numbers,
$\Sigma_g^+$. Any static potential that is different from the 
$\Sigma_g^+$ ground state
will be referred to as a ``hybrid''
potential. Since these potentials are gluonic excitations
they can be thought of as being hybrids between pure ``glueballs''
and a pure static-static state;
indeed, high hybrid excitations are
unstable and will decay into lower lying potentials via radiation of
glueballs. We will address the question of hybrid potentials in detail in
Sections~\ref{sec:hybrid} and \ref{sec:gluino}.

\subsection{Exact results}
We identify the static potential, $V({\mathbf r})$,
with the ground state energy,
$E_0({\mathbf r})$, of Eq.~(\ref{eq:specw}) that can be extracted from the
Wilson loop of Eq.~(\ref{eq:wl}) via Eq.~(\ref{eq:limit}).
By exploiting the symmetry of a Wilson loop under an interchange of space and
time directions, it can be proven that the static potential cannot
rise faster
than linearly as a function of the distance $r$ in the limit,
$r\rightarrow\infty$~\cite{Seiler:1978ur}.
Moreover, reflection
positivity of Euclidean $n$-point
functions~\cite{Osterwalder:1973dx,Osterwalder:1975tc} implies
convexity of the static
potential~\cite{Bachas:1986xs},
\begin{equation}
V''(r)\leq 0.
\end{equation}
The proof also applies to ground state
potentials between sources in
non-fundamental representations.
However, it does not apply to hybrid potentials
since in this case the required creation operator
extends into
spatial directions orthogonal to the direction of ${\mathbf r}$.
Due to positivity, the potential is bound from below\footnote{
The potential that is determined from Wilson loops depends on the
lattice cut-off, $a$, and can be factorised into a ``physical''
potential $\hat{V}(r)$ and a (positive) self energy contribution:
$V(r,a)=\hat{V}(r)+V_{\mbox{\scriptsize self}}(a)$.
The latter diverges in the continuum limit (see Section~\ref{sec:pert}).
While the ``physical'' potential, $\hat{V}(r)$, will become negative 
at small distance, $V(r,a)$ is indeed non-negative.}.
Therefore,
convexity implies that $V(r)$ is a monotonically rising function of $r$,
\begin{equation}
V'(r)\geq 0.
\end{equation}

In Ref.~\cite{Simon:1982yv}, which in fact preceded Ref.~\cite{Bachas:1986xs},
somewhat more strict upper and lower limits on Wilson loops,
calculated on a lattice, have been derived:
let $a_{\sigma}$ and $a_{\tau}$
be temporal and spatial lattice resolutions.
The main result for rectangular Wilson loops
in representation $D$ and $d$ space-time dimensions then
is,
\begin{equation}
\langle W(a_{\sigma},a_{\tau})\rangle^{rt/(a_{\sigma}a_{\tau})}\leq
\langle W(r,t)\rangle
\leq(1-c)^{r/a_{\sigma}+t/a_{\tau}-2},
\end{equation}
with
$c=\exp[-4(d-1)D\beta]$.
The resulting bounds
on $V(r)$ for $r>a_{\sigma}$ read,
\begin{equation}
\label{eq:lowbound}
-\ln(1-c)\leq a_{\tau}V(r)\leq -\frac{r}{a_{\sigma}}\ln\langle W(a_{\sigma},a_{\tau})\rangle;
\end{equation}
in consistency with Ref.~\cite{Seiler:1978ur},
the potential (measured in lattice units, $a_{\tau}$)
is bound from above by a linear function of $r$ and
it takes positive values everywhere.

\subsection{Strong coupling expansions}
Expectation
values, Eq.~(\ref{eq:vacexp}), can be approximated by expanding
the exponential of the action, Eq.~(\ref{eq:wilglue}), in terms of $\beta$, 
$\exp(-\beta S)=1-\beta S+\cdots$.
This strong coupling expansion is similar to a
high temperature expansion in statistical mechanics. When the Wilson
action is used each factor, $\beta$, is accompanied by a plaquette and
certain diagrammatic rules can be
derived~\cite{Wilson:1974sk,Wilson:1976zj,Balian:1975xw,Creutz:1978yy,Drouffe:1983fv}.
Let us consider a strong coupling expansion of the Wilson loop,
Eq.~(\ref{eq:wl}).
Since the integral over a single group element vanishes,
\begin{equation}
\int dU\,U=0,
\end{equation}
to zeroth
order, we have, $\langle W\rangle = 0$. To the next order in $\beta$, it
becomes possible to cancel the link variables on the contour, $\delta C$,
of the Wilson
loop by tiling the whole minimal enclosed
(lattice) surface, $C$,
with plaquettes. Hence, one
obtains the expectation
value~\cite{Creutz:1978yy,Creutz:1984mg,Montvay:1994cy},
\begin{equation}
\langle W(C)\rangle=\left\{\begin{array}{l}
\left[\beta/4\right]^{-\mbox{\scriptsize area}(\delta C)}+\cdots, N=2\\
\left[\beta/2N^2\right]^{-\mbox{\scriptsize area}(\delta C)}+\cdots,
N>2\end{array}\right.,
\end{equation}
for $SU(N)$ gauge theory. $\mbox{area}(\delta C)$ denotes the
area of the minimal lattice world sheet that is enclosed by the
contour $\delta C$.

 If we now consider the case of a rectangular
Wilson loop that extends $r/a$ lattice points into a spatial
and $t/a$ points into the temporal direction,
we find the area law,
\begin{equation}
\label{eq:areal}
\langle W({\mathbf r},t)\rangle = \exp\left[-\sigma_d rt\right]+
\cdots,
\end{equation}
with a string tension,
\begin{equation}
\label{eq:str}
\sigma_da^2 = -d\,\ln \frac{\beta}{18}.
\end{equation}
The numerical value of the denominator applies to
$SU(3)$ gauge theory;
the potential is linear with slope, $\sigma_d$,
and colour sources are confined at strong coupling.
$d=(|r_1|+|r_2|+|r_3|)/r\geq 1$ denotes the ratio between lattice and
continuum norms and deviates from $d=1$ for source separations,
${\mathbf r}$, that are not parallel to a lattice axis. The string tension
of Eq.~(\ref{eq:str}) depends on $d$ and, therefore,
on the lattice direction; $O(3)$ rotational
symmetry is broken down to the cubic subgroup $O_h$. The extent of violation
will eventually be reduced as one increases $\beta$ and considers
higher orders of the expansion.
Such high order strong coupling expansions have indeed been performed
for Wilson loops~\cite{Munster:1980vk} and glueball
masses~\cite{Munster:1981es}.
Unlike standard perturbation theory,
whose convergence is known to be at best
asymptotic~\cite{Dyson:1952tj,Zinn-Justin:1981uk}, the strong coupling
expansion is analytic around
$\beta=0$~\cite{Osterwalder:1978pc} and, therefore, has
a finite radius of convergence.

Strong coupling $SU(3)$ gauge theory results 
seem to converge for~\cite{Creutz:1984mg}
$\beta < 5$. One would have hoped
to eventually identify a crossover region of finite extent
between the validity regions of the strong and weak coupling
expansions~\cite{Kogut:1980pm}, or at least a transition point between
the leading order strong coupling behaviour, $a^2\propto-\ln(\beta/18)$,
of Eq.~(\ref{eq:str})
and the weak coupling limit, $a^2\propto\exp[-2\pi\beta/(3\beta_0)]$,
of Eq.~(\ref{eq:weakc}).
However, even after re-summing the strong coupling series
in terms of improved expansion parameters and applying sophisticated
Pad\'e approximation techniques~\cite{Smit:1982fx}, nowadays
such a direct crossover region does not appear to exist, necessitating
one to employ Monte Carlo simulation techniques.
One reason for the break down of the strong
coupling expansion around $\beta\approx 5$ seems to be the roughening
transition that is e.g.\
discussed in Refs.~\cite{Luscher:1981ac,Drouffe:1981dp};
while at strong coupling the dynamics is
confined to the minimal
area spanned by a Wilson loop (plus small ``bumps'' on top of this surface),
as the coupling decreases, the colour fields between the sources can penetrate
over several lattice sites into the vacuum.

We would like to remark that the area law of Eq.~(\ref{eq:areal}) is a rather
general result for strong coupling expansions in the fundamental representation
of compact gauge groups. In particular, it also applies to $U(1)$ gauge theory
which we do not expect to confine in the continuum. In fact,
based on
duality arguments, Banks, Myerson and Kogut~\cite{Banks:1977cc} have
succeeded in proving
the existence of a confining phase in the four-dimensional theory and
suggested the existence of a phase transition while
Guth~\cite{Guth:1980gz} has proven that, at least in the
non-compact formulation
of $U(1)$, a Coulomb phase exists.
Indeed, in numerical simulations of (compact) $U(1)$ lattice gauge theory two
such distinct phases were found~\cite{Creutz:1979zg,Lautrup:1980xr},
a Coulomb phase
at weak coupling and a confining phase at strong coupling.
The question whether the confinement one finds in $SU(N)$
gauge theories in the strong coupling limit
survives the continuum limit, $\beta\rightarrow\infty$,
can at present only be answered by means of
numerical simulation.

\subsection{String picture}
The infra-red properties of QCD might be reproduced by effective theories of
interacting strings. String models share many aspects with
the strong coupling expansion.
Originally, the string picture of confinement has
been discussed by Kogut and Susskind~\cite{Kogut:1975ag}
as the strong coupling limit
of the Hamiltonian formulation of lattice QCD. The strong coupling
expansion of a Wilson loop can be cast into a sum of weighted random
deformations of the minimal area world sheet. This sum can
then be
interpreted to represent
a vibrating string. The physical picture behind such
an effective
string description is that of the electric flux between two colour
sources being squeezed into
a thin, effectively one-dimensional, flux tube or Abrikosov-Nielsen-Olesen
(ANO)
vortex~\cite{Abrikosov:1957sx,Nielsen:1973ve,'tHooft:1974jz,Migdal:1984gj}.
As a consequence, this yields a constant energy density
per unit length and a static
potential that is linearly rising as a function of the distance.

One can study the spectrum of such a vibrating string in simple
models~\cite{Luscher:1980fr,Luscher:1981ac,Luscher:1981iy}. Of course,
the string action is not {\em a priori} known. The simplest possible
assumption, employed in the above references,
is that the string is described by the Nambu-Goto
action~\cite{Goto:1971ce,Nambu:1974zg}
in terms of ($d-2$) free bosonic fields
associated to the transverse degrees of freedom of the string.
In this picture, the static potential
is~\cite{Luscher:1980fr,Arvis:1983fp} (up to a constant term) given by,
\begin{equation}
\label{eq:stringe}
V(r)=\sigma r\sqrt{1-\frac{(d-2)\pi}{12\,\sigma\,r^2}}=
\sigma\, r-\frac{(d-2)\pi}{24\,r}-\frac{(d-2)^2\pi^2}{1152\,\sigma\,r^3}
-\cdots,
\end{equation}
while for a fermionic string~\cite{Caselle:1987ek}
one would expect
the coefficient of the correction term to the linear behaviour
to be only one quarter as big as the Nambu-Goto one above.
In the bosonic string picture,
excited levels are separated from the ground state by,
\begin{equation}
\label{eq:excit}
V_n^2(r)=V^2(r)+(d-2)\pi n\,\sigma=\left[V(r)+\frac{(d-2)\pi n}{2\,r}-
\cdots\right]^2,
\end{equation}
with $n$ assuming integer values. 
It is clear from Eq.~(\ref{eq:stringe}) that
the string picture at best applies to distances,
\begin{equation}
r\gg r_c=\sqrt{\frac{(d-2)\pi}{12\,\sigma}}.
\end{equation}
In four dimensions one obtains,
$r_c\approx 0.33$~fm, from the value, $\sqrt{\sigma}\approx 430$~MeV, from
the $\rho,a_2,\ldots$ Regge trajectory.

The expectation of Eq.~(\ref{eq:stringe}) has been very accurately
reproduced in numerical simulations of $Z_2$ gauge theory in $d=3$ space-time
dimensions~\cite{Caselle:1997ii}.
In a recent study of $d=4$ $SU(3)$ gauge theory~\cite{Morningstar:1998da}
the hybrid potentials have been
found to group themselves
into various bands that are separated by approximately equi-distant gaps
at large $r$. However, up to distances as large as 3~fm
these gaps seem to be inconsistent with $\pi/r$, the expectation of
Eq.~(\ref{eq:excit}).
These newer data contradict earlier findings
in $SU(2)$ gauge theory~\cite{Perantonis:1989uz} where good agreement
with the Nambu-Goto string picture has been reported, such that we do not
regard this issue as finally settled.
The consistency of lattice data
with Eq.~(\ref{eq:excit}) at large separations
would support the
existence of a bosonic string description of confining gauge theories in the
very low energy
regime~\cite{Akhmedov:1996mw,Polyakov:1997nc,Chernodub:1998ie,Antonov:1998wi,Baker:1999xn}.
Of course, in $d<26$, the string Lagrangian is not
renormalisable but only effective and higher order correction terms 
like torsion and rigidity will in
general have to be added~\cite{Polchinski:1991ax}.

It is hard to disentangle in $d=4$
the (large distance) $1/r$ term, expected from string vibrations, from the
perturbative Coulomb term at short distances. Therefore, three-dimensional
investigations (where perturbation theory yields a logarithmic contribution)
have been suggested~\cite{Ambjorn:1984yu}. Another way out is to determine
the mass of a
closed string, encircling a boundary of the lattice~\cite{'tHooft:1979uj}
with a spatial
extent, $l=L_{\sigma}a$ (a torelon~\cite{Michael:1987cj};
for details see Appendix~\ref{centres}),
which is not polluted by a perturbative tail.
The bosonic string expectation in this case would be~\cite{Ambjorn:1984yu},
\begin{equation}
\label{eq:tor}
E_n(l)=
\sigma\, l-\frac{(d-2)\pi}{6\,l}+\cdots.
\end{equation}
The na\"{\i}ve range of validity of the picture is
$l\gg l_c=2\,r_c\approx 0.66$~fm. The numerical value applies to
$d=4$. An investigation of the finite size dependence of the
torelon mass in $d=4$ $SU(2)$ gauge theory has been done by
Michael and Stephenson~\cite{Michael:1994ej} who found excellent agreement
with the bosonic string picture
already for distances, $1\,\mbox{fm}\leq l\leq 2.4$~fm, quite close
to $l_c$,
on the 3~\% level. Qualitative agreement has also been reported by
Teper~\cite{Teper:1998te} from simulations of
$SU(2)$, $SU(3)$, $SU(4)$ and $SU(5)$ gauge theories in three dimensions.

The bosonic string picture for $r\gg\beta=aL_{\tau}$ predicts a behaviour 
similar to Eq.~(\ref{eq:tor})
for the finite temperature potential, calculated from Polyakov line
correlators~\cite{deForcrand:1985cz},
\begin{equation}
-\frac{1}{\beta}\ln\langle P^*(r)P(0)\rangle
=\sigma(\beta)r+\cdots,\quad\sigma(\beta)=
\sigma-\frac{(d-2)\pi}{6\beta^2}+\cdots.
\end{equation}
The Polyakov line is defined as [Eq.~(\ref{eq:pl})],
\begin{equation}
\label{eq:pl2}
P({\mathbf x})=\Tr\,\left\{
{\mathcal T}\left[\exp\left(
i\int_0^{aL_{\tau}}\!dx_4\,A_4(x)\right)\right]\right\}=
\Tr\,\left(\prod_{x_4=0}^{aL_{\tau}}U_{x,4}\right),
\end{equation}
where ${\mathcal T}$ denotes time ordering of the argument.
The dependence of the effective string tension on the
temperature has recently been checked
for rather low $T^{-1}=\beta<1.13\,\beta_c\approx 0.85\,\mbox{fm}$
in a study of $SU(3)$ gauge theory~\cite{Kaczmarek:1999mm}.
Although the sign of the leading correction term
to the zero temperature limit
is correct, the difference comes out to be bigger than predicted.
It would be interesting to check whether the result will converge towards
the string expectation at lower temperatures.

\subsection{The potential in perturbation theory}
\label{sec:pert}
Besides the strong coupling expansion, which is specific to the lattice
regularisation, the expectation value of a Wilson loop can be
approximated using standard perturbative techniques.

We will discuss the leading order weak coupling result that
corresponds to single gluon exchange between the static
colour sources which, although we neglect the spin structure,
we will call ``quarks'' for convenience. From the Lagrangian,
${\mathcal L}_{YM}=\frac{1}{2g^2}\tr F_{\mu\nu}F_{\mu\nu}$, one can easily
derive the propagator of a gluon with four-momentum, $q$,
\begin{equation}
\label{eq:gluonprop}
G^{ab}_{\mu\nu}(q)=g^2\frac{\delta^{ab}\delta_{\mu\nu}}{q^2},
\end{equation}
where $\mu,\nu$ are Lorentz indices and $a,b=1,\ldots N_A=N^2-1$ label the
colour generators. The same calculation can be done, starting from a
lattice discretised action. The Wilson action, Eq.~(\ref{eq:wilglue}),
yields the result of Eq.~(\ref{eq:gluonprop}), up to the
replacement,
\begin{equation}
q_{\mu}\rightarrow \hat{q}_{\mu}=\frac{2}{a}\sin
\left(\frac{aq_{\mu}}{2}\right).
\end{equation}
Other lattice actions yield slightly different results but
they all approach Eq.~(\ref{eq:gluonprop}) in the continuum limit,
$a\rightarrow 0$.
Up to order $\alpha_s^2$, the
momentum space potential can be obtained from the
on-shell static quark anti-quark  scattering amplitude:
the gluon interacts with two static external currents pointing into
the positive and negative time directions,
$A^a_{\mu,\alpha\beta}=\delta_{\mu,4}T^a_{\alpha,\beta}$
and $A_{\nu,\alpha\beta}^{\prime b}=-\delta_{\nu,4}T^b_{\gamma,\delta}$.
Hence, we obtain the tree level interaction kernel,
\begin{equation}
\label{eq:kern}
K_{\alpha\beta\gamma\delta}(q)=-\frac{g^2}{q^2}
T^a_{\alpha\beta}T^a_{\gamma\delta}.
\end{equation}

For sources in the fundamental representation,
the Greek indices run from $1$ to $N$ and the quark anti-quark state can
be decomposed into two irreducible representations of $SU(N)$,
\begin{equation}
{\mathbf N}\otimes{\mathbf N^*}={\mathbf 1}\oplus{\mathbf N_A}.
\end{equation}
We can now either start from a singlet or
an octet\footnote{We call the state ${\mathbf N_A}$ an ``octet'' state,
having the group $SU(3)$ in mind.} initial
$\Phi_{\beta\gamma}=Q_{\beta}Q^*_{\gamma}$ 
state,
\begin{eqnarray}
\Phi^{\mathbf 1}_{\beta\gamma}&=&\delta_{\beta\gamma},\\
\Phi^{\mathbf N_A}_{\beta\gamma}&=&
\Phi_{\beta\gamma}-\frac{1}{N}\delta_{\beta\gamma},
\end{eqnarray}
where the normalisation is such that
$\Phi^i_{\alpha\beta}\Phi^{j}_{\beta\alpha}=\delta^{ij}$.
A contraction with the group generators of Eq.~(\ref{eq:kern}) yields,
\begin{eqnarray}
\Phi^{\mathbf 1}_{\beta\gamma}T_{\alpha\beta}^aT_{\gamma\delta}^a
&=&C_F\Phi^{\mathbf 1}_{\alpha\delta},\\
\Phi^{\mathbf N_A}_{\beta\gamma}T_{\alpha\beta}^aT_{\gamma\delta}^a
&=&-\frac{1}{2N}\Phi^{\mathbf N_A}_{\alpha\delta},
\end{eqnarray}
where
$C_F=N_A/(2N)$
is the quadratic Casimir charge of the fundamental representation.

We end up with the potentials in momentum space,
\begin{equation}
V_s(q)=-C_Fg^2\frac{1}{q^2},\quad V_o(q)=\frac{g^2}{2N}\frac{1}{q^2}=
-\frac{1}{N_A}V_s(q),
\end{equation}
governing interactions between fundamental charges coupled to a
singlet and to an octet,
respectively: the force in the singlet channel is attractive while that
in the octet channel is repulsive and smaller in size.

How are these potentials related to the static 
position space inter-quark potential,
defined non-perturbatively through the Wilson loop,
\begin{equation}
\label{eq:vdef}
V({\mathbf r})=-\lim_{t\rightarrow\infty}\frac{d}{dt}
\ln\langle W({\mathbf r},t)\rangle?
\end{equation}
The quark anti-quark state creation operator, $\Psi_{\mathbf r}$,
within the Wilson loop 
contains a gauge transporter and couples to the gluonic degrees of freedom.
Thus, in general,
it will have overlap with both, $Q Q^*$ singlet and octet
channels\footnote{Of
course, for quark and anti-quark being at different 
spatial positions, the singlet-octet classification
should be consumed with caution in a non-perturbative
context.}.
Since the singlet channel is energetically preferred, $V_s<V_o$,
we might expect the static potential to correspond to the singlet
potential.

To lowest order in perturbation theory, the Wilson loop is given by
the Gaussian integral,
\begin{equation}
\label{eq:wile}
\langle W({\mathbf r},t)\rangle=
\exp\left\{-\frac{1}{2}\int\! d^4x\,d^4y
J^a_{\mu}(x)
G^{ab}_{\mu\nu}(x-y)J^b_{\nu}(y)\right\},
\end{equation}
where $J_{\mu}^a=\pm T^a$ if $(x,\mu)\in\delta C$ and $J_{\mu}^a=0$,
elsewhere\footnote{Note that this formula that automatically
accounts for multi-photon exchanges is exact in non-compact QED to any order of
perturbation theory. However, in theories containing more complicated vertices,
like non-Abelian gauge theories or
compact lattice $U(1)$ gauge theory, correction terms have to be
added at higher orders in $g$.}.
Eq.~(\ref{eq:wile}) implies for $t\gg r$,
\begin{equation}
\label{eq:wilint}
\langle W({\mathbf r},t)\rangle=
\exp\left(C_Fg^2t\int_{-t/2}^{t/2}\!dt'\ [G({\mathbf r},t')-G({\mathbf 0},t')]
\right).
\end{equation}
We have omitted gluon exchanges between the spatial closures
of the Wilson loop from the above formula. Up to order $\alpha_s^3$
(two loops), such contributions result in terms
whose exponents are proportional to $r$ and $r/t$
and, therefore, do not affect the potential
of Eq.~(\ref{eq:vdef}).
$G^{ab}_{\mu\nu}(x)$, the Fourier transform of
$G^{ab}_{\mu\nu}(q)$, contains the function,
\begin{equation}
G(x)=\int\frac{d^4q}{(2\pi)^4}\frac{e^{iqx}}{q^2},
\quad\int_{-\infty}^{\infty}\!dx_4\,G(x)=\frac{1}{4\pi}\frac{1}{r}.
\end{equation}
After performing the $t$-integration, we obtain,
\begin{equation}
\label{eq:vfundpert}
V({\mathbf r},\mu)=-C_F\frac{\alpha_s}{r} + V_{\mbox{\scriptsize self}}(\mu),
\end{equation}
where $\alpha_s=g^2/(4\pi)$. The piece,
\begin{equation}
\label{eq:self1}
V_{\mbox{\scriptsize self}}(\mu)=C_Fg^2\int_{q\leq\mu}
\frac{d^3q}{(2\pi)^3}\frac{1}{q^2}=C_F\alpha_s\frac{2}{\pi}\mu,
\end{equation}
that linearly diverges with the ultra-violet cut-off, $\mu$,
results from self-inter\-act\-ions of the static (infinitely heavy)
sources. Beyond tree level, $g^2$ will depend on $q$, such that
$\alpha_s$ in momentum space has to be replaced by $\alpha_s(q^*)$
with some effective $q^*(\mu)$.
We find,
\begin{equation}
\label{eq:equal}
V(q)=V_s(q),
\end{equation}
where
\begin{equation}
V({\mathbf q},0)=
\int\!d^3r\,e^{i{\mathbf q}\cdot{\mathbf r}} \hat{V}({\mathbf r}),\quad
\hat{V}({\mathbf r}) = V({\mathbf r},\mu)
-V_{\mbox{\scriptsize self}}(\mu).
\end{equation}
This self-energy problem is well known on the lattice
and has recently received attention in continuum QCD, in the context of
renormalon ambiguities in quark mass
definitions~\cite{Bigi:1994em,Beneke:1998rk}.

At order $\alpha_s^4$ a class of diagrams appears in a perturbative
calculation of the Wilson loop that results in
contributions to the static potential that diverge logarithmically
with the interaction time~\cite{Appelquist:1977tw}.
In Ref.~\cite{Brambilla:1999qa}, within the framework of effective
field theories, this effect has been related to ultra-soft
gluons due to which an extra scale, $V_o-V_s$,
is generated. Moreover, a systematic procedure has been
suggested to isolate and subtract such terms to obtain
a finite
interaction potential between heavy quarks. However, one would wish
to understand and regulate such contributions not only
for heavy quarks but also in the static case.
At present it is not clear
whether the interaction potential within a heavy quark bound
state whose effective Hamiltonian contains a kinetic term
will, in the limit of infinite quark masses, approach
the static potential that is defined through the Wilson loop.
Hence, one should carefully distinguish between the static and heavy 
quark potentials.
We shall discuss
a physically motivated reason for the breakdown of standard
high order perturbative
calculations of the Wilson loop in Section~\ref{sec:beyond}. In our opinion
the presence of a low energy
scale, which we shall identify with the gap between ground state potential and
hybrid excitations, results in problems within perturbation
theory in the limit of large $t$.

That something in the position space derivation of the perturbative
potential might be
problematic is reflected in Eq.~(\ref{eq:wilint}) that contains
an integration over the interaction time.
We know for instance from the spectral decomposition
of Section~\ref{sec:mass} that
for any fixed distance $r$, Wilson loops will decay exponentially
in the limit of large $t$. However, the tree level propagator in position space
is proportional to, $(r^2+t^2)^{-1}$, i.e.\ asymptotically
decays with $t^{-2}$ only. We notice that
the integral receives significant contributions from the region of large $t$
as demonstrated by the finite $t\gg r$ tree level result,
\begin{equation}
-\ln\langle W(r,t)\rangle=-\frac{C_F\alpha_s}{r}t\frac{2}{\pi}
\left\{
\arctan\frac{t}{r}-\frac{r^2}{2t}\left[
\ln\left(1+\frac{t^2}{r^2}\right)\right]\right\}
+(r+t)V_{\mbox{\scriptsize self}}.
\end{equation}
Ignoring this problem for the moment,
one finds the weak coupling equality, Eq.~(\ref{eq:equal}),
to hold up to two loops (order $\alpha_s^3$) in perturbation theory.
Some of the hybrid potentials of Section~\ref{sec:hybrid}
that can be extracted from generalised
Wilson loops, $\langle W_{\Psi}\rangle$,
in which the wave function, $\Psi$, transforms
non-trivially under the cylindrical rotation group $D_{\infty h}$, however,
receive leading order octet contributions. This is because the
creation operator, $\Psi$, explicitly couples to the gluonic background.

The tree level lattice potential can easily be obtained by replacing
$q_{\mu}$ by $\hat{q}_{\mu}$ and (in the case of finite lattice volumes)
the integrals by discrete sums over lattice momenta,
\begin{equation}
q_i=\frac{2\pi}{L_{\sigma}}\,\frac{n_i}{a},\quad
n_i=-\frac{L_{\sigma}}{2}+1,\ldots,\frac{L_{\sigma}}{2}.
\end{equation}
The lattice potential reads,
\begin{equation}
\label{eq:latpot}
V({\mathbf r})=V_{\mbox{\scriptsize self}}(a)-C_F\alpha_s\left[\frac{1}{{\mathbf r}}\right],
\end{equation}
where
\begin{equation}
\label{eq:latpr}
\left[\frac{1}{{\mathbf r}}\right]=\frac{4\pi}{L_{\sigma}^3a^3}
\sum_{{\mathbf q}\neq {\mathbf 0}}\frac{e^{i{\mathbf qR}}}{\sum_i
\hat{q}_i\hat{q}_i},
\end{equation}
and $V_{\mbox{\scriptsize self}}(a)=C_F\alpha_s\left[1/{\mathbf 0}\right]$.
We have neglected the zero mode contribution that is suppressed
by the inverse volume, $(aL_{\sigma})^{-3}$. In the continuum limit,
$[1/{\mathbf r}]$ approaches $1/r$ up to quadratic lattice artefacts
whose coefficients depend on the direction of ${\mathbf r}$ while
$V_{\mbox{\scriptsize self}}(a)$ with $n_f$ flavours of Wilson fermions
diverges like~\cite{Heller:1985hx,Martinelli:1998vt},
\begin{equation}
\label{eq:self2}
V_{\mbox{\scriptsize self}}(a)=C_F\alpha_s a^{-1}\left[3.1759115\ldots+
\left(16.728\ldots-0.423\ldots n_f\right)\alpha_s\right].
\end{equation}
The numerical values apply to the limit, $L_{\sigma}\rightarrow\infty$
and, in the case of the one loop coefficient, $N=3$.
Note that under the substitution, $\mu\approx 1.5879557\,\pi/a$, the
tree level term of
Eq.~(\ref{eq:self2}) is identical to Eq.~(\ref{eq:self1}).
A one loop computation of on-axis lattice
Wilson loops in pure gauge theories can be
found in Ref.~\cite{Heller:1985hx}.
The tree level
form, Eq.~(\ref{eq:latpot}), is often employed to parameterise
lattice artefacts.

Besides defining the static potential from Wilson loops, on a volume with
temporal extent, $\beta=aL_{\tau}$, and periodic boundary conditions
it can be extracted from Polyakov line correlators\footnote{The
Polyakov line is defined in Eq.~(\ref{eq:pl2}).},
\begin{equation}
V(r)=-\lim_{\beta\rightarrow\infty}\frac{d}{d\beta}
\langle P^*(r)P(0)\rangle:
\end{equation}
at any given time the pair of Polyakov lines has the gauge transformation
properties of a static quark anti-quark pair and, thus, the
ground state is the same as that of a
Wilson loop\footnote{This statement is not entirely correct on a finite
spatial volume as we
shall see in Section~\ref{sec:lattfse}.
However, for distances, ${\mathbf r}$, with $r_i\leq aL_{\sigma}/2$,
the ground state is indeed the same.}.
In the Polyakov line correlator, no projection is
made onto the $\Sigma_g^+$ ground state of the flux tube.
Therefore, one might
expect~\cite{Nadkarni:1986cz},
\begin{equation}
\label{eq:potaver}
\langle P^*(r)P(0)\rangle\approx\frac{1}{N^2}
\left[
e^{-\beta V_s({\mathbf r})}+N_Ae^{-\beta V_o({\mathbf r})}\right],
\end{equation}
where the ``octet'' potential,
$V_o$, can be thought to be related to hybrid excitations of the
inter-quark string.
At small $\beta$ (high temperature) the exponentials
can be expanded and the term proportional to $g^2$ vanishes due
to $V_s=-N_AV_o$: the leading order
$r$ dependent
contribution to the correlation
function requires two gluons to be exchanged,
\begin{equation}
\label{eq:potexp}
\langle P^*(r)P(0)\rangle
=\left(1+\frac{N_A}{8N^2}\alpha_s^2\frac{\beta^2}{r^2}\right)e^{-\beta
V_{\mbox{\tiny self}}}.
\end{equation}
The above result can also be produced by
a direct perturbative evaluation of the Polyakov
line correlator in position space: the correlation
function contains two disjoint colour traces, therefore, single
gluon exchanges only contribute to the self-energy.
The colour factor
that accompanies two gluon exchanges
is,
$\frac{1}{N}\tr(T^aT^b)\delta^{ac}\delta^{bd}\frac{1}{N}\tr(T^cT^d)
=\frac{N_A}{4N^2}$.
Hence, we indeed reproduce
Eq.~(\ref{eq:potexp}).
By assuming the singlet channel ($V_s< V_o$)
to dominate Eq.~(\ref{eq:potaver})
in the asymptotic limit
of large $\beta$ one obtains the result of
Eq.~(\ref{eq:vfundpert}), i.e.\
the same potential as from Wilson loops. 
However, if we insist on perturbation theory
to hold for the correlation function itself at large $\beta$, i.e.\ at low
temperature, a misleading (and divergent) result is obtained.
We have demonstrated that extra information how to treat 
the limit $\beta\rightarrow\infty$ has to be provided to
obtain the correct zero temperature tree level potential
from Polyakov line correlation functions. We take this as an
indication that in three loop calculations of the Wilson loop
the $t\rightarrow\infty$ limit should be performed with caution too.

\subsection{Potential models}
Several parametrisations of the QCD potential
have been suggested in the past, either QCD inspired or purely
phenomenological. One should keep in mind
that one would not necessarily expect a potential that reproduces the
observed quarkonia levels to coincide with the static potential
calculated from QCD,
due to the approximations involved, namely the adiabatic and non-relativistic
approximations.

A purely phenomenological logarithmic potential, $V(r)=C\ln(r/r_0)$,
has been suggested as an easy way to produce identical
spin-averaged charmonia and bottomonia  level splittings~\cite{Quigg:1977dd}.
This idea has been incorporated into the Martin
potential~\cite{Martin:1980jx,Martin:1981rm},
$V(r)=C+(r/r_0)^{\alpha}$, with $\alpha\approx 0.1$.
Potentials that have QCD-like behaviour built in at small
distances have been suggested for instance in
Refs.~\cite{Eichten:1975af,Richardson:1979bt,Buchmuller:1981su}.
We have already discussed the prototype Cornell
potential~\cite{Eichten:1975af}, $\hat{V}(r)=-e/r+\sigma r$, that interpolates
between perturbative one gluon exchange for small distances and a
linear confining behaviour for large distances.
Another elegant interpolation between the two domains,
containing the one loop running
of the QCD coupling,
\begin{equation}
\alpha_V(q)=\frac{1}{\beta_0 t_V},\quad
 t_V=\ln\left(\frac{q^2}{\Lambda_V^2}\right),
\end{equation}
has been suggested by Richardson~\cite{Richardson:1979bt}:
in momentum space, $t_V$ is substituted by, $t_V'=\ln(1+q^2/\Lambda_V^2)$,
which does not affect the perturbative ultra-violet domain since
$t_V'\rightarrow t_V$ as $q^2\rightarrow\infty$.
However, the Landau pole at $q^2=\Lambda_V^2$
is regulated and the low energy behaviour of the resulting potential,
\begin{equation}
\label{eq:ric1}
V(r)=-\frac{4\pi C_F}{\beta_0}
\int\frac{d^3q}{(2\pi)^3}
\frac{e^{i{\mathbf q}{\mathbf r}}}
{q^2\ln(1+q^2/\Lambda_V^2)},
\end{equation}
is given by,
\begin{eqnarray}
\label{eq:ric}
V(r)&\rightarrow& \sigma r \quad(r\rightarrow\infty),\\
\label{eq:ric2}
\sigma&=&\frac{C_F}{2\beta_0}\Lambda_V^2,
\end{eqnarray}
i.e.\ the ansatz connects the QCD scale parameter, $\Lambda_V$
to the string tension, $\sigma$.
We have neglected an infinite additional constant
from Eq.~(\ref{eq:ric}) that can be eliminated by adding an appropriate
counter term to the integrand of Eq.~(\ref{eq:ric1}).

From Eq.~(\ref{eq:ric2}) and the relation,
\begin{equation}
\Lambda_V=\Lambda_{\overline{MS}}e^{a_1/(2\beta_0)},
\end{equation}
with~\cite{Fischler:1977yf,Appelquist:1977tw},
\begin{equation}
\label{eq:a0}
a_1=\left(\frac{31}{3}-\frac{10}{9}n_f\right)\frac{1}{4\pi},
\end{equation}
we find
$\Lambda_{\overline{MS}}/\sqrt{\sigma}\approx 0.71639$ for $n_f=0$ or
$\Lambda_{\overline{MS}}/\sqrt{\sigma}\approx 0.70253 (0.70048)$
for $n_f=3(4)$, respectively.
This has to be compared to the value,
$\Lambda_{\overline{MS}}/\sqrt{\sigma}=0.52\pm 0.05$,
determined by lattice simulations~\cite{Capitani:1998mq,Schilling:1993bk}
for $n_f=0$.
Experimental results from $e^+e^-$ scattering experiments at LEP and
SLAC indicate somewhat bigger
ratios~\cite{Caso:1998tx}, $\Lambda_{\overline{MS}}^{(4)}/\sqrt{\sigma}
=0.88(12)$~MeV, for $n_f=4$,
where we have assumed, $\sqrt{\sigma}=(430\pm 20)$~MeV:
while the Richardson potential overestimates the $\Lambda$-parameter
in the quenched case it might approximate the experimental
$n_f=3$ situation quite well.
However, this coincidence is rather accidental.

Many so-called QCD potentials have been suggested that incorporate two loop
perturbation theory at short distances, with varying interpolation
prescriptions to different assumptions on the
large distance behaviour. The most popular
potential within this class is probably
the Buchm\"uller-Tye parametrisation~\cite{Buchmuller:1981su} that, like
the Richardson potential, is formulated in momentum space.
For collections of various parametrisations, we refer to
Refs.~\cite{Kuhn:1988ty,Lucha:1991vn}. 
While phenomenological potentials like a logarithmic as well as
the Martin potential are ruled out at large and intermediate distances
by lattice data and at short distances by pQCD,
such parametrisations may still serve to explore the sensitivity
of the heavy quark spectrum towards QCD.

Basically, all potentials that yield a
correct description for the spin-averaged quarkonia spectra are only slight
variations around the Cornell potential in the relevant region,
$0.2\,\mbox{fm}<r<1$~fm. Unfortunately, the top quark
is too heavy
to form stable hadronic states and basically only the production rate
of $t\bar{t}$ in $e^+e^-$ or $\mu^+\mu^-$ collisions as a function of the
energy will directly depend on the potential at very short distances.
Decay rates and fine structure splittings of quarkonia
in principle can probe the
potential at short distances too and predictions of these quantities
indeed depend very sensitively on the underlying ansatz~\cite{Eichten:1994gt}.
As we will see
in Sections~\ref{relativistic} and \ref{sec:uncert},
the predictive power of quarkonium physics on the short range
potential is reduced by
theoretical uncertainties in the matching of
an effective field theory to QCD.
A big part of the (multiplicative) uncertainty
in the fine structure, however, cancels from ratios of
such splittings.

\subsection{Lattice results}
The static QCD potential has been determined to high accuracy
in quenched
lattice studies with
Wilson~\cite{Lang:1982tj,Stack:1983wb,Griffiths:1983ah,Otto:1984qr,Hasenfratz:1984gc,Barkai:1984ca,Sommer:1985du,deForcrand:1985cz,Huntley:1986ts,Hoek:1987uy,Ford:1988ki,Perantonis:1989uz,Perantonis:1990dy,Michael:1992az,Bali:1992ab,Booth:1992bm,Bali:1993ru,Schilling:1993bk,Bali:1995de,Edwards:1997xf}
as well as various improved lattice
actions~\cite{Itoh:1986gy,Iwasaki:1997sn,Pennanen:1997ni,Beinlich:1997ia,Morningstar:1998da}
in $SU(2)$ and $SU(3)$ gauge theories.
Results for QCD with sea quarks have been obtained in
Refs.~\cite{Born:1994cq,Heller:1994rz,Glassner:1996xi,Bali:1997bj,Aoki:1998sb,Bali:2000vr,Bernard:2000gd}.
After discussing the methods most commonly used we will present results
on the potential in QCD, without and with sea quarks.

\subsubsection{Evaluation method}
\label{sec:lattsmear}
The relative statistical errors of Wilson loop expectation values
turn out to increase exponentially fast
with the Euclidean time extent, $t$, of the loop.
Therefore, after some pioneering
studies~\cite{Lang:1982tj,Stack:1983wb,Otto:1984qr,Hasenfratz:1984gc,Barkai:1984ca},
replacement of the straight spatial connection within the Wilson loop
by operators with improved overlap to the physical ground state turned out to
be essential for a reliable determination of the potential at large distances
from data at moderate $t$ separations.
For this purpose, in
Refs.~\cite{Griffiths:1983ah,Sommer:1985du,Huntley:1986ts}, linear
combinations of certain spatial paths connecting quark and anti-quark
were employed. Subsequently, iterative smearing
techniques~\cite{Teper:1987wt,Albanese:1987ds,Perantonis:1989uz,Bali:1992ab}
turned out to be extremely successful in optimising the
ground state overlap.

Among all algorithmic and technical tricks employed in
lattice simulations smearing is certainly the most important one.
The underlying concept somewhat resembles that of
cooling techniques that are applied to extract classical properties of
quantum field configurations~\cite{Teper:1986ek} with the difference
that, since smearing is a purely spatial procedure, the spectrum of the theory
remains unaffected: fat links are iteratively constructed
by replacing a given link by the sum of itself and the neighbouring
six (in $d=3+1$ dimensions)
spatial staples with some weight parameter, $\alpha>0$,
\begin{equation}
\label{eq:smear}
U_{x,i}\rightarrow P_{SU(N)}\left(U_{x,i}+\alpha\sum_{j\neq i}
U_{x,j}U_{x+\hat{\jmath},i}U^{\dagger}_{x+\hat{\imath},j}\right).
\end{equation}
$P_{SU(N)}$ denotes a projection operator, back onto the $SU(N)$ manifold.
One possible definition is~\cite{Bali:1992ab}, $U=P_{SU(N)}(A)\in SU(N)$,
$\re\tr UA^{\dagger}=\max$. The procedure, Eq.~(\ref{eq:smear}),
can be iterated several times
over the whole lattice. The number of iterations and $\alpha$ represent free
parameters which can be varied to optimise the overlap of an operator,
constructed from the fat links, with the physical ground state in question.
Several variations of the algorithm
exist. For example, all links within a given
timeslice can be replaced at once
or several subgroups can be replaced, subsequently.
Blocking or fuzzing algorithms can be used, in which
a fat link of smearing level $n$ extends over more lattice sites
than
the previous links of level $n-1$. Smearing and fuzzing
can be combined etc..
All smearing and fuzzing methods have in common that the expectation value of
a plaquette built from fat spatial links is increased during the
iterations, similar to cooling, which means that the contribution
to the gauge action from spatial links is reduced:
the movement of the magnetic field through colour space
under a change of the spatial position is minimised.
Operators, built from such fat links, are likely to effectively
decouple from excitations since the ground state
wave function is always the smoothest wave function within any given channel.
Smearing or fuzzing methods can be combined
with variational minimisation techniques when determining a
correlation matrix between a set of different
operators~\cite{Perantonis:1989uz}, to achieve further improvement.

The potential is finally extracted from expectation values of
smeared Wilson loops, $W({\mathbf r},t)$, where the spatial transporters are
constructed from fat links,
\begin{eqnarray}
V({\mathbf r})&=&\lim_{t\rightarrow\infty}V({\mathbf r},t),\\
V({\mathbf r},t)&=&-\frac{d}{dt}\ln\langle W({\mathbf r},t)\rangle
\approx a_{\tau}^{-1}
\ln\frac{\langle W({\mathbf r},t)\rangle}{\langle
W({\mathbf r},t+a_{\tau})\rangle}.
\end{eqnarray}
$a_{\tau}$ denotes the
temporal lattice spacing. On the lattice, the limit of large
temporal separation is approximated by a single- or multi-exponential fit
to Wilson loops for a range, $t>t_{\min}({\mathbf r})$.
Positivity of the transfer matrix implies that $V({\mathbf r},t)$
converges towards the asymptotic value,
$V({\mathbf r})$, monotonically from above, a feature
that is  essential for the reliable detection of saturation of effective
masses, $V({\mathbf r},t)$, into a
plateau. In general, within given statistics,
$t_{\min}({\mathbf r})$ will depend on the distance, ${\mathbf r}$.
Within the typical window of lattice spacings, 0.2~fm~$\geq a\geq
0.05$~fm, in pure gauge theories and standard smearing and simulation
techniques, this dependence happens to be weak.
However, this does not necessarily have to be so but depends very much
on the interplay between the dynamics of the underlying theory,
smearing methods and statistical errors.

In order to illustrate the importance of a careful analysis
of the $t$ dependence of the lattice data we consider the
case of an unsmeared on-axis Wilson loop
on an isotropic lattice, $\langle W(r,t)\rangle=\langle
W(t,r)\rangle$. For $t\gg r$, we expect $\langle W(r,t)\rangle\propto
e^{-V(r)t}$. The symmetry under interchange of $r$ and $t$
implies, $\langle
W(r,t)\rangle\propto e^{-V(t)r}$ for $r\gg t$. This means,
\begin{equation}
V(r,t)=\sigma_{\mbox{\scriptsize eff}}(t)r,\quad 
\sigma_{\mbox{\scriptsize eff}}(t)=V'(t)
\end{equation}
Thus, approximating $V(r)$ by an effective
potential, $V(r,t_{\min})$, with an $r$-independent
value of $t_{\min}$ automatically 
implies a linear rise~\cite{Bali:1995de,Diakonov:1998rk}
within the region, $r\gg t_{\min}$, for any potential
with non-vanishing derivative. This illustrates the importance
of separately investigating the approach to
the plateau for each distance.
Let us examine closely the situation for the Cornell potential,
$V(r)=V_{\mbox{\scriptsize self}}+\sigma r-e/r$. In this case, taking
one and the same $t$-value for all separations we find,
\begin{equation}
\sigma_{\mbox{\scriptsize eff}}(t)=\sigma+\frac{e}{t^2};
\end{equation}
even a pure Coulomb potential, $\sigma=0$, implies a non-vanishing
$\sigma_{\mbox{\scriptsize eff}}$ at finite $t\ll r$.
Of course, the symmetry of the Wilson loop under interchange of $r$ and $t$
also implies that no plateau in $V(r,t)$ can be found,
unless $t\gg r$.
For smeared Wilson loops, one would still expect
a similar $1/t^2$ approach (with a different coefficient)
of $\sigma_{\mbox{\scriptsize eff}}$
towards the asymptotic limit, while effective masses,
$V(r,t)$, will approach
$V(r)$ exponentially fast at any $r$.

\subsubsection{The quenched potential}
\label{sec:latt}
\begin{figure}[thb]
\centerline{\epsfxsize=10truecm\epsffile{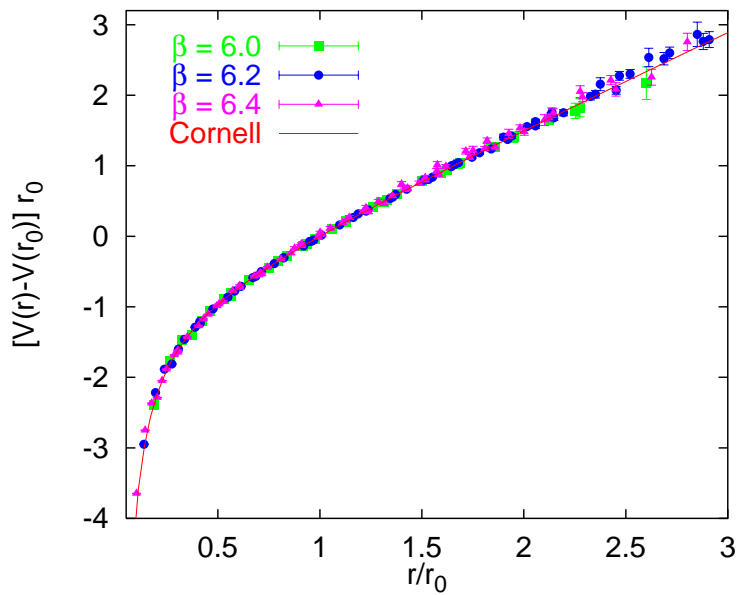}}
\caption{The quenched Wilson action $SU(3)$ potential, normalised
to $V(r_0)=0$.}
\label{figpotsu3}
\end{figure}
In Figure~\ref{figpotsu3}, we display the quenched potential, obtained at three
different $\beta$ values in units of $r_0\approx 0.5$~fm
from the data of Refs.~\cite{Bali:1993ru,Bali:1997am}. The lattice
spacings, determined from $r_0$, correspond to $a\approx 0.094$~fm,
0.069~fm and 0.051~fm, respectively. The curve represents the Cornell
parametrisation with $e=0.295$. At small
distances the data points lie somewhat
above the curve, indicating a weakening of the
effective coupling and, therefore, asymptotic freedom. We will discuss this
observation later. All data points for $r>4a$ collapse onto a universal curve,
indicating that for $\beta\geq 6.0$ the scaling
region is effectively reached for the static potential.
Moreover, continuum rotational symmetry is restored:
in addition to on-axis separations, many off-axis distances of the sources
have been realised and the corresponding data points are well parameterised
by the Cornell fit for $r> 0.6\,r_0$. Prior to comparison between the
potential at various $\beta$, the additive self-energy contribution,
associated with the static
sources, that diverges in the continuum limit has been removed. This is
achieved by the parametrisation-independent normalisation of
the data to $V(r_0)=0$.

\begin{figure}[thb]
\centerline{\epsfxsize=10truecm\epsffile{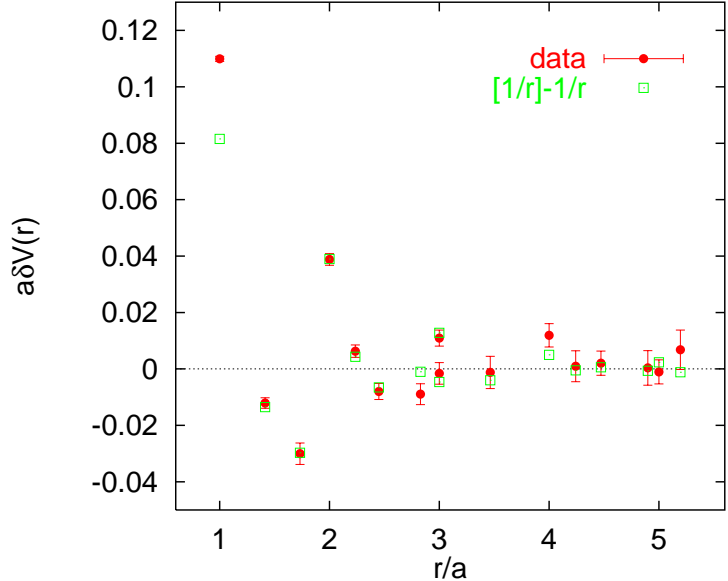}}
\caption{Comparison between the tree level expectation of violations of
  rotational invariance and lattice data at $\beta=6.4$.}
\label{figpotrot}
\end{figure}

The lattice potential
at $\beta\leq 6.5$ is  well described
by the functional form~\cite{Booth:1992bm},
\begin{equation}
\label{eq:v1fit}
V({\mathbf r})=V_{\mbox{\scriptsize cont}}(r)-l\delta V({\mathbf r}),
\quad V_{\mbox{\scriptsize cont}}(r)=V_0+\sigma\,r-\frac{e}{r}+\frac{af}{r^2},
\end{equation}
for separations as small as $r\geq\sqrt{3}a$.
The $1/r^2$ term is not physically motivated but effectively
parameterises the weakening of the coupling with
the distance while 
the difference between tree level lattice
and continuum perturbation theory results,
\begin{equation}
\label{eq:v2fit}
\delta V({\mathbf r})=\left[\frac{1}{\mathbf r}\right]-\frac{1}{r},
\end{equation}
is used to quantify lattice artefacts. 
In Figure~\ref{figpotrot}, we compare the theoretical difference
$\delta V({\mathbf r})$ to
$\delta V({\mathbf r})=
[V_{\mbox{\scriptsize cont}}(r)-V({\mathbf r})]/l$, as calculated from the
lattice data
after determination of the fit parameters,
$V_0,\sigma,e,f$ and $l$,
at $\beta=6.4$, in lattice units~\cite{Schilling:1993bk}.
The Figure demonstrates
that at the level of precision achieved, deviations from the
continuous fit curve are statistically significant for $r\leq 4a$.
Moreover, deviations from $V_{\mbox{\scriptsize cont}}(r)$ are
qualitatively indeed very well parameterised by a multiple of
the tree level difference.

\subsubsection{Finite size effects}
\label{sec:lattfse}
In lattice simulations the potential is determined on a torus with finite
volume and the question of finite size effects (FSE) arises.
Obviously, the ground state
potential is affected by the infra-red cut-off.
For instance, by exploiting the $r\leftrightarrow t$ symmetry
of Euclidean space-time, it is clear from
Appendix~\ref{centres} that
for the extreme case of a spatial extent smaller than
the critical temperature of the deconfinement phase transition,
any asymptotic string tension will disappear.
The other source of FSE on Wilson loops
is related to un-wanted interactions of source and
anti-source around the periodic boundaries that will become negligible
as $L_{\sigma}a\rightarrow\infty$: by unwrapping the spatial
torus onto
an infinite hyper-cubic lattice of cells with spatial periods, $L_{\sigma} a$,
it becomes obvious that each
charge at position ${\mathbf r}$ is accompanied by an infinite set
of mirror
charges at ${\mathbf r}_{\mathbf n}=
{\mathbf r}+{\mathbf n}L_{\sigma}a$, $n_i$ integer.
For on-axis geometries, ${\mathbf r}=r\hat{\mathbf\imath}$,
for instance, the
closest mirror charge will be separated by a distance $L_{\sigma}a-r$
from the origin,
followed by another charge at $L_{\sigma}a+r$. Therefore,
in this case one would na\"{\i}vely expect $V(r)=V(L_{\sigma}a-r)$.
The symmetry, $V({\mathbf r})=V({\mathbf r}_{\mathbf n})$,
is indeed reflected in the tree level weak coupling expansion results of
Eqs.~(\ref{eq:latpot}) and (\ref{eq:latpr}) and
in fact holds to any order of perturbation theory
for the singlet and octet potentials.

General considerations, based on the centre symmetry of the action
which is discussed in
Appendix~\ref{centres}, however, lead us to expect the potential to
be more robust against FSE than perturbation theory suggests.
We will assume the source to be at position ${\mathbf 0}$ and the
anti-source to be at ${\mathbf r}$.
The spatial connection, $\Psi_{\mathbf r}^{\dagger}$,
built into a (smeared) Wilson loop
is a linear combination
of products of link variables along paths
that have trivial winding number
around the periodic boundaries. 
$\Psi_{\mathbf r}^{\dagger}$
has definite
eigenvalues, $z_i\in Z_N$, with respect to the centre transformations
associated with the three spatial directions. If we place the hyperplanes at
which centre transformations, $z$, are applied at position ${\mathbf x}$ with
$x_i\geq r_i$, i.e.\ such that they do not interfere with the shortest
connection between the two test charges, we have $z_i=0$. However,
a gauge transporter to a mirror charge at ${\mathbf r}_{\mathbf n}$
has the centre transformation property, $z_i=z^{n_i}$.
Since the centre symmetry is both, a symmetry of the action as well as of the
path integral measure, $z_i$ are conserved quantum
numbers:
the creation operator, $\Psi_{\mathbf r}$,
only couples to mirror charges at distances in which all $n_i$'s are multiples
of $N$. For on-axis
separations in $SU(N)$ this means that
the closest mirror charge contributing to the Wilson loop will be at a distance
$NL_{\sigma}a-r$~\cite{Huntley:1986ts,Bali:1995de}, rather than
$L_{\sigma}a-r$ as the geometric argument alone or perturbation theory would
have suggested. This suppression
of FSE does obviously not work for Polyakov loop correlators in which the
state of the gluonic flux tube remains unspecified.
While in the standard weak coupling expansion the gauge group only
influences the group theoretical pre-factors the centre
charge affects
the zero mode sector~\cite{Coste:1985mn}, ${\mathbf q}={\mathbf 0}$
(that is suppressed
by a power of the volume).

Numerical simulations of $SU(2)$ and $SU(3)$ gauge
theories~\cite{Bali:1995de,Bali:1992ab} suggest that
FSE on the static potential determined from Wilson loops,
even at distances as big as
$r=\sqrt{3}/2\,L_{\sigma}a$,
are numerically undetectable on the 1 -- 2~\% level
for $L_{\sigma}a>3\,r_0$. A reason besides the protection
due to the centre symmetry for this finite size
friendliness is the
rather rapid on-set of the deconfinement phase transition which is
first order in $SU(3)$ gauge theory. Full QCD, however, is less
well explored yet and one might expect somewhat bigger FSE, at
least for light sea quarks since the fermionic part of the action
explicitly breaks
centre symmetry. In particular, it might be hard to discriminate
between breaking of the flux tube due to screening by sea quarks
and FSE as reasons for an eventually flattening
potential at large distances.

\subsubsection{Sea quark effects}
\label{sec:seaq}
\begin{figure}[thb]
\centerline{\epsfxsize=10truecm\epsffile{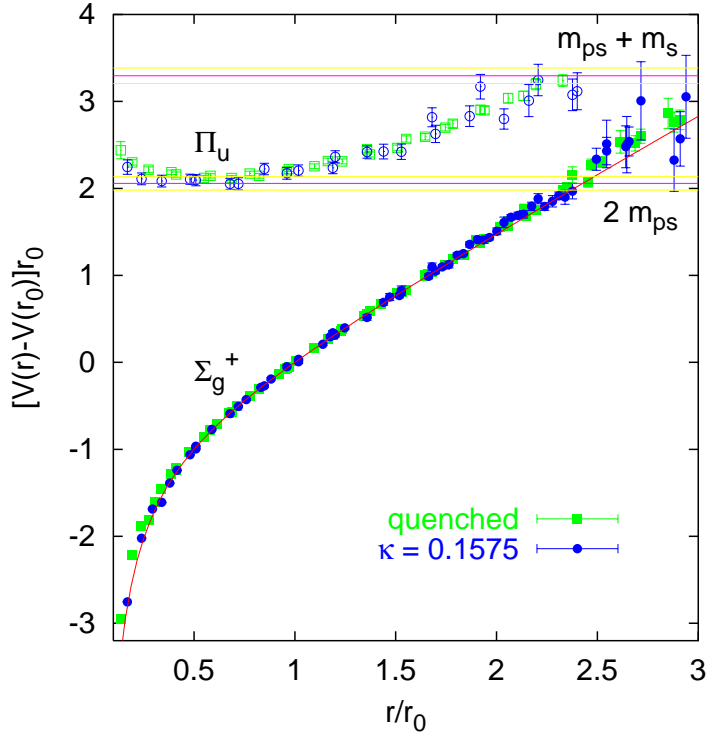}}
\caption{Comparison between quenched ($\beta=6.2$) and un-quenched $n_f=2$
  ($\beta=5.6, \kappa=0.1575$) ground state and $\Pi_u$ potentials
(from Ref.~\cite{Bali:2000vr}).}
\label{figunq}
\end{figure}
When including sea quarks, one would expect two physical effects,
one at large distances and one at small distances.
While within the quenched approximation the number of quarks and anti-quarks
are separately conserved, with sea quarks, only the difference (the baryon
number) is a conserved quantity. Light quark anti-quark pairs can be created
from the vacuum and in general transitions between a colour ``string'' state,
spanned between two static sources, and
two static-light mesons can occur.
If the energy stored in the colour string between the sources
exceeds a certain critical value at some distance, $r=r_c$,
the string will ``break'' and decay into two static-light mesons,
separated by a distance, $r$.
Therefore, in the limit, $r\rightarrow \infty$, the ground state
energy will stop rising with the distance and
saturate at a constant level: the static sources will be completely
screened by light quarks that pop up out of the vacuum.

The other effect will change the potential at short distances.
While the vacuum polarisation due to gluons has an anti-screening effect
on fundamental sources, sea quarks result in screening.
Therefore, the running of the QCD coupling 
with the distance is slowed down with respect to the quenched approximation.
This is for instance
reflected in the factor, $11-2n_f$, within the perturbative $\beta$-function
coefficient, $\beta_0$, of Eq.~(\ref{beta0}).
When running the coupling from an infra-red
hadronic reference scale down to short distances, the effective Coulomb
strength in presence of sea quarks should, therefore, remain
at a higher value than in the quenched case. It should be possible to detect
this effect in the coefficient, $e$, within the Cornell parametrisation.

\begin{figure}[thb]
\centerline{\epsfxsize=10truecm\epsffile{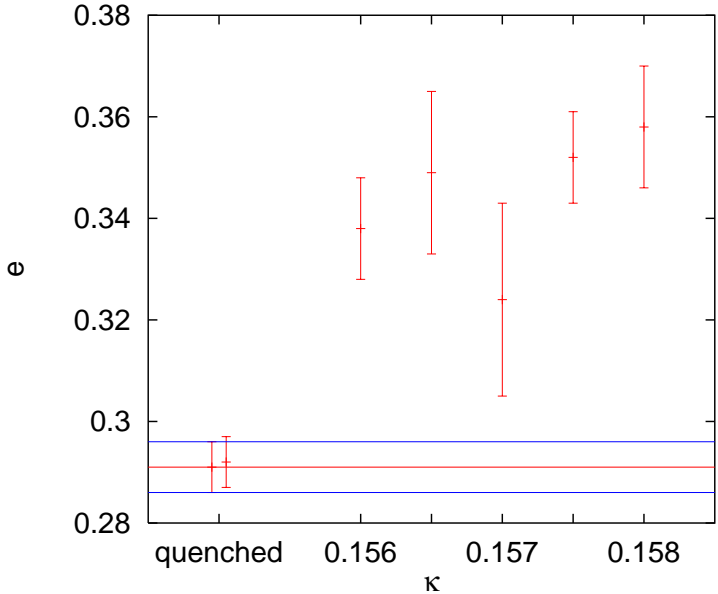}}
\caption{Sea quark mass dependence of the effective Coulomb strength of the
static inter-quark potential.}
\label{figcoul}
\end{figure}

In Figure~\ref{figunq}, a recent comparison between
the quenched ($\beta = 6.2$) and $n_f=2$ static potential ($\beta=5.6$,
$\kappa = 0.1575$)
by the $T\chi L$ collaboration
at a sea quark mass, $m_{ud}\approx m_s/2$, is
displayed~\cite{Bali:1999gq,Bali:2000vr}.
Besides the ground state potential, $\Sigma^+_g$, the lowest  lying
hybrid potential, $\Pi_u$, is shown.
Estimates of masses of pairs of static-light mesons ($2m_{ps}$ and
$m_{ps}+m_s$)
into which the static-static systems can decay are also included
into the Figure.
The potentials have been matched to each other at a
distance, $r=r_0 \approx 0.5$~fm.
Around $r\approx 2.3\,r_0\approx 1.15$~fm,
both un-quenched potentials, $\Sigma_g^+$ and $\Pi_u$, are
expected to become
unstable. However, the data are not yet precise enough to resolve this
effect. A similar comparison between the static potential
and $2m_{ps}$ has first been performed by the
UKQCD collaboration~\cite{Allton:1998gi}.

At small $r$ the un-quenched data points
are found to be
somewhat below their quenched counterparts: the effective Coulomb
force indeed remains stronger. To quantify this effect,
we fit the
potentials~\cite{Glassner:1996xi,Bali:1997ec,Bali:1997bj,Bali:2000vr}
(quenched at $\beta=6.0$ and $\beta=6.2$ and un-quenched 
at $\beta=5.6$ and various quark masses) for identical
fit range in physical units, $0.4\,r_0<r<2\,r_0$,
to the parametrisation
of Eqs.~(\ref{eq:v1fit}), (\ref{eq:v2fit}), with $f=0$. The resulting
effective $e$ values from these four-parameter fits are
displayed in Figure~\ref{figcoul}. Larger $\kappa$ values correspond to
smaller quark masses.
With two flavours of sea quarks of masses slightly larger than that
of the strange quark, down to $m_{ud}\approx m_s/3$,
the effective Coulomb strength is
increased by 17 to 22~\% which is not too far off from
the most na\"{\i}ve expectation
of 14~\%, from the ratio, $\beta_0^{(n_f=0)}/\beta_0^{(n_f=2)}=33/29$.
Within given errors the dependence on the sea quark mass cannot be
resolved. However, one would expect this to be rather
weak as the simulated quark masses
are all much smaller than infra-red reference scales like $r_0^{-1}$.
Similar results have been reported by the CP-PACS
collaboration~\cite{Aoki:1998sb}.

\subsection{Beyond perturbation theory at short distances}
\label{sec:beyond}
The singlet potential,
\begin{equation}
V_s(q)=-C_F\frac{4\pi\alpha_V(q)}{q^2},
\end{equation}
has been calculated to one loop long
ago~\cite{Fischler:1977yf,Appelquist:1977tw} and now the two loop
result
is also known\footnote{The leading log contribution to the
three loop result has been derived by Brambilla and
collaborators~\cite{Brambilla:1999qa} and confirmed in
Ref.~\cite{Manohar:1999xd} while a two loop result for the case of
massive quarks has recently
been obtained by
Melles~\cite{Melles:2000dq}.}~\cite{Schroder:1998vy,Peter:1997ig,Peter:1997me}.
It is,
\begin{equation}
\label{eq:vq}
\alpha_{V,{\mbox{\scriptsize pert}}}(q)=\alpha_{\overline{MS}}(q)\left(1+
a_1\alpha_{\overline{MS}}(q)
+a_2\alpha_{\overline{MS}}^2+\cdots\right).
\end{equation}
$a_1$ is defined in Eq.~(\ref{eq:a0}) and,
\begin{eqnarray}
a_2&=&\left[\frac{4343}{18}+36\pi^2-\frac{9}{4}\pi^4+66\zeta(3)\right.
\\\nonumber
&-&\left.\left(\frac{1229}{27}+\frac{52}{3}\zeta(3)\right)n_f+
\frac{100}{81}n_f^2\right]\frac{1}{16\pi^2}.
\end{eqnarray}
$\zeta(3)=1.2020569\ldots$ denotes the Riemann $\zeta$-function.
For $n_f=5$ the numerical values are, $a_1=0.3802034\ldots$,
$a_2=0.9868211\ldots$. Since $a_2\gg a_1$, perturbation theory seems
to be rather slowly or badly convergent. 

Na\"{\i}vely, one might expect the perturbative calculation of the
static QCD potential to be reliable
in the limit of large energy scales, $q\simeq 1/r$, i.e\ at
short distances. However, unlike in QED,
the QCD potential is the ground state energy
of a bound state composed of the two static
colour sources and gluons. Bound state
properties are associated with
a characteristic scale, $\lambda$,
which plays the r\^ole of
an inverse gluonic coherence length.
We identify $\lambda$ with
the gap between ground state
and first excitation. As we shall see in Section~\ref{sec:hybrid},
for large $r$ this gap corresponds to the difference between a hybrid state
and the ground state potential which, from the bosonic string picture,
we expect to decrease at large distances like $\pi/r$. However,
in the limit $r\rightarrow 0$ the gap will not diverge but
saturate at a constant level  that corresponds to
the scalar glueball mass, $\lambda\approx
1.7$~GeV. Note that in QCD with light sea quarks it will be even smaller,
of the order of
the mass of two pions.
The presence of such a low energy scale, affecting the
short distance behaviour,
can result in differences between the perturbatively calculated
singlet potential and the static potential. 

Non-perturbative $\Lambda^4/q^4$ power corrections to $\alpha_V$ that
are due to the gluon condensate
are indeed expected
from the
standard operator product
expansion~\cite{'tHooft:1977am,Lautrup:1977hs,Shifman:1979cg,Mueller:1985vh}.
Recently, this picture has been challenged by
several
authors~\cite{Beneke:1992mn,Brown:1992ic,Zakharov:1992bx,Vainshtein:1994ff,Baker:1998jw,Akhoury:1998by,Grunberg:1998ix,Gubarev:1999zq,Simonov:1999gk,Gubarev:1999ie}
who found various arguments in support of 
a term, proportional to
$\Lambda^2/q^2$,
\begin{equation}
\alpha_V(q)=\alpha_{V,\mbox{\scriptsize pert}}(q)
+c_V\frac{\Lambda^2_{\overline{MS}}}
{q^2}+\cdots.
\end{equation}
Lorentz invariance implies that no power law corrections of even lower order
in $1/q$ exist.
For the quenched case, where precise data exist down to
lattice spacings as small as $a^{-1}\approx 5.5$~GeV,
the lattice potential has been compared to perturbation theory and,
indeed, a non-vanishing value,
$c_V=4.8\pm 1.4$,
has been found
in Ref.~\cite{Bali:1999ai}
for $n_f=0$, after subtracting one loop perturbation theory.
We briefly summarise this analysis below.

A Fourier transform of the momentum space potential yields,
\begin{equation}
\label{Vrun}
V(r)=-C_F\frac{\alpha_R(1/r)}{r},
\end{equation}
with~\cite{Billoire:1980ih,Peter:1997me,Jezabek:1998wk,Bali:1999ai},
\begin{equation}
\label{alphaR}
\alpha_R(1/r)=\alpha_{V,\mbox{\scriptsize pert}}(\mu)
\left(1+\frac{\pi^2\beta_0^2}{3}\alpha_{V,\mbox{\scriptsize pert}}^2+\cdots\right)
-2c_V\Lambda_{\overline{MS}}r^2+\cdots,
\end{equation}
where $\mu=\exp(-\gamma_E)/r$ and
$\gamma_E=0.5772156\ldots$ denotes the
Euler constant. While in the ultra-violet the effect of the $1/q^2$
power correction to $\alpha_V$ on $V(q)$ is suppressed by a factor, $1/q^4$,
this suppression is proportional to $r$ only
in position space [Eqs.~(\ref{Vrun}) -- (\ref{alphaR})].

\begin{figure}[thb]
\centerline{\epsfxsize=10truecm\epsffile{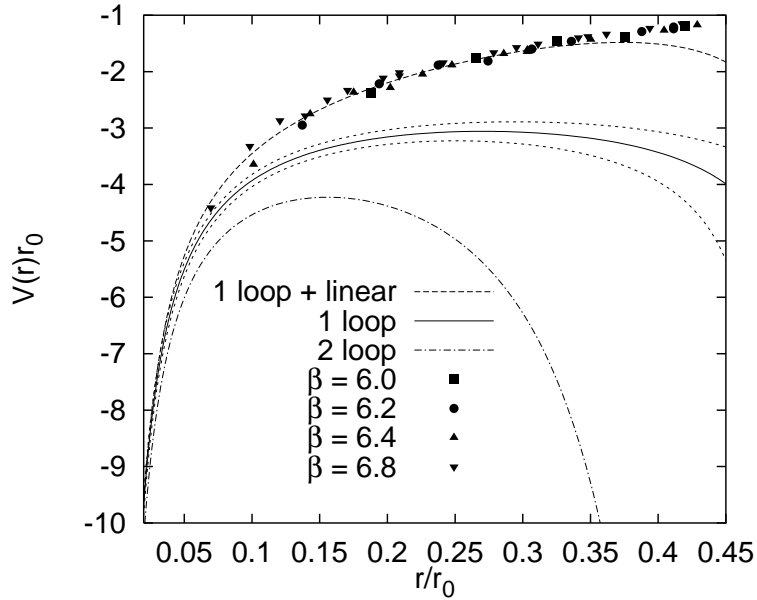}}
\caption{The quenched lattice potential and
perturbation theory.
The lattice data have been normalised such that $V(r_0)=0$.
The values $0.94\,r_0^{-1}$ and $0.77\,r_0^{-1}$ have been
subtracted from the one  and two loop formulae, respectively,
to allow for comparison.}
\label{figpertpot}
\end{figure}

By employing a recursive lattice finite size
technique~\cite{Luscher:1992an,Luscher:1994gh},
the ALPHA collaboration
has recently obtained a value for the running coupling in quenched
$SU(3)$ QCD~\cite{Capitani:1998mq}. They quote the result,
\begin{equation}
\label{eq:lms0}
\Lambda_{\overline{MS}}^{(0)}=0.602(48)/r_0,
\end{equation}
in units of the Sommer scale $r_0$.
We use this result as an input to determine $\alpha_{\overline{MS}}$
at a high energy scale (1000~$r_0^{-1}$) and run it down from there to scales
$\mu$, using the four loop renormalisation group equation,
Eq.~(\ref{rge}).
Subsequently, the resulting $\alpha_{\overline{MS}}(\mu)$
is converted into $\alpha_R(e^{\gamma_E}\mu)$ to one
and two loops via Eqs.~(\ref{eq:vq}) and (\ref{alphaR}) (with $c_V=0$),
and the perturbative potential is determined.
In Figure~\ref{figpertpot}, we compare the result to lattice data.
The disagreement is increased when going to higher order perturbation theory.
Furthermore, the difference between perturbation theory and the
non-perturbative determination is consistent with a linear
term~\cite{Bali:1999ai},
as expected from Eqs.~(\ref{Vrun}) and (\ref{alphaR}).

Because of the significant size of the
coefficient, $a_2$, different ways of re-summing the
series or performing the Fourier transform
can result in somewhat different results and, therefore, in a
different coefficient of the linear term,
$-2C_Fc_V\Lambda_{\overline{MS}}^2=(3.4\pm 1.0)\sigma$.
Due to the non-convergent character of the perturbative series,
subtracting the perturbative tail from a physical operator
is never a well-defined procedure anyway. The ambiguity involved is
related to renormalons that result from the interplay between
perturbation theory and non-perturbative contributions --- between the
ultra-violet and the infra-red.
By subtracting two loop perturbation theory, we find the
linear term to have a slope about six times as big as the string
tension. In contrast, tree level perturbation theory,
with the coupling being treated as a free parameter,
is compatible with the Cornell potential, i.e.\ results
in the same linear slope, $\sigma$,
for small and large distances.

\subsection{String breaking}
\label{sec:stringb}
Having discussed the potential at very
short distances, we shall re-examine the large
distance behaviour. While a linear rise is expected in pure $SU(N)$ gauge
theory, in full QCD the
coupling of gluons to fundamental matter fields will
result in a screening of inter-quark forces at large distances. However,
this behaviour has not been detected so far in simulations involving
sea quarks (cf.\ Figure~\ref{figunq}).
One reason might be that smeared Wilson loops
are highly optimised to achieve enhanced overlap with the lowest lying
string state and might, therefore, almost completely
decouple~\cite{Bali:1997ec} from the physical ground state at
large $r$ that consists of two disjoint
static-light mesons. Arguments based on the strong coupling
expansion~\cite{Drummond:1998ar,Drummond:1998eh} as well as
on the bosonic string picture~\cite{Gliozzi:1999wq}
support this 
suggestion.
Investigating string breaking in full QCD
is computationally very
expensive as high statistics are required to resolve the
potential in the large $r$ region of interest.

While string breaking has not been detected in the Wilson loop,
the finite temperature potential, extracted from Polyakov line correlators
at temperatures close to the deconfinement phase transition,
exhibits a flattening, once sea quarks are included into the
action~\cite{DeTar:1998qa,Buerger:1993bq}.
First indications
of this effect have been reported as early as in 1988~\cite{Faber:1988pi}.
Unlike Wilson
loops, Polyakov line correlators automatically have non-vanishing overlap
with any excitation, containing static quark and
anti-quark, separated by a distance, $r$; in particular the static
quarks can be accompanied by two disjoint sea quark loops, encircling the
temporal boundaries, while in the Wilson loop case, co-propagating sea quarks
are terminated by the spatial transporters at $x_4=0$ and $x_4=t$.
The difference becomes perhaps most obvious in the loop expansion
of Ref.~\cite{Gliozzi:1999wq}.
Since for non-zero temperatures the finite temperature potential
(or free energy)
extracted from Polyakov line correlators will in general
differ from the
potential, extracted from Wilson loops, the situation is not
yet satisfying.

\begin{figure}[thb]
\centerline{\epsfxsize=9truecm\epsffile{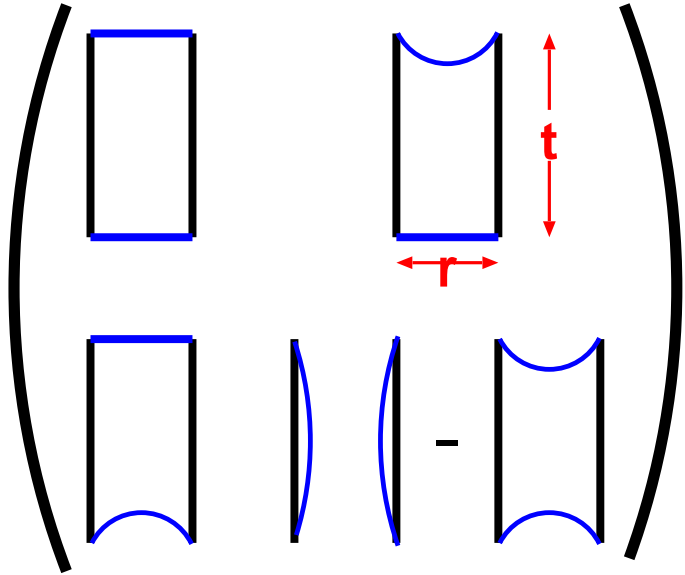}}
\caption{Graphical representation of the correlation matrix
that is relevant for an investigation of string breaking in full QCD.}
\label{figcorr}
\end{figure}

The first ambitious studies of string breaking in QCD using operators with
better
projection on the broken string state are at present
being performed (see
e.g.~\cite{Allton:1998gi,Michael:1999nq,Pennanen:2000yk,Schilling:1999mv,Bali:2000vr}). Such
calculations
involve diagonalisation of a two by two correlation matrix
between string states and two pairs of static-light states.
This matrix is visualised in Figure~\ref{figcorr},
where straight lines correspond to
gauge transporters while curved lines represent light quark propagators
that are obtained by inverting the fermionic matrix, $M$, of
Eq.~(\ref{eq:wilsonf}).
The off-diagonal elements encode transitions between a string state
with fixed ends
and two static-light mesons while the bottom-right element 
represents two interacting static-light mesons, separated by a distance $r$.
The second of the two contributions to this element corresponds
to the exchange of a light meson. The situation becomes slightly more involved
when the Dirac spin structure is taken into account.

Light quark propagators are normally just calculated
for one source point to reduce the
effort in terms of computer time and memory to a tolerable size.
However, pairs of quark propagators emanating from different sites are
required to calculate the bottom right element of the correlation
matrix for various distances and times.
Moreover, for a precise determination
of expectation values of Wilson loops one usually exploits self-averaging:
the average of the Wilson loop is not only taken over
the Monte-Carlo generated ensemble of gauge configurations but also within
each configuration;
Wilson loops with different corner point coordinates are averaged,
exploiting translational invariance.
This practice is essential to reduce statistical fluctuations to an acceptable
level. By use of refined stochastic estimator techniques~\cite{Michael:1998sg}
to calculate the all-to-all
light quark propagators required for this purpose one will eventually
be able to confirm string breaking at distances $r\approx 2.3\,r_0$.

\begin{figure}[thb]
\centerline{\epsfxsize=10truecm\epsffile{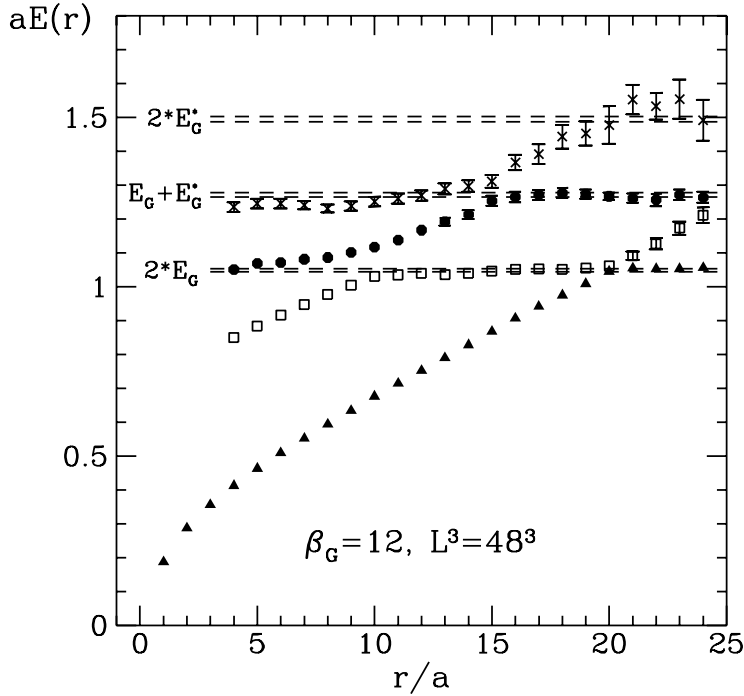}}
\caption{The breaking of the adjoint string in $2+1$-dimensional $SU(2)$ gauge
  theory at a gauge coupling, $\beta=12$, in lattice units (from
  Ref.~\cite{Philipsen:1999wf}).}
\label{figadjbreak}
\end{figure}

In addition to the potential, breaking of closed strings (torelons)
in QCD with sea
quarks has been investigated~\cite{Bali:1997bj,Bali:1997ec,Bali:2000vr}.
Although the results suggest an effect,
its statistical significance of $2.5$ standard deviations
is not yet entirely convincing.
While the situation in the case of interest is
not settled yet, toy models have been investigated
in three and four space-time dimensions.
Similar to the situation
of fundamental QCD colour sources being
screened by sea quarks, one expects the string between adjoint sources in pure
gauge theories to decay into a pair of gluelumps (or glueballinos),
bound states
between a static adjoint source (that can be thought to approximate a
heavy gluino) and gluons (for a detailed discussion see e.g.\
Ref.~\cite{Jorysz:1988qj}). Until recently, the situation was controversial:
while in some early studies indications of the
breaking of the adjoint string in four-dimensional $SU(2)$ and $SU(3)$
gauge theories have been reported~\cite{Griffiths:1985ip,Campbell:1986kp},
in a later simulation of three-dimensional $SU(2)$
the broken string state has not been detected within
Wilson loop correlators~\cite{Poulis:1997nn}.

\begin{figure}[thb]
\centerline{\epsfxsize=10truecm\epsffile{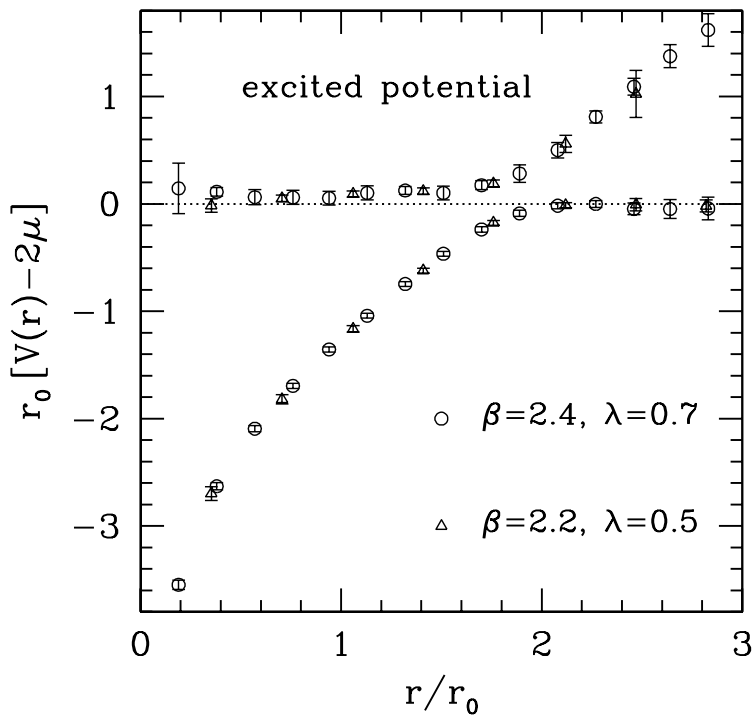}}
\caption{The breaking of the fundamental string in the
$3+1$-dimensional $SU(2)$-Higgs model in physical units (from
  Ref.~\cite{Knechtli:1999av}). $\mu$ corresponds to the mass of
a Higgs particle, bound to the static source.}
\label{fighiggsbreak}
\end{figure}

In other studies a correlation matrix similar to that
of Figure~\ref{figcorr} between the string state and a two gluelump basis
has been investigated. This was first done in four-dimensional
$SU(3)$ gauge theory~\cite{Michael:1992nc},
followed by simulations of
three-dimensional~\cite{Stephenson:1999kh,Philipsen:1999wf}
and four-dimensional~\cite{deForcrand:1999kr,Kallio:2000jc}
$SU(2)$ gauge
theories.
The main result of the $d=3$ study of
Ref.~\cite{Philipsen:1999wf} is depicted in
Figure~\ref{figadjbreak}. In addition to the ground state the first
three excitations are included into the Figure. At small $r$,
these resemble the first radial
excitation of the string, two ground state gluelumps
and one ground state gluelump plus an excited state gluelump, respectively.
The horizontal lines indicate masses of pairs of isolated gluelumps where
$E_G$ stands for the ground state and $E_G'$ for the first excitation.
A similar breaking pattern of the adjoint string has also been confirmed
in four dimensional $SU(2)$ gauge
theory~\cite{deForcrand:1999kr,Kallio:2000jc}.
Studies of the breaking of the adjoint string have been
preceded by investigations~\cite{Bock:1990kq}
of the breaking of the fundamental string in
$SU(2)$ gauge
theory with a
Higgs field in the fundamental representation.
The latest results from such simulations obtained in
three~\cite{Philipsen:1998de} and four
dimensions~\cite{Knechtli:1998gf,Knechtli:1999av}.
confirm the expected screening of the static sources
(see Figure~\ref{fighiggsbreak}).

In QCD one would expect a cross-over of the string state and the
two static-light meson levels similar to the observations within
the toy models considered above, even in the
quenched approximation~\cite{Sommer:1996fr,Michael:1999nq}. How then
can one distinguish the quenched scenario from the un-quenched one where
the string is allowed to break?
In the quenched case, the separate conservation of baryon and anti-baryon
numbers implies that an open string state creation operator
is orthogonal to a creation operator for the two static-light state.
Therefore, each operator has zero overlap with the respective
other state and only the assignment of the ground state to a particular
operator will become interchanged around
the would-be string breaking distance at which
the two energy levels cross. Unlike the behaviour depicted in
Figures~\ref{figadjbreak} and \ref{fighiggsbreak}, no energy gap at
this distance will occur.
Both, the string breaking distance and the associated energy gap,
which is related to a phase shift in the mixing matrix of the
un-quenched case, are
relevant for an understanding of decay rates such as that
of the $\Upsilon(4S)$ or $\Upsilon(5S)$ mesons into pairs of $B\overline{B}$
mesons.

\subsection{Colour confinement}
We have established the linearly rising potential in pure Yang-Mills gauge
theories as a numerical fact and have made this behaviour plausible
from strong coupling and string arguments. However, the dynamical question
of how $SU(N)$ gauge theory as the theory of asymptotic freedom results
in the formation of colour flux tubes with constant energy density per unit
length remains un-answered.
In the past decades, many explanations of the confinement mechanism
have been proposed, most of which share the feature that topological
excitations of the vacuum
play a major r\^ole. These pictures include, among others, the
dual superconductor scenario of
confinement~\cite{Mandelstam:1974pi,'tHooft:1975pr,'tHooft:1981ht}
and the centre vortex
model~\cite{Nielsen:1979xu,'tHooft:1979uj,Ambjorn:1980xi,Mack:1980zr,Cornwall:1979hz,Mack:1982gy}.
Depending on the underlying scenario, the excitations
giving rise to confinement are thought to be
magnetic monopoles, instantons, dyons,
centre vortices, etc.. Different ideas are not necessarily
exclusive. For instance,
all fore-mentioned excitations are
found to be correlated with each other
in numerical as well as in some analytical studies, such that at present
it seems to be rather
a matter of personal preference which one to consider
as more fundamental.

\begin{figure}
\epsfxsize=10.truecm
\centerline{\epsfbox{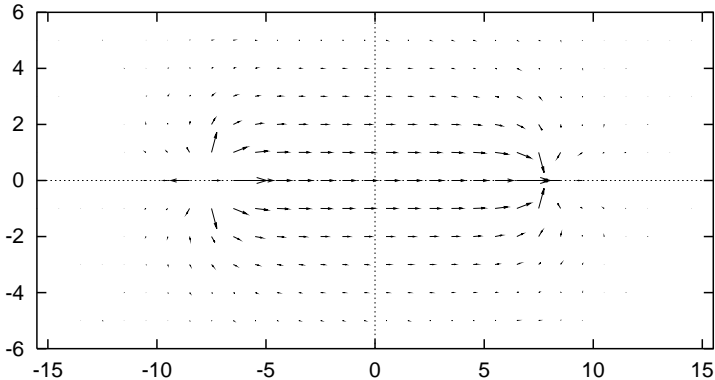}}
\caption{Electric field distribution between two static $SU(2)$
  sources
in the MA projection, in lattice units, $a\approx
0.081$~fm.
The sources are located at the coordinates $(-7.5,0)a$ and
$(7.5,0)a$.} 
\label{fig1}
\end{figure}

Recently, the centre vortex model has enjoyed renewed
attention~\cite{DelDebbio:1998uu}. In this picture,
excitations that can be classified in accord with
the centre group provide the disorder required to
produce an area law of the Wegner-Wilson loop and, therefore,
confinement. One striking
feature is that --- unlike monopole currents --- centre vortices
form gauge invariant
two-dimensional objects, such that in four space-time dimensions,
a linking number between a Wegner-Wilson loop and a centre vortex can
unambiguously be defined, providing a geometric interpretation of
the confinement mechanism~\cite{Chernodub:1998vk}.

We will only discuss the
superconductor picture,
which is based on the concept of electro-magnetic
duality after an Abelian gauge projection that has originally been
proposed by 't~Hooft and Mandelstam~\cite{Mandelstam:1974pi,'tHooft:1975pr}.
The QCD vacuum is thought to behave analogously
to an electrodynamic superconductor
but with the r\^oles of electric and magnetic fields being interchanged:
a condensate of magnetic monopoles expels electric fields
from the vacuum. If one now puts electric charge and anti-charge
into this medium, the electric flux that forms between them
will be squeezed into a thin, eventually
string-like, Abrikosov-Nielsen-Olesen vortex which
results in linear confinement.

In all quantum field theories in which confinement has been proven,
namely in compact $U(1)$ gauge theory,
the Georgi-Glashow model and ${\mathcal N}=2$ SUSY Yang-Mills
theories, this scenario is indeed realised.
However, before one can apply this simple
picture to QCD or $SU(N)$ gluodynamics
one has to identify the relevant dynamical variables: it is not straight
forward to generalise the electro-magnetic duality of a $U(1)$ gauge theory
to $SU(N)$ where gluons carry colour charges. How can one define
electric fields and dual fields in a gauge invariant way?

In the Georgi-Glashow model, the $SO(3)$ gauge symmetry is broken down to
a residual $U(1)$ symmetry as the vacuum expectation value of
the Higgs field becomes finite. It is currently unknown whether QCD
provides a similar mechanism
and various reductions of the $SU(N)$
symmetry have been conjectured.
In this spirit,
it has been proposed~\cite{'tHooft:1981ht}
to identify the monopoles in a $U(1)^{N-1}$ Cartan subgroup
of the $SU(N)$ gauge theory after gauge fixing with respect to the
off-diagonal $SU(N)/U(1)^{N-1}$ degrees of freedom.
After such an Abelian gauge fixing QCD
can be regarded as a theory of interacting
photons, monopoles and matter fields (i.e.\ off-diagonal gluons and quarks).
One might assume that the off-diagonal
gluons do not affect long range interactions.
This conjecture is known as {\em Abelian dominance}~\cite{Ezawa:1982bf}.
Abelian as well as monopole dominance are
qualitatively realised in lattice studies of
$SU(2)$ gauge theory~\cite{Bali:1996dm,Hioki:1991ai}
in the maximally Abelian (MA) gauge projection~\cite{Kronfeld:1987vd},
which 
appears to be a suitable gauge fixing condition.

\begin{figure}
\epsfxsize=10.0truecm
\centerline{\epsfbox{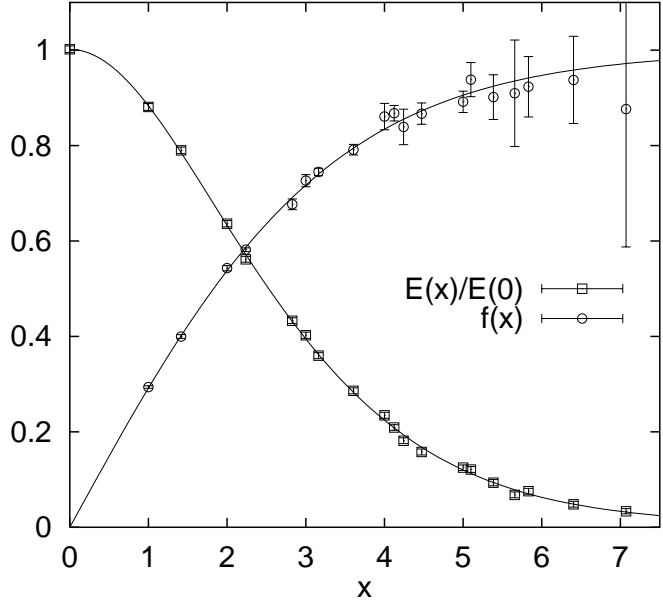}}
\caption{Longitudinal electric field, $E=E_{\parallel}=|{\mathbf E}|$,
and density of superconducting magnetic monopoles, $f$,
in the centre plane between the sources.} 
\label{fig2}
\end{figure}

In Figure~\ref{fig1}, the electric field distribution between
$SU(2)$ quarks, separated by a distance, $r=15a\approx 1.2$~fm,
is displayed~\cite{Bali:1997cp}. This distribution has been obtained
within the MA gauge projection.
The physical 
scale, $a\approx 0.081$~fm, derived from
the value, $\sqrt{\sigma}=440$~MeV, for the string tension,
is intended to serve as
a guide to what one might expect in ``real'' QCD.
Clearly,
an elongated Abrikosov-Nielsen-Olesen vortex forms between the charges.
In Figure~\ref{fig2}, a cross section through the centre plane
of this vortex is displayed. While the electric field
strength decreases with the
distance from the core, the modulus of the dual Ginsburg-Landau (GL)
wave function,
$f$, i.e.\ the density of superconducting magnetic monopoles, decreases
towards the centre of the vortex where superconductivity breaks down.
In this study
the values $\lambda=0.15(2)$~fm and $\xi=0.25(3)$~fm have been
obtained~\cite{Bali:1997cp,Bali:1998de}
for penetration depth and GL coherence length, respectively.
The ratio $\lambda/\xi=0.59(13)<1/\sqrt{2}$ classically
corresponds to a
type I superconductor very close to the border of type II behaviour, i.e.\
QCD flux tubes appear to weakly attract each other.
However, for a final settlement
on which side of the 
Abrikosov limit $SU(2)$ gauge theory lies,
quantum corrections should be considered.
A recent analysis of the same lattice data in terms of
the classical four-dimensional Abelian Higgs model
has resulted in similar conclusions~\cite{Gubarev:1999yp}. 
For a more detailed discussion the reader is referred to
Ref.~\cite{Bali:1998de}.

\section{More static potentials}
\label{potential2}
We will discuss a variety of excitations of the pure gauge vacuum,
such as hybrid potentials, glueballs, gluelumps, potentials between
charges in non-fundamental representations and three-body potentials.
In particular hybrid potentials, whose short range behaviour is related
to the glueball and gluelump spectra, turn out to be relevant for quarkonia
as they give rise to extra states that are not expected from the quark
model. They are also related to relativistic correction terms to
the static potential and determine the validity range of the adiabatic
approximation as we shall see in Sections~\ref{sec:potper} and
\ref{sec:transi}. Prior to discussing hybrid potentials, we shall
introduce hybrid mesons.

\subsection{Hybrid mesons}
At the same time that QCD was invented it has been
noticed~\cite{Fritzsch:1973pi} that the spectrum of
this theory should in principle contain bound states without constituent
quark content, the so-called glueballs, in addition to the
mesons and baryons of the quark model. The question, however, arises
what constitutes the difference between a flavour singlet meson that
contains ``sea'' gluons and a glueball that contains sea quarks.
In general such hypothetically pure states
will mix with each other to yield the observed particle spectrum.
Still, the possibility of gluonic excitations will
result in extra levels within certain mass regions
that would not have been expected from
simplistic
pure constituent quark model arguments.
Moreover, glueballs with exotic, quark
model forbidden, quantum numbers should exist.
While in QCD the difference at least between
non-exotic glueballs and flavour singlet mesons
is somewhat obscured, the quenched approximation contains
``pure'' glueballs and the spectrum of such states may be used as an input
for mixing models~\cite{Lee:1999kv}.

Another non-trivial spectroscopic consequence of the QCD vacuum
structure are so-called hybrid
mesons~\cite{Jaffe:1976fd,Hasenfratz:1978dt,Horn:1978rq}, i.e.\ mesons
with ``constituent'' glue; by considering excitations of the glue,
mesons can acquire exotic quantum numbers too\footnote{The possibility
of mixing with such
exotic hybrids as well as four quark ($q\bar{q}q\bar{q}$) molecules
in fact
renders the notion even of a spin-exotic glueball fuzzy in full QCD.}.
There is a slight problem with the notion
of ``constituent'' glue. Neither the number of gluons is conserved,
nor do they have a non-vanishing rest mass. How then can one define the
difference between ``constituent'' and ``sea'' glue?
Do not all mesons include a gluonic component? Even in the quenched
approximation, where a glueball is a perfectly well defined object,
we cannot easily switch off
``sea'' gluons to identify hybrids.
What a
``hybrid'' is can only be understood 
within certain models like
bag models~\cite{Chodos:1974je,Hasenfratz:1978dt}, 
the strong coupling lattice model~\cite{DeGrand:1975cf,Kogut:1976zr}
or the flux tube model~\cite{Isgur:1983wj} that distinguish
between hybrids and standard quark model states.
Such models offer extensions of the quark model that
help in classifying the observed hadron spectrum and can
guide lattice simulations as well as sum rule calculations.
In Section~\ref{sec:transi} we shall also see that in the framework
of a semi-relativistic expansion the classification can be made
more precise.

In full QCD an operator that is bilinear in the quark fields
with given $J^{PC}$ content and flavour related quantum
numbers isospin, $I$ and $I_3$, strangeness, charm
and beauty will in general couple to all mesonic states
within the given channel: in particular QCD makes no clear distinction between
states with identical quantum numbers such as the flavour singlet states
$\eta$, $\eta'$,
$\eta_c$ and $\eta_b$ and, for instance, pseudo-scalar glueballs:
in the flavour singlet sector
even the notion of a valence quark as opposed to a sea quark
is, strictly speaking, ill defined.
However,
the $\eta_c$ is experimentally clearly
distinct from an $\eta$, containing light
constituent quarks; almost no mixing between the
would-be pure $c\bar{c}$
and the corresponding pure light quark state of the na\"{\i}ve quark model
occurs: while this is not an exact symmetry
the assumption that the number of valence charm quarks
and anti-quarks are separately conserved is a very good approximation of the
physical situation.
In this sense, one can still assign a flavour
to individual constituent quarks.

We intend to create a meson, i.e.\ a state containing a
quark, $q$, and an anti-quark, $\bar{q}'$, with given $J^{PC}$ assignment.
The most general creation operator that is
bilinear in the quark fields is,
\begin{equation}
\sum_{{\mathbf x}}
\bar{q}^{\prime\,\alpha}_{{\mathbf x},\mu}\Gamma_{\mu\nu}
O^{\alpha\beta}_{\mathbf x}[U]
q^{\beta}_{{\mathbf 0},\nu},
\end{equation}
where we have chosen the coordinates such that
the quark is at the origin.
$\alpha,\beta=1,2,3$ are colour and $\mu,\nu=1,\ldots,4$ Dirac
indices.
While $\Gamma$ determines the internal spin symmetry of the state,
the function of the gauge fields, $O$,
generates both, relative angular momentum of the quarks as well as
excitations of the gluonic degrees
of freedom. For trivial $O^{\alpha\beta}_{\mathbf
  x}=\delta_{\alpha\beta}\delta({\mathbf x})$, $\bar{q}\gamma_5 q$ 
creates a pseudo-scalar meson ($J^{PC}=0^{-+}$) and $\bar{q}\gamma_{\mu} q$
a vector ($1^{--}$): in a colour and flavour singlet state
without relative angular momentum of the quarks,
Fermi statistics implies, $P=-1$
and $C=(-1)^s$ (in this special case ${\mathbf J}={\mathbf S}$).
When one allows
$q$ to differ from
$\bar{q}'$, the creation operator is no longer a
charge eigenstate and
scalars $\bar{q}'q$ ($0^+$) as well as axial-vectors
$\bar{q}'\gamma_5\gamma_{\mu}q$ ($1^+$) can easily be created too.

In general, each $O_{\mathbf x}$ is a
combination of gauge connections
between ${\mathbf 0}$ and ${\mathbf x}$.
If one allows $O$ to have a non-trivial spatial distribution,
angular momentum, ${\mathbf L}$, can be introduced. This can be achieved by
the choice, $O_{\mathbf x}=Y_{ll_3}(\theta,\phi)U({\mathbf x})$, where
$U({\mathbf x})$ denotes a Schwinger line connecting
${\mathbf 0}$ with ${\mathbf x}$ and $Y_{ll_3}$ are the
familiar spherical harmonics\footnote{On the lattice the continuum
$O(3)$ rotation group is broken down to the discrete point
group, $O_h$, associated with the cubic symmetry plus inversions,
and the spherical harmonics will be replaced by functions
that are designed to project onto irreducible representations
of the latter subgroup, rather than onto continuum $l$. The
necessary group theory has been worked out in
Refs.~\cite{Berg:1983kp,Michael:1990ry} for glueballs and in
Ref.~\cite{Lacock:1996vy} for hybrid mesons. Since $O_h$ is a subgroup of
$O(3)$, irreducible representations of the point group can be subduced from
a spin representation of the continuous group.
See also Refs.~\cite{Landau:1987qm,Hamermesh:1962xx}.}.
By combining such angular excitations with the
pseudo-scalar creation operator, all $q\bar{q}$ states,
$J^{PC}=0^{-+},1^{+-},2^{-+},\ldots$, can be created while
combination with a vector results in, $J^{PC}=0^{++},1^{--},1^{++},
2^{--},2^{++},\ldots$. Note that $P=(-1)^{l+1},
C=(-1)^{l+s}$.

\begin{figure}
\epsfxsize=10.0truecm
\centerline{\epsfbox{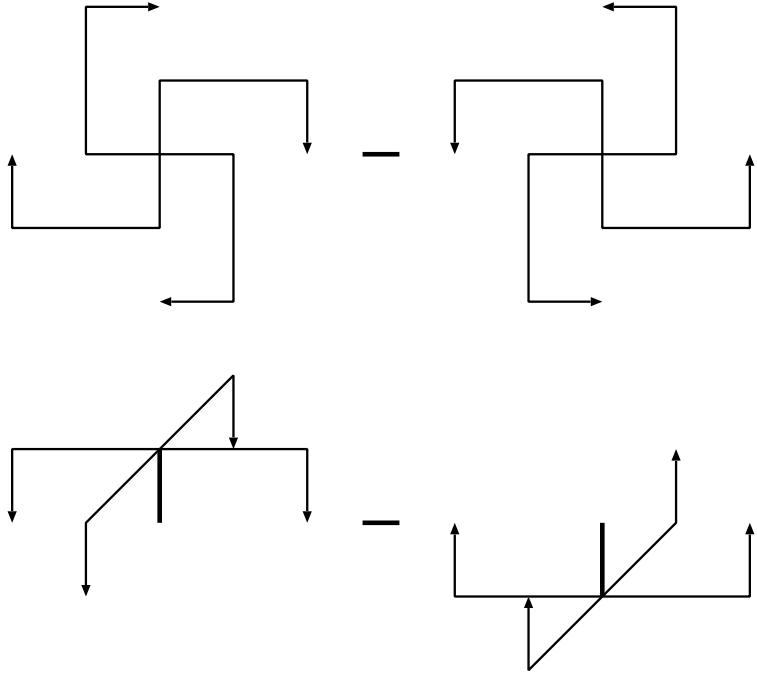}}
\caption{Lattice paths, $O$, with $O_h\otimes C$ quantum numbers,
$T_1^{+-}$ and $T_1^{-+}$, respectively.}
\label{hybcre}
\end{figure}

Let us now investigate the case, $O_{\mathbf x}^{\alpha\beta}=
\delta({\mathbf x})B_{{\mathbf 0},i}^{\alpha\beta}$. The chromo-magnetic
field, ${\mathbf B}=\sum_a {\mathbf B}^aT^a$,
transforms like an octet under gauge transformations and
is traceless. It has the internal quantum numbers of an axial
vector, $1^{+-}$, while the electric field ${\mathbf E}$
is a vector, $1^{--}$. Obviously, the combination,
$\bar{q}^{\alpha}\gamma_5B_i^{\alpha\beta}q^{\beta}$, in which
the quarks couple to a colour octet and interact with the magnetic
gluon, shares
the vector $J^{PC}=1^{--}$ spin assignment with
the quark singlet state,
$\bar{q}^{\alpha}\gamma_iq^{\alpha}$, while both,
$\bar{q}^{\alpha}\gamma_iB_i^{\alpha\beta}q^{\beta}$
and $\bar{q}^{\alpha}\gamma_5q^{\alpha}$ are pseudo-scalars:
It appears plausible to assume
that the colour singlet operators have a better overlap with the physical
ground state while the colour octet operators show an improved coupling
with would-be hybrid excitations.
We finally note that
$\epsilon_{ijk} \bar{q}^{\alpha}\gamma_jB_k^{\alpha\beta}q^{\beta}$
results in a spin-exotic $1^{-+}$ assignment.

In general, one will employ a spatially extended creation operator.
Two examples of such lattice operators, $O$, that incorporate bended
gauge transporters (staples) which result in a non-trivial gluonic state
are depicted in Figure~\ref{hybcre}.
The first one corresponds to a lattice spin content, $T_1^{+-}$,
while the second one is within the $T_1^{-+}$ representation
of $O_h\otimes C$. The lowest lying continuum spin from which
$T_1$ can be subduced is, $l=1$. In combining the above paths with
various possible quark bilinears~\cite{Lacock:1996vy},
the first operator projects onto mesons
with
$J^{PC}=0^{-+},{\mathbf 1^{-+}},1^{--},2^{-+},\ldots$ while the
second operator yields,
$J^{PC}={\mathbf 0^{+-}},1^{+-},1^{++},{\mathbf 2^{+-}},\ldots$.
Spin-exotic states have been indicated in bold. 
The lightest spin exotic mesons
come out to have $J^{PC}=1^{-+}$
in studies of both, quenched QCD and QCD
with two flavours of sea
quarks~\cite{Lacock:1997ny,Bernard:1997ib,Lacock:1998be}.
As a next step mixing effects with possible $\pi f_1$ spin-exotic four-quark
molecules should be considered.

\subsection{Hybrid potentials}
\label{sec:hybrid}
\begin{figure}
\epsfxsize=10.0truecm
\centerline{\epsfbox{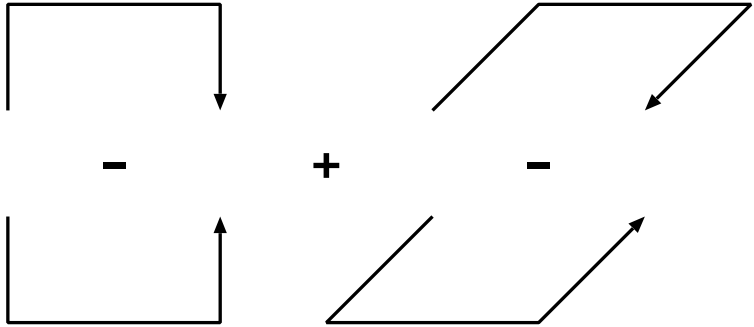}}
\caption{Creation operator for the $E_u$ hybrid potential.}
\label{hybcre2}
\end{figure}
While the distinction between a hybrid meson and an ordinary meson is
not well defined, a hybrid potential with quantum numbers other than
$\Sigma_g^+$ between static colour sources,
separated by a distance ${\mathbf r}$
is clearly distinct from the ground state potential or its
radial excitations of the ground state potential. 
Hybrid potentials can be classified in analogy to
excitations of homonuclear diatomic
molecules~\cite{Landau:1987qm,Hamermesh:1962xx}.
The relevant symmetry group is $D_{\infty h}$ in the continuum
and $D_{4h}$ on a cubic lattice (for on-axis separation of the sources).
An angular momentum $\Lambda_{\hat{\mathbf r}}$ about the
molecular axis can be assigned to the state. In addition,
the state might transform
evenly (gerade, g) or oddly (ungerade, u) under the combined parity of
a charge inversion and a
reflection about the midpoint of the axis, $\eta$.
Finally, reflections with respect to
a plane that includes the axis can be performed.
For $\Lambda=|\Lambda_{\hat{\mathbf r}}|\neq 0$ such reflections just transform
one state within a $\Lambda$-doublet into the
other: $\Lambda_{\hat{\mathbf r}} \rightarrow -\Lambda_{\hat{\mathbf r}}$.
However, for
$\Lambda=0$, the transformation property under this reflection
gives rise to an extra parity index, $\sigma_v$. Conventionally, the
angular momentum
is labelled by a capital Greek letter, $\Lambda=0,1,2,3\ldots
=\Sigma,\Pi,\Delta,\Phi\ldots$. The straight line 
connection transforms in accord
with the representation, $\Sigma_g^+$. In Figure~\ref{hybcre2},
we have visualised a creation operator for
the lattice $D_{4h}$ state, $E_u$, that can be subduced from
the continuum representation, $\Pi_u$. 
The fact that staples pointing into positive and
negative directions are subtracted from each other
reflects the spin one nature of the state.
Note that the combinations of Figure~\ref{hybcre} contain similar
elementary paths.
The necessary group theory and lattice operators have been
worked out in Ref.~\cite{Griffiths:1983ah}.

\begin{figure}
\centerline{\epsfysize=7.7truecm\epsfbox{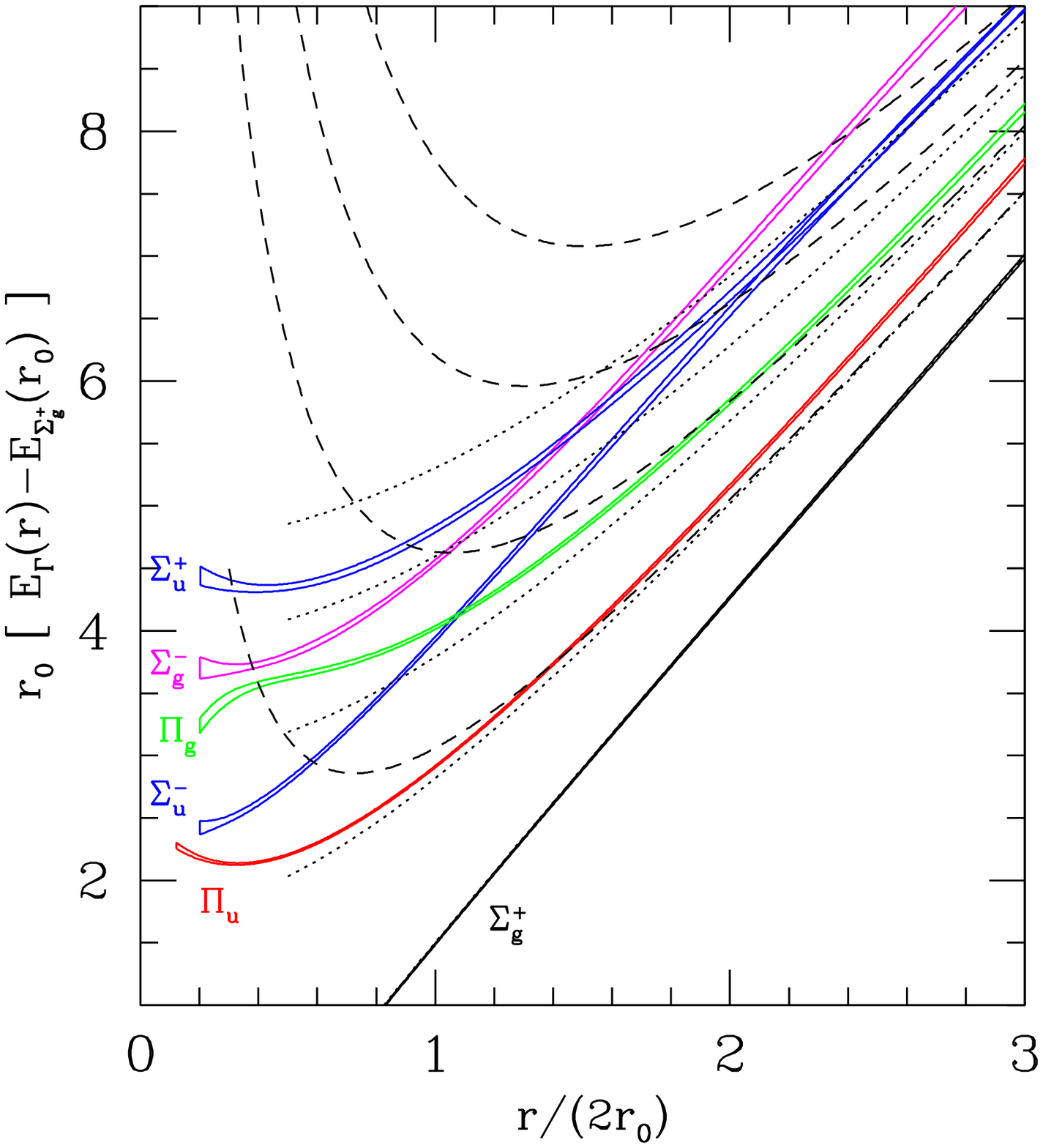}\epsfysize=7.7truecm\epsfbox{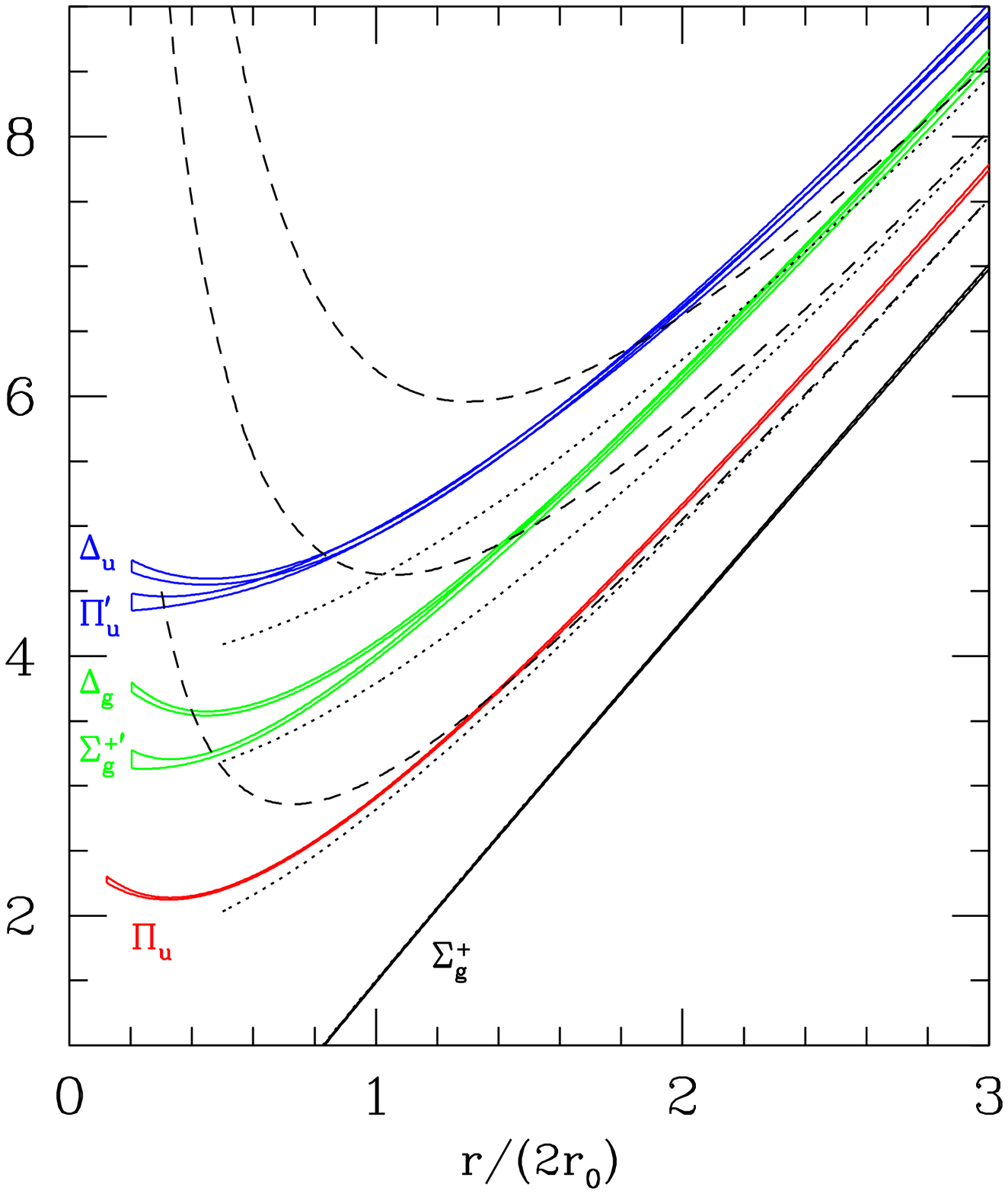}}
\caption{Hybrid excitations of the static $SU(3)$ potential
(from Ref.~\cite{Morningstar:1998da}).}
\label{hybrids}
\end{figure}

Lattice results for hybrid potentials have been obtained
in $SU(2)$~\cite{Griffiths:1983ah,Perantonis:1989uz,Michael:1992az}
and
$SU(3)$~\cite{Campbell:1984fe,Ford:1989as,Perantonis:1990dy,Collins:1997cb,Morningstar:1998da,Juge:1999ie}
gauge theories as well as in QCD with two flavours of sea
quarks~\cite{Bali:1997bj,Bali:2000vr}. For a recent review, see
Ref.~\cite{Michael:1998sm}. Employing the adiabatic and non-relativistic
approximations for heavy quarks, one can estimate possible hybrid
charmonia and bottomonia levels by solving the Schr\"odinger equation
with such hybrid potentials. The only peculiarity is that
the angular momentum, ${\mathbf K}={\mathbf L}+{\mathbf S}_g$,
that couples to the spin of the quarks, ${\mathbf S}={\mathbf S}_1+
{\mathbf S}_2$,
to produce the total spin, ${\mathbf J}={\mathbf K}+{\mathbf S}$,
differs from the angular momentum due to the relative motion of the quarks,
${\mathbf L}$. ${\mathbf S}_g$
denotes the spin of the gluonic flux tube whose projection onto the axis
is, $\Lambda_{\hat{\mathbf r}}={\mathbf S}_g\hat{\mathbf r}$.
Thus, $\langle k\Lambda|{\mathbf S}_g^2|k\Lambda\rangle
\geq \Lambda(\Lambda+1)$ and $k\geq\Lambda$.
Within the leading order Born-Oppenheimer approximation, ${\mathbf K}$ and
$\Lambda$ are conserved, but not ${\mathbf L}$ or
${\mathbf S}_g$.
The centrifugal term, $l(l+1)$ that appears in the radial Schr\"odinger
equation, Eq.~(\ref{eq:radial}), has to be substituted by the
correct factor~\cite{Landau:1987qm},
$\langle {\mathbf L}^2\rangle=
k(k+1)-2\Lambda^2+\langle{\mathbf S}_g^2\rangle$.

Mass estimates of hybrid bottomonia, obtained in this way from
hybrid potentials, can be found in
Refs.~\cite{Perantonis:1990dy,Collins:1997cb,Juge:1999ie}. Like
in the case of light mesons the $1^{-+}$, $0^{+-}$ and $2^{+-}$
quarkonium spin-exotica,
that are governed by the $\Pi_u$ potential
in the adiabatic approximation,
turn out to be the lightest ones.
Within the quenched, non-relativistic and leading order Born-Oppenheimer
approximations
bottomonia hybrids
come out to lie only slightly above the $B\overline{B}$ threshold.
To this order in the semi-relativistic expansion, which does not yet
incorporate
spin sensitive terms, the masses of hybrid
${\mathbf 0^{+-}}$, $0^{-+}$, ${\mathbf 1^{-+}}$,
$1^{--}$, $1^{+-}$, $1^{++}$, ${\mathbf 2^{+-}}$ and  $2^{-+}$ states
are degenerate. It is clear, however, that for
the non-exotic hybrids the use of an excited state potential
within the Born-Oppenheimer approximation is at best dubious.

In Figure~\ref{hybrids},
the spectrum of  hybrid potentials from the most comprehensive
study so far~\cite{Morningstar:1998da} is displayed.
Continuum limit extrapolated lattice results are indicated by pairs of solid
curves while dotted curves correspond to the classical
Nambu-Goto string expectation in four dimensions, Eq.~(\ref{eq:excit}).
Dashed curves indicate $n\pi/r$ gaps, added to the ground state potential,
the leading order contribution of the bosonic string picture.
To guide the eye, the lowest lying states, $\Sigma_g^+$ and $\Pi_u$, are
included into both plots.
Note that a $\Phi_u$ interpretation of the $\Pi_u'$
state cannot be excluded from the lattice data. However, as we shall see
at the end of Section~\ref{sec:gluino}, other evidence speaks in favour
of the $\Pi_u'$ assignment.
Most states are in 
clear disagreement with the simple model expectation up to
distances as big as 3~fm where sub-leading terms of the string picture
are rather small as the differences between dashed and dotted curves show.
While this contrasts the findings of Ref.~\cite{Michael:1994ej}
for closed strings (torelons) and those of Ref.~\cite{Perantonis:1989uz}
for hybrid potentials, investigations of the ground state
flux tubes between
static sources indicate half widths of
about 1~fm~\cite{Bali:1995de,Bali:1998de}.
Thus, although 3~fm is big in comparison with typical hadronic scales,
the amplitude of string fluctuations is still
quite large in relation to the longitudinal extent.
Therefore, in an effective string representation
the possibility of higher dimensional correction terms
to the Nambu-Goto action might have to be considered.

The small distance behaviour exhibits a rich structure too and
some states appear to try to become degenerate.
In particular the change of curvature of the $\Pi_g$
potential at small $r$ appears puzzling.
In the limit, $r\rightarrow 0$, the quarks combine to
an octet or a singlet colour representation.
The
octet channel in which the sources explicitly couple to
gluons should have
relevance for the hybrid potentials that differ from the ground
state by excitations of the gluonic flux tube. One might
therefore assume that the short distance
behaviour~\cite{Foster:1998wu,Brambilla:1999xf}
is determined by the perturbative octet potential $V_o(r)=-1/8V_s(r)$.
This is in agreement with the observation that
the curvature of all potentials (with the exception
of $\Pi_g$) is smaller than
and opposite in sign to the one of the ground state
potential. Note that in the framework of potential NRQCD (pNRQCD),
the hybrid potentials have also been predicted to follow
$V_o$ to leading
order, up to non-perturbative constants~\cite{Brambilla:1999xf}.

We would like to mention that in QED potentials can be classified
in exactly the same way. Nonetheless, in the deconfined phase, that
is realised in nature,
the spectrum of excitations above the ground state Coulomb
potential is continuous since
photons of arbitrary momentum can be emitted.
This is not so in QCD. However, the spectrum of QCD potentials
will become continuous too above glueball pair\footnote{Due to momentum
conservation radiation of a
single glueball is forbidden.} radiation
thresholds or, when allowing for light sea quarks,
meson pair radiation thresholds.

\subsection{Glueballs, glueballinos and hybrid potentials}
\label{sec:gluino}
In the limit, $r\rightarrow 0$, the cylindrical symmetry of
a (hybrid) potential creation operator is enlarged to
that of the full rotational group in three dimensions,
$D_{\infty h}\subset O(3)\otimes C$ (or, on a cubic lattice,
$D_{4h}\subset O_h\otimes C$). Irreducible representations
of the subgroup with spin $\Lambda$ can be subduced from
irreducible representations of the rotational group
with spin $J\geq\Lambda$, as illustrated in Table~\ref{tab:subduce}.
Note that $P=\sigma_v$, $C=\eta\sigma_v$.
Moreover, states can be classified as singlets and octets
in accord with their local gauge transformation properties.
While a singlet state decouples from the temporal
transporters within an $r=0$ ``Wilson loop'',
an octet state couples to a temporal Schwinger line
in the adjoint representation.
In the infinite mass limit, where spin can be neglected, the
temporal transporter can be interpreted as the propagator
of a static gluino, in analogy to fundamental lines representing
a static quark.
Consequently, the octet state is called a glueballino or
gluelump~\cite{Campbell:1986kp,Jorysz:1988qj,Michael:1992nc,Foster:1998wu}
while the singlet state that, neglecting quark pair creation,
contains nothing but glue
represents a glueball.

\begin{figure}
\centerline{\epsfysize=7truecm\epsfbox{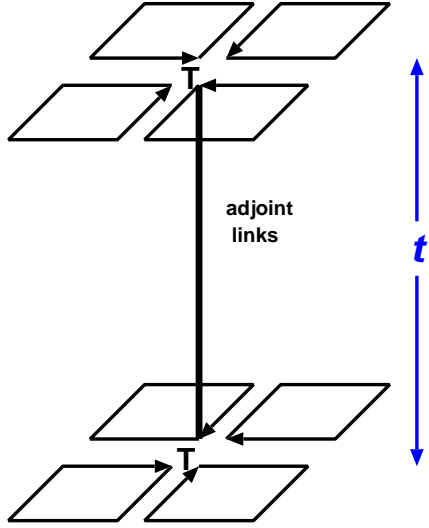}}
\caption{The gluelump correlation function,
Eq.~(\ref{eq:corgl}).}
\label{fig:glcreate}
\end{figure}

Gluelump masses can be extracted from the decay of the
correlation function,
\begin{equation}
\label{eq:corgl}
C(t)=
\frac{1}{2N}
\left\langle H^a_{{\mathbf 0},t}[U^A_{\mathbf 0}(t)]^{ab}
H^b_{{\mathbf 0},0}\right\rangle,
\end{equation}
in Euclidean time.
$U^A_{\mathbf x}(t)$ denotes an adjoint Schwinger line connecting the
point $({\mathbf x},0)$ with $({\mathbf x},t)$ and $H$ is a local
operator in the adjoint representation. The simplest example
is, $H^a\propto B^a_3$, where
$2\,\tr(H^FT^a)=\sum_b H^b 2\,\tr(T^bT^a)=H^a$.
This operator corresponds to an axial-vector, $J^{PC}=1^{+-}$,
from which the $D_{\infty h}$ hybrid potentials,
$\Pi_u$ and $\Sigma_u^-$, can be
subduced in the limit, $r\rightarrow 0$.
The three possible orthogonal choices of the
direction $i$ of $B^A_i$ correspond to the dimensionality, $2J+1$,
of the $J=1$
representation which is identical to the sum of dimensions of the
subduced representations, $\Pi_u$ and $\Sigma_u^-$: $2+1$.
From Eqs.~(\ref{eq:ebconv}), (\ref{eq:fieldstr}) and (\ref{eq:fields2}),
we obtain lattice definitions of magnetic and electric
field strength operators,
\begin{equation}
\label{eq:field1}
gB_{x,i}=\frac{1}{2ia^2}\epsilon_{ijk}\Pi_{x,jk},\quad
gE_{x,i}=\frac{1}{2ia^2}\left(\Pi^t_{x,i}-
\Pi^{t\dagger}_{x,i}\right),
\end{equation}
that approximate the continuum limit up to ${\mathcal O}(a)$ lattice
artefacts [${\mathcal O}(a^2)$ in
$SU(2)$ gauge theory].
In $SU(3)$ gauge theory one would preferably
modify the above definitions,
\begin{equation}
B_{x,i}\rightarrow B'_{x,i}= B_{x,i}-\Tr(B_{x,i}){\mathbf 1},\quad
E_{x,i}\rightarrow E'_{x,i}= E_{x,i}-\Tr(E_{x,i}){\mathbf 1},
\end{equation}
to eliminate order $a$
scaling violations.
\begin{equation}
\Pi_{x,ij}=\frac{1}{4}\left(U_{x,i,j}+U_{x,-i,j}
+U_{x,-i,-j}+U_{x,i,-j}\right)
\end{equation}
denotes a ``clover leaf'' sum of four elementary plaquettes,
Eq.~(\ref{eq:plaq}), while,
\begin{equation}
\label{eq:field3}
\Pi^t_{x+\frac{a}{2}\hat{4},i}=\frac{1}{2}\left(U_{x,i,4}+U_{x,-i,4}\right),
\end{equation}
is defined
at half-integer values of the lattice time, $t/a$.
Note that $\Pi_{x,ij}=\Pi^{\dagger}_{x,ji}$.

\begin{table}
\caption{What $O(3)\otimes C$ representation contains
what $D_{\infty h}$ representations?}
\label{tab:subduce}

\begin{center}
\begin{tabular}{c|c}
$\Lambda^{\sigma_v}_{\eta}$&$J^{PC}$\\\hline
$\Sigma_g^+$&$0^{++},1^{--},2^{++},3^{--},\ldots$\\
$\Sigma_g^-$&$0^{--},1^{++},2^{--},3^{++},\ldots$\\
$\Sigma_u^+$&$0^{+-},1^{-+},2^{+-},3^{-+},\ldots$\\
$\Sigma_u^-$&$0^{-+},1^{+-},2^{-+},3^{+-},\ldots$\\
$\Pi_g$&$1^{++},1^{--},2^{++},2^{--},\ldots$\\
$\Pi_u$&$1^{+-},1^{-+},2^{+-},2^{-+},\ldots$\\
$\Delta_g$&$2^{++},2^{--},3^{++},3^{--},\ldots$\\
$\Delta_u$&$2^{+-},2^{-+},3^{+-},3^{-+},\ldots$
\end{tabular}
\end{center}
\end{table}

The correlation function of Eq.~(\ref{eq:corgl}) is visualised
in Figure~\ref{fig:glcreate}.
$1^{--}$ states can be created
by operators,
$H_i\propto E^A_i$, or by the operators, $H_i\propto\epsilon_{ijk}D^A_jB^A_k$.
The latter operator is local in time
and would preferably be used in
lattice simulations.
The five operators,
$D_i^AB_j^A-\frac{1}{3}\delta_{ij}D_i^AB_j^A$,
couple to $2^{--}$ states etc..
A table containing continuum
creation operators for various quantum numbers can be found
for instance in Ref.~\cite{Brambilla:1999xf}.

The correlation function, Eq.~(\ref{eq:corgl}), can be rewritten
in terms of operators in the fundamental representation
by use of the completeness relation,
\begin{equation}
2\sum_a T^a_{\alpha\beta}T^a_{\gamma\delta}=\delta_{\alpha\delta}
\delta_{\beta\gamma}-
\frac{1}{N}\delta_{\alpha\beta}\delta_{\gamma\delta},
\end{equation}
and the identity $(U^A)^{ab}=2\,\tr(UT^aU^{\dagger}T^b)$.
The result reads,
\begin{equation}
\label{eq:corgl2}
C(t)=
\left\langle \Tr_F \left[H^F_{{\mathbf 0},t}U_{\mathbf 0}(t)
H^{F,\dag}_{{\mathbf 0},0}U^{\dagger}_{\mathbf 0}(t)
\right]\right\rangle,
\end{equation}
where the disconnected part,
$-\left\langle\Tr\, H^F\Tr\, H^{F,\dag}
\right\rangle$,
vanishes due to, $\Tr\, H^F=H^a\Tr\, T^a=0$.
The above correlation function
resembles a ``hybrid'' Wilson loop in the limit, $r\rightarrow 0$.
In this limit, the Wilson loop can be factorised into
singlet and octet components,
\begin{equation}
\langle W_{\Psi}(r,t)\rangle=
c_1e^{-m_{\mbox{\scriptsize gluelump}}(a)t}+
c_2e^{-[m_{\mbox{\scriptsize glueball}}+V_{\Sigma_g^+}(r,a)]t}+\cdots
\quad(r\rightarrow 0),
\end{equation}
where on the lattice, $V_{\Sigma_g^+}(0,a)=\hat{V}_{\Sigma_g^+}(0)
+V_{\mbox{\scriptsize self}}(a)=0$.
At $r\gg a$, $\hat{V}_{\Sigma^+_g}(r)$ will approach the
continuum potential. From the above representation
we expect certain groups of
hybrid potentials to become degenerate with each
other as $r\rightarrow 0$
and to assume the mass of the lightest glueball or gluelump
within the sector of allowed $J^{PC}$ quantum numbers that have
overlap with the hybrid string creation operator, $\Psi^{\dagger}$.

\begin{figure}[thb]
\centerline{\epsfxsize=10truecm\epsfbox{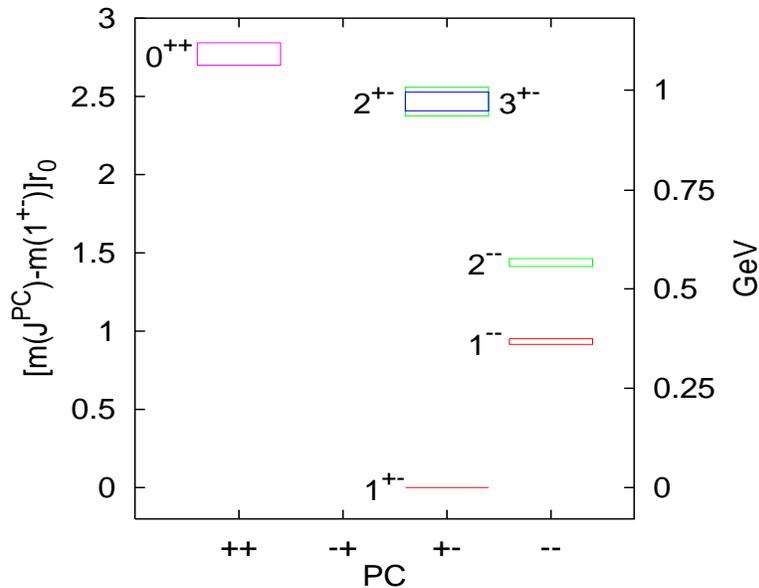}}
\caption{The lowest six $SU(3)$ gluelump states, extrapolated
to the continuum limit~\cite{Foster:1998wu}.}
\label{fig:glspect}
\end{figure}

Like static potentials, any gluelump mass will contain
a finite contribution and a ($J^{PC}$ independent)
contribution due to the self-energy
of the static sources that will diverge in the continuum limit,
\begin{equation}
\label{eq:fac}
m_{\mbox{\scriptsize gluelump}}(a)=m_{\mbox{\scriptsize finite}}
+
m_{\mbox{\scriptsize self}}(a).
\end{equation}
In order to obtain predictions on (hypothetical)
glueballino masses, one has to substitute the (unphysical) self-energy
by the rest mass of the constituent gluino in some appropriate scheme.
Keeping this in mind,
without additional input,
only splittings of glueballino masses with respect to the
ground state can be determined from lattice simulations of
glueballino correlation functions.
In analogy to Eq.~(\ref{eq:self2}), we obtain the
tree level result,
\begin{equation}
\label{eq:glself}
m_{\mbox{\scriptsize self}}(a)=\frac{C_A}{C_F}
\frac{V_{\mbox{\scriptsize self}}(a)}{2}
=\frac{N^2}{N^2-1}V_{\mbox{\scriptsize self}}(a)
>V_{\mbox{\scriptsize self}}(a):
\end{equation}
the self-energy associated with the adjoint static source diverges faster
than that of the two fundamental sources
within the static potential. 
In view of this observation, it is clear that in the continuum limit,
the glueball within Eq.~(\ref{eq:fac})
will be the lighter state and that the level ordering of
the hybrid potentials at zero distance will be determined by the
glueball spectrum. Increasing the separation
a bit such that breaking of the rotational symmetry still remains
small the generalised Wilson loop will contain a contribution
which resembles the correlation
function of a gluelump with a self-energy that is
reduced as the adjoint source becomes smeared out
into two fundamental sources. Depending on the size of gluelump level
splittings in relation with the glueball spectrum, it is therefore quite
possible that
at small distances the spectrum of hybrid potentials will be guided
by the ordering of
gluelump levels before, towards $r\rightarrow 0$, the glueballs
finally take over.
Note that if we allow for sea quarks, flavour singlet
mesons and meson pairs will become lighter than the respective
glueball
levels and determine the short distance behaviour.

\begin{figure}[thb]
\centerline{\epsfxsize=10truecm\epsfbox{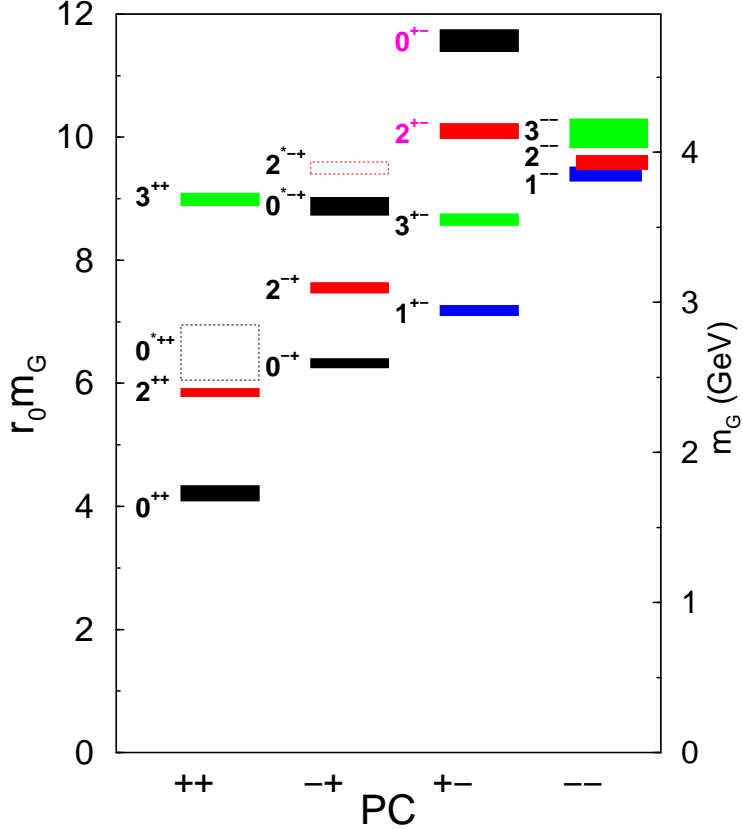}}
\caption{The glueball spectrum of $SU(3)$ gauge theory
(from Ref.~\cite{Morningstar:1999rf}).}
\label{glueballs}
\end{figure}

What ordering of gluelump and glueball states do we expect?
In the MIT bag
model~\cite{DeGrand:1975cf} for instance
the lightest gluonic mode  is the TE mode~\cite{Jaffe:1976fd}
($J^P=1^+$), followed by the TM mode ($J^P=1-$). Hence,
one might expect the axial-vector gluelump to be lighter than
the vector gluelump.
Such concepts have
been generalised~\cite{Jaffe:1986qp} by assuming that
masses of particles increase with the lowest possible
dimension of an operator with which the state in question can be created.
While derivatives, $D$,
have dimension $m$, quark creation operators, $q$,
carry dimension
$m^{3/2}$ and chromo electro-magnetic field operators, $E$ and $B$,
dimension
$m^2$. Only the $1^{+-}$ and the $1^{--}$ gluelumps can be created
by operators of dimension two; all other states require
derivatives or additional fields. Based on this simple picture,
one would expect $1^{+-}$ to be the lighter
state since a magnetic operator ($B$) excites a TE field.
The next state would be $1^{--}$ ($E$), followed by
$2^{--}$ ($DB$) and $2^{+-}$ ($DE$) and eventually
states containing two derivatives ($3^{+-}, 3^{--}$) or
two gluonic fields ($0^{++}, 2^{++}, 0^{-+},2^{-+}$) etc..
Indeed, the gluelump spectrum~\cite{Foster:1998wu}
of Figure~\ref{fig:glspect} seems to follow this qualitative pattern
that has also been predicted in Refs.~\cite{Jorysz:1988qj,Brambilla:1999xf}.

The lowest dimensional operator that can be used to create a glueball
has dimension four. Here, we would expect the lowest states to be made
up from two TE gluons ($BB$), coupling to $0^{++}$ and
$2^{++}$, followed by $0^{-+}$ and $2^{-+}$, containing
a TE plus a TM excitation, followed by $1^{++}$ and $3^{++}$ from
dimension five $BDB$ operators (or an excited $0^{++}$ from two TE modes)
etc.. However, as is revealed by
Figure~\ref{glueballs}~\cite{Morningstar:1999rf},
this simple picture fails after the first 3--4 states:
the $1^{+-}$ is too light.
The strong coupling model~\cite{Kogut:1976zr}, in which one would expect
the ordering $0^{++}$, $2^{++}$, $1^{+-}$ from the perimeter of the minimal
loop required to create the state in question on the lattice,
in contrast, fails to predict
the low mass of the pseudo-scalar glueball.
Of course an abundance of alternative qualitative and quantitative
pictures of the QCD vacuum
exists that result in somewhat different expectations. A detailed
discussion of such models and the underlying assumptions is beyond the scope
of the present article.

From the spectrum of glueballs\footnote{When allowing for light sea quarks,
due to mixing with flavour singlet mesons, the level ordering will be
completely different, starting with the pseudo-scalar
$\Sigma_u^-$.}
and Table~\ref{tab:subduce} we expect
the $\Sigma_g^{+\prime}$ potential to be
separated from the ground state
by a scalar glueball mass $m(0^{++})$ at small distances, followed by
three degenerate potentials $\Sigma_g^{+\prime\prime}$, $\Pi_g$ and $\Delta_g$
which will be separated from the ground state
by $m(2^{++})$, $\Sigma_u^-$ separated by $m(0^{-+})$,
another $m(0^{++\prime})$ triplet of potentials and
a set of $\Pi_u$ and $\Sigma_u^{-\prime}$ states, separated by $m(1^{+-})$.
In the regime of somewhat bigger $r$, which is dominated by
gluelumps, we expect a low, almost degenerate
pair of hybrid potentials, $\Pi_u$ and $\Sigma_u^-$,
corresponding to $1^{+-}$, followed by a $\Pi_g,\Sigma_g'$ ($1^{--}$) pair
and a $\Sigma_g^-,\Pi_g',\Delta_g$ ($2^{--}$) triplet.
Indeed, Figure~\ref{hybrids} reveals
that the $\Sigma_u^-$ and $\Pi_u$ potentials are the lowest
excitations at small $r$, and approaching each other. With $r\rightarrow 0$
we would expect the levels to cross as the value of
$\Sigma_u^-$ will tend towards
the ground state potential plus a
pseudo-scalar glueball mass. Confirmation of this effect, however,
requires lattice spacings that are sufficiently small to yield
a gluelump mass exceeding that of the glueball in question\footnote{
Moreover, some hybrid Wilson loops are constructed in such a way that
one would expect them to better project onto states determined
by the gluelump spectrum rather than
the glueballs which will complicate numerical
studies of the expected level crossings at short distance.}.
All the remaining levels are in complete agreement with the ordering
and degeneracy expectations from the gluelump
considerations too, with the exception of
the $\Delta_u$ that comes out to be somewhat higher than its
degenerate $2^{+-}$ $\Sigma_u^+$ and $\Pi_u'$ partners.
Unfortunately, no data on $\Pi_g'$ exists, which we would have expected
to become degenerate with
$\Delta_g$ and
$\Sigma_g^-$ at small $r$.

Lattice simulations~\cite{Foster:1998wu} reveal that at spacings,
$a^{-1}>2$~GeV, the sum of the scalar glueball mass and
the ground state
potential at the shortest accessible distance, $V_{\Sigma_g^+}(a)$,
becomes smaller than the mass of the lightest ($1^{+-}$) gluelump.
In the framework of effective field theories (see Section~\ref{relativistic})
a cut-off on gluon momenta
is imposed. We conclude that as long as this cut-off
does not exceed about 2~GeV hybrid related interactions
are governed
by the spectrum of gluelumps at short distance
while when allowing for harder gluons, glueball channels will
become increasingly important.

\subsection{Casimir scaling}
\label{sec:casimir}
It is possible to determine the potential between colour sources
not only in the fundamental representation (quarks) but in any
representations of the gauge group. We have already discussed bound states
between static adjoint sources (gluinos) and relativistic gluons above.
Despite the availability of
a wealth of information on fundamental potentials,
only few lattice investigations of forces 
between sources in higher
representations of gauge groups, $SU(N)$, exist.
Most of these studies have been performed in $SU(2)$ gauge theory in
three~\cite{Ambjorn:1984dp,Poulis:1997nn,Stephenson:1999kh,Philipsen:1999wf}
and
four~\cite{Bernard:1982pg,Ambjorn:1984mb,Michael:1985ne,Griffiths:1985ip,Trottier:1995fx,deForcrand:1999kr,Kallio:2000jc}
space-time dimensions.
Zero temperature results for $SU(3)$ can be found in
Refs.~\cite{Campbell:1986kp,Michael:1992nc,Michael:1998sm,Deldar:1998ne,Bali:1999hx,Deldar:1999vi}
while four-dimensional determinations of Polyakov line correlators in
non-fundamental representation
have been performed at finite temperature by
Bernard~\cite{Bernard:1982pg,Bernard:1983my} for $SU(2)$
and Refs.~\cite{Ohta:1986pc,Markum:1988na,Muller:1991xj,Buerger:1993bq}
for $SU(3)$ gauge
theory. 
We have already discussed the $SU(2)$ results of
Refs.~\cite{Stephenson:1999kh,Philipsen:1999wf,deForcrand:1999kr}
and the finite temperature
results in the context of string breaking in Section~\ref{sec:stringb}
and shall focus on $d=4$ $SU(3)$ zero
temperature simulations below.

\begin{table}[hbt]
\caption{Group factors for $SU(3)$. $D$ is the dimension of the
  representation, $(p,q)$ are the weight factors, $z=\exp(2\pi i/3)$,
  and $d_D=C_D/C_F$ denotes ratios of Casimir factors.}
\label{tab:reps}

\begin{center}
\begin{tabular}{c|r|c|c|l}
$D$&$(p,q)$&$z^{p-q}$&$p+q$&$d_D$\\\hline
3&$(1,0)$&$z$&1&1\\
8&$(1,1)$&1&2&2.25\\
6&$(2,0)$&$z^*$&2&2.5\\
$15a$&$(2,1)$&$z$&3&4\\
10&$(3,0)$&1&3&4.5\\
27&$(2,2)$&1&4&6\\
24&$(3,1)$&$z^*$&4&6.25\\
$15s$&$(4,0)$&$z$&4&7
\end{tabular}
\end{center}
\end{table}

For the static potential in the singlet channel in position space, tree level
perturbation theory yields the result\footnote{In fact this relation
turns out to hold to at least two loops
(order $\alpha_s^3$)~\cite{Schroder:1998vy}.},
\begin{equation}
\label{eq:pothigh}
V(r,\mu)=-C_D\frac{\alpha_s}{r}
+d_DV_{\mbox{\scriptsize self}}(\mu),
\end{equation}
in analogy to Eq.~(\ref{eq:vfundpert}).
$D=1,3,6,8,10,\ldots$ labels the representation of $SU(3)$.
$D=3$ corresponds to the fundamental representation, $F$, and
$D=8$ to the adjoint representation, $A$. $C_D$ labels the corresponding
quadratic Casimir operator, $C_D=\mbox{Tr}_DT_a^DT_a^D$, with
the generators $T^D_a$ fulfilling the commutation
relations of Eq.~(\ref{eq:commu}), $[T_a^D,T_b^D]=if_{abc}T_c^D$.
Table~\ref{tab:reps} contains all representations $D$,
the corresponding weights $(p,q)$ for $p+q\leq 4$ and
the ratios of Casimir factors, $d_D=C_D/C_F$. In $SU(3)$ we have $C_F=4/3$,
and $z=\exp(2\pi i/3)$ denotes a third root of 1.

We denote group elements in the fundamental
representation by $U$. 
The traces of $U$ in various representations,
$W_D=\tr U_D$, can easily be worked out,
\begin{eqnarray}
\label{eq:reps}
W_3&=&\tr U,\\
W_8&=&\left(|W_3|^2-1\right),
\\
W_6&=&\frac{1}{2}\left[(\tr U)^2+\tr U^2\right],
\\
W_{15a}&=&\tr U^*\,W_6-\tr U,\\
W_{10}&=&\frac{1}{6}\left[(\tr U)^3+3\,\tr U\,\tr U^2+2\,\tr U^3\right],\\
W_{24}&=&\tr U^*\,W_{10}-W_6,\\
W_{27}&=&|W_6|^2-|W_3|^2,\\
W_{15s}&=&\frac{1}{24}\left[
(\tr U)^4+6(\tr U)^2\tr U^2
+3(\tr U^2)^2\right.\\\nonumber
&+&\left. 8\,\tr U\,\tr U^3+6\,\tr U^4\right].
\label{eq:reps15s}
\end{eqnarray}
Note the difference, $\mbox{Tr}_DU_D=\frac{1}{D}\tr U_D=\frac{1}{D}W_D$:
the normalisation of $W_D$ differs from that of the
Wilson loop of Eq.~(\ref{eq:wl})
by a factor $D$.
Under the replacement, $U\rightarrow z\,U$, $W_D$ transforms like,
$W_D\rightarrow z^{p-q}W_D$. 

In Section~\ref{sec:latt} we have seen that
for distances $r\geq 0.6\,r_0\approx$~0.3~fm the
fundamental potential is well described by the
Cornell parametrisation,
\begin{equation}
\label{eq:param}
V_F(r)=V_{0,F}-\frac{e_F}{r}+\sigma_Fr.
\end{equation}
Perturbation theory [Eq.~(\ref{eq:pothigh})] tells us,
$V_{0,D}\approx d_DV_{0,F}$ and $e_D\approx d_De_F$.
While the fundamental potential in pure gauge theories linearly
rises {\em ad infinitum}, the adjoint potential
will be screened by gluons and,
at sufficiently large distances, decay into two disjoint
gluelumps.
This string breaking has indeed been
confirmed in numerical
studies~\cite{Markum:1988na,Muller:1991xj,Michael:1992nc,Stephenson:1999kh,Philipsen:1999wf,deForcrand:1999kr,Kallio:2000jc}.
Therefore,
strictly speaking, the adjoint string tension is {\em zero}. 
In fact, all charges in higher than the fundamental representation
will be at least partially screened by the background gluons. For instance,
${\mathbf 6}\otimes {\mathbf 8}={\mathbf 2}{\mathbf 4}\oplus
{\mathbf 1}{\mathbf 5}{\mathbf a}^*
\oplus {\mathbf 6}\oplus {\mathbf 3}^*$:
in interacting with the glue, the sextet potential obtains a fundamental
(${\mathbf 3}^*$) component. 
A simple rule, related to the centre of
the group, is reflected in Eqs.~(\ref{eq:reps}) -- (\ref{eq:reps15s}):
wherever $z^{p-q}=1$, the source will be reduced into a singlet component
at large distances while, wherever
$z^{p-q}=z (z^*)$, it will be screened, up to a residual (anti-)\-triplet
component, i.e.\
one can easily read off the asymptotic string tension ($0$ or $\sigma_F$) from
the third column of Table~\ref{tab:reps}, rather than having to multiply
and reduce representations. As a result, the self adjoint representations,
${\mathbf 8}$ and ${\mathbf 2}{\mathbf 7}$, as well as the representation,
${\mathbf 1}{\mathbf 0}$, will be completely
screened while in all other representations with $p+q\leq 4$ a residual
fundamental component survives. The same argument, applied to
$SU(2)$, results in the prediction that all odd-dimensional (bosonic)
representations
are completely screened while all even-dimensional (fermionic)
representations will tend towards the fundamental string tension at large
distances.

\begin{figure}[thb]
\centerline{\epsfysize=12.5truecm\epsffile{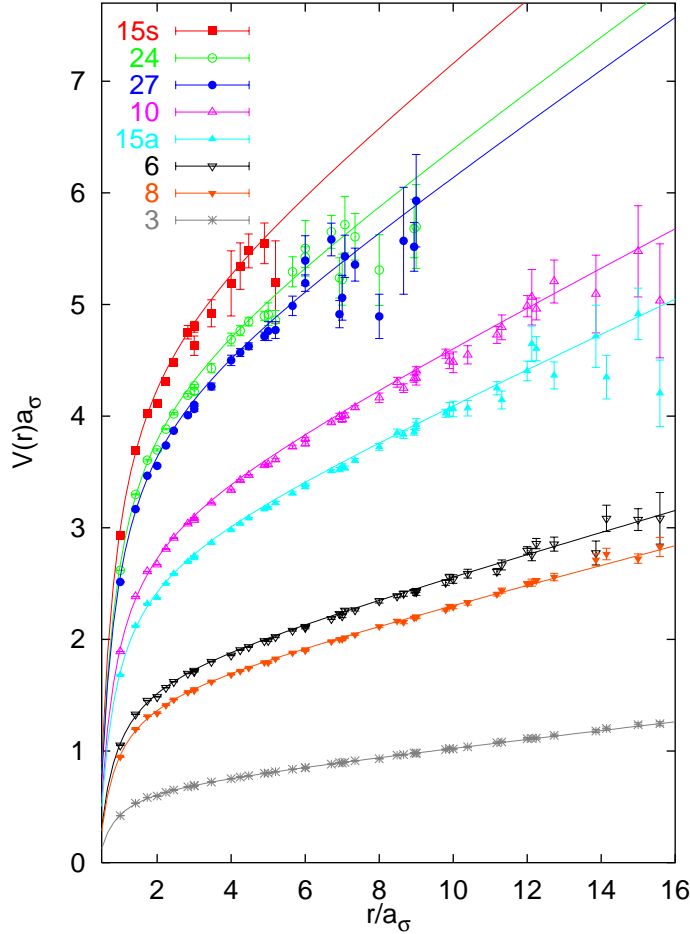}}
\caption{Static potentials between sources in various representations
of $SU(3)$ in lattice units, $a_{\sigma}\approx 0.085$~fm.}
\label{fig:Casimir}
\end{figure}

While the string tension approaches either $0$ or $\sigma_F$ at very 
large distances,
in an intermediate range an approximate linear behaviour is
found~\cite{Michael:1985ne,Griffiths:1985ip,Campbell:1986kp,Muller:1991xj,Michael:1992nc,Michael:1998sm,Deldar:1998ne,Stephenson:1999kh,Philipsen:1999wf,Bali:1999hx}, such that
one might speculate
whether in this region
the Casimir scaling hypothesis~\cite{Ambjorn:1984dp},
$\sigma_D\approx d_D\sigma_F$, that is exact in two-dimensional QCD, holds.
This hypothesis has been challenged by the fact that in all but
two~\cite{Bali:1999hx,Deldar:1999vi}
lattice simulations
the expected Casimir slope is under-estimated. Motivated by this
observation other
models have been suggested, like scaling in proportion with the number
of fundamental flux tubes embedded into the higher representation
vortex\footnote{Some other alternatives have been suggested in the past.
In a bag model calculation, for instance, the result,
$\sigma_D=\sqrt{C_D/C_F}\,\sigma_F$, was obtained~\cite{Johnson:1976sg}.}
[$p+q$ in $SU(3)$]~\cite{Michael:1998sm,Bali:1999hx}, which
happens to coincide with Casimir scaling in the large $N$ limit.
Casimir scaling and flux counting predictions, at least for the lower
dimensional representations, are close to each
other, such that discriminating between them
represents a numerical challenge. 

The latest lattice results for $SU(3)$
gauge theory from Ref.~\cite{Bali:1999hx}
are displayed in Figure~\ref{fig:Casimir} in lattice units, $a_{\sigma}\approx
r_0/6\approx 0.085$~fm. Note, that
the raw lattice data are displayed and no self-energy pieces have been
subtracted. The data have been obtained on
lattices with an anisotropic Wilson action and tiny temporal lattice spacing,
$a_{\tau}^{-1}\approx 4a_{\sigma}^{-1}\approx 24\, r_0^{-1}\approx 9.5$~GeV.
The fundamental potential for distances, $r\geq 0.6~r_0$, has been fitted
to Eq.~(\ref{eq:param}). The expectations on the potentials, $V_D(r)$,
which are displayed in the Figure, correspond to the resulting fit curve,
multiplied by the factors $d_D$.
As one can see, up to distances where the signal disappears into noise or the
string might break, the data are well described by the
Casimir scaling assumption. Since this study has been performed on the
finest
lattice resolution so far, it can very well be that the underestimation
of the Casimir scaling prediction of previous studies is a lattice artefact
that will disappear after an extrapolation to the continuum limit.
Indeed, a close inspection of data for the three-dimensional $SU(2)$
case~\cite{Poulis:1997nn,Stephenson:1999kh,Philipsen:1999wf} shows
the tendency that the Casimir scaling expectation is approached from below
with decreasing lattice spacing.
It is still an open question whether Casimir
scaling only holds approximately or if it is exact for distances
smaller than the corresponding string breaking scales.

\subsection{Three-body potentials}
\label{sec:multi}
Although weak decays turn the phenomenology of hadrons composed
of three or more heavy quarks experimentally unpromising,
predicting properties of such heavy quark systems is the starting point
for understanding multi-quark bound states from QCD and, eventually,
nuclear physics. The first steps into the latter direction
of including light quarks
have been done by Michael and Pennanen who investigate systems
composed of two light and two heavy quarks~\cite{Michael:1999nq} or
the even more ambitious study of the $uuddss$ $H$-dibaryon, containing six
light valence quarks by Wetzorke and collaborators~\cite{Wetzorke:1999rt}.
Forces between three and more static sources are not only interesting
to guide the phenomenology of multi-quark states and to develop and
test the lattice methodology required in this context
but also for model
builders: can multi-quark interactions be understood in terms of
two-body interactions or have genuine three- and many-body effects
to be considered?
Hadronisation models for instance, which intend to explain the
formation of hadronic jets in high energy scattering experiments, crucially
rely on a factorisation hypothesis.

In the past years the Helsinki group has made extensive investigations
of systems composed of four static $SU(2)$
sources~\cite{Green:1993yw,Green:1995cg,Green:1996df,Green:1998nt,Pennanen:1998nu}, where
no distinction between quarks and anti-quarks exists.
These systems, therefore, have the capacity
to approximate both,
meson-meson and baryon-baryon interactions in QCD. The lack of difference
between baryons and mesons is of course a serious limitation
when trying to understand multi-quark interactions. For instance,
unlike in $SU(3)$ where just two different pairings within four quark
systems are possible, in $SU(2)$
three different ways of dividing the system into two
colour singlets are viable: combinatorially, the four quark system
finds its generalisation in $SU(3)$ systems composed of six quarks;
however, geometrically, $q\bar{q}q\bar{q}$ systems come closer.
Here, we will restrict our discussion
to the simpler case of three
quarks in QCD.

\begin{figure}[thb]
\centerline{\epsfxsize=8truecm\epsffile{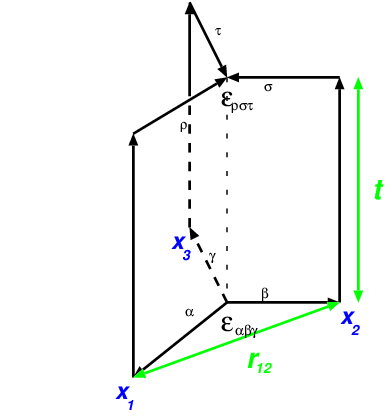}}
\caption{A baryonic Wilson loop.}
\label{fig:bwl}
\end{figure}

In analogy to the standard (mesonic) Wilson loop, in $SU(N)$ 
($N\geq 3$) gauge theories gauge invariant baryonic Wilson loops, $W_{Nq}$,
can be defined. for the case of $SU(3)$
the binding energy of a system
of three static quarks at positions ${\mathbf x}_1$,
${\mathbf x}_2$, and ${\mathbf x}_3$ (baryonic potential, $V_{3q}$)
can be extracted in the limit, $t\rightarrow\infty$. The baryonic
Wilson loop
is composed of three staples, $U^i$, $i=1,2,3$, whose colour indices are
contracted at Euclidean times $0$ and $t$ by completely
antisymmetric tensors,
\begin{equation}
W_{3q}({\mathbf x}_1,{\mathbf x}_2,{\mathbf x_3};t)
=\frac{1}{3!}\epsilon_{\alpha\beta\gamma}\epsilon_{\rho\sigma\tau}
U^1_{\alpha\rho}U^2_{\beta\sigma}U^3_{\gamma\tau}.
\end{equation}
The definition of the staples, $U^i$, is evident from Figure~\ref{fig:bwl}.
The spatial parts of the baryonic loop will in general be composed
of fat or smeared links for enhanced overlap with the physical
ground state. The contraction of the colour indices
can take place at a spatial coordinate at time $t$ that differs from that at
time $0$. Moreover, the contraction points do not necessarily
have to differ from the static quark positions, ${\mathbf x}_i$.
In the limit, ${\mathbf r}_{12}\rightarrow 0$, two quarks combine to
an anti-triplet, ${\mathbf 3}\otimes {\mathbf 3} = {\mathbf 3}^*
\oplus {\mathbf 6}$, that interacts with the remaining quark at position
${\mathbf x}_3$: in this limit the baryonic Wilson loop becomes a
mesonic Wilson loop.

\begin{figure}[thb]
\centerline{\epsfxsize=10truecm\epsffile{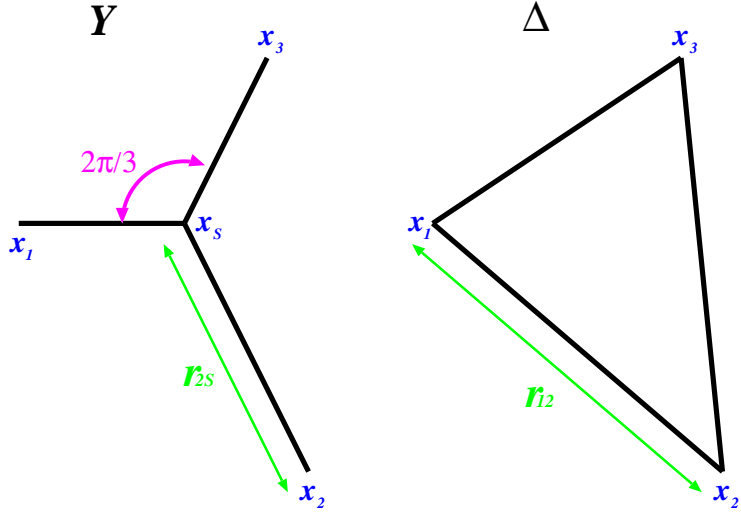}}
\caption{The star geometry with, $r_{Y}=\sum_i r_{iS}$, and the
$\Delta$ geometry with, $r_{\Delta}=\sum_{i>j} r_{ij}$.}
\label{fig:merced}
\end{figure}

The colour factor that accompanies the tree level perturbative result
reads,
\begin{equation}
\frac{1}{3!}\epsilon_{\alpha\beta\gamma}\epsilon_{\alpha\sigma\tau}
T^a_{\beta\sigma}T^a_{\gamma\tau}=-\frac{1}{2}\Tr\left(
T^aT^a\right)=-\frac{C_F}{2}.
\end{equation}
With another minus sign due to the different relative orientation
of the quark lines with respect to a mesonic Wilson loop
we arrive at the tree level\footnote{In fact, this relation
turns out to hold at least to order $\alpha_s^2$ in perturbation theory.}
relation,
\begin{equation}
\label{eq:v3qdelt}
V_{3q}({\mathbf x}_1,{\mathbf x}_2,{\mathbf x}_3)=
\frac{1}{2}\left[V({\mathbf r}_{12})+V({\mathbf r}_{23})+V({\mathbf
    r}_{31})\right],
\end{equation}
between baryonic and mesonic potentials.
For the non-perturbative long range part, two models compete with
each other, which we shall refer to as the star (or $Y$) and
the $\Delta$ laws. The first model originates from strong coupling
and area minimisation
considerations~\cite{Artru:1975zn,Dosch:1976gf,Isgur:1985bm,Brambilla:1994zw,Kalashnikova:1997px}.
The solution of the problem of finding the shortest connecting path
is well known for the case of three points, i.e.\ for a planar geometry;
three straight lines emanating from the quarks
will meet at an angle of $2\pi/3$ at a central Steiner
point, ${\mathbf x}_S$ (Figure~\ref{fig:merced}). Unless one of the
angles within the baryonic triangle exceeds the value
$2\pi/3$, in which case 
a linear geometry will be preferred, the resulting
minimal area configuration resembles a Mercedes
star shape. In this case, we expect the baryonic potential
to be described by parameters extracted from a
Cornell fit, Eq.~(\ref{eq:param}), to the mesonic potential, in the
following way,
\begin{equation}
V_{3q}({\mathbf x}_1,{\mathbf x}_2,{\mathbf x}_3)
\approx \frac{3}{2}V_0-\frac{e}{2}\sum_{i>j}\frac{1}{r_{ij}}
+\sigma\, r_Y,
\end{equation}
where we have approximated the terms associated with the short
range behaviour by the perturbative expectation\footnote{One might, however,
argue that at least at large $r$ the $1/r$ term is related to
Gaussian string fluctuations around the minimal area
string world
sheet~\cite{Luscher:1980fr,Luscher:1981ac} and try a somewhat
different ansatz.}.

\begin{figure}[thb]
\centerline{\epsfxsize=10truecm\epsffile{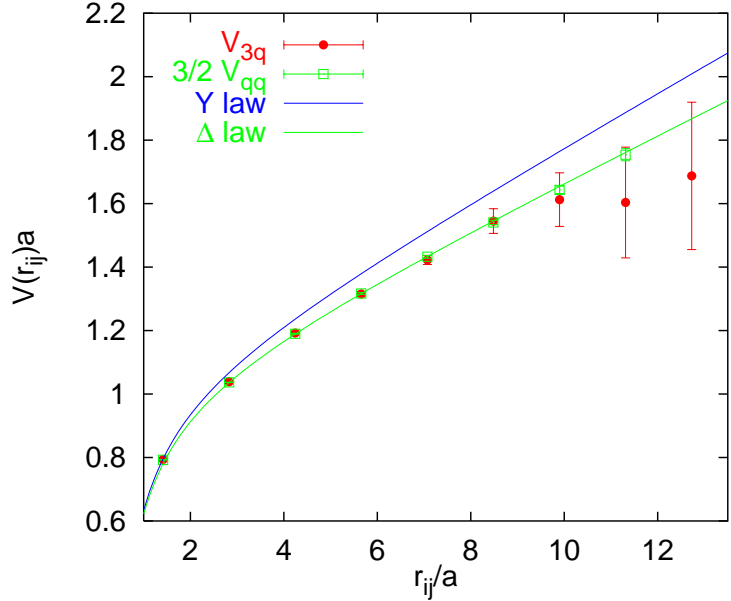}}
\caption{Three quark potential obtained on an equilateral triangle,
$r_{ij}=r_{12}=r_{23}=r_{31}$, in $SU(3)$ gauge theory with
Wilson action at $\beta=6.0$~\cite{Bali:2000ab}.}
\label{fig:3pot}
\end{figure}

The competing contender is the $\Delta$
law~\cite{Cornwall:1977xd,Cornwall:1996xr},
\begin{equation}
V_{3q}({\mathbf x}_1,{\mathbf x}_2,{\mathbf x}_3)
\approx \frac{3}{2}V_0-\frac{e}{2}\sum_{i>j}\frac{1}{r_{ij}}
+\frac{\sigma}{2}r_{\Delta}:
\end{equation}
although obviously, $r_{\Delta}> r_Y$,
in this case each static quark line is shared
by two surfaces which, depending on the underlying
model~\cite{Cornwall:1996xr}, can
result in a pre-factor, $1/2$, to avoid over-counting.
Since $r_{\Delta}/2\leq r_Y$, this might then be the dominant configuration.
Obviously, whenever the three quarks belong to a straight line,
the two models yield identical predictions. The biggest difference is
encountered for the case of an equilateral triangle where
the predictions disagree by about 15~\%,
\begin{equation}
r_Y=\sqrt{3}\,r_{ij}=\frac{2}{\sqrt{3}}\frac{r_{\Delta}}{2}.
\end{equation}
Only very few lattice results~\cite{Sommer:1986da,Flower:1986ex,Thacker:1987aq}
on baryonic potential existed
so far, with statistical errors too big to rule out either possibility.
In an as yet unpublished
study~\cite{Bali:2000ab}, however, clear evidence in support of the
$\Delta$ law has been found. The result for an equilateral
triangle in lattice units is displayed
in Figure~\ref{fig:3pot} for $SU(3)$  gauge theory
with Wilson action at $\beta=6.0$ ($a\approx 0.094$~fm): the data
perfectly agree with the simple expectation,
$V_{3q}=(3/2) V_{q\bar{q}}(r_{12})$,
which, like the
ratios between potentials between charges in different
representations of the gauge group presented above,
happens to coincide with
tree level (and higher order) perturbation theory,
Eq.~(\ref{eq:v3qdelt}). 

It is interesting to notice that phenomenological fits
of the baryon spectrum, for instance
in the framework of relativised quark models~\cite{Godfrey:1985xj}
in which the $Y$ law is assumed, yield a string tension that is reduced
by about 20~\%~\cite{Capstick:1986bm},
compared to the corresponding mesonic result.
This is fairly consistent with the lattice results for the
configuration of the equilateral triangle presented above.
We conclude that while the agreement of the lattice data with the
$\Delta$ law is appealing the question
whether the $\Delta$ law holds or an approximate $Y$ law with
a reduced string tension
is satisfied cannot conclusively be answered until
additional source geometries have been investigated.

\section{Relativistic corrections}
\label{relativistic}
In this Section we attempt to bridge the gap between QCD dynamics of
heavy quark bound states and potential models.
We will sketch the derivation of a quantum mechanical Hamiltonian,
containing the static potential as well as semi-relativistic correction
terms. To leading order this has been pioneered by
Wilson,
Brown and Weisberger~\cite{Wilson:1974sk,Brown:1979ya}
some 20 years ago. As soon as the approach was
generalised to higher
orders~\cite{Eichten:1979pu,Eichten:1981mw,Gromes:1984pm} in
the inverse heavy quark mass, $m^{-1}$, or,
better, relative heavy quark velocity, $v$, certain inconsistencies appeared
between the non-perturbatively derived general form of the interaction
and a direct perturbative evaluation~\cite{Pantaleone:1986uf}
of the potential between two heavy
quark sources at order $\alpha_s^2/m^2$.

A lot of progress in the understanding of effective theories, in particular
in the matching of low energy theories to QCD
has been achieved since then, and the problem is now
understood~\cite{Chen:1995dg}
and removed. Motivated by these developments, we choose
to start our discussion from non-relativistic QCD (NRQCD)
in the continuum and on the lattice, before we address
relativistic corrections to the heavy quark potential.
Special emphasis is put on the matching problem.
We shall also see that the validity of the adiabatic approximation
is very closely tied to that of the non-relativistic expansion. Finally,
lattice results on the heavy quark interaction will be presented.

\subsection{NRQCD}
\label{sec:nrqcd}
\subsubsection{The problem}
We wish to consider mesonic bound states that contain two heavy quarks,
namely the $J/\psi$, $\Upsilon$ and $B_c$ quarkonia families.
Typical binding energies, $\overline{\Lambda}$,
turn out to be a few hundred
MeV, similar to
systems that are entirely composed of light constituent quarks.
The quark mass, $m$, however, is much larger.
This difference in scales results in complications when
evaluating physical properties. In a standard lattice computation
for instance
one has to work at lattice cut-offs, $a\ll m^{-1}$, in
order to resolve the heavy quark
while at the same time the box size has to be kept
sufficiently large to resolve the scales that are relevant for the
dynamics of the bound state like the binding energy,
$L_{\sigma}a\gg \overline{\Lambda}^{-1}$.
This results in a prohibitively large number of
lattice sites that seems physically unnecessary
since the scale, $m\gg\overline{\Lambda}$,
appears to be rather irrelevant for
the quarkonia level splittings (cf.~Section~\ref{sec:quark}).
Indeed, closer inspection shows that
only the temporal lattice spacing, $a_{\tau}\ll m^{-1}$, is limited by the
quark mass. The computational effort becomes tolerable
when an anisotropy,
$a_{\tau}/a_{\sigma}\approx \overline{\Lambda}/m$, is introduced.

The two scale
problem can even be turned into a virtue within an effective field theory
formalism. The strategy would be to integrate out the
ultra-violet behaviour at scales, $\mu\approx m$,
into local Wilson coefficients of an effective low
energy action that encodes the information relevant for bound
state properties. Heavy quark effective
theory~\cite{Isgur:1989vq,Eichten:1990zv,Georgi:1990um} (HQET) for instance
is a very effective framework for the calculation of properties
of systems containing one heavy quark.
The strategy is to write down an effective action
that approximates QCD to a given power $\nu$ in $m^{-1}$.
In general, the effective Lagrangian will then contain all operators
of dimensions smaller than or equal to that of $m^{\nu+4}$.
The coefficients that accompany these terms
can be determined by matching
on-shell Green functions, calculated in the effective theory,
to those calculated in QCD, in the ultra-violet.
This can be done for example in perturbation theory which is supposed to be
applicable as long as, $\mu\gg\Lambda_{QCD}$.
The tree level matching coefficients can be obtained
by formally expanding the Dirac Lagrangian in terms of $m^{-1}$.

Although NRQCD~\cite{Caswell:1986ui}, that applies to
systems containing two heavy quarks, is somewhat more involved it has in fact
been formulated earlier than HQET. The power counting scheme required
for quarkonia differs from the one used in heavy-light systems.
This is related to the fact that
in the lowest order HQET Lagrangian, heavy quarks with non-vanishing
relative velocity decouple from each other. In order to allow
for interactions, a kinetic term, $p^2/2m$, is required that causes changes
of the relative quark velocity, $v$. Therefore, the lowest order
effective Lagrangian depends explicitly on the quark mass, $m$,
in a way that cannot be absorbed into simple field redefinitions:
the HQET power counting is obscured and
a different expansion parameter is required.

As an alternative it has been suggested to expand the effective Lagrangian
in terms of the quark velocity, $v$. One consequence is that in NRQCD
a hierarchy of scales, $m\gg mv\gg mv^2\gg\ldots$, is introduced.
The binding energy, $\overline{\Lambda}$, is of the order of
the ultra-soft scale, $mv^2$, while the typical three-momenta
exchanged, $mv$, are soft.
The hard scale, $m$, is integrated out into matching coefficients,
$c_i(\mu/m,\alpha_s)$,
at a scale $m\geq\mu\geq mv$. With the hierarchy of scales
comes the possibility of a hierarchy of effective theories: after
integrating out the soft scale, $mv$, another effective field theory,
potential NRQCD (pNRQCD), can be
formulated~\cite{Pineda:1997bj,Beneke:1999ff,Brambilla:1999xf}.

\begin{figure}[thb]
\centerline{\epsfxsize=10truecm\epsffile{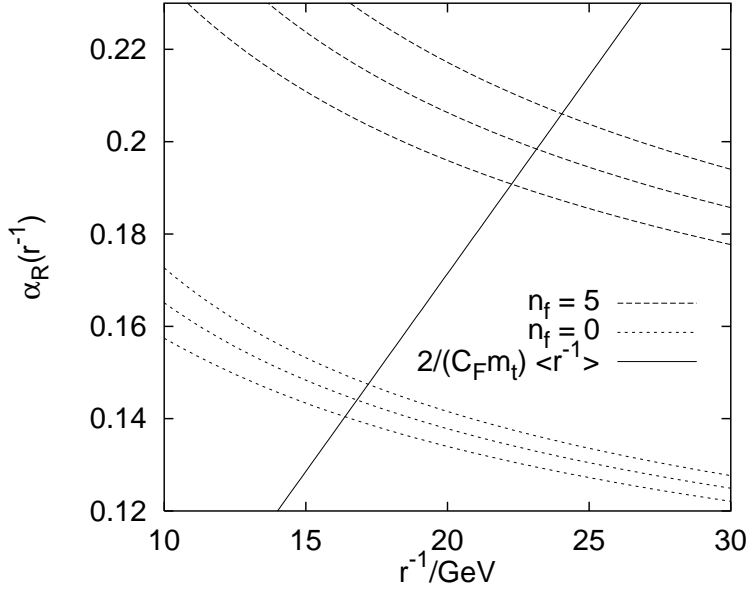}}
\caption{Determination of the relevant coupling for
hypothetical $t\bar{t}$ bound states.}
\label{fig:alphatop}
\end{figure}

\begin{table}
\caption{Hard, soft and ultra-soft scale estimates.}
\label{tab:scales}

\begin{center}
\begin{tabular}{c|c|c|c|c}
&$J/\psi$&$\Upsilon$&$t\bar{t}$&$e^+e^-$\\\hline
$m$&1.4~GeV&4.7~GeV&175~GeV&511~keV\\
$mv$&0.7~GeV&1.3~GeV&45~GeV&3.7~keV\\
$mv^2$&0.4~GeV&0.4~GeV&12~GeV&0.027~keV\\
$v$&0.5&0.29&0.26&0.007
\end{tabular}
\end{center}
\end{table}

In Table~\ref{tab:scales}, estimates of the three scales for
the charmonium, bottomonium and (unstable and therefore hypothetical)
toponium ground states from
a potential calculation are listed. For comparison, we include
the corresponding estimates for positronium.
The top quark is
so heavy that for our estimate only a Coulomb like potential
needs to be considered.
In this case, $V(r)=-C_F\alpha_R/r$, and one easily obtains from the
virial theorem [Eq.~(\ref{eq:virial}) with $\nu=1$], 
$\langle V\rangle = -2\langle T\rangle=2E_1$, using $E_1=-mC_F^2
\alpha_R^2/4$,
\begin{equation}
\label{eq:vir2}
\frac{\langle v^2\rangle}{c^2}=C_F^2\alpha_R^2,
\quad\frac{\langle r^{-1}\rangle}{c}=\frac{1}{2}m C_F\alpha_R.
\end{equation}
In Figure~\ref{fig:alphatop}, $2/(C_Fm_t)\langle r^{-1}\rangle$
and $\alpha_R$ are
plotted as functions of $r$.
The upper $n_f=5$
curve has been
obtained from the input value, $\alpha_{\overline{MS}}(m_Z)=0.1214(31)$,
from $e^+e^-$ experiments at LEP and SLAC~\cite{Caso:1998tx},
by use of Eqs.~(\ref{eq:vq})
and (\ref{alphaR}). The quenched, $n_f=0$, estimate
has been calculated by use of
the lattice result of Ref.~\cite{Capitani:1998mq}, displayed in
Eq.~(\ref{eq:lms0}).
The intersects correspond to the values,
$\alpha_R^{(0)}(17\,\mbox{GeV})\approx 0.145$,
and,
$\alpha^{(5)}_R(23\,\mbox{GeV})\approx 0.20$.
The estimates
quoted in Table~\ref{tab:scales} were obtained
using the latter ($n_f=5$)
result. 
The matching coefficients between QCD and NRQCD are calculable in
perturbation theory as long as $m\gg\Lambda_{QCD}$ while perturbative
matching between NRQCD and pNRQCD can be performed
whenever $mv\gg\Lambda_{QCD}$.
However, given the numbers in the Table, a reliable
perturbative determination
of the matching coefficients for charmonia states
appears to be doubtful while
for top quarks even perturbative
pNRQCD should be applicable, up to power
corrections~\cite{Gubarev:1999ie,Bodwin:1998mn,Caswell:1986ui}.

\subsubsection{The NRQCD Lagrangian and power counting}
In order to derive an effective field theory
that includes a kinetic term in its leading order Lagrangian,
we introduce a second dimension $[v]=c$, in addition to
the mass, and expand the Lagrangian formally in terms
of~\cite{Grinstein:1998xb,Grinstein:1998gv} $1/c$.
As a result, time is measured in different units than space:
$t=x_4/c$, $\partial_t=c\partial_4$
and $D_t=cD_4=\partial_t-iA_4$. The spatial covariant
derivative reads, $D_i=\partial_i-\frac{i}{c}A_i$.
The kinetic term, $\tr F_{\mu\nu}F_{\mu\nu}$, has dimension\footnote{
Note that $[t]=1/(mc^2), [x_i]=1/(mc)$ and,
$gE_i=-i[D_i,D_4], gB_i=-\frac{i}{2}\epsilon_{ijk}c\,[D_j,D_k]$.}
$m^4c^6$. We define,
\begin{equation}
\label{eq:sdef}
S=\frac{1}{c}\int\!d^3x\,\int\!dt\,{\mathcal L},
\end{equation}
where,
\begin{equation}
\label{eq:Ldef}
{\mathcal L} = \frac{1}{2g^2}\tr F_{\mu\nu}F_{\mu\nu}
+\bar{q}[{\mathbf \gamma}\cdot{\mathbf D}c
+\gamma_4D_t+mc^2]q.
\end{equation}
In order to make the leading order NRQCD Lagrange density independent
of $c$, we have rescaled ${\mathcal L}$ by an overall
factor, $c$ [Eq.~(\ref{eq:sdef})],
which is compensated in the fermionic part by rescaling
the fermion fields, $q$ and $\bar{q}$, by factors, $\sqrt{c}$.

At tree level a classical derivation of the NRQCD Lagrangian
from the Dirac
Lagrangian is possible by means of the Foldy-Wouthuysen-Tani (FWT)
rotation~\cite{Foldy:1950wa,Itzykson:1980xx}: one starts from the Dirac basis
in which $\gamma_4$ is diagonal and then, order by order in
${\mathbf p}/(mc)$, the fermion kernel is iteratively
block diagonalised into separate non-interacting quark and anti-quark
sectors.
The first order transformation takes the form,
\begin{equation}
q\rightarrow\exp\left(\frac{{\mathbf \gamma}\cdot{\mathbf p}}{2mc}\right)q,
\end{equation}
which happens to be the correct expression
to all orders in the free field case.
Subsequently, the rest mass can be removed
by rescaling the (anti-)fermion fields with the factor, $\exp(\pm mc^2t)$.
For details see e.g.\ Ref.~\cite{Lepage:1992tx}.
We choose the decomposition,
\begin{equation}
q=\left(\begin{array}{c}\psi\\\chi\end{array}\right),
\end{equation}
of the Dirac spinor in the FWT basis
into quark and anti-quark Pauli spinors.

The resulting effective
Lagrangian to order $c^{-2}$ for the two-particle sector reads,
\begin{equation}
\label{eq:lagnrqcd}
{\mathcal L}_{\psi}+{\mathcal L}_{\chi}=-
\psi^{\dagger}\left[D_4+H_{\psi}\right]\psi
-
\chi^{\dagger}[D_4-H^{\dagger}_{\chi}]\chi
\end{equation}
with
\begin{eqnarray}
\label{eq:ef1quark}
H_{\psi}&=&-\frac{{\mathbf D}^2}{2m_{\psi}}-
c_F\frac{g{\mathbf\sigma}\cdot{\mathbf B}}{2m_{\psi}c}
-\frac{({\mathbf D}^2)^2}{8m_{\psi}^3c^2}\\\nonumber
&-&ic_D\frac{g({\mathbf D}\cdot{\mathbf E}
-{\mathbf E}\cdot{\mathbf D})}{8m_{\psi}^2c^2}
+c_S\frac{g{\mathbf\sigma}\cdot
({\mathbf D}\wedge {\mathbf E}
-{\mathbf E}\wedge {\mathbf D})}{8m_{\psi}^2c^2}
+\frac{1}{c^3}{\mathcal O}(c^3).
\end{eqnarray}
Note that we are using the conventions of Eq.~(\ref{eq:ebconv}) to
relate the field strength tensor and electric/magnetic fields.
The well known Fermi term of the Pauli equation that is responsible for the
hyperfine splittings in atomic physics is accompanied by a matching
coefficient, $c_F$, the Darwin term by $c_D$ and the
spin-orbit (Thomas) term by $c_S$. The normalisation within
Eq.~(\ref{eq:ef1quark}) is
such that in the free field limit,
$c_F=c_D=c_S=1$. The kinetic term defines the mass,
$m_{\psi}=m+{\delta m}$, where $\delta m$ accounts for the
difference in the self-energy subtractions between effective
theory and QCD. In the classical limit, $\delta m =0$.
The coefficient of the relativistic correction to the kinetic energy
is fixed by Lorentz [or, in Euclidean space, $O(4)$] symmetry.
If this is broken, as for example on the lattice, it can also obtain
a non-trivial value~\cite{Morningstar:1994qe}.

\begin{table}
\caption{NRQCD operators and dimension.}
\label{tab:dim}

\begin{center}
\begin{tabular}{c|c|c|l}
operator&dimension&effect&description\\\hline
${\mathcal L}$&$m^4c^6$&$m^4v^5$&Lagrange density\\
$\psi$&$m^{3/2}c^2$&$(mv)^{3/2}$&quark creation operator\\
$\chi$&$m^{3/2}c^2$&$(mv)^{3/2}$&anti-quark creation operator\\
$D_t$&$mc^2$&$mv^2$&covariant time derivative\\
${\mathbf D}$&$mc$&$mv$&covariant spatial derivative\\
$g{\mathbf E}$&$m^2c^3$&$m^2v^3$&chromo-electric field\\
$g{\mathbf B}$&$m^2c^3$&$m^2v^4$&chromo-magnetic field\\
$g^2$&1&$v$&strong coupling ``constant''
\end{tabular}
\end{center}
\end{table}

In Table~\ref{tab:dim}, we list the na\"{\i}ve dimensions of various operators
as they result from the above equation. 
By considering the fermionic part of the action
and writing down the equations
of motion in Coulomb gauge, phenomenological scaling laws
have been derived~\cite{Thacker:1991bm} that somewhat differ
from the power counting
in $c$ but should closely
resemble the relative numerical importance
of a given operator with respect to the resulting quarkonium spectrum\footnote{
To add to the confusion, yet another
set of counting rules that arises from a multipole expansion
of the gluon field has recently been suggested~\cite{Luke:1999kz}.}.
According to the analysis of Ref.~\cite{Thacker:1991bm},
the coupling $g^2$ is expected to
scale in proportion to the velocity [which is the
case for a Coulomb potential, Eq.~(\ref{eq:vir2})], making the rescaling
of the quark fields by factors, $\sqrt{c}$,
in Eq.~(\ref{eq:Ldef}) superficial.
The results of Ref.~\cite{Thacker:1991bm}
are included in the third column of
the Table. In what follows, we will distinguish between ``dimension''
in terms of $c$ of a given operator and the 
``effect'' on energy levels of a bound state in terms of $v$.
The above order $c^{-2}$ Lagrangian, without radiative corrections
($\alpha_s\propto v$), corresponds to order $v^4$ in terms of
these (original) NRQCD power counting rules.

We find
the $v$ power classification of Ref.~\cite{Thacker:1991bm}
useful for phenomenological purposes.
However, we believe that for a consistent construction of an effective field
theory, formal expansion in terms of a dimensionful parameter, $c$, is
more illuminating~\cite{Grinstein:1998xb}.
As long as it is not possible to cleanly
disentangle soft ($mv$) and ultra-soft ($mv^2$) degrees of freedom,
each operator will receive additional contributions that are
sub-leading in $v$; the velocity scaling arguments
are not exact but have to be interpreted at leading order. Moreover,
the effective $v$ will depend on both, the state
under consideration and the operator in question.
While an expansion in terms of $c^{-1}$ can be
performed on the quark-gluon level,
the velocity size classification is based on bound state properties.
For instance, we find the Fermi term, ${\mathbf \sigma}\cdot {\mathbf B}$,
of Eq.~(\ref{eq:ef1quark})
to be of order $c^{-1}$ while according to the velocity classification it
has relative size $v^2$. Within a bound state, the spin variable
within this term has
to be saturated by a second spin, such that its leading
effect on the energy levels is suppressed by an additional power of
$c^{-1}$, in accord with the velocity size counting.
Taking such bound state arguments into account, it appears
favourable to always truncate
the Lagrangian at an even order in $c^{-1}$.
The Darwin and Thomas terms have the same dimension
($c^{-2}$ and $v^2$) in both counting schemes.
In what follows, we will assign the orders $1$ and $v^2$ in the
$c^{-1}$ and $v$ power counting schemes, respectively, to the lowest
order Lagrangian.

For completeness of the effective Lagrangian
we have to consider the two-particle sector:
\begin{eqnarray}
{\mathcal L}_{\psi\chi}
&=&\frac{d_{ss}}{m_{\psi}m_{\chi}c^2}\psi^{\dagger}\psi\chi^{\dagger}\chi
+\frac{d_{sv}}{m_{\psi}m_{\chi}c^2}\psi^{\dagger}{\mathbf \sigma}
\psi\chi^{\dagger}{\mathbf \sigma}\chi\label{eq:lanr}\\\nonumber
&+&\frac{d_{vs}}{m_{\psi}m_{\chi}c^2}\psi^{\dagger}T^a\psi\chi^{\dagger}T^a\chi
+\frac{d_{vv}}{m_{\psi}m_{\chi}c^2}\psi^{\dagger}T^a{\mathbf \sigma}\psi
\chi^{\dagger}T^a{\mathbf\sigma}\chi+\frac{1}{c^3}{\mathcal O}(c^7).
\end{eqnarray}
Note that by
means of a Fiertz transformation an alternative basis can be chosen,
\begin{eqnarray}
{\mathcal L}'_{\psi\chi}
&=&\frac{d^c_{ss}}{m_{\psi}m_{\chi}c^2}\psi^{\dagger}\chi\chi^{\dagger}\psi
+\frac{d^c_{sv}}{m_{\psi}m_{\chi}c^2}\psi^{\dagger}{\mathbf \sigma}
\chi\chi^{\dagger}{\mathbf \sigma}\psi\\\nonumber
&+&\frac{d^c_{vs}}{m_{\psi}m_{\chi}c^2}
\psi^{\dagger}T^a\chi\chi^{\dagger}T^a\psi
+\frac{d^c_{vv}}{m_{\psi}m_{\chi}c^2}\psi^{\dagger}T^a{\mathbf \sigma}\chi
\chi^{\dagger}T^a{\mathbf\sigma}\psi+\frac{1}{c^3}{\mathcal O}(c^7).
\end{eqnarray}
The coefficients are related to each other~\cite{Pineda:1998kj},
\begin{eqnarray}\label{eq:rela1}
d_{ss}&=&\frac{1}{2N}\left[-d_{ss}^c-3d_{sv}^c-C_Fd_{vs}^c-3C_Fd_{v}^c\right],\\
d_{sv}&=&\frac{1}{2N}\left[-d_{ss}^c+d_{sv}^c-C_Fd_{vs}^c+C_Fd_{v}^c\right],\\
d_{vs}&=&\frac{1}{2N}\left[-2Nd_{ss}^c-6Nd_{sv}^c+d_{vs}^c+3d_{v}^c\right],\\
d_{vv}&=&\frac{1}{2N}\left[-2Nd_{ss}^c+2Nd_{sv}^c+d_{vs}^c-d_{v}^c\right].
\label{eq:rela4}
\end{eqnarray}
For $m_{\psi}\neq m_{\chi}$ the tree level coefficients are all zero
while in the equal mass case the coefficients in the standard basis
are proportional to $\alpha_s$, due to an annihilation diagram
that results in, $d_{vv}^c=-\pi\alpha_s+\cdots$.
According to the $v$ power counting rules the first two terms are
only suppressed by a factor $v$, relative to the
leading kinetic term. However, the matching
coefficients guarantee a suppression by an additional factor of
$\alpha_s\simeq v$, in the equal mass case and of $\alpha_s^2\simeq v^2$
for $m_{\psi}\neq m_{\chi}$.
Therefore, the effect of the
contact terms is of combined orders $v^4$ and $v^5$, for equal
and non-equal quark flavours, respectively.

Finally, we consider corrections to the gauge action~\cite{Manohar:1997qy}, 
\begin{eqnarray}
g^2{\mathcal L}_{YM}&=&\frac{b_1}{2}\tr F_{\mu\nu}F_{\mu\nu}\nonumber\\
\label{eq:lagnrqcdend}
&-&\left(\frac{b_{2,\psi}}{m_{\psi}^2c^2}
+\frac{b_{2,\chi}}{m_{\chi}^2c^2}\right)\tr F_{\mu\nu}{\mathbf D}^2
F_{\mu\nu}\\\nonumber
&-&2\left(\frac{b_{3,\psi}}{m_{\psi}^2c^2}
+\frac{b_{3,\chi}}{m_{\chi}^2c^2}\right)\tr F_{\mu i}[D_i,D_j]F_{j\mu}
+\frac{1}{c^3}{\mathcal O}(c^7).
\end{eqnarray}
We have adopted the notation, ${\mathbf D}^2=D_i^aD_i^a$.
The tree level values are, $b_{2,\psi}=b_{2,\chi}=
b_{3,\psi}=b_{3,\chi}=0$. It turns out
that the radiative corrections to the
tree level value, $b_1=1$, compensate
the effect of the heavy quarks on the running
of the QCD coupling~\cite{Pineda:1998kj}
in the effective theory, between the matching scale,
$\mu$, and the quark masses, $m_{\psi}$ and $m_{\chi}$.
Heavy quark loops are subsequently explicitly reintroduced into the
NRQCD Lagrangian
via the terms containing derivatives
that are proportional to $b_2$ and $b_3$.
As long as we are only concerned with the quenched approximation to QCD
the corrections to the gauge action should be ignored. In an un-quenched
world they should, however, be included for consistency. It is clear
though that such heavy-quark un-quenching effects are numerically tiny
in comparison to the error one makes when ignoring light sea quarks
for example.

\subsubsection{Matching NRQCD to QCD}
We wish to construct
an effective theory which is applicable to gluon momenta, $q\leq\mu$,
where
$mv<\mu<m$, and which reproduces QCD up to corrections that
are of higher order in terms of the expansion parameter ($m^{-1}$ in the case
of HQET and $c^{-1}$ in the case of NRQCD). Differences between the
effective theory and the correct QCD behaviour that would otherwise
arise for
momenta, $q>\mu$, have to be compensated for by an adequate choice
of the Wilson coefficients, $c_i(\mu/m,\alpha_s)$, $d_i(\mu/m,\alpha_s)$
and $b_i(\mu/m,\alpha_s)$, that
encode the high energy behaviour. The full relativistic Poincar\'e symmetry
[or $O(4)$ plus translations for the Euclidean space-time conventions
adopted here] is not evident from the NRQCD Lagrangian of
Eqs.~(\ref{eq:lagnrqcd}) --
(\ref{eq:lagnrqcdend})
in which only Galilean invariance is explicitly manifest.
By imposing the full four-dimensional invariance to the respective order
of the non-relativistic expansion, matching coefficients
that accompany operators of different dimensions become related.
This re-parametrisation invariance~\cite{Luke:1992cs,Chen:1993sx} is discussed
in Refs.~\cite{Finkemeier:1997re,Manohar:1997qy}.
One result is the relation,
\begin{equation}
\label{eq:repa}
c_S(\mu/m,\alpha_s)=2c_F(\mu/m,\alpha_s)-1,
\end{equation}
which can be derived by imposing invariance under
an infinitesimal Lorentz boost, $v\rightarrow v+\delta v$.
Furthermore, the relativistic dispersion relation implies that
the matching coefficients accompanying the kinetic terms are unity.

Those coefficients that are not determined by fundamental symmetries
can be obtained by matching amplitudes, calculated in the effective theory,
to their QCD counterparts. This can be done
non-perturbatively on the lattice or to a given order in $\alpha_s$
in perturbation theory.
In the first case one would calculate a set of quantities that are
sensitive to the choice of the matching coefficients in 
lattice NRQCD at finite values of the lattice spacings,
$a_{\sigma}\leq \pi/q$, for a range of
quark masses $a_{\sigma}^{-1}\pi\geq m\geq a_{\sigma}^{-1}$ and,
ideally,
match them to their continuum QCD counterparts. The difficulty
of obtaining the continuum QCD result, which requires
simulations at small lattice spacings, $a_{\tau}\ll m^{-1}$,
has turned out to be prohibitive so far.
The exception is the mass
renormalisation,
$\delta m$, that can be fixed
by demanding that the rest mass of an $\Upsilon$ meson must equal
its kinetic mass. This is done by comparing finite
momentum $\Upsilon$ masses with the expected dispersion
relation\footnote{$\delta m$ has also been computed
to one loop perturbation theory
in one version of lattice NRQCD~\cite{Morningstar:1993de}.
On 
the lattice, where $O(3)$ rotational symmetry is broken,
the (non-trivial) coefficients accompanying the kinetic terms
can in principle be fixed by imposing the continuum dispersion
relation.}~\cite{Davies:1994pz,Davies:1994mp,Davies:1998im}.
The only other attempt
into this direction was an estimation of the
coefficients of
the order $\alpha_s$ correction terms to $c_F=1$ and $\delta m=0$
from small volume simulations~\cite{Trottier:1997bn,Dimm:1995fy}.

As an alternative to matching to continuum QCD, one could in principle
treat
all coefficients as free parameters and fix them by demanding
$\Upsilon$ splittings, determined in lattice NRQCD, to match
experimental input values. This, of course, would severely limit
the predictive power of (NR)QCD calculations\footnote{For a prediction
of $B$ meson properties at the $1~\%$ level, it appears
to be sufficient to consider the order $m^{-1}$ HQET/NRQCD 
Lagrangian~\cite{AliKhan:1996ub,Collins:1999ff}.
To this order, the only parameter that requires continuum QCD
input is $c_F$, such that the reduction in predictive power
by using $\Upsilon$ fine structure splittings as an input is not
great.}. Moreover,
it is hard to combine experimental input and
lattice NRQCD in a conceptually clean way;
experiment has $2+1+1$ flavours of
sea quarks of the right physical masses
built in while lattice QCD calculations in
general require extrapolations to the physical sea quark masses and,
eventually, the relevant number
of sea quark flavours.

The general procedure of matching an effective theory to QCD
is to start with the lowest dimensional operators, the dimension
of which we assume
to be $n$ in terms of an expansion parameter, $\lambda$,
and to determine their Wilson coefficients in one or another
scheme. Since the theory only has to reproduce QCD
to the given order, $n$, of the expansion,
the coefficients
are ambiguous: corrections of order $\lambda$ can always be added.
In the next step one would examine the set of operators of the
next available dimension, $n+1$.
These terms will not only undergo mixing with
each other under renormalisation group transformations but also
with lower dimensional operators.
The resulting set of coefficients to this order will depend on
the conventions used to determine the lower dimensional ones:
order $\lambda$ terms added to the coefficients
that accompany dimension $n$ operators have to be cancelled
by operators of dimension $n+1$. This freedom
of re-shuffling power corrections between ultra-violet
Wilson coefficients and
infra-red operators of higher dimension is nothing but the
well known
renormalon
ambiguity~\cite{'tHooft:1977am,Lautrup:1977hs,Shifman:1979cg,Mueller:1985vh}.
In conclusion, any matching scheme can be used but it has to be
employed consistently.

\begin{figure}[thb]
\centerline{\epsfxsize=10truecm\epsffile{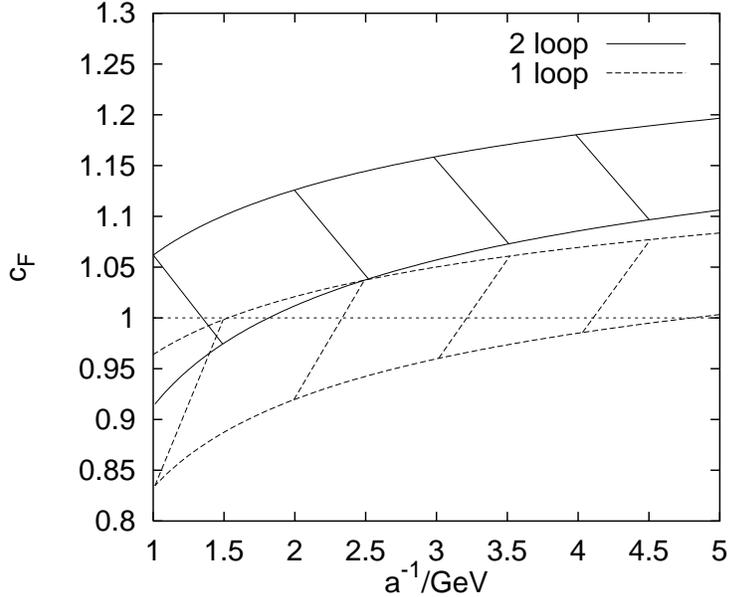}}
\caption{Continuum inspired estimates of the
matching coefficient, $c_F(\mu)$, as a function of
the lattice spacing for bottom quarks. The upper curves correspond
to $\mu=\pi/a$, the lower ones to $\mu=1/a$.}
\label{fig:cfbottom}
\end{figure}

\begin{figure}[thb]
\centerline{\epsfxsize=10truecm\epsffile{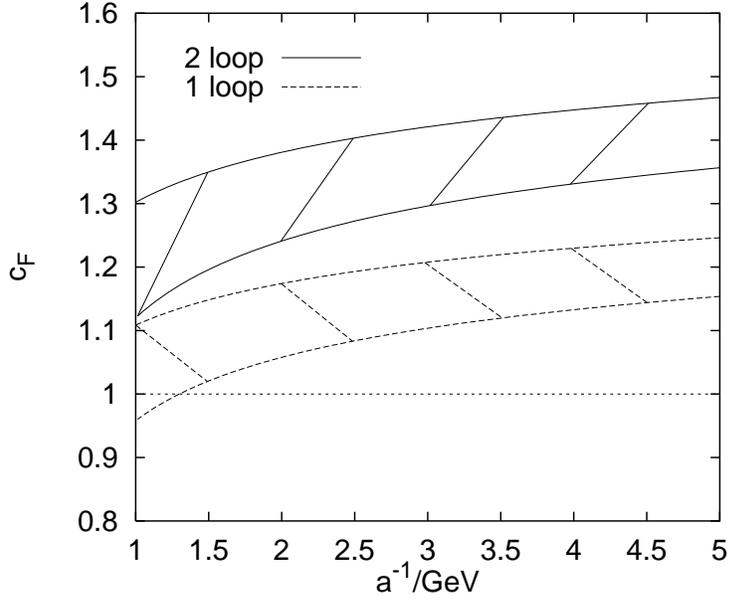}}
\caption{The same as Figure~\ref{fig:cfbottom} for charm quarks.}
\label{fig:cfcharm}
\end{figure}

\begin{figure}[thb]
\centerline{\epsfxsize=10truecm\epsffile{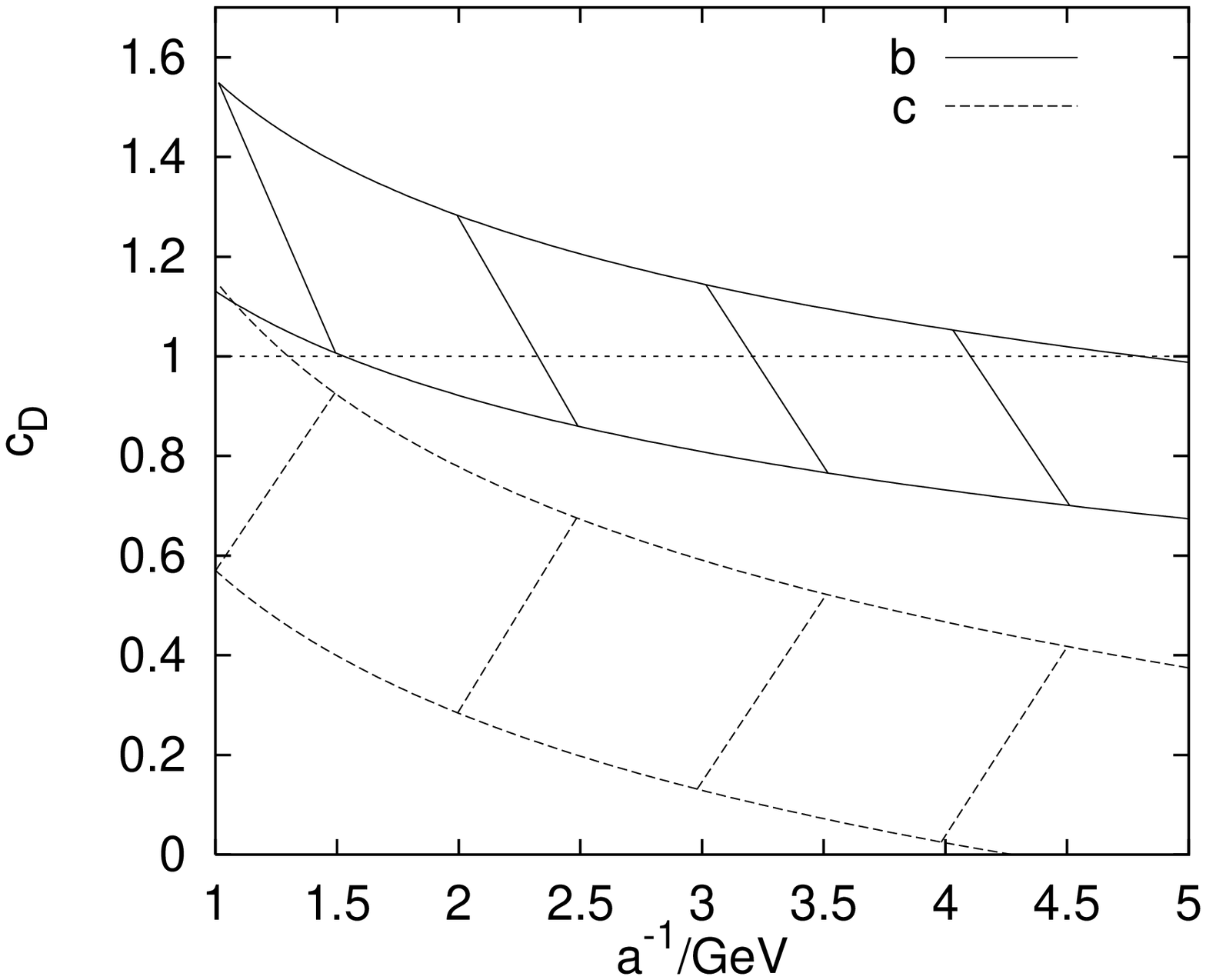}}
\caption{Continuum inspired estimates of $c_D(\mu)$, as a function of
the lattice spacing for charm and bottom quarks. The lower curves correspond
to $\mu=\pi/a$, the upper ones to $\mu=1/a$.}
\label{fig:cdbotcharm}
\end{figure}

In Refs.~\cite{Manohar:1997qy,Pineda:1998kj} it has been argued that
despite of the fact that the
leading order NRQCD Lagrangian differs from its HQET counterpart by
the kinetic term it incorporates, HQET delivers a viable prescription
for determining the NRQCD matching coefficients, at least up to order
$m^{-3}$. In HQET the Fermi coefficient, $c_F$, is known to two loops
and all other coefficients at the one loop level in the $\overline{MS}$
scheme of dimensional regularisation. We display the results in
Appendix~\ref{app:match}.
As discussed above, the coefficients are specific to the prescription used.
Other regularisation schemes or ways of
organising the expansion, e.g.\
in powers of $c^{-1}$, will in general yield different results.

Unfortunately, no lattice NRQCD perturbation theory results exist for
the matching coefficients, $c_F$ and $c_D$.
Large contributions from lattice  tadpole
diagrams~\cite{Lepage:1993xa} in general result in
big renormalisations between non-spectral
quantities, calculated by use of lattice
regularisation, with respect to continuum schemes such
as the $\overline{MS}$ scheme.
This is for instance reflected in the
ratio~\cite{Hasenfratz:1980kn,Weisz:1981pu,Dashen:1981vm,Kawai:1981ja},
$\Lambda_{\overline{MS}}/\Lambda_L\approx 28.81\Lambda_L$,
for the pure gauge Wilson action.

It has been suggested~\cite{Parisi:1980pe,Lepage:1993xa} 
to (partially) cancel
tadpole contributions by dividing each lattice link
that appears in a given operator by the fourth root of the
measured expectation value of a plaquette,
$U_{P}=\re\langle\Tr\, U_{x,\mu\nu}\rangle$.
Other prescriptions using the average link in Landau gauge or
expressions containing the logarithm of the plaquette have been suggested as
alternatives~\cite{Lepage:1993xa}.
It is argued that the diagrams that result in large renormalisations
also cause the plaquette (or the average gauge fixed link) to
substantially deviate from the free field
expectation of unity. Moreover, such (ultra-violet) renormalisation effects
might commute with the infra-red physics of interest.
Based on these ideas, the following replacement has been
suggested~\cite{Parisi:1980pe},
$\beta S=\beta_{\mbox{\scriptsize MF}}S_{ir}$,
with $S_{ir}=S/U_{P}$. This yields the
relation, $\alpha_{\mbox{\scriptsize MF}}=\alpha_L/U_{P}$.
Another popular choice of an
``improved'' coupling is~\cite{Lepage:1993xa},
$\alpha_{\mbox{\scriptsize FNAL}}=-\ln U_{P}/(\pi C_F)$. From the perturbative
expansion of the plaquette expectation
value~\cite{DiGiacomo:1981wt,DiGiacomo:1982dp}, one finds
$\Lambda_{\overline{MS}}\approx
2.63\Lambda_{\mbox{\scriptsize MF}}
\approx 4.19\Lambda_{\mbox{\scriptsize FNAL}}$ for $n_f=0$;
indeed, the coefficients appearing in the one loop perturbative
matching between $\alpha_{\overline{MS}}(\pi/a)$ and
$\alpha_{\mbox{\scriptsize MF}}(a)$
or $\alpha_{\mbox{\scriptsize FNAL}}(a)$ are much
smaller than for the bare lattice coupling.

After obtaining a ``tadpole improved'' version of lattice NRQCD
by following the above recipe, one might hope that the running of the
coefficients with the quark mass closely resembles that of continuum NRQCD.
Unlike dimensional regularisation, the lattice imposes a hard cut-off, $\pi/a$,
on gluon momenta, $q$, such that
it is not entirely clear what matching scale corresponds to the $\mu$
of the $\overline{MS}$ formulae of Appendix~\ref{app:match}.
It is reasonable, however, to assume~\cite{Lepage:1993xa},
$\pi/a\geq \mu\geq 1/a$. In Figure~\ref{fig:cfbottom},
we display the resulting estimates for $c_F$ for the bottom quark
as a function of the inverse lattice spacing, based on
Eq.~(\ref{eq:cfper}). The widths of the two
bands correspond to the above scale uncertainty.
$\alpha_{\overline{MS}}$ as a function of the scale has been
obtained by running down the quenched result~\cite{Capitani:1998mq}
of Eq.~(\ref{eq:lms0}) from a high energy scale by means of the
four loop $\beta$ function, Eq.~(\ref{rge}).

Within the region,
1.5 GeV~$<a^{-1}<3$~GeV, $c_F$ can easily deviate from the tree level value
by as much as 15~\%. Moreover, the two loop result significantly deviates
from the one loop prediction, indicating a slow convergence of the
perturbative series. In Figure~\ref{fig:cfcharm}, it is convincingly
demonstrated that for lattice resolutions better than 1~GeV
a perturbative estimation of $c_F$ for charmonia is unreliable.
On the other hand lattice spacings, $a^{-1}\ll 2$~GeV,
are too big to sample the relevant
bound state dynamics and would result in huge scaling
violations.
Finally, in Figure~\ref{fig:cdbotcharm} we plot our estimates,
Eq.~(\ref{eq:cdper}),
of the cut-off
dependence of the Darwin coefficients, $c_D$, for both,
bottom and charm quarks. We find $c_D$ to vary much more
with the quark mass than $c_F$. Unfortunately,
no two loop calculation for this quantity is available.

In conclusion, the perturbative calculation exhibits a significant
dependence of the coefficients on the quark mass (or lattice spacing).
While at a given lattice spacing $c_F$ decreases with
increasing quark mass, $c_D$ shows the opposite behaviour.
Perturbation theory seems to be slowly convergent. Moreover, in general
power corrections can contribute to the coefficients.

\subsection{Lattice NRQCD}
It is straight forward to discretise the Lagrangian, Eqs.~(\ref{eq:lagnrqcd})
-- (\ref{eq:lanr}), and to simulate it directly on a lattice. Let us
start with the leading order continuum NRQCD Lagrangian,
\begin{equation}
\label{eq:leading}
{\mathcal L}_{NRQCD,v^2}=\psi^{\dagger}\left(-D_4+
\frac{{\mathbf D}^2}{2m_{\psi}}\right)\psi.
\end{equation}
Note that from now on we use, $c=1$.
We define the heavy quark propagator, $K=\langle\psi\psi^{\dagger}\rangle$.
$K$ is the
direct product of a $2\times 2$ matrix acting on the Pauli spinor
space, a $3\times 3$ matrix acting on colour space and a
$L_{\sigma}^3L_{\tau}\times L_{\sigma}^3L_{\tau}$ matrix
acting on space-time.
From Eq.~(\ref{eq:leading})
it follows that the evolution of $K$ with time is governed
by the Hamiltonian,
\begin{eqnarray}
H_0&=&-\frac{{\mathbf D}^2}{2m_{\psi}}:\\
-\partial_4K&=&\left(igA_4+H_0\right)K.\label{eq:hlnrqcd}
\end{eqnarray}
By formally solving the above differential equation,
we obtain the evolution equation,
\begin{equation}
\label{eq:formal2}
K({\mathbf x},t+a)
=\sum_{\mathbf y}
\int_{{\mathbf s}(0)={\mathbf y}}^{{\mathbf s}(t)={\mathbf x}}\!\!\!
D{\mathbf s}\,{\mathcal P}\left\{\exp\left[-\int_t^{t+a}\!\!\!dt'\,
(igA_4+H_0)\right]\right\}K({\mathbf y},t),
\end{equation}
where we have assumed the sum over all paths to be appropriately
normalised.
The initial condition reads,
\begin{equation}
\left.K(x)\right|_{x_4=0}=\delta^3({\mathbf x}).
\end{equation}
Note that we have suppressed the dependence of the propagator
on the source point, $K(x)=K(x,y=0)$.

A natural discretisation of Eq.~(\ref{eq:formal2})
is~\cite{Lepage:1992tx},
\begin{equation}
K(t+a)=\left(1-\frac{aH_0(t+a)}{2n}\right)^nU^{\dagger}_{t,4}
\left(1-\frac{aH_0(t)}{2n}\right)^nK(t).
\end{equation}
We have omitted the dependence on the spatial coordinates
from the above equation. The temporal link, $U^{\dagger}_{t,4}$,
is diagonal in space.
The covariant Laplacian within $H_0(t)$ can be written as,
\begin{equation}
\label{eq:lapla}
{\mathbf D}^2_{{\mathbf x}{\mathbf y}}(t)
=a^{-2}\sum_{i=1}^3\left[U_{({\mathbf x},t),\hat{\imath}}\delta_{{\mathbf x}+
a\hat{\mathbf\imath},{\mathbf y}}+
U^{\dagger}_{({\mathbf x}-a\hat{\mathbf\imath},t),\hat{\imath}}
\delta_{{\mathbf x}-
a\hat{\mathbf\imath},{\mathbf y}}-2\delta_{{\mathbf x}{\mathbf y}}\right],
\end{equation}
up to ${\mathcal O}(a^2)$ lattice artefacts. 
For the na\"{\i}ve $n=1$ discretisation,
the evolution equation might become numerically unstable
as $1-aH_0$ becomes negative
for momenta larger than the quark mass; 
lighter quarks try to travel faster than they are
allowed by the evolution equation.
Introducing the stabilisation parameter, $n$, improves the
spatial propagation and
relaxes this criterion to
$\max(aH_0)<n$.
In the free field case,
the maximal eigenvalue of the
Laplacian, Eq.~(\ref{eq:lapla}),
is $\sum_i\max{\hat{p}_i\hat{p}_i}=3a^{-2}$, such that
$ma>3/(2n)$ has to be maintained. When switching on
interactions, the factor 3/2 is reduced somewhat.

Working with an
anisotropy, $a_{\tau}<a_{\sigma}$, offers an alternative
to introducing the parameter, $n$.
In this case, the free field stability criterion relaxes to,
$ma_{\sigma}>(3/2)a_{\tau}/a_{\sigma}$. It is amusing to see that
in lattice NRQCD simulations,
discretisation effects become more pronounced in light quark
propagators,
rather than for heavy quarks as in relativistic lattice QCD.
While in the latter case, heavier quarks can be realised
by reducing $a_{\tau}$, in NRQCD lighter quarks require smaller
$a_{\tau}$ (or larger $n$).
Of course one would not rely on results obtained
for quark masses, $m<a_{\sigma}^{-1}$, as the non-relativistic
expansion breaks down
for a cut-off on gluon momenta larger than the quark mass.
On the other hand one would also not want to simulate quarks much
heavier than the lattice resolution to keep the scale, $mv$, separated
from the lattice cut-off, $m\ll (a_{\sigma}v)^{-1}$.
Otherwise, the matching coefficients between
lattice NRQCD and QCD would explode and their behaviour could 
no longer reliably be estimated.

The above evolution equation approximates the continuum equation
only up to ${\mathcal O}(a_{\tau})$ lattice artefacts.
These can be removed at tree level by the substitution,
$H_0\rightarrow H_0\left[1+(a_{\tau}/4n)H_0\right]$ (or
reduced by increasing $n$).
${\mathcal O}(v^4)$ correction terms, $\delta H_{v^4}$, can be
included too,
\begin{eqnarray}
K(t+a)&=&\left(1-\frac{a\delta
    H_{v^4}}{2}\right)\left(1-\frac{aH_0}{2n}\right)^n\\\nonumber
&\times& U^{\dagger}_{t,4}
\left(1-\frac{aH_0}{2n}\right)^n
\left(1-\frac{a\delta H_{v^4}}{2}\right)
K(t).
\end{eqnarray}
Details can be found in Ref.~\cite{Lepage:1992tx}.
Although a method to incorporate the four fermion terms
of Eq.~(\ref{eq:lanr}) is suggested in this Reference too,
these have not been included into any lattice simulation so far.
Typically, tadpole improvement is employed in NRQCD simulations,
i.e.\ link variables are divided by factors $U_P^{1/4}$ or
equivalent quantities that approach unity
in the continuum limit.

The NRQCD evolution equation has also been applied to the heavy
quark within heavy-light systems~\cite{AliKhan:1996ub}.
$H_0$ does not only
consist of the static propagator but also incorporates
the kinetic term, while the Fermi term, that is of the same order
in $m^{-1}$, appears within
$\delta H$. The main
advantages of this procedure over a na\"{\i}ve discretisation
of HQET lie in smaller wave function renormalisations and in a reduction
of statistical fluctuations. Both effects are related to the use of
a propagator that samples gauge fields over an extended spatial region.
The disadvantages in applying lattice ``NRQCD'' with HQET like power counting
to heavy-light mesons is a loss in conceptual
clarity as
the wave function renormalisation depends on the expansion
parameter, $m^{-1}$, in a way that cannot be absorbed into
multiplicative field redefinitions.

By contracting quark and anti-quark propagators with suitable combinations of
gauge transporters and Pauli matrices, particular
${}^sl_J$ states can be realised whose ground state masses can be extracted
from the asymptotic decay of two-particle Green functions in
Euclidean time in the usual way. Like in all direct spectrum
evaluations, radial
excitations present a major problem. Thus, it is a
tremendous achievement
that the $3S$ as
well as the $2P$
states
have been determined,
with statistical errors of about 100~MeV~\cite{Davies:1998im,Davies:1994mp,Eicker:1998vx,Spitz:1999tu,Manke:1997gt}.
Precision results exist for
$2S$, $1P$ and $1S$ states.

\subsection{The potential approach}
\label{sec:gluon}
\subsubsection{Deriving a bound state Hamiltonian}
We wish to derive a Hamiltonian that governs the evolution of a
quarkonium state from the order $c^{-2}$ (or $v^4$)
NRQCD Lagrangian of Eqs.~(\ref{eq:lagnrqcd}) -- (\ref{eq:lagnrqcdend})
that is formulated on
the quark-gluon level. As a first step in this direction,
we calculate a heavy quark propagator in a representation that will turn
out to be suitable for our purpose.
The time evolution of the Pauli propagator, $K$,
is controlled by the equation,
\begin{equation}
\label{eq:evelove}
-\partial_4K=H_1K,
\end{equation}
where the Hamiltonian,
\begin{equation}
H_1=m+igA_4+H_{\psi},
\end{equation}
can be read off from Eq.~(\ref{eq:ef1quark}).
Unlike in Eq.~(\ref{eq:hlnrqcd})
we decide not to eliminate the heavy quark
rest mass, $m=m_{\psi}-\delta m$,
and not to rescale the (anti-)fermion fields by factors,
$\exp(\pm mt)$.
For the initial condition,
\begin{equation}
\left.K(x,y)\right|_{x_4=y_4}=\delta^3({\mathbf x}-{\mathbf y}),
\end{equation}
Eq.~(\ref{eq:evelove}) can be formally solved by summing over all
possible paths connecting $y$ with $x$,
\begin{equation}
\label{eq:formal}
K(x,y)=\int_{{\mathbf z}(y_4)={\mathbf y}}^{{\mathbf z}(x_4)={\mathbf x}}
D{\mathbf z}\,D{\mathbf p}\exp\left\{\int_{y_4}^{x_4}dt
\left[{\mathbf p}\dot{\mathbf z}-H_1({\mathbf z},{\mathbf p})\right]\right\},
\end{equation}
where the dot denotes a derivative with respect to the time coordinate.
The correct normalisation is assumed to be included into the definitions
of $D{\mathbf z}$ and $D{\mathbf p}$.

\begin{figure}[thb]
\centerline{\epsfxsize=10truecm\epsffile{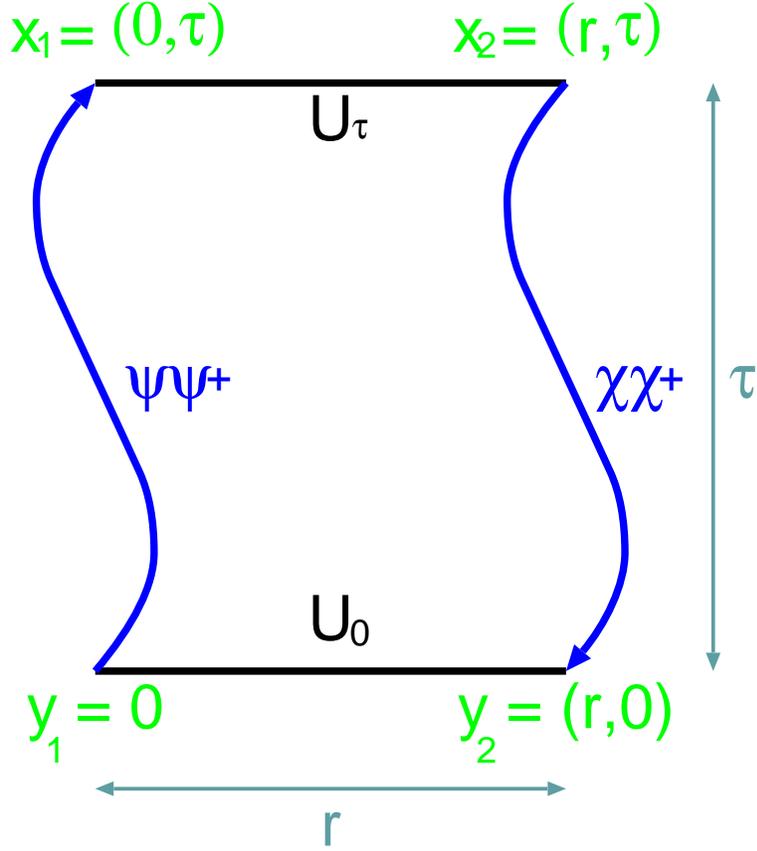}}
\caption{The four point function of Eq.~(\ref{eq:4point}).}
\label{fig:4point}
\end{figure}

We can now combine two such propagators into a generalised (fluctuating)
rectangular Wilson loop,
\begin{equation}
\label{eq:4point}
G_{i'j'ij}(r,\tau)
=\left\langle\Tr\left[U_0K_{i'i}(y_1,x_1)U_{\tau}
K_{j'j}(x_2,y_2)\right]\right\rangle,
\end{equation}
where the indices $i,j$ and $i',j'$ represent the spins of the
initial and final states. Note that $G$, unlike the argument
of the expectation
value,
is real in Euclidean space-time.
For the case of quark and anti-quark having different masses, two different
propagators, $K_{\psi}$ and $K_{\chi}$,
must be used within the above formula.
We denote the temporal extent by, $\tau=y_{1,4}-x_{1,4}=y_{2,4}-x_{2,4}$,
and the spatial separation by ${\mathbf r}={\mathbf y}_2-{\mathbf y}_1=
{\mathbf x}_2-{\mathbf x}_1$. 
The situation is visualised in Figure~\ref{fig:4point}.
For the sake of simplicity, we switch to leading order
NRQCD with equal quark masses. In this case,
\begin{equation}
\label{kern}
H_1=m+igA_4+\frac{p^2}{2m_{\psi}}.
\end{equation}
To lowest order in $v/c$, the exponent within
Eq.~(\ref{eq:formal}) can be
approximated by the value it takes
along the shortest path~\cite{Barchielli:1988zs,Barchielli:1990zp}. Thus,
\begin{equation}
G({\mathbf r},\tau)=\exp\left[
\int_0^{\tau}\!dt\,\sum_{j=1}^2\left({\mathbf p}_j\dot{\mathbf x}_j
  -m-
\frac{{\mathbf p}_j^2}{2m_{\psi}}\right)\right]\langle W({\mathbf r},\tau)\rangle.
\end{equation}
In higher orders of the $v/c$ expansion fluctuations of the propagators
around the classical paths have to be taken
into account that result in additional terms.
From the spectral decomposition of the Wilson loop,
\begin{equation}
\label{eq:hh1}
\langle W({\mathbf r},\tau)\rangle
\propto \exp[-V_0({\mathbf r})\tau]\quad(\tau\rightarrow\infty),
\end{equation}
we arrive at,
\begin{equation}
\label{eq:rr1}
-\frac{d}{dt}G = HG,
\end{equation}
with
\begin{equation}
\label{eq:hh3}
H=2m+\frac{p^2}m_{\psi}+V_0(r),
\end{equation}
in the limit of large $\tau$:
the result is a Schr\"odinger equation that governs the evolution
of a quark anti-quark state in a gluonic background whose average effect
is contained in the static potential, $V_0$.
The validity of the instantaneous approximation
is tied to that of the na\"{\i}ve quark model:
if quarkonium states
can be completely classified by the quantum numbers of the constituent
quarks, then the spectrum and wave functions can be obtained by solving the
quantum mechanical
equation,
\begin{equation}
\label{eq:rr2}
H\psi_{nll_3}({\mathbf r})=E_{nl}\psi_{nll_3}({\mathbf r}).
\end{equation}

We have discussed above that $m$ will in general differ
from the ``kinetic'' mass of the quark, $m_{\psi}=m(\mu)+\delta m(\mu)$.
Furthermore, this difference will
depend on the matching scale, $\mu$. In Section~\ref{sec:pert}
we have also seen that the potential, $V_0(r,\mu)=\hat{V}_0(r)+
V_{\mbox{\scriptsize self}}(\mu)$, can be factorised into a
physical and a self-energy part. This observation results in the relation,
\begin{equation}
\delta m(\mu')=\delta m(\mu)+
\frac{1}{2}\left[ V_{\mbox{\scriptsize self}}(\mu')-
V_{\mbox{\scriptsize self}}(\mu)\right],
\end{equation}
i.e.\ $\delta m(\mu)$ diverges as $\mu\rightarrow\infty$.

In lattice NRQCD, it is straight forward to calculate masses
of quarkonia states, $E_{\Upsilon}(p)$,
projected onto non-vanishing momentum, $p$. We use the convention
that $E_{\Upsilon}$ is the sum of the bare quark masses, $2m$, and
the energy shift due to the interaction terms of the Hamiltonian.
By requiring the correct dispersion relation to the given order
of the expansion,
\begin{equation}
E_{\Upsilon}(p)-E_{\Upsilon}(0)=\frac{p^2}{2m_{\Upsilon}}
-\frac{p^4}{8m_{\Upsilon}^3}\quad\left(\quad+\cdots\quad\right),
\end{equation}
the $\Upsilon$ rest mass, $m_{\Upsilon}$, can be determined. The mass shift,
then, is given by, $2 \delta m =E_{\Upsilon}(0)-m_{\Upsilon}$.
Within the potential approach, the zero point energy at first
appears to be difficult to determine in a similar way.
However, in principle it should be possible to calculate potentials
governing finite
momentum quarkonia states too, by starting the derivation
from a boosted NRQCD Lagrangian.

\subsubsection{Relativistic corrections}
The Hamiltonian Eqs.~(\ref{eq:hh1}) -- (\ref{eq:hh3})
was first obtained in Ref.~\cite{Brown:1979ya} in a systematic way
from continuum QCD where the static
Dirac equation, $\left(\gamma_4D_4+m\right)q=0$,
is solved by a Schwinger line times a factor, $e^{-mt}$,
after projecting
onto quark and anti-quark states.
Starting from QCD, Eichten, Feinberg and Gromes derived spin
dependent correction terms~\cite{Eichten:1979pu,Eichten:1981mw,Gromes:1984pm}
(see also the article by Peskin~\cite{Peskin:1983up}). Finally,
Brambilla and collaborators
(BBP)~\cite{Barchielli:1988zs,Barchielli:1990zp} found an additional
relativistic correction term
to the central potential that had previously been ignored and 
added further velocity (or momentum) dependent
terms by taking fluctuations of
the heavy quark propagators into account.
In general, apart from the one-particle Lagrangians,
${\mathcal L}_i, i=\psi,\chi$, [Eqs.~(\ref{eq:lagnrqcd}) -- (\ref{eq:ef1quark})]
the two-particle Greens function receives contact term
contributions from the two-particle sector Lagrangian of
Eq.~(\ref{eq:lanr})~\cite{Chen:1995dg},
${\mathcal L}_{\psi\chi}$.
Taking these into account too,
the complete result to this order in $c^{-1}$,
with the NRQCD matching coefficients
included~\cite{Chen:1995dg,Bali:1997am,Brambilla:1998vm},
in the centre of mass frame (${\mathbf p}={\mathbf p_1}=-{\mathbf p_2}$
and ${\mathbf L}=
{\mathbf L_1}={\mathbf L_2}$), for $m_1\geq m_2$ is,
\begin{eqnarray}
H&=&\sum_{i=1}^2\left(m_i-\delta m_i
+\frac{p^2}{2m_i}-\frac{p^4}{8m_i^3}\right)\nonumber\\\label{ham}
&+&V(r,{\mathbf p},{\mathbf L},{\mathbf S_1},{\mathbf S_2}),
\end{eqnarray}
where the potential,
\begin{equation}
V(r,{\mathbf p},{\mathbf L},{\mathbf S_1},{\mathbf S_2})=
V_0(r)+
V_{\mbox{\scriptsize C}}+
V_{\mbox{\scriptsize SD}}(r,{\mathbf L},{\mathbf S_1},{\mathbf S_2})+
V_{\mbox{\scriptsize MD}}(r,{\mathbf p}),\label{prp}
\end{equation}
contains corrections to the central potential (C) as well
as spin dependent (SD) and
momentum dependent (MD) corrections. $V_0(r)$ denotes the static potential
while,
\begin{eqnarray}
V_{\mbox{\scriptsize C}}(r)&=&\frac{d_s}{m_1m_2}4\pi C_F\alpha_s
\delta^3({\mathbf r})\nonumber\\
\label{cepo}
&+&
\sum_{i=1}^2\frac{1}{8m_i^2}\left\{
c_D^{(i)}\left[\nabla^2V_0(r)+\nabla^2V_a^E(r)\right]
+{c_F^{(i)}}^2\nabla^2V_a^B(r)\right\}\\\nonumber
&+&\left(\frac{1}{m_1}+\frac{1}{m_2}\right)V_{\prime}(r),
\end{eqnarray}
\begin{eqnarray}
V_{\mbox{\scriptsize SD}}(r,{\mathbf L},{\mathbf S_1},{\mathbf S_2})
&=&\left(\frac{{\mathbf S_1}}{m_1^2}
+ \frac{{\mathbf S_2}}{m_2^2}\right){\mathbf L}
\frac{(2c_+-1)V_0'(r)+2c_+V_1'(r)}{2r}\nonumber\\
&+&\frac{{\mathbf S_1} + {\mathbf S_2}}{m_1m_2}{\mathbf L}
\frac{c_+V_2'(r)}{r}\nonumber\\\label{sdpo}
&+&\left(\frac{{\mathbf S_1}}{m_1^2}
- \frac{{\mathbf S_2}}{m_2^2}\right){\mathbf L}
\frac{c_-[V_0'(r)+V_1'(r)]}{r}\nonumber\\\nonumber
&+&\frac{{\mathbf S_1} - {\mathbf S_2}}{m_1m_2}{\mathbf L}
\frac{c_-V_2'(r)}{r}\\
&+&\frac{S_1^iS_2^j}{m_1m_2}c_F^{(1)}c_F^{(2)}R_{ij}V_3(r)
\\\nonumber
&+&\frac{{\mathbf S_1}\cdot{\mathbf S_2}}{3m_1m_2}
\left[c_F^{(1)}c_F^{(2)}V_4(r)-12 d_v 4\pi C_F\alpha_s
\delta^3({\mathbf
  r})\right],
\end{eqnarray}
and
\begin{eqnarray}
V_{\mbox{\scriptsize MD}}(r,{\mathbf p})
&=&-\frac{1}{m_1m_2}
\left\{p_i,p_j,[\delta_{ij}V_b(r)-R_{ij}V_c(r)]\right\}_{\mbox{\scriptsize
Weyl}}\label{mdpo}\\
&+&\sum_{k=1}^2\frac{1}{m_k^2}\left\{p_i,p_j,[\delta_{ij}V_d(r)-R_{ij}V_e(r)]
\right\}_{\mbox{\scriptsize Weyl}},\nonumber
\end{eqnarray}
with
\begin{eqnarray}
\label{eq:rij}
R_{ij}&=&\frac{r_ir_j}{r^2}-\frac{\delta_{ij}}{3},\\
\delta m_i&=&\delta m(m_i,\mu),\\
c_{F,D}^{(i)}&=&c_{F,D}(m_i,\mu),\\
c_{\pm}&=&c_{\pm}(m_1,m_2,\mu)=\frac{1}{2}\left(c_F^{(1)}\pm
c_F^{(2)}\right),\\
d_s&=&\frac{1}{4\pi C_F\alpha_s}\left[
d_{ss}(m_1,m_2,\mu)+C_Fd_{vs}(m_1,m_2,\mu)\right],\\
d_v&=&\frac{1}{4\pi C_F\alpha_s}
\left[d_{sv}(m_1,m_2,\mu)+C_Fd_{vv}(m_1,m_2,\mu)\right].
\end{eqnarray}
The symbol $\{a,b,c\}_{\mbox{\scriptsize Weyl}}=
\frac{1}{4}\{a,\{b,c\}\}$
denotes Weyl ordering of the three arguments.
Note that in the equal mass case, that has been considered 
in Ref.~\cite{Eichten:1981mw}, where $c_-$ assumes its
tree level value, $c_-=0$, two of the spin-orbit terms vanish.
The term proportional to $V_{\prime}$ in Eq.~(\ref{cepo}) has been
identified very recently~\cite{Brambilla:2000gk} and in principle
additional $1/m^2$ corrections to $V_{\mbox{\scriptsize C}}$
should exist~\cite{Brambilla:2000gk}, albeit to higher order in the
$c^{-1}$ power counting than order $c^{-2}$ considered above.

The last term of Eq.~(\ref{sdpo}) has been written in a
somewhat suggestive way that is motivated by the expectation,
$V_4(r)\approx 8\pi C_F\alpha_s\delta^3({\mathbf r})$.
$c_S$ has been eliminated from the above formulae by using the
re-parametrisation invariance relation, Eq.~(\ref{eq:repa}).
Note that neither $d_{vv}^c$ or $d_{vs}^c$ nor
$d_{ss}$ or $d_{sv}$ contribute to $d_s$ or $d_v$. This means that
even in the equal mass case, where $d_{vv}^c=-\pi\alpha_s+\cdots$,
$d_s$ and $d_v$ are of order $\alpha_s$. The one loop results
in the $\overline{MS}$ scheme are displayed
in Eqs.~(\ref{eq:ds2}) and (\ref{eq:dv}) of Appendix~\ref{app:match}.

$V_0$, $V_{\prime}$,
$\nabla^2V_a^E$, $\nabla^2V_a^B$, $V_1',\ldots, V_4$ and $V_b,\ldots,
V_e$ can be computed from lattice correlation functions
(in Euclidean time) of Wilson loop like operators.
The functions
$V_1',\ldots, V_4$ are related to spin-orbit and spin-spin interactions.
The MD potential
gives rise to correction terms of the
form $\frac{1}{r}{\mathbf L}^2$, $\frac{1}{r^3}{\mathbf L}^2$,
$\frac{1}{r}p^2$, $\frac{1}{r}$ and $\delta^3(r)$, and
the correction to the central potential
includes the expected Darwin term,
$\nabla^2V_0$, as well as $\nabla^2
V_a^E$ and $\nabla^2 V_a^B$. 

\subsubsection{Scale dependence}
The SD potentials,
$V_1',\ldots, V_4$, as well as $\nabla^2V_a^E$
and $\nabla^2V_a^B$ depend on the matching scale, $\mu$.
The potentials $V_0$, $V_{\prime}$ as well as
$V_b,\ldots,V_e$ can contain additive, $\mu$ dependent self energy
contributions; however, their derivatives are scale independent\footnote{
$V_0$ is a spectral quantity while $V_{\prime}$ and the MD potentials
$V_b,\ldots,V_e$ originate from the terms $D_4$ and ${\mathbf D}^2/(2m)$
of the NRQCD action
that are protected by reparametrisation invariance. Therefore, these
potentials do not undergo multiplicative renormalisation.}.

Due to Lorentz invariance, certain pairs of potentials are related
to the static
potential by the Gromes~\cite{Gromes:1984ma} and BBP~\cite{Barchielli:1990zp}
relations,
\begin{eqnarray}
V_2'(\mu;r)-V_1'(\mu;r)&=&V_0'(r),\label{grom}\\
V_b(r;\mu)+2V_d(r;\mu)&=&\frac{r}{6}V_0'(r)-\frac{1}{2}V_0(r;\mu),
\label{bram2}\\
V_c(r)+2V_e(r)&=&-\frac{r}{2}V_0'(r)\label{bram},
\end{eqnarray}
such that three potentials, e.g.\ $V_2'$, $V_d$ and $V_e$ can be
eliminated from the Hamiltonian. 
Note that Eq.~(\ref{grom}) implies Eq.~(\ref{eq:repa}).
Given the structure of the Hamiltonian,
Eqs.~(\ref{ham}) -- (\ref{sdpo}), and the Gromes
relation, Eq.~(\ref{grom}),
we can deduce~\cite{Bali:1995yz,Bali:1997am} the following
relations between potentials, evaluated at cut-off
scales $\mu$ and $\mu'$,
by demanding\footnote{The $\delta$ function
within $V_{\mbox{\tiny C}}$ represents a problem: no
spin- and momentum-independent counter term is known
that has the right
mass dependence to cancel the running of the
coefficient, $d_s$. This might hint at further, not yet
discovered, relations.}
$dH/d\ln\mu=0$,
\begin{eqnarray}
\nabla^2V_a^E(\mu';r)&=&\frac{1}{c_D(\mu')}
\left\{c_D(\mu)\nabla^2V_a^E(\mu;r)+[c_D(\mu)-c_D(\mu')]\nabla^2V_0(r)\right.
\nonumber\\
&+&
\left.
c_F^2(\mu)\nabla^2V_a^B(\mu;r)+c_F^2(\mu')\nabla^2V_a^B(\mu';r)\right\},
\label{vabsca}\\
V_1'(\mu';r)&=&V_1'(\mu;r)-
\left[1-\frac{c_F(\mu')}{c_F(\mu)}\right]V_2'(\mu;r),\label{v1sca}\\
V_2'(\mu';r)&=&\frac{c_F(\mu)}{c_F(\mu')}V_2'(\mu;r),\label{v2sca}\\
V_3(\mu';r)&=&\frac{c_F^2(\mu)}{c_F^2(\mu')}
V_3(\mu;r),\label{v3sca}\\\label{v4sca}
V_4(\mu';r)&=&\frac{1}{c_F^2(\mu')}\left\{c_F^2(\mu)V_4(\mu;r)\right.
\nonumber\\
&-&\left.
12[d_v(\mu)-d_v(\mu')]4\pi C_F\alpha_s\delta^3({\mathbf r})\right\}.
\end{eqnarray}
Since the potentials, appearing in the above relations, do not depend on the
quark mass, the ratios $c_{F,D}(m,\mu)/c_{F,D}(m,\mu')$ must not depend on
$m$. Therefore, the matching coefficients can always be factorised into
two separate functions,
$c_i(m,\mu)=f_i(m)g_i^{-1}(\mu)$.

\subsubsection{Integrating out gluons}
We have managed to
separate the time dependence of the interaction
into coefficient functions of various interaction terms,
$V_i$, which we shall call the potentials. These potentials can be computed
as expectation values in presence of a gauge field
background~\cite{Eichten:1981mw,Barchielli:1988zs,Bali:1997cj,Bali:1997am},
\begin{eqnarray}
\label{ce_1}
\nabla^2{V}_a^E({\mathbf r})&=&2\,g^2\lim\limits_{\tau\to\infty}
\int_0^\tau
\!dt\, \langle\langle {\mathbf E}({\mathbf 0},0)
\cdot{\mathbf E}({\mathbf 0},t)\rangle\rangle_W^c,\\
\label{ce_2}
\nabla^2{V}_a^B({\mathbf r})&=&2\,g^2\lim\limits_{\tau\to\infty}
\int_0^\tau 
\!dt\, \langle\langle {\mathbf B}({\mathbf 0},0)\cdot
{\mathbf B}({\mathbf 0},t)\rangle\rangle_W,\\
V_{\prime}({\mathbf r})&=&-\frac{g^2}{2}\lim\limits_{\tau\to\infty}
\int_0^\tau
\!dt\, t\,\langle\langle {\mathbf E}({\mathbf 0},0)
\cdot{\mathbf E}({\mathbf 0},t)\rangle\rangle_W^c,\label{ce_3}
\end{eqnarray}
where the superscript ``$c$'' denotes the connected part\footnote{
Both, electric and magnetic fields transform oddly under
charge conjugation. Therefore, in $SU(2)$ gauge theory,
where all traces are real,
$\langle\langle {\mathbf E}\rangle\rangle_W=\langle\langle {\mathbf
  B}\rangle\rangle_W = {\mathbf 0}$. Under $PC$ transformations
the electric field transforms evenly. However, the magnetic field
has $PC=-1$. Therefore, 
$\langle\langle {\mathbf B}\rangle\rangle_W={\mathbf 0}$ 
still holds for $SU(3)$ gauge theory. However,
components of $\langle\langle {\mathbf E}\rangle\rangle_W$ that
are not orthogonal to ${\mathbf r}$ do not have to
vanish (cf.\ Table~\ref{tab:subduce}).},
\begin{equation}
\langle\langle AB\rangle\rangle_W^c
=
\langle\langle AB\rangle\rangle_W-
\langle\langle A\rangle\rangle_W
\langle\langle B\rangle\rangle_W.
\end{equation}
For the SD potentials one finds,
\begin{eqnarray}
\label{ef_1}
\frac{r_k}{r}{V}_1'({\mathbf r}) &=& 
\epsilon_{ijk}g^2
\lim\limits_{\tau\to\infty}\!
\int_0^\tau 
\!\!dt\,t \langle\langle {B}_i({\mathbf 0},0){E}_j({\mathbf
0},t)
\rangle\rangle_W, \\
\label{ef_2}
\frac{r_k}{r}{V}_2'({\mathbf r}) &=& \frac{1}{2}\epsilon_{ijk}
g^2\lim\limits_{\tau\to\infty}\!
\int_0^\tau
\!dt\, t\langle\langle {B}_i({\mathbf
0},0){E}_j({\mathbf r},t)
\rangle\rangle_W, \\
\label{ef_3}
R_{ij}V_3({\mathbf R}) &=& 2\,g^2\lim\limits_{\tau\to\infty}
\int_0^\tau 
\!dt\,\left[\langle\langle {B}_i({\mathbf 0},0)
{B}_j({\mathbf r},t)\rangle\rangle_W\right.\\\nonumber
&-&\frac{\delta_{ij}}{3}\left.
\langle\langle {\mathbf B}({\mathbf 0},0)\cdot{\mathbf
B}({\mathbf r},t)
\rangle\rangle_W\right],\\
\label{ef_4}
{V}_4({\mathbf R}) &=& 2\,g^2\lim\limits_{\tau\to\infty}
\int_0^\tau 
\!dt\, \langle\langle {\mathbf B}({\mathbf 0},0)\cdot
{\mathbf B}({\mathbf r},t)\rangle\rangle_W,
\end{eqnarray}
where $R_{ij}$ is defined in Eq.~(\ref{eq:rij}).
Finally, the MD potentials are,
\begin{eqnarray}
\label{md_1}
{V}_b({\mathbf r})&=&-\frac{1}{3}g^2\,\lim\limits_{\tau\to\infty}
\int_0^\tau 
\!dt\,t^2\langle\langle {\mathbf E}({\mathbf 0},0)\cdot
{\mathbf E}({\mathbf r},t)\rangle\rangle_W^c,\\
\label{md_2}
R_{ij}{V}_c({\mathbf r}) &=& g^2\lim\limits_{\tau\to\infty}
\int_0^\tau 
\!dt\,t^2\left[\langle\langle {E}_i({\mathbf 0},0)
{E}_j({\mathbf r},t)\rangle\rangle_W^c\right.\\\nonumber
&-&\frac{\delta_{ij}}{3}\left.
\langle\langle {\mathbf E}({\mathbf 0},0)\cdot{\mathbf
E}({\mathbf r},t)
\rangle\rangle_W^c\right],\\
\label{md_3}
{V}_d({\mathbf r})&=&\frac{1}{6}g^2\lim\limits_{\tau\to\infty}
\int_0^\tau 
\!dt\,t^2\langle\langle {\mathbf E}({\mathbf 0},0)\cdot
{\mathbf E}({\mathbf 0},t)\rangle\rangle_W^c,\\
R_{ij}{V}_e({\mathbf r}) &=&-\frac{1}{2}g^2
\lim\limits_{\tau\to\infty}
\int_0^\tau
\!dt\,t^2\left[\langle\langle {E}_i({\mathbf 0},0)
{E}_j({\mathbf 0},t)\rangle\rangle_W^c\right.\label{md_4}\\\nonumber
&-&\frac{\delta_{ij}}{3}\left.
\langle\langle {\mathbf E}({\mathbf 0},0)\cdot{\mathbf
E}({\mathbf 0},t)
\rangle\rangle_W^c\right].
\end{eqnarray}
While $V_0,V_b,\ldots, V_e$ have the dimension $m$,
$V_{\prime}$, $V_{1}'$ and $V_2'$ have dimension $m^2$ and
$V_{3}$, $V_4$, $\nabla^2V_a^{E}$ and $\nabla^2V_a^{B}$
have dimension $m^3$.

\begin{figure}[thb]
\centerline{\epsfxsize=10truecm\epsffile{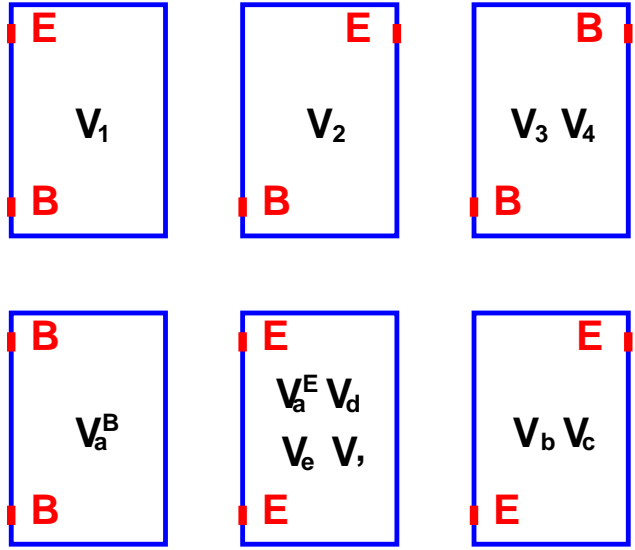}}
\caption{The nominators of the arguments of
the integrals within
Eqs.~(\ref{ce_1}) -- (\ref{md_4}).}
\label{fig:spinpot}
\end{figure}

Throughout the previous equations,
the expectation value, $\langle\langle F_1F_2\rangle\rangle_W$, is 
defined as,
\begin{equation}
\label{dex}
\left\langle\langle F_1F_2\rangle\right\rangle_{W(C)}=
\frac{\langle\mbox{Tr}\,{\mathcal P}
\left[\exp\left(ig\int_{\delta C} dx_\mu\,A_\mu\right)F_1F_2\right]\rangle}{
\langle\mbox{Tr}\,{\mathcal P}
\left[\exp\left(ig\int_{\delta C} dx_\mu\,A_\mu\right)\right]\rangle},
\end{equation}
where $\delta C$ represents a closed path [the contour of a Wilson loop,
$W({\mathbf r},T)$, $T\geq\tau$].
The nominators of Eq.~(\ref{dex}) that are required to compute the
potentials are depicted in Figure~\ref{fig:spinpot}. The correlators
appearing within the coefficient
functions of the spin-orbit potentials, $V_1'$ and $V_2'$, involve
electric and magnetic fields, the latter originating from the angular
movement. Correlators between two magnetic fields are required
in the spin-spin potentials, $V_3$ and $V_4$, which arise from
interactions between the two Fermi terms
$\propto g\,{\mathbf S_i}\cdot{\mathbf B}/m_i$. The corrections to the central
potential, $\nabla^2V_a^E$ and $\nabla^2V_a^B$, involve
electric-electric and magnetic-magnetic interactions, respectively, while
$V_{\prime}$ and
all MD corrections involve two electric field insertions.
The latter arise from re-expressing derivatives acting
on the static propagators in terms of field strength insertions.

In principle similar results that would include Wilson loops with more
than two
field strength insertions can be obtained from the order $c^{-4}$ (or $v^6$)
NRQCD Lagrangian of Ref.~\cite{Lepage:1992tx}. This tedious work
has not been done yet since the dominant sources of error at present are
the uncertainties of the matching coefficients and certain
transition matrix elements
(cf.\ Sections~\ref{sec:uncert}, \ref{sec:pnrqcd} and
\ref{sec:transi}), rather than
higher order relativistic corrections.

\subsubsection{The potentials as perturbations}
\label{sec:potper}
In Refs.~\cite{Bali:1997cj,Bali:1997am} spectral decompositions of the
above potentials have been derived.
The results can be written as follows,
\begin{eqnarray}
\label{eq:spectra1}
V_{3,4}(r)
&=&\sum_{n>0}D_n^{(3,4)}(r)\int_0^{\tau}dt\,e^{-\Delta V_n(r)t}=
\sum_{n>0}\frac{D_n^{(3,4)}(r)}{\Delta V_n(r)},\\
\nabla^2V_{a}^{E,B}(r)
&=&\sum_{n>0}D_n^{(E,B)}(r)\int_0^{\tau}dt\,e^{-\Delta V_n(r)t}=
\sum_{n>0}\frac{D_n^{(E,B)}(r)}{\Delta V_n(r)},\\
V'_{1,2}(r)
&=&\sum_{n>0}D_n^{(1,2)}(r)\int_0^{\tau}dt\,t\,e^{-\Delta V_n(r)t}=
\sum_{n>0}\frac{D_n^{(1,2)}(r)}{[\Delta V_n(r)]^2},\\
V_{\prime}(r)
&=&\sum_{n>0}D_n'(r)\int_0^{\tau}dt\,t\,e^{-\Delta V_n(r)t}=
\sum_{n>0}\frac{D_n'(r)}{[\Delta V_n(r)]^2},\\
V_{b,c,d,e}(r)
&=&\sum_{n>0}D_n^{(b,c,d,e)}(r)\int_0^{\tau}dt\,\frac{t^2}{2}
\,e^{-\Delta V_n(r)t}=
\sum_{n>0}\frac{D_n^{(b,c,d,e)}(r)}{[\Delta V_n(r)]^3},\label{eq:spectra4}
\end{eqnarray}
where $\Delta V_n(r)=V_n(r)-V_0(r)$ denotes the difference between
the $n$th hybrid excitation and the ground state $\Sigma_g^+$ potential.
The coefficients, $D_n(r)$, are real parts of products of two transition
amplitudes and can easily be read off from Eqs.~(\ref{ce_1})
-- (\ref{md_4}). 
For instance, in the case of $V_4$, one obtains,
\begin{equation}
\label{eq:specco}
D_n^4(r)=
2\,g^2\re\left[\langle\Phi_{{\mathbf r},0}|{\mathbf B}({\mathbf 0})|
\Phi_{{\mathbf r},n} \rangle
\langle \Phi_{{\mathbf r},n}|{\mathbf B}({\mathbf r})|
\Phi_{{\mathbf r},0} \rangle\right],
\end{equation}
where $|\Phi_{{\mathbf r},n}\rangle$ denotes the $n$th excitation
of a quark anti-quark state at separation, ${\mathbf r}$, and
the states are thought to be normalised, 
$\langle\Phi_{{\mathbf r},i}|\Phi_{{\mathbf r},i}\rangle=1$.
Note that, $D_n^E=-4D_n'=6D_n^d$.

Physically, the above result can be interpreted as
follows~\cite{Michael:1985wf,Bali:1997cj}:
at time $0$ the spin of the quark at position, ${\mathbf 0}$,
interacts with the
background glue and excites the gluonic vortex until, at time $\tau$
a second interaction with the spin of the anti-quark at ${\mathbf r}$ takes
place that returns the flux tube into its ground state: the non-perturbative
analogue of a gluon exchange! From
Table~\ref{tab:subduce} one can read off that in general the intermediate
state will be a superposition of excitations within the $\Sigma_u^-$ and
$\Pi_u$ channels in the particular cases of $V_3, V_4$ and $\nabla^2V_a^B$.

We add the term proportional
to $\nabla^2V^B_a$ of Eq.~(\ref{cepo}) to the
two terms proportional to $V_3$ and $V_4$ of
Eq.~(\ref{sdpo}). The result reads,
\begin{eqnarray}
\label{eq:vss}
V_{\mbox{\scriptsize ss}}(r)&=&c_F^{(1)}c_F^{(2)}
\frac{S_1^iS_2^j}{3m_1m_2}\left[3R_{ij}V_3(r)
+\delta_{ij}V_4(r)\right]\\\nonumber
&+&
\left(\frac{{c_F^{(1)}}^2}{8m_1^2}+\frac{{c_F^{(2)}}^2}{8m_2^2}\right)
\nabla^2V_a^B(r).
\end{eqnarray}
By inserting the spectral decomposition
with the correct coefficients, $D_n$, determined from
Eqs.~(\ref{ce_2}), (\ref{ef_3}) and (\ref{ef_4}), into
Eq.~(\ref{eq:vss}) one obtains,
\begin{eqnarray}
\label{eq:pert}
V_{\mbox{\scriptsize ss}}(r)&=&
\sum_{n>0}\frac{1}{V_n(r)-V_0(r)}\\\nonumber
&\times&
\left|\left
\langle\Phi_{{\mathbf r},0}\left|\left[
c_F^{(1)}\frac{g{\mathbf S}_1\cdot{\mathbf B}
({\mathbf 0})}{m_1}+c_F^{(2)}
\frac{g{\mathbf S_2}\cdot{\mathbf B}({\mathbf r})}{m_2}\right]\right|
\Phi_{{\mathbf r},n} \right\rangle\right|^2,
\end{eqnarray}
where we have exploited the fact that ${\mathbf B}^2=4
({\mathbf S}\cdot{\mathbf B})^2$.
The result is exactly the energy shift one would have expected
in second order perturbation theory from the Fermi terms
for quark and anti-quark
within Eq.~(\ref{eq:ef1quark}),
\begin{equation}
\label{eq:per2}
V_{\mbox{\scriptsize ss}}(r)=\Delta E_{\mbox{\scriptsize ss}}=
\sum_{n>0}\frac{\langle \Phi_{{\mathbf r},0}
|\Delta H_{\mbox{\scriptsize ss}}|\Phi_{{\mathbf r},n}\rangle
\langle \Phi_{{\mathbf r},n}|\Delta H_{\mbox{\scriptsize ss}}|
\Phi_{{\mathbf r},0}\rangle}{V_n(r)-V_0(r)},
\end{equation}
with
\begin{equation}
\label{eq:pertu}
\Delta H_{\mbox{\scriptsize ss}}({\mathbf x}) = 
\frac{c_F^{(1)}g{\mathbf \sigma_1}\cdot{\mathbf B}({\mathbf x})}{2m_1}
\delta^3({\mathbf x})
+\frac{c_F^{(2)}g{\mathbf \sigma_2}\cdot{\mathbf B}({\mathbf x})}{2m_2}
\delta^3({\mathbf x}-{\mathbf r}),
\end{equation}
where ${\mathbf \sigma}_i=2{\mathbf S}_i$.

Other potentials in their spectral representation
can be interpreted as perturbations too.
However, relating these to the NRQCD Lagrangian
requires somewhat more involved formal
manipulations. From the considerations above it is obvious that
the formalism cannot readily be applied to spin dependent
interactions of hybrid quarkonia where the $\Sigma^+_g$
ground state would appear as an intermediate state:
since $\exp[-(V_n-V_m)t]$ diverges with $t$
for $m>n$; the
matrix elements corresponding to Eq.~(\ref{eq:per2}),
for an external state, $|\Phi_{{\mathbf r},m}\rangle$, cannot be
obtained from a simple time integral over double bracket expectation values.

\subsection{Model expectations}
We discuss expectations for the potentials and the resulting
Hamiltonian,
and discuss the Lorentz structure of the effective
interaction kernel.

\subsubsection{The potentials}
We will present simple model expectations for the above potentials.
The double bracket expectation values of colour field operators
can be obtained from
infinitesimal deformations of a generalised, non-static
Wilson loop\footnote{For the definition of the functional
derivative acting on a Wilson loop with
respect to a surface element, $S_{\mu\nu}(x)$,
see e.g.\ Ref.~\cite{Migdal:1984gj}.}~\cite{Brambilla:1994zw},
\begin{equation}
g^2\langle F_{\mu\nu}(x)F_{\rho\sigma}(y)\rangle\rangle^c_W=
-\frac{\delta^2\ln\langle W\rangle}{\delta S_{\mu\nu}(x)\delta
  S_{\rho\sigma}(y)}.
\end{equation}
If the functional dependence of the Wilson loop expectation value
on its contour is known, the above formula can be applied to
calculate the corresponding long distance behaviour of the
potentials. This has been done for the stochastic vacuum model (SVM),
dual QCD and the area law assumption in
Refs.~\cite{Baker:1996mu,Brambilla:1997aq}.
A variety of predictions on SD and MD potentials exists in the literature
that are based on effective modified one gluon exchanges or Bethe
Salpeter kernels. Ref.~\cite{Ebert:1999xv} represents
a recent example\footnote{Note, however, that their result
is incompatible with Eqs.~(\ref{md_1}) -- (\ref{md_4}).}.
Given these different suggestions,
lattice results with a precision that is sufficient to discriminate between
them are highly desirable.

Here, we shall only discuss the area law
expectations~\cite{Barchielli:1988zs,Barchielli:1990zp},
combined with tree level
perturbation theory and constraints from the
renormalisation group mixing between the potentials~\cite{Bali:1997am},
Eqs.~(\ref{vabsca}) -- (\ref{v4sca}),
\begin{eqnarray}
V_0(r;\mu)&=&V_{\mbox{\scriptsize self}}(\mu)
-\frac{e}{r}+\sigma r,\label{v0pa}\\
\nabla^2V_a^E(r;\mu)&=&C_a^E(\mu)-
\frac{2\sigma+b(\mu)}{r},\label{vaepa}\\
\nabla^2V_a^B(r;\mu)&=&C_a^B(\mu),\label{vabpa}\\
V_1'(r;\mu)&=&-\frac{h(\mu)}{r^2}-
\sigma,\label{v1pa}\\
V_2'(r;\mu)&=&\frac{e-h(\mu)}{r^2},\label{v2pa}\\
V_3(r;\mu)&=&3\frac{e-h(\mu)}{r^3},\label{v3pa}\\
V_4(r;\mu)&=&8\pi [e-h(\mu)]\delta^3(r),\label{v4pa}\\
V_{\prime}(r)&=&0,\label{vprimepa}\\
V_b(r;\mu)&=&C_b(\mu)+\frac{2}{3}\frac{e}{r}-\frac{\sigma}{9}r,\label{vbpa}\\
V_c(r)&=&-\frac{1}{2}\frac{e}{r}-\frac{\sigma}{6}r,\label{vcpa}\\
V_d(r;\mu)&=&C_d(\mu)-\frac{\sigma}{9} r,\label{vdpa}\\
V_e(r)&=&-\frac{\sigma}{6} r,\label{vepa}
\end{eqnarray}
with\footnote{In one loop perturbation theory
one obtains~\cite{Brambilla:2000gk},
$V_{\prime}(r)=-C_FC_A\alpha_s^2/(4r^2)$.} $e-h(\mu)\approx C_F\alpha_s$.
The above formulae conform with the
Gromes and BBP relations,
Eqs.~(\ref{grom}) -- (\ref{bram}).
While $V_0$, $V_{\prime}$ and the MD potentials
do not undergo multiplicative renormalisation,
$V_0$, $V_b$ and $V_d$ still contain additive self-energy contributions
that will diverge as $\mu\rightarrow\infty$ and whose $\mu$ dependence
has to
be cancelled by the quark mass shifts\footnote{We ignore the
possibility of self energy contributions to $V_{\prime}$,
$V_c$ and $V_e$ that vanish at least
to lowest order perturbation theory
in the parametrisation.
In lattice determinations of
$V_c$ and $V_e$ these have indeed been found to
agree with zero within errors~\cite{Bali:1997am}.}, $\delta m_i$.
The constants, $V_{\mbox{\scriptsize self}}(\mu)$, $C_a^E(\mu)$,
$C_a^B(\mu)$, $C_b(\mu)$  and $C_d(\mu)$ as well as all terms
proportional to $e$ originate from perturbation theory, while all
terms proportional to the string tension, $\sigma$, are due to the
area law ansatz, with the exception of $\nabla^2V_a^E$.
We allow for terms proportional to $h$ in $V_1'$ and
$V_2'$ that are thought to originate from the mixing between
these two potentials
under renormalisation group transformations, Eq.~(\ref{v1sca}).
In perturbation theory as well as in vector exchange models,
one obtains, $V_3(r)=V_2'/r-V_2''$ and $V_4=2\nabla^2V_2$. Therefore,
replacing $C_F\alpha_s$ by $e-h$ within these potentials appears
to be reasonable. However, we remark that further corrections
to this ansatz must exist since the scaling behaviours
under $\mu\rightarrow\mu'$ of $V_2$ [Eq.~(\ref{v2sca})],
$V_3$ [Eq.~(\ref{v3sca})] and $V_4$ [Eq.~(\ref{v4sca})] are incompatible
with each other.
Finally, the expectation, $2\sigma+b(\mu)$, within $\nabla^2V_a^E$
is motivated by the lattice results to be presented in
Section~\ref{sec:latsdres} as well
as by dual QCD and SVM
calculations~\cite{Baker:1996mu,Brambilla:1997aq}.
One would expect additional $\delta$-like contributions to
$\nabla^2V_a^E$ and $\nabla^2V_a^B$ from
Eq.~(\ref{vabsca}),
which we ignore for the moment.

Interestingly, by adding a perimeter term to the Wilson loop
area law~\cite{Barchielli:1988zs,Barchielli:1990zp},
one obtains a non-vanishing $C_d=-V_{\mbox{\scriptsize self}}/4$,
which agrees with the expectation from perturbation theory. However,
the perimeter term does not contribute to
$C_a^E$, $C_a^B$ or $C_b$.
In continuum perturbation theory
as well as in lattice perturbation theory in the
infinite volume limit, one obtains
the tree level results~\cite{Bali:1997am},
\begin{equation}
C_b(\mu)=0,\quad
C_d(\mu)=-\frac{1}{4}V_{\mbox{\scriptsize self}}(\mu),
\end{equation}
where the lattice perturbation theory result for
$V_{\mbox{\scriptsize self}}(a)$ with the Wilson action is given
in Eq.~(\ref{eq:self2}).
By using the lattice field definitions
of Eqs.~(\ref{eq:field1}) -- (\ref{eq:field3}), we obtain the
lattice perturbation theory results~\cite{Bali:1997am},
\begin{eqnarray}
C_a^E(a)&=&-C_F\alpha_s a^{-3}\times 7.91084\ldots,\\
C_a^B(a)&=&C_F\alpha_s a^{-3}\times 14.89413\ldots,
\end{eqnarray}
for the other two self-energy contributions.

\subsubsection{The Hamiltonian}
The Hamiltonian that results from the ansatz Eqs.~(\ref{v0pa}) -- (\ref{vepa}),
in the equal mass
case, $m=m_1=m_2$, takes the form,
\begin{eqnarray}
H&=&H_0+\delta H_{\mbox{\scriptsize kin}}+\frac{1}{m^2}\left(\delta H_{\delta}
+\delta H_{\mbox{\scriptsize MD}}+\delta H_{\mbox{\scriptsize SD}}\right),\\
H_0&=&2(m-\delta m)+V_{\mbox{\scriptsize self}}+\frac{1}{4m^2}
\left(c_DC_a^E+c_F^2C_a^B\right)\nonumber\\
&+&\left[1-\frac{1}{2m}\left(V_{\mbox{\scriptsize self}}
+4C_b\right)\right]\frac{p^2}{m}\label{eq:h0}\\\nonumber
&-&
\left[e+\frac{3c_Db+2\sigma}{12m^2}\right]\frac{1}{r}
+\sigma r,\\
\delta H_{\mbox{\scriptsize kin}}&=&-\frac{p^4}{4m^3},\\\label{eq:hdelta}
\delta H_{\delta}&=&
\left(\frac{3}{4}+d_s\right)4\pi e\delta^3({\mathbf r}),\\\label{eq:hmd}
\delta H_{\mbox{\scriptsize MD}}&=&-\frac{\sigma}{6r}{\mathbf L}^2
-\frac{e}{r}\left(p^2-\frac{{\mathbf L}^2}{2r^2}\right),\\
\delta H_{\mbox{\scriptsize SD}}&=&
\left[-\frac{\sigma}{r}+\frac{4c_F(e-h)-e}{r^3}\right]
\frac{{\mathbf L}\cdot{\mathbf S}}{2}\nonumber\\\label{eq:hsd}
&+&\frac{3c_F^2(e-h)}{r^3}T+\left[2c_F^2(e-h)-12d_v e\right]4\pi
\delta^3({\mathbf r})\frac{{\mathbf S}_1\cdot{\mathbf S}_2}{3},
\end{eqnarray}
with
\begin{eqnarray}
{\mathbf L}^2&=&l(l+1)\\
\frac{{\mathbf S}_1\cdot{\mathbf
    S}_2}{3}&=&\frac{1}{6}\left[s(s+1)-\frac{3}{2}\right],\\
{\mathbf L}\cdot{\mathbf S}&=&\frac{1}{2}\left[
J(J+1)-l(l+1)-s(s+1)\right],\\
T&=&R_{ij}S_1^iS_2^j=-\frac{6({\mathbf L}\cdot{\mathbf S})^2
+3{\mathbf L}\cdot{\mathbf S}-2s(s+1)l(l+1)}{6(2l-1)(2l+3)}.
\end{eqnarray}
For a discussion of the non-equal mass case we refer the reader 
to Ref.~\cite{Bali:1997am}. 
The parametrisations of the potentials that enter the above Hamiltonian,
can of course be improved
in several ways,
for example~\cite{Brambilla:1998vm,Brambilla:1998qg}
by including the known one loop perturbative
results for the spin dependent terms~\cite{Gupta:1982kp,Pantaleone:1986uf}
and the two loop result for the static
potential~\cite{Peter:1997me,Schroder:1998vy}. Note that
all terms containing the low-energy parameter, $\sigma$,
are independent of the matching scale, $\mu\gg\sqrt{\sigma}$.

We have eliminated $C_d$ from
the above Hamiltonian by use of Eq.~(\ref{bram2}),
$V_{\mbox{\scriptsize self}}=-2C_b-4C_d$. The subscripts of
the correction terms, $\delta H_i$, do not necessarily relate to
the potentials of origin.
$H_0$ contains contributions from $V_0$, $V_{\mbox{\scriptsize C}}$
as well as from $V_{\mbox{\scriptsize MD}}$ while
$\delta H_{\delta}$ contains terms due to $V_{\mbox{\scriptsize C}}$
and $V_{\mbox{\scriptsize MD}}$.
We have used the relation $2\pi\langle \delta^3({\mathbf r})\rangle
=-i\langle r^{-3}{\mathbf r}\cdot{\mathbf p}\rangle$ to cast
a term that appears within $V_{\mbox{\scriptsize MD}}$ into
a $\delta$ function.
The radial Schr\"odinger equation, Eq.~(\ref{eq:radial}),
can be solved numerically
for $H_0$ and, subsequently, the $\delta H_i$ terms can conveniently be
treated as perturbations.

We substitute,
\begin{equation}
\label{eq:h00}
\tilde{m}=m+\frac{V_{\mbox{\scriptsize self}}+4C_b}{2}
\end{equation}
into $H_0$. To order $m^{-2}$ this yields,
\begin{eqnarray}
\label{eq:h02}
H_0&=&C_0+2\tilde{m}+\frac{p^2}{\tilde{m}}+\tilde{V}(r),\\
\tilde{V}(r)&=&-\frac{\tilde{e}}{r}+\sigma r,\\
\label{eq:h03}
C_0&=&-2\delta m
-4C_b+\frac{1}{4{\tilde{m}}^2}\left(c_DC_a^E+c_F^2C_a^B\right),\\
\label{eq:h04}
\tilde{e}&=&e+\frac{3c_Db+2\sigma}{12\tilde{m}^2}:
\end{eqnarray}
the static quark self-energy shift, $V_{\mbox{\scriptsize self}}$,
is eliminated from the Hamiltonian.
This was first noticed in Ref.~\cite{Barchielli:1990zp}.
The remaining scale dependence of $\delta m(\mu)$
has to compensate that of the sum of the (small) term
$C_b(\mu)$, which vanishes in tree level perturbation
theory, and the term containing
$C_a^E$ and $C_a^B$, which is suppressed by a factor $\tilde{m}^{-2}$.
Moreover, $C_a^E$ and $C_a^B$ have different relative signs, such that
partial cancellations occur.
Of course, the above substitution is only valid for quark masses,
$m\gg V_{\mbox{\scriptsize self}}\propto \alpha_s\mu$. This relation,
however, is automatically fulfilled for matching scales, $\mu<m$.
In conclusion: the mass shift,
$\delta m$, which is related to the wave function renormalisation,
becomes reduced as
relativistic
corrections are taken into account.

Some of the correction terms are well known from atomic physics, others
are specific to non-Abelian gauge theories. One piece of
$\delta H_{\delta}$ [Eq.~(\ref{eq:hdelta})] as well as
the term
proportional to $b/r$ within Eq.~(\ref{eq:h0}) stem from
the Darwin interaction.
A string whose energy density, $\sigma$,
is carried by a constant longitudinal electric
field~\cite{Buchmuller:1982fr}, gives rise to the
(classical) orbit-orbit interaction
term, $-\sigma/(6r){\mathbf L}^2$, that appears
within Eq.~(\ref{eq:hmd}).
$\delta H_{\mbox{\scriptsize SD}}$ [Eq.~(\ref{eq:hsd})]
contains a spin-orbit (Thomas) interaction term
that, unlike its QED counterpart, only falls off like
$r^{-1}$ at large distances.
In addition, it contains two spin-spin interaction terms
that
take very much the same form as in QED,
the first of which does not affect $S$ waves and the second of which
only affects $S$ waves to the order in $\alpha_s$ considered above.

\subsubsection{The Lorentz structure of the effective interaction}
The general form of a Hamiltonian governing relativistic two-particle
bound states has been derived within the Bethe-Salpeter formalism
(see e.g.~\cite{Lucha:1991vn} and references therein), under the assumption
that the interaction kernel only depends on the transfer momentum, $q^2$:
The momentum space kernel can be decomposed into the five
Lorentz invariants,
\begin{equation}\label{eq:lorinv}
\tilde{I}=\tilde{V_S}\,{\mathbf 1}\otimes {\mathbf 1}
+\tilde{V_V}\,\gamma_{\mu}\otimes\gamma_{\mu}+
\tilde{V_T}\frac{1}{2}\sigma_{\mu\nu}\otimes\sigma_{\mu\nu}
+\tilde{V_A}\,\gamma_{\mu}\gamma_5\otimes\gamma_{\mu}\gamma_5
+\tilde{V_P}\,\gamma_5\otimes\gamma_5,
\end{equation}
where the form factors, $\tilde{V}_i$, only depend on $q^2$.
In the QED case, within the ladder approximation, only
$V_V$ assumes a non-trivial value and
the resulting Hamiltonian has the Breit-Fermi form, well known
from atomic physics. In the most general case~\cite{Lucha:1991vn}
the equal mass Hamiltonian reads,
\begin{eqnarray}
V(r)&=&V_V(r)+V_S(r)
+4\left[V_T(r)-V_A(r)\right]{\mathbf S}_1\cdot
{\mathbf S}_2\nonumber\\\label{eq:convo}
&+&\frac{1}{m^2}\left\{\frac{1}{4}\nabla^2 V_V(r)+\frac{{\mathbf L}\cdot
{\mathbf S}}
{2r}\left[3V_V'(r)-V_S'(r)\right]\right.\\\nonumber
&+&T\left[\frac{V_V'(r)-V_P'(r)}{r}+V_P''(r)-V_V''(r)\right]\\\nonumber
&+&\left.\frac{{\mathbf S}_1\cdot{\mathbf S}_2}{3}
\left[2\nabla^2V_V(r)+\nabla^2V_P(r)\right]
\right\},
\end{eqnarray}
where we have ignored momentum dependent terms as well as
possible $m^{-2}$ corrections to $V_A(r)$ and $V_T(r)$.

Since QCD interactions are spin-independent to leading order,
$V_A(r)=V_T(r)$ must be satisfied.
Moreover, in comparing the above formula with the potential
of Eqs.~(\ref{prp}) -- (\ref{mdpo}), with tree level matching constants,
$c_i=1, d_i=0$, one finds, 
\begin{eqnarray}
V_0&=&V_V+V_S,\\
V_1&=&-(1-\eta)V_S,\\
V_2&=&V_V+\eta V_S,\label{eq:v2co}\\
V_3&=&\frac{V_V'-V_P'}{r}-(V_V''-V_P''),\label{eq:v3co}\\
V_4&=&2\nabla^2V_V+\nabla^2V_P,\label{eq:v4co}
\end{eqnarray}
where we have also used the relation, Eq.~(\ref{grom}).
There are indications that the linear term, $\sigma\,r$, within $V_0$
should be
purely scalar since vector type potentials
are thought to rise at most
logarithmically in $r$~\cite{Gromes:1988zx}.
The Darwin term appearing within Eq.~(\ref{cepo}) implies
that $b=0$, i.e.\ any scalar contribution to $\nabla^2V_0$
has to be cancelled by
the combination, $\nabla^2V_a^E+\nabla^2V_a^B$.
It is clear that the picture becomes more involved when
the matching coefficients assume non-trivial values. Moreover,
the assumption that the interaction kernel
only depends on the momentum transfer does not necessarily apply.

\subsection{Beyond the adiabatic approximation}
\label{sec:retard}
We shall briefly discuss the interrelation between local potentials,
sum rules and the stochastic vacuum model. Following this, we shall
describe pNRQCD which is a systematic and conceptionally
attractive approach to quarkonia bound state problems. Subsequently,
we will discuss some consequences that arise from including
MD potentials. Finally, we incorporate hybrid states
and transitions between different gluonic excitations of the string
into the potential approach.

\subsubsection{Are potentials enough?}
The local potential picture of heavy quark bound states has often been
challenged. Voloshin~\cite{Voloshin:1979hc}
and Leutwyler~\cite{Leutwyler:1981tn}
for instance investigated the effect of the gluon
condensate on quarkonia levels and found a dependence
proportional to $n^6\langle \alpha_s F^2\rangle$, on the
principal quantum number, $n$. From this they concluded that
this effect could not be reproduced by a local potential.
However, a term growing that rapidly would certainly dominate
the spectrum, if not for $n=2$, then for $n=3$, in contradiction
to experiment. In this light, it appears questionable whether
all non-perturbative physics relevant for excited state quarkonia
can be approximated by the gluon condensate alone or if other infra-red scales
play a r\^ole. The gluon condensate does not result in a linear
contribution to the static potential
but will only add a short distance
term, proportional~\cite{Balitsky:1985iw} to $r^2$,
to the perturbative result. Thus, the gluon condensate alone is
not sufficient for an understanding of
non-perturbative physics at large (as well as small,
cf.~Section~\ref{sec:beyond}) distances.
Based on somewhat different arguments this has also been pointed out in
Refs.~\cite{Gromes:1982su,Bertlmann:1983pf}.

One instructive
extension of the sum rule approach is the stochastic vacuum model (SVM)
by
Dosch and Simonov~\cite{Dosch:1987sk,Dosch:1988ha,Simonov:1988rn} in which
non-local
condensates, i.e.\ correlators of field strength tensors
at different space-time points,
\begin{equation}
\label{eq:ds}
D(x)=\langle \alpha_s F_1(x)
U^A(x)F_2(0)\rangle,
\end{equation}
are introduced. $F_i$ symbolise
linear combinations of electric or magnetic fields.
Calculating a Wilson loop in this approach indeed yields
a linear contribution to the static potential at
large distances~\cite{Brambilla:1997aq}.
In order to achieve gauge
invariance of the correlation function, the 
adjoint Schwinger line, $U^A$, has been included into the
definition, Eq.~(\ref{eq:ds}).
Note that the above non-local condensate resembles the gluelump
correlator of Eq.~(\ref{eq:corgl}). It is not
entirely clear how to cancel the self-energy contribution that is due
to the Schwinger line and how to interprete the possible path dependence
of the result. Putting these problems aside for the moment,
lattice determinations of such
correlators by use of two different methods
exist~\cite{DelDebbio:1994zn,Bali:1998aj,Ilgenfritz:1999tg}.

The correlator will decay exponentially for large Euclidean
separations, $D(x)\propto\exp(-|x|/T_G)\quad (|x|\rightarrow\infty)$,
with the gluon correlation time, $T_G$, being a second dimensionful
infra-red scale. Let us further introduce the characteristic
time scale, associated with a quark in a bound state,
$T_{nl}\propto \bar{\Lambda}_{nl}^{-1}\propto
1/(mv_{nl}^2)$.
One can now distinguish between two limits. In the case,
$T_G\gg T_{nl}$, the non-local condensate can be well
approximated by a local condensate. Therefore,
the Leutwyler~\cite{Leutwyler:1981tn} result is
reproduced and no local potential that describes the long distance
behaviour can be found. This is not too surprising, though,
as one would expect the adiabatic approximation to be violated
if the characteristic time scale of the gluon dynamics becomes
larger than that associated to the heavy quarks.
On the other hand, for gluons harder than the bound state energies,
$T_G\ll T_{nl}$, the effect of the non-local condensate cannot be
neglected and under certain additional model assumptions
one indeed finds level splittings to scale like~\cite{Marquard:1987pe},
$\Delta E_{nl}\propto T_G\langle r^2\rangle$. This would
imply the local potential itself to be proportional
to $r^2$ at small distances, in contradiction to the lattice
results but in agreement with sum rule expectations\footnote{
The static potential differs
from
the interaction potential between moving quarks of finite mass.
Sum rules predict the latter to be proportional to $r^3$ at short
distances~\cite{Leutwyler:1981tn}.}
on the static potential~\cite{Balitsky:1985iw}. 
However, one would not expect the SVM to reproduce the correct behaviour
for distances, $r<T_G$, anyway.

\subsubsection{Potential NRQCD}
\label{sec:pnrqcd}
A more systematic approach to the bound state problem is
potential NRQCD (pNRQCD)~\cite{Pineda:1997bj,Beneke:1999ff,Brambilla:1999xf},
the QCD generalisation of
pNRQED~\cite{Caswell:1986ui,Labelle:1992hd,Labelle:1998en}
in which on top of the NRQCD Lagrangian, an expansion in terms of
the quark separation, $r\propto 1/(mv)$, is performed. The remaining
colour fields  are living at the centre of
mass coordinate, ${\mathbf 0}$. By means of a multipole expansion,
$A_{\mu}({\mathbf r},t)$ can be obtained from $A_{\mu}({\mathbf 0},t)$
and derivatives thereof.
The resulting Lagrangian is~\cite{Brambilla:1999xf},
\begin{eqnarray}
{\mathcal L}_{pNRQCD}&=&-\Tr\left\{S^{\dagger}\left[\partial_4+V_s(r)-
\frac{{\mathbf \nabla}^2}{2\mu_R}+\cdots\right]S\right.\nonumber\\
&+&O^{\dagger}\left[D_4+V_o(r)-\frac{{\mathbf D}^2}{2\mu_R}+
\cdots\right]O\label{eq:pnrq}\\\nonumber
&+&gV_A(r)\left(O^{\dagger}{\mathbf r}\cdot{\mathbf E}S+S^{\dagger}
{\mathbf r}\cdot{\mathbf E}O\right)\\\nonumber\
&+&\left.
g\frac{V_B(r)}{2}\left(O^{\dagger}{\mathbf r}\cdot{\mathbf E}O+O^{\dagger}
O{\mathbf r}\cdot{\mathbf E}\right)\right\}+\cdots,
\end{eqnarray}
where $V_s(r)$, $V_o(r)$, $V_E(r)$ and $V_B(r)$ represent
(infinitely many) matching coefficients that have to be determined
by some prescription.
Apart from $r$ the coefficients depend on the scale $\mu$ and, to
higher orders of the expansion, spins and momenta.
Since all $r$ dependence has been separated from the
interaction terms, these can be factorised according to
their properties under local gauge transformations.
$V_s(r)$ and $V_o(r)$ can be identified with the singlet and
octet potentials of Section~\ref{sec:pert} in the case that
no relevant physical
scale exists between $mv$ and $mv^2$; $S$ is the colour
singlet contribution to the wave function while $O$ represents the
colour octet part.

Interestingly, in the situation, $\bar{\Lambda}\approx
\Lambda_{QCD}$, a non-perturbative $r^2$ contribution
to $V_s$ is obtained, in agreement with Ref.~\cite{Balitsky:1985iw}.
For details we refer the reader to
Ref.~\cite{Brambilla:1999xf}. We also remark that in Ref.~\cite{Luke:1999kz}
vNRQCD is introduced which is based on a similar multipole expansion
in momentum, rather than in position space.

While in pNRQCD local and non-local terms are clearly separated,
unfortunately, it is not clear how to arrange for such a factorisation
in lattice simulations. Moreover, once the matching coefficients, $V_i(r)$, are
determined, all remaining dynamics are ultra-soft, requiring lattice
resolutions, $a^{-1}$, of order $mv$ or smaller. This would result in
intolerably large
discretisation errors, unless one is interested
in top quarks. However, the form of the pNRQCD Lagrangian
with its transitions between singlet and octet states is quite instructive.

\subsubsection{Consequences of momentum dependence}
We will briefly discuss an effect that is sometimes mistaken as
a violation of the adiabatic approximation:
let us
assume that the spectrum, $E_N$, $N=\{nll_3\}$, and
Coulomb gauge wave functions, $\psi_N({\mathbf r})=
\langle {\mathbf r}|\psi_N\rangle$, of a quarkonium bound state
are known. In this case one might attempt to determine the interaction
potential from the Schr\"odinger equation,
\begin{equation}
H|\psi_N\rangle=E_N|\psi_N\rangle.
\end{equation}

In the non-relativistic case, we have $d{\mathbf r}/dt={\mathbf p}/{\mu_R}$.
Therefore,
\begin{equation}
\label{eq:easy}
[H,{\mathbf r}]=-i\frac{d{\mathbf r}}{dt}=-\frac{i}{\mu_R}{\mathbf p}.
\end{equation}
Let us consider a Hamiltonian of the form,
\begin{equation}
H=\frac{p^2}{2\mu_R}+H_i.
\end{equation}
From the canonical commutation relation, $[{\mathbf p},{\mathbf r}]=-i$,
and Eq.~(\ref{eq:easy}), one can easily see that,
\begin{equation}
\label{eq:commute}
[H_i,{\mathbf r}]=0,
\end{equation}
i.e.\ the interaction term, $H_i=V_0({\mathbf r})$,
is only a function of the distance and does not
depend on the momentum.
In this case,
the potential can be obtained, wherever $\psi_N(r)\neq 0$,
\begin{equation}
\label{eq:vext}
V_0(r)=E_N-\frac{1}{2\mu_R}
\frac{\langle r|p^2|\psi_N\rangle}{\langle r|\psi_N\rangle},
\end{equation}
where we have assumed rotational symmetry. Note that $V_0$ does not depend
on the state $|\psi_N\rangle$ under consideration!

To higher orders of the non-relativistic expansion
not only spins and angular momentum have to be included into
the set of canonical coordinates but also
Eq.~(\ref{eq:commute}) will in general be violated:
the interaction Hamiltonian contains
the explicitly 
momentum dependent terms of
Eq.~(\ref{mdpo}).
Ignoring SD terms
as well as the correction to the
kinetic energy to keep the expressions simple, we have,
\begin{equation}
\label{eq:mdquas}
H_i=V_0(r)+V_{\mbox{\scriptsize MD}}(r,{\mathbf p}).
\end{equation}
Na\"{\i}vely applying Eq.~(\ref{eq:vext})
will result in the effective interaction potential (due
to being  forced to depend only on
the position variables)
to change with the state under consideration,
\begin{equation}
V_N(r)=V_0(r)+
\frac{\langle r|V_{\mbox{\scriptsize MD}}(r,{\mathbf p})
|\psi_N\rangle}{\langle r|\psi_N\rangle}+\cdots;
\end{equation}
this dependence of $V_N(r)$ on the state has nothing to
do with the Lamb shift of
QED since the
(MD) potential,
$V(r,{\mathbf p})=V_0(r)+V_{\mbox{\scriptsize MD}}(r,{\mathbf p})$,
of Eq.~(\ref{eq:mdquas}) does of course not dependent on the quantum numbers
$N$.

\subsubsection{What is the effect of hybrid states?}
\label{sec:transi}
From the discussion of Section~\ref{sec:hybrid} it is clear that
gluonic excitations can play an important r\^ole in bound state
problems. In general, the
total angular momentum will be the sum of the angular momentum due to
the relative movement of the quarks within the
bound state, ${\mathbf L}={\mathbf r}\wedge{\mathbf p}$, and the spin of
the gluons, ${\mathbf S}_g$: ${\mathbf K}={\mathbf L}+{\mathbf S}_g$.
$\Lambda_{\hat{\mathbf r}}={\mathbf S}_g\hat{\mathbf r}$ denotes
the projection of the gluon spin onto the inter-quark axis and
$\Lambda=|\Lambda_{\hat{\mathbf r}}|$.
${\mathbf K}^2$ has eigenvalues, $k(k+1)$, $k\geq\Lambda$.
${\mathbf K}$ will
couple to the quark spin to give the total spin of the state,
${\mathbf J}={\mathbf K}+{\mathbf S}$. We also
recall that the gluonic string
could be classified with respect to, $\Lambda^{\sigma_v}_\eta$,
where $\eta$ denotes the combined parity under charge inversion and
reflection about the midpoint of the axis, and $\sigma_v$
denotes the symmetry under reflection with respect
to a plane, containing the axis.
$\Sigma$ states with $\sigma_v=\pm 1$ fall into two
different irreducible representations of the relevant symmetry group,
$D_{\infty h}$, while irreducible representations,
$\Lambda\geq 1$, which are two-dimensional,
contain both
$\sigma_v$ parities: $|\Lambda_{\pm}\rangle=2^{-1/2}\left(
|\Lambda\rangle\pm |\!\!-\!\!\Lambda\rangle\right)$, with
$\sigma_v|\Lambda_{\pm}\rangle=\pm |\Lambda_{\pm}\rangle$.
The resulting (hybrid-) quarkonium state has the symmetries,
\begin{equation}
P=\sigma_v(-1)^{k+\Lambda+1},\quad
C=\sigma_v\eta(-1)^{k+\Lambda+s}.
\end{equation}

\begin{table}
\caption{Combinations of spins and angular momenta that can couple
to $J^{PC}=1^{--}$.}
\label{tab:jpc}

\begin{center}
\begin{tabular}{c|c|c}
$k$&$s$&$\Lambda^{\sigma_v}_{\eta}$\\\hline
$S$&1&${\mathbf \Sigma_g^+}$\\
$P$&0&$\Sigma_u^-$\\
$P$&0&$\Pi_u$\\
$P$&1&$\Sigma_g^-$\\
$P$&1&$\Pi_u$\\
$D$&1&${\mathbf \Sigma_g^+}$\\
$D$&1&$\Pi_u$\\
$D$&1&$\Delta_g$
\end{tabular}
\end{center}
\end{table}

In general, many possibilities exist to realise a given $J^{PC}$ assignment.
In Table~\ref{tab:jpc}, we illustrate this by listing all combinations that
yield a vector, $J^{PC}=1^{--}$. Note that even without considering
hybrids, the state can either be an $S$ ($k=0$) wave or a $D$
($k=2$) wave.
In a direct lattice NRQCD simulation of the spectrum, all the
above combinations
will share the same $1^{--}$ ground state and none of the quantum numbers,
$s,k,\Lambda$, are strictly
conserved. However, we shall see that mixing between
$S$ and $D$ waves for instance is likely to be small, such that almost
pure $S$ or $D$ states, that can be created by different almost
orthogonal operators, should still be distinguishable.
In the potential approach mixing effects have been completely
neglected so far and they may matter, at least for high radial excitations.
Dipole transitions are suppressed by order $c^{-1}$ in the
NRQCD velocity expansion while quadrupole transitions
are accompanied by pre-factors, $c^{-2}$. Dipole induced
mixing effects will be suppressed by order $c^{-2}$ with respect to the
leading order NRQCD Lagrangian and should, therefore, be included
into an order $c^{-2}$ spectrum calculation.
$k_3$ will not be affected by magnetic dipole transitions,
however, $s_3$ and $\eta$ are changed. Magnetic transitions
also alter the $D_{\infty h}$ representation:
the ${}^3S_1$ state in the Table
can mix with hybrid ${}^1P_1$ states, which contain a flux tube
in the $\Sigma_u^-$ or in the $\Pi_u$ representation.

Electric dipole transitions cannot affect $s_3$ or $\eta$
but change $k_3$. As the Table reveals, only the mixing
of ${}^3S_1$ $\Sigma_g^+$ states
with ${}^3P_1$ $\Sigma_g^-$ states is possible in this case. We shall,
however, see that the corresponding transition amplitude vanishes
identically. Either a quadrupole transition or two separate
dipole transitions connect the $S$ and $D$ wave $\Sigma_g^+$ states.
Therefore, mixing effects between these channels
only have to be considered from order $c^{-4}$ onwards.

\begin{figure}[thb]
\centerline{\epsfxsize=8truecm\epsffile{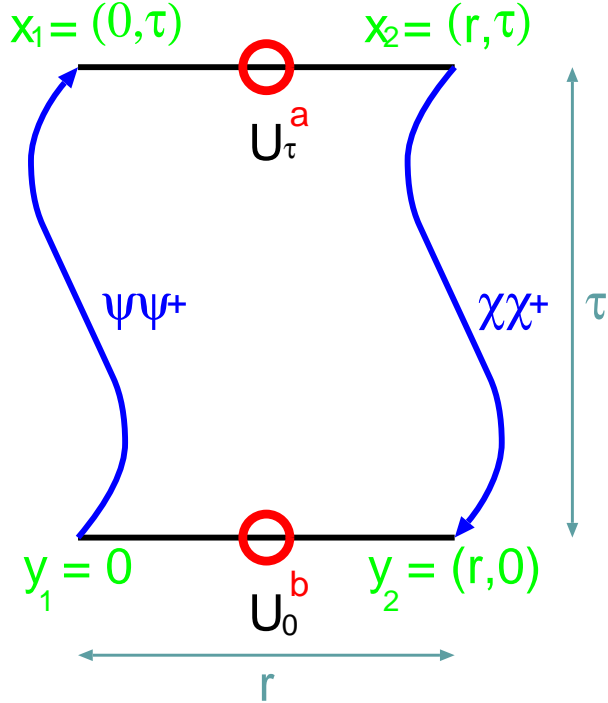}}
\caption{The four point function, $G_{ab}$.}
\label{fig:4point2}
\end{figure}

In the derivation of the Schr\"odinger type bound state equation,
Eq.~(\ref{eq:rr2}), from Eq.~(\ref{eq:rr1}) we have assumed that
quarkonia can be completely classified by the quantum numbers
of the constituent quarks. If this is not the case,
the two-particle Green function, $G$, of Eq.~(\ref{eq:4point})
and Figure~\ref{fig:4point} has to be generalised to the $G_{ab}$
of Figure~\ref{fig:4point2}, where the indices, $a$
and $b$, run over all excitations that will contribute to the $J^{PC}$
of interest.
To account for energy level shifts of
$S$ wave vector mesons,
$\Upsilon(nS)$, to order $c^{-2}$, clearly only
$a,b=\Sigma_g^+,\Sigma_u^-,\Pi_u$ are relevant. All other channels
decouple to this order in $c^{-1}$.
The Hamiltonian, $H$, acts on $G_{ab}$,
\begin{equation}
\label{eq:rr3}
-\frac{d}{dt}G_{ac} = \sum_b H_{ab}G_{bc},
\end{equation}
and the resulting
Schr\"odinger equation reads,
\begin{equation}
\label{eq:schr4}
\sum_b H_{ab}\psi_{nJ^{PC}}^b({\mathbf r})=E_{nJ^{PC}}
\psi_{nJ^{PC}}^a({\mathbf r}).
\end{equation}
The normalisation is such that,
$\sum_a\langle \psi_a|\psi_a\rangle=1$.
Note that the state vector, $(\psi^a_{nJ^{PC}})$,
now contains information about gluonic excitations too.

The ${\mathcal O}(1)$ Hamiltonian is diagonal
in the space of hybrid excitations and, to this order,
${\mathbf K}$, ${\mathbf S}$ and $\Lambda^{\sigma_v}_{\eta}$
are separately conserved. However, to ${\mathcal O}(c^{-2})$,
off-diagonal elements appear and the direction of $\psi$ will
change with time. To compute the off-diagonal elements of
$H$ we introduce $W_{ab}$, Wilson loops where the spatial
transporter at $t=0$ is in representation $b$ of $D_{\infty h}$
and at $t=\tau$ in representation $a$.
The orthogonality of states within different representations
of $D_{\infty h}$ implies,
$\langle W_{ab}\rangle=\delta_{ab}\langle W_a\rangle$.
We now intend to relate the
generalised four point function,
$G_{ab}$, to the expectation value of $W_{ab}$, with appropriate colour
field insertions on the temporal lines.
Let us first consider the corrections from fluctuations around the
static propagator,
${\mathbf x}_2(t)={\mathbf r}+{\mathbf v}(t-\tau/2)$. The expectation value
of the perturbed Wilson loop, $W^{\mathbf v}_{ab}$, can be related to
that of the static Wilson loop,
\begin{equation}
\langle W^{\mathbf v}_{ab}\rangle=\langle W_{ab}\rangle+
{\mathbf v}g\int_{-\tau/2}^{\tau/2}\!dt\,t\,\langle {\mathbf E}({\mathbf r},t)
W_{ab}\rangle:
\end{equation}
the integral vanishes, unless the expectation value
is negative under time reversal, $CP=T=-1$, in which case
the correction matrix element itself
disappears. We conclude that to the lowest non-trivial
order, electrically
mediated transitions between different hybrid potentials do not exist.

Next, we consider magnetic transitions. The relevant perturbation term,
$\Delta H_{\mbox{\scriptsize ss}}({\mathbf x})$, is given in
Eq.~(\ref{eq:pertu}). In analogy to Eq.~(\ref{eq:per2}), we obtain,
\begin{equation}
H_{ab}=\frac{\langle a|\Delta H_{\mbox{\scriptsize ss}}|b\rangle}{(\langle
a|a\rangle\langle b|b\rangle)^{1/2}}.
\end{equation}
We have introduced the denominator, such that $|a\rangle$ and
$|b\rangle$ do not need to be normalised.
In the equal mass case
the matrix element can be expressed in terms of Wilson loops in the following
way, where we have exploited the fact that
${\mathbf B}$ is even under parity inversions and, ${\mathbf S}=
{\mathbf S}_1+{\mathbf S}_2$,
\begin{eqnarray}
H_{ab}(r)&=&\frac{c_F(m)S_i}{m}V_{ab,i}(r),\\
V_{ab,i}(r)&=&g\,\lim_{\tau\rightarrow\infty}
\langle\langle B_i({\mathbf 0},\tau/2)\rangle\rangle_{ab},\label{v_ab}
\end{eqnarray}
where,
\begin{equation}
\langle\langle F\rangle\rangle_{ab}=
\frac{\langle \Tr(U_{ab}F)\rangle}{(\langle W_a\rangle
\langle W_b\rangle)^{1/2}}.
\end{equation}
$U_{ab}$ is a path ordered product of $SU(N)$ matrices, starting from
and ending at the
space-time position of $F$, with $\Tr\, U_{ab}=W_{ab}$.
Note that $V_{ab,i}=V_{ba,i}$ have dimension $m^2$.

If we are interested in corrections to $\Sigma_g^+$ states only,
it is sufficient to consider the leading order diagonal elements,
\begin{equation}
H_{0,a}=2(\tilde{m}-\delta m)+\frac{p^2}{\tilde{m}^2}+V_a(r),
\end{equation}
where $V_a$ denotes the respective hybrid potential.
We can start from the unperturbed (diagonal) Hamiltonian, $H_0$,
and determine the spectrum in all hybrid channels,
\begin{equation}
H_{0,a}\psi^{0,a}_N=E^{0,a}_N\psi^{0,a}_N.
\end{equation}
Subsequently, the order $c^{-2}$ corrections
to the $\Sigma_g^+$ levels can be determined in perturbation theory,
the corrections to the diagonal part, $H_{0,{\Sigma_g^+}}$ in first (and,
for spin-spin interactions as well as MD corrections, second) order,
the corrections due to mixing with hybrids, $\Delta E_N^{mix}$,
in second order,
\begin{equation}
\label{eq:per7}
\Delta E_N^{mix}=\sum_{M,a\neq\Sigma_g^+}
\frac{\left|\langle\psi^{0,\Sigma_g^+}_N|H_{\Sigma_g^+,a}|\psi^{0,a}_M\rangle
\right|^2}{E_M^{0,a}-E_N^{0,\Sigma_g^+}}.
\end{equation}
Note that radial excitations like $3S$ and $4S$ whose energy levels
are close to those of hybrid states, will be more strongly affected
by the mixing than $1S$ or $2S$ states, that are separated
from the hybrids by substantial energy gaps. Also note that although the
above equation very much resembles the general form of Eq.~(\ref{eq:per2}),
in Eq.~({\ref{eq:per7}}) static hybrid state creation operators
are substituted by
wave functions of quarkonia bound states, and hybrid potentials
by ($r$-independent) quarkonia energy levels.

In QED similar mixing effects between $|e^+e^-\rangle$ states and
$|e^+e^-\gamma\rangle$ states
exist~\cite{Labelle:1992hd,Labelle:1998en,Pineda:1998kn}.
In QCD such effects are na\"{\i}vely enhanced
by factors, $\alpha_s v^2_{\Upsilon}/
(\alpha_{fs}v^2_{e^+e^-})$, with respect to QED, however, the denominator
of Eq.~(\ref{eq:per7})
guarantees an additional suppression;
the lowest hybrid level is well separated from the ground state
and the spectrum of hybrid potentials is discrete, rather
than continuous. In addition to transitions between the ground state
string and
hybrid excitation, glueball creation can be considered. However,
with masses of 3 -- 4 GeV, the vector and axial-vector glueballs
will only play a minor r\^ole while the scalar glueball will only
enter the scenario
at order $c^{-4}$, when quadrupole transitions
have to be considered. In the case of QCD with sea quarks, additional
flavour singlet meson channels
open up, however, these particles are rather heavy too.
Another possibility is the (OZI suppressed) radiation of
three $\pi$s.
The main change with respect to the quenched approximation
is related to the spectra of static
potentials at large $r$, where string breaking becomes possible.
This will give rise to
mixing effects with $B\overline{B}$ states.

It is interesting to observe that Eq.~(\ref{eq:schr4}),
which corresponds to the Lagrangian,
\begin{eqnarray}
{\mathcal L}&=&-\left(\psi_{\Sigma_g^+}^{\dagger}
H_{\Sigma_g^+}\psi_{\Sigma_g^+}
+\psi_{\Sigma_u^-}^{\dagger}
H_{\Sigma_u^-}\psi_{\Sigma_u^-}+\cdots\right.\nonumber\\
&+&\left.
\psi_{\Sigma_g^+}^{\dagger}H_{\Sigma_g^+,\Sigma_u^-}\psi_{\Sigma_u^-}^{\dagger}
+\psi_{\Sigma_u^-}^{\dagger}H_{\Sigma_g^+,\Sigma_u^-}\psi_{\Sigma_g^+}+\cdots\right),
\end{eqnarray}
somewhat resembles the general form of the pNRQCD Lagrangian,
Eq.~(\ref{eq:pnrq}). In our case, $\psi_{\Sigma_g^+}$ replaces the
singlet wave function, $S$, while the octet finds its
analogue in various hybrids. An important difference is that,
unlike in Eq.~(\ref{eq:pnrq}), the leading order mixing elements
contain magnetic fields while electric
contributions proportional to, ${\mathbf r}\cdot{\mathbf E}$, 
have been found to vanish. Of course
in higher orders of pNRQCD similar magnetic terms will appear too.

The potential approach not only allows
all sorts of effects to be systematically incorporated
but also enables the determination of many
quantities that are not directly observable, for example
the spectra of
would-be hybrid states and the mixing matrix elements between these states
and quark model states. This information is hidden in a direct lattice
simulation. The results can readily
be translated into languages commonly used in the context of the quark model
and flux tube extensions thereof and put otherwise
only heuristically defined concepts onto a firm basis.
It also becomes obvious that the heavy quark interaction potential will
only converge towards the static potential
in the limit $v/c\rightarrow 0$, rather
than $m\rightarrow\infty$ as one na\"{\i}vely might have assumed, ignoring
the kinetic term in the NRQCD Lagrangian. However,
unlike in heavy-light systems,
$v/c$ is not proportional to $m$
but $v/c\propto\alpha_R(r)$ [Eq.~\ref{eq:vir2}]:
the desired limit $v/c\rightarrow 0$ will be approached
logarithmically slowly
as the spatial extent $r$ of the bound state wave function vanishes.
This freezing of $v/c$
as a function of the quark mass $m$ at
large $m$ is also illustrated by the estimates
in the last row of Table~\ref{tab:scales}.

\subsection{Lattice determinations of the potentials}
\label{sec:latsdres}
The potentials, Eqs.~(\ref{ce_1}) -- (\ref{md_4}), are given in a
form in which they can be easily evaluated on the lattice.
Spin dependent potentials have been computed
in $SU(2)$ gauge
theory~\cite{Michael:1985wf,Michael:1986rh,Huntley:1987de,Bali:1995yz,Bali:1997cj},
$SU(3)$ gauge
theory~\cite{deForcrand:1985zc,Campostrini:1986ki,Campostrini:1987hu,Huntley:1987de,Koike:1987jh,Ford:1989as,Bali:1997am}
and in exploratory studies of
QCD with sea quarks~\cite{Koike:1989jf,Born:1994cp}.
In Refs.~\cite{Bali:1995yz,Bali:1997am} the momentum dependent
corrections in $SU(2)$ and $SU(3)$ gauge theories, respectively,
have been considered too. The correction to the central potential,
$V_{\prime}$, of
Eq.~(\ref{ce_3})~\cite{Brambilla:2000gk} as well as
the transition potentials, $V_{ab,i}$,
of Eq.~(\ref{v_ab}),
however, have not been calculated so far.

\subsubsection{The method}
The simplest discretisations of magnetic and electric field
insertions, $g{\mathbf B}$ and $g{\mathbf E}$, are the clover leaf
definitions of Eqs.~(\ref{eq:field1}) -- (\ref{eq:field3}).
Alternative discretisations have been investigated in the first lattice
study~\cite{Michael:1985wf} of spin dependent potentials.
Since the temporal lattice extent is always finite, the limit,
$\tau\rightarrow\infty$, cannot be performed exactly.
Moreover, the arguments of the
integrals within Eqs.~(\ref{ce_1}) -- (\ref{md_4}) can only be obtained
on a discrete set of $t$ values.
Spectral representations of the potentials,
Eqs.~(\ref{eq:spectra1}) -- (\ref{eq:spectra4}), however, are extremely
useful to guide and control interpolations and extrapolations as well as in
improving the lattice operators used.

\begin{figure}[thb]
\centerline{\epsfxsize=10truecm\epsffile{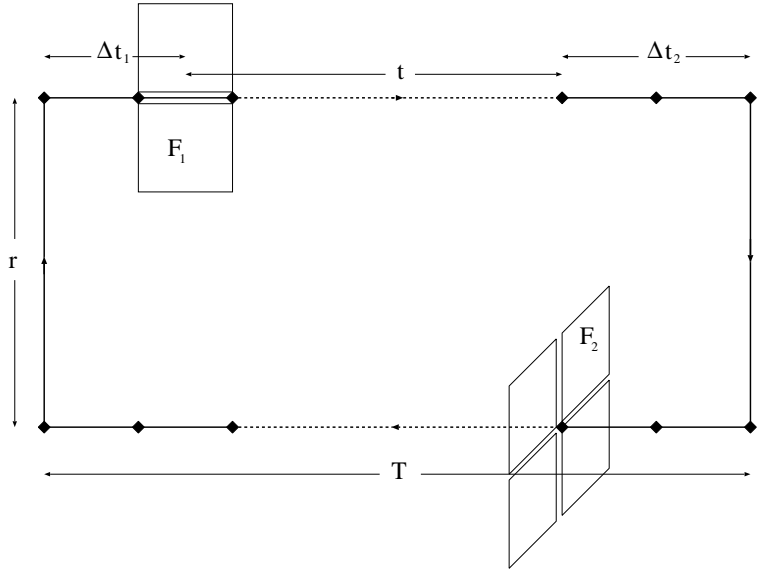}}
\caption{Lattice definition of the operator in the argument of the
nominator of
Eq.~(\ref{dex})
for the example of $F_1$ being an electric and $F_2$ being a magnetic
field.}
\label{fig:latvdep}
\end{figure}

Relative
statistical errors explode with the temporal extent of a Wilson loop,
$T\geq\tau$,
while the distances between the field strength insertions and the
spatial closures of the loop, $\Delta t_1$ and $\Delta t_2$, determine
the degree of pollution from excited states.
Therefore, adapting the size of the Wilson loop within the double bracket
expectation value, Eq.~(\ref{dex}), to the distance between the
two field insertions, $t$,
$T(t)=\Delta t_1+\Delta t_2 +t$, turns out to be the optimal
choice in terms of statistical errors as well as in terms of
a fast convergence to the asymptotic limit of interest~\cite{Bali:1997cj}.
The resulting lattice operator is depicted in  Figure~\ref{fig:latvdep}.
In addition to keeping $\Delta t_i$ large, the overlap with the ground state
can be enhanced by smearing the spatial connections
within the Wilson loop (cf.\ Section~\ref{sec:lattsmear}).
In the first lattice studies~\cite{deForcrand:1985zc} the integrals,
Eqs.~(\ref{ce_1}) -- (\ref{md_4}),
were replaced by discrete lattice sums. By parameterising
the arguments as continuous functions of $t$~\cite{Ford:1989as,Bali:1997am},
prior to the integration, discretisation errors can be reduced and the
effects of
the region of large $t$ (where statistical errors dominate the signal)
can still be incorporated.
If the hybrid potentials are known, the exponents of
such multi-exponential fits to
Eqs.~(\ref{eq:spectra1}) -- (\ref{eq:spectra4}) can be
determined independently~\cite{Ford:1989as,Michael:1986rh,Huntley:1987de}.

\subsubsection{Matching to the continuum}
In all lattice studies, based on na\"{\i}ve discretisations of the
continuum expressions,
the potentials $V_2', V_3$ and $V_4$
have been found to be
much smaller than one would have expected from perturbative
arguments or quarkonia phenomenology.
In Ref.~\cite{Michael:1985wf} this has been attributed to
the anomalous dimension of the
magnetic moment while in Ref.~\cite{deForcrand:1985zc} this
has been interpreted as a lattice artefact.
As we shall see, both suggestions are true in parts.
In particular the difference, $V_2'-V_1'$,
has been found to
be a factor of three to four times
smaller~\cite{Michael:1986rh,Campostrini:1987hu,Huntley:1987de}
than the inter-quark force
$V_0'$, in violation of the Gromes relation, Eq.~(\ref{grom}).

Nowadays, we know that such behaviour is
caused by large renormalisations between lattice operators
and their continuum counterparts~\cite{Bali:1997am}. In addition,
the matching coefficients between NRQCD and QCD,
discussed in Section~\ref{sec:nrqcd}, will affect
quarkonium spectrum predictions.
We can separately perform
two matchings: lattice NRQCD to continuum NRQCD and continuum NRQCD to QCD.
In Ref.~\cite{Michael:1986rh} a procedure reminiscent
of ``tadpole improvement''~\cite{Lepage:1993xa}
has been suggested to reduce the former renormalisation factor:
the lattice operators are improved
by dividing out factors, $U_{P}^2$, from the
double bracket correlation functions. This prescription does not affect
the continuum limit and still the leading
order lattice artefacts are proportional to $a^2$.
However, in perturbation
theory all lattice specific
one loop self-interactions of the field insertions
are cancelled. This procedure can be
refined by the Huntley-Michael (HM) construction~\cite{Huntley:1987de},
in which additional un-wanted higher order graphs cancel
too. This becomes possible
by taking the relative position of the field insertions with
respect to the Wilson loop into account. This HM
scheme has
been employed in the simulations of
Refs.~\cite{Bali:1995yz,Bali:1997cj,Bali:1997am},
and as a result the Gromes and BBP relations Eqs.~(\ref{grom}) -- 
(\ref{bram}) are found to be respected within the achieved numerical
accuracy of a few per cent.

\begin{figure}[thb]
\centerline{\epsfxsize=10truecm\epsffile{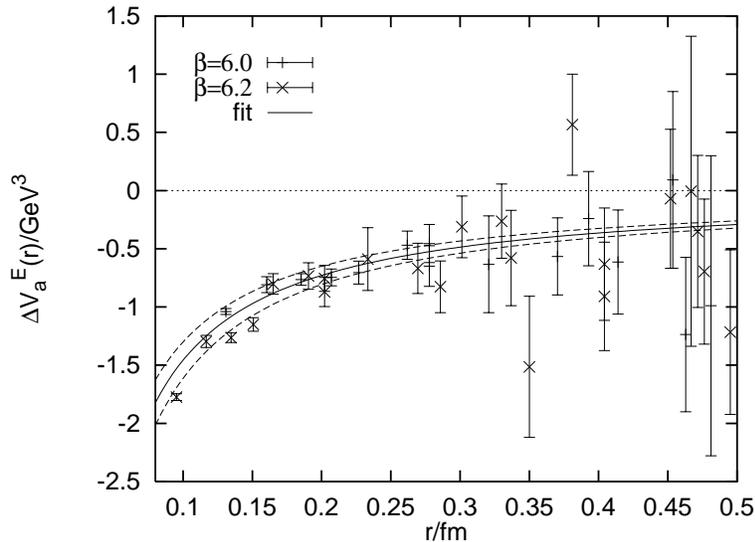}}
\caption{The potential $\nabla^2V_a^E$, together with a fit curve,
$\nabla^2V_a^E=-(2\sigma+b)/r$.}
\label{fig:va}
\end{figure}

It has been suggested~\cite{Campostrini:1987hu} to  fix the
lattice renormalisation factors
non-pertur\-ba\-tive\-ly from the Gromes relation
at distances, $r\gg a_{\sigma}$, where rotational symmetry is effectively
restored on the lattice.
From Eqs.~(\ref{eq:spectra1}) -- (\ref{eq:spectra4}) it is evident that
the relativistic corrections to the static potential, rather than being
spectral quantities themselves, are proportional to
amplitudes, $D_n^{(i)}({\mathbf r})$, which will undergo renormalisation.
Ratios of these amplitudes for different $n$, however,
should approach the continuum ratios, up to order $a^2$ scaling violations.
Let us define the renormalisation constants, $Z_{\mathbf B}({\mathbf r})$ and
$Z_{\mathbf E}({\mathbf r})$,
\begin{eqnarray}
\label{eq:zb}
\langle \Phi_{{\mathbf r},0}|{\mathbf B}({\mathbf 0})|\Phi_{{\mathbf r},n}
\rangle &=& Z_{\mathbf B}
({\mathbf r})\langle \Phi_{{\mathbf r},0}|{\mathbf B}^L
({\mathbf 0})|\Phi_{{\mathbf r},n}\rangle\\
\langle \Phi_{{\mathbf r},0}|{\mathbf E}({\mathbf 0})|\Phi_{{\mathbf r},n}
\rangle &=& Z_{\mathbf E}
({\mathbf r})\langle \Phi_{{\mathbf r},0}|{\mathbf E}^L
({\mathbf 0})|\Phi_{{\mathbf r},n}\rangle,\label{eq:ze}
\end{eqnarray}
where $n$ should be chosen such that the
corresponding amplitude does not vanish.
From considerations analogous to Eq.~(\ref{eq:specco}) it is
obvious that $V_1'$ and $V_2'$, measured on the lattice, have to be
multiplied by factors $Z_EZ_B$,
$V_3,V_4$ and $\nabla^2V_a^B$
by factors $Z_B^2$ and all other
potentials by $Z_E^2$ to make contact
with the potentials in a continuum scheme. 

\begin{figure}[thb]
\centerline{\epsfxsize=10truecm\epsffile{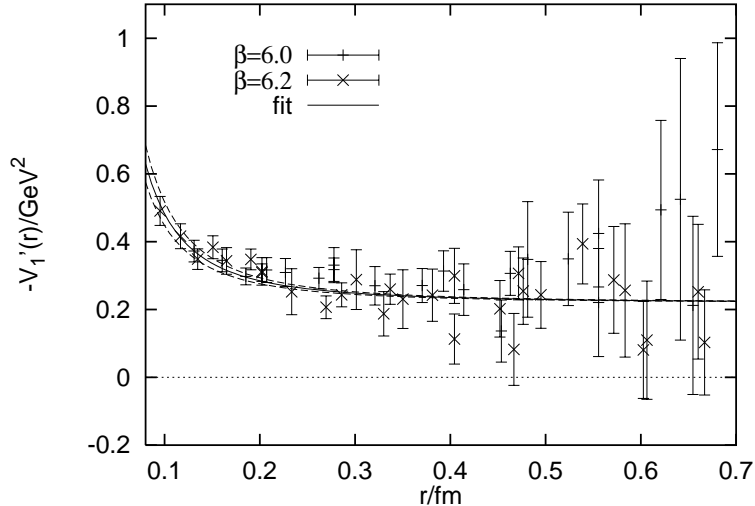}}
\caption{The spin-orbit potential $V_1'$, with a fit, $-V_1'=\sigma+h/r^2$.}
\label{fig:v1}
\end{figure}

The definitions of $Z_{\mathbf B}$ and $Z_{\mathbf E}$ are ambiguous; any term 
that vanishes at least like $a^2$ can be added.
Since the left hand sides of Eqs.~(\ref{eq:zb}) and (\ref{eq:ze})
are approached by the right
hand sides in the continuum limit,
$Z_{\mathbf B}$ and $Z_{\mathbf E}$
are $r$ independent, up to order $a^2$ lattice artefacts,
and can be defined from the value at $r = r_0$, for instance.
Therefore, we only have to distinguish between four independent
renormalisation factors, $Z_{B,\perp}$, $Z_{B,\|}$,
$Z_{E,\perp}$ and $Z_{E,\|}$, where $\perp$ refers to a component
orthogonal to the inter-quark axis and $\|$ parallel to the axis.
By demanding the Gromes relation, Eq.~(\ref{grom}), to hold for
$r\approx r_0$, different linear combinations of products between
$Z_{\mathbf E}$ and $Z_{\mathbf B}$ components can be determined.
By varying the
direction of ${\mathbf r}$,
the three combinations,
$Z_{B,\|}Z_{E,\perp}$, $Z_{B,\perp}Z_{E,\|}$ and $Z_{B,\perp}Z_{E,\perp}$,
can be fixed. From the BBP relations,
Eqs.~(\ref{bram2}) and (\ref{bram}),
all $Z_EZ_E$ products can be over-determined. Therefore, a completely
non-perturbative evaluation of the renormalisations required to restore
the continuum Lorentz symmetry
is viable. From the rotational symmetry
of the relativistic correction potentials, observed in
Ref.~\cite{Bali:1997am}, one can conclude $Z_{B,\parallel}\approx Z_{B,\perp}$
as well as $Z_{E,\parallel}\approx Z_{E,\perp}$.
In fact, up to the inherent order $a^2$ ambiguity, one would
expect such (approximate) equalities
if one considers that the renormalisation between
lattice and continuum NRQCD is an ultra-violet effect
and, therefore, should be primarily related to properties of the
local field strength insertions themselves,
rather than to their interaction with
the ultra-soft background of bound state gluons.
Moreover, using the same argument, on an isotropic lattice, $a_{\sigma}=
a_{\tau}$,
one would expect, $Z_E\approx Z_B$.

\subsubsection{Results}
We conclude this section by reviewing the lattice results obtained in
the most concise and precise study so far~\cite{Bali:1997am}.
The $SU(3)$
potentials have been determined by use of the quenched
Wilson action on isotropic lattices at $\beta=6.0$ and
$\beta=6.2$, that correspond to lattice spacings, $a^{-1}\approx 2.14$~GeV
and $a^{-1}\approx 2.94$~GeV, respectively.
In this reference, the HM renormalisation procedure~\cite{Huntley:1987de}
has been employed. Subsequently, the 
continuum Gromes and BBP relations were found to be satisfied
within the statistical accuracy of the study.

\begin{figure}[thb]
\centerline{\epsfxsize=10truecm\epsffile{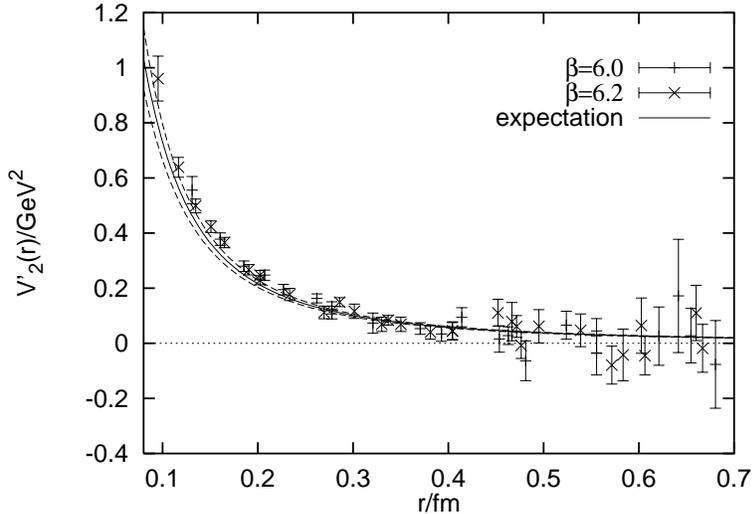}}
\caption{The spin-orbit potential $V_2'$, in comparison with the expectation,
Eq.~(\ref{v2pa}).}
\label{fig:v2}
\end{figure}

In Figure~\ref{fig:va}, we display the result for $\nabla^2V_a^E$,
together with a fit of the form of Eq.~(\ref{vaepa}).
$\nabla^2V_a^B$ was found to be consistent with a constant.
In the Figure, we have subtracted the fitted self-energy constants,
$C_a^Ea$, from the data points. The resulting $c_DC_a^Ea\approx -1$ seemed
to cancel $c_F^2C_a^Ba\approx 1$ almost perfectly.
At $\beta=6.0$ and $\beta=6.2$, values,
$b=(1.13\pm 0.45)\sigma$ and $b=(1.83\pm 0.61)\sigma$, have been
found, respectively. The sign of the difference, although not statistically
significant, coincides with the expectation that the matching
coefficient $c_D$ decreases with the lattice spacing
[Figure~\ref{fig:cdbotcharm}].
Note that a value, $b\neq 0$, is incompatible
with Eq.~(\ref{eq:convo}) that results from the assumption that
the form factors, $\tilde{V}_i$, within the interaction kernel,
Eq.~(\ref{eq:lorinv}), only depend on the momentum transfer, $q^2$.

The observation, $3c_Db+2\sigma>0$, means that
besides the $\delta$ like Darwin term,
another $1/r$
like mass- (and, therefore, flavour-) dependent correction to the central
potential, with a coefficient of approximate size, $-2\,\sigma/(4m^2)$, exists.
However, this correction, together with an additional
$-\sigma/(6m^2)$ term from the MD potentials, yields an increase in
the effective Coulomb coefficient of the
Cornell potential of less than 2.5~\% for bottomonium.
In the case of charmonium the situation is less clear: the uncertainty
in $c_D$ can result in
an increase of the effective Coulomb coefficient of
anything from
8~\% to 18~\%.
The effective Coulomb coupling within
the static potential will
weaken at short distances
as soon as one goes beyond the tree level inspired
Cornell parametrisation.
This is, however, not the case for the coefficient
of the mass dependent 
corrections proportional to,
$\nabla^2 (V_0-V_{0,{\mbox{\scriptsize pert}}})\approx 2\sigma/r$
whose relative weight will, thus, increase
at very short distances.
Considering the discussion of the potential at very short
distances in Section~\ref{sec:beyond},
such contributions could turn out to be more important
than one would have assumed, guided by the Cornell parametrisation alone.

\begin{figure}[thb]
\centerline{\epsfxsize=10truecm\epsffile{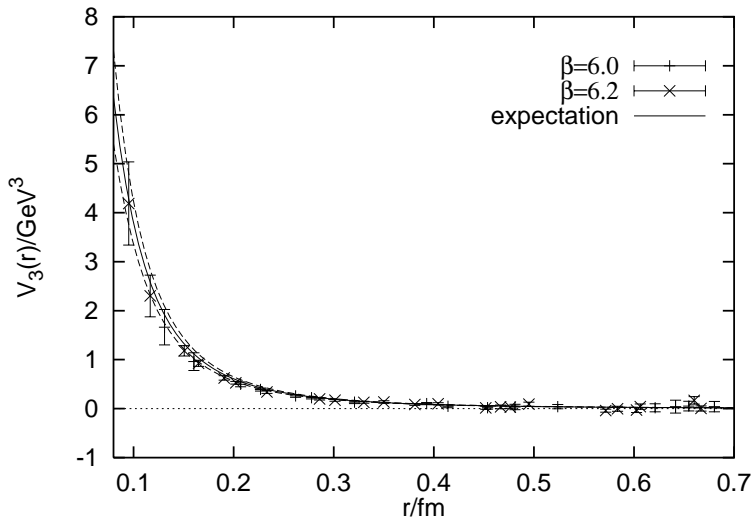}}
\caption{The spin-spin potential, $V_3$, in comparison with the expectation,
Eq.~(\ref{v3pa}).}
\label{fig:v3}
\end{figure}

In Figure~\ref{fig:v1}, the long range spin-orbit potential $V_1'$ is displayed,
together with a fit of the form Eq.~(\ref{v1pa}), where
the string tension has been taken from a fit to the central potential.
The values, $h=0.071\pm 0.013$ and $h=0.065\pm 0.009$, have been
found at the two lattice spacings, respectively.
Therefore, the dimensionless parameter
$h$ turns out to be somewhat bigger than
one fifth of the Coulomb coefficient, $e\approx 0.3$. Since $c_F$
increases with decreasing $a$, we expect $h$ to decrease slightly
as a function of the lattice spacing. We can ask ourselves at what lattice
spacing, $a'\approx 1/\mu'$, we would expect $h$ to assume its (unmixed) value,
$h=0$. From Eq.~(\ref{v1sca}), we can derive the relation,
$c_F(\mu')=[(e-h)/e]\,c_F(\mu)\approx 0.78\, c_F(\mu)$: a decrease of
$c_F$ by more than 20~\% is required which, as can be seen from
Figure~\ref{fig:cfbottom}, will correspond to a scale (much)
smaller than 1~GeV.

In Figures~\ref{fig:v2} and \ref{fig:v3}, we display $V_2'$ and $V_3$,
together with the model expectations of Eqs.~(\ref{v2pa}) and (\ref{v3pa}).
After having determined $e$ from the static potential and $h$ from $V_1'$
there are no free parameters in the function displayed. 
Excellent agreement between the data and the predictions is found.
$V_2'$ does not contain any
long distance contribution and therefore can be identified with the
vector potential, $V_V$, within models that are based on an
interaction kernel that only depends on the momentum transfer,
i.e.\ $\eta=0$, within Eq.~(\ref{eq:v2co}).
The fact that $V_3\approx V_2'/r-V_2''$ implies [Eq.~(\ref{eq:v3co})]
$V_P(r)\approx c r^2$
and, therefore,
$V_P\approx 0$.

\begin{figure}[thb]
\centerline{\epsfxsize=10truecm\epsffile{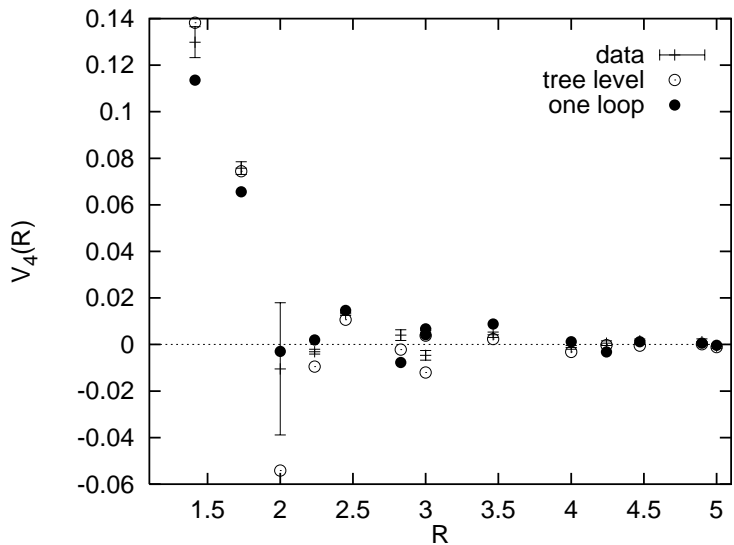}}
\caption{The spin-spin potential, $V_4$, in lattice units at $\beta=6.2$,
in comparison with the expectation,
Eq.~(\ref{v4pa}), and a one loop perturbative improvement.}
\label{fig:v4}
\end{figure}

In Figure~\ref{fig:v4}, we display the potential $V_4$, determined
at $\beta = 6.2$ in lattice units. We decided not to plot the
potential in physical units since the behaviour expected
from Eq.~(\ref{v4pa}) is a $\delta$ function. Hence,
the result will be cut-off and discretisation dependent.
For the clover leaf definition of the magnetic fields, employed in the
study, the lattice $\delta$ function has been calculated (indicated
as ``tree level'' in the plot).
Indeed, the data are described well by this expectation.
A one loop improved version brings
the data even more in line with the expectation.

\begin{figure}[thb]
\centerline{\epsfxsize=10truecm\epsffile{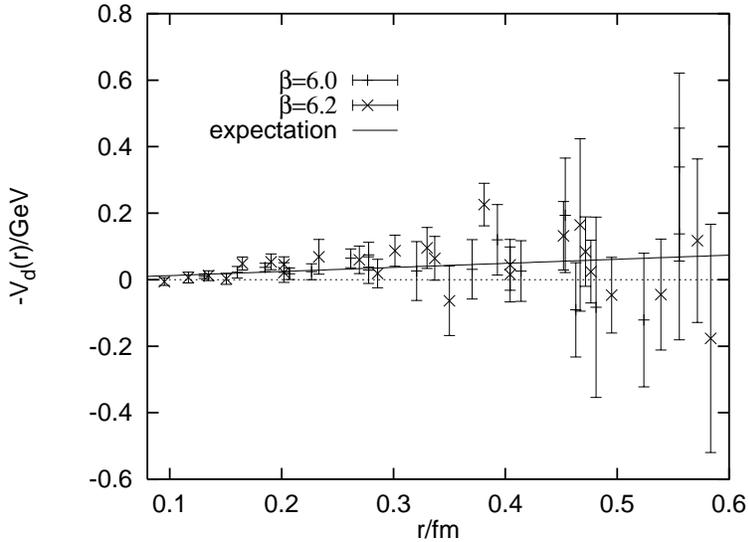}}
\caption{The potential, $V_d$, together with the curve $V_d(r)=-\sigma/9\,r$.}
\label{fig:vd}
\end{figure}

The errors in the lattice determination
of the relativistic correction potentials
are much bigger than those on the static potential of
Figure~\ref{figpotsu3}. However, one should keep in mind that
the effect of these terms on the spectrum is suppressed
by factors of $v^2/c^2$ with respect to the static potential;
even an error as big as 10~\% on a 10~\% correction term
is completely tolerable for most
phenomenological purposes as the induced uncertainty of
a few MeV on the $\Upsilon$ spectrum
will still be smaller than the effect of neglecting
higher order relativistic or
radiative corrections.
In general, operators involving electric field insertions
result in stronger statistical fluctuations than
magnetic fields. Therefore, $V_3$ and $V_4$ are the most precisely
determined potentials, followed by $V_1'$ and $V_2'$
while $\nabla^2V_a^E$ as well as the MD corrections
are subject to big statistical uncertainties.

In the case of the MD potentials, this is particularly
disappointing
as the expectations, Eqs.\ (\ref{vbpa}) -- (\ref{vepa}),
all contain a long range part and are all numerically small, in comparison
with the other potentials. This means of course that these potentials are
not of prime phenomenological interest. However, being dominated by
non-perturbative effects, they are needed to discriminate between
competing predictions arising from different assumptions on the QCD
vacuum~\cite{Baker:1996mu,Brambilla:1997aq}. As an example of a MD potential,
$V_d(r)$ is depicted in Figure~\ref{fig:vd}, together with the
expectation, Eq.~(\ref{vdpa}). The fitted self-energies $C_d$
have been subtracted from the data sets. Note that the BBP relation,
Eq.~(\ref{bram2}), has been confirmed to hold for the self-energies,
$2C_b+4C_d+V_{\mbox{\scriptsize self}}=0$.
Clearly, further improved numerical simulations are required to
arrive at definite conclusions about the functional form
of the MD potentials.

\section{Application to the quarkonium spectrum}
\label{quarkonia}
After having determined the potentials, quarkonia spectra and wave functions
can readily be predicted, within the limitations of the non-relativistic and
adiabatic approximations.
Vice versa, quarkonium spectra can in principle be used
as an input to fix parameters that have not yet been determined accurately,
in particular the matching coefficients appearing within the effective
action. The same values could then be taken in lattice NRQCD studies
or HQET calculations of heavy-light bound states and their decay
matrix elements. In particular, the $S$ and $P$ state fine splittings
react in a very sensitive way towards variations of these coefficients.
Unfortunately, the $\eta_b$ whose splitting with respect
to $\Upsilon$ states would
yield the cleanest information has not been discovered yet.
Moreover, the fine structure as well as decay rates,
that are proportional to the wave function (or in the case of
$P$ states, its derivative) at the origin, probe the heavy quark interaction
at very small distances.
Here we will restrict our discussion to
spectrum determinations and estimations of the systematic errors
inherent in order $v^4$ (or $c^{-2}$) continuum and lattice
NRQCD as well as uncertainties from neglecting sea quarks.

\subsection{Solving the Schr\"odinger equation}
\label{quarkprop}
Once the interaction potentials are determined,
the Schr\"odinger equation, Eq.~(\ref{eq:schroe}), can be
solved numerically on any personal computer,
either on a discrete lattice~\cite{Bali:1998pi} or
in the continuum~\cite{Eichten:1975af,Quigg:1979vr,Eichten:1980ms}.
In the latter case, one would start by integrating the radial equation,
Eq.~(\ref{eq:radial}) or
Eq.~(\ref{eq:radial2}), for the Hamiltonian $H_0$
of Eqs.~(\ref{eq:h0}) and (\ref{eq:h02}),
\begin{equation}
H_0|nll_3\rangle=E_{nl}^0|nll_3\rangle.
\end{equation}
Subsequently, the
$1/m^2$ corrections, Eqs.~(\ref{eq:hdelta}) -- (\ref{eq:hsd}),
can be treated as perturbations,
\begin{equation}
E_{nJls}=E_{nl}^0+\frac{1}{\tilde{m}^2}\sum_i\langle {nll_3}|\delta
H_i(r,J,l,s,{\mathbf p})|{nll_3}\rangle.
\end{equation}
By use of the identities~\cite{Quigg:1979vr,Bali:1998pi},
\begin{eqnarray}
\langle n ll_3|f(r)p^2|nll_3\rangle&=&\tilde{m}\left\langle nll_3\left|
f(r)\left[E_{nl}^0-\tilde{V}(r)\right]\right|nll_3
\right\rangle,\\
4\pi\langle nll_3|\delta^3(r)|nll_3\rangle&=&\left|\psi_{nll_3}(0)\right|^2\\
\nonumber
&=&\frac{\tilde{m}\sigma}{\pi}\left(1+\frac{\tilde{e}}{\sigma}\langle nll_3|r^{-2}|nll_3
\rangle\right),
\label{eq:psio}
\end{eqnarray}
all perturbations can readily be computed from expectation values,
$\langle r^{\alpha}\rangle$,
$\alpha = -4,\ldots,1$. Note that
$\tilde{m}$, $\tilde{V}(r)$ and $\tilde{e}$
are defined in Eqs.~(\ref{eq:h00}) -- (\ref{eq:h04}).

\begin{figure}[thb]
\centerline{\epsfysize=13truecm\epsffile{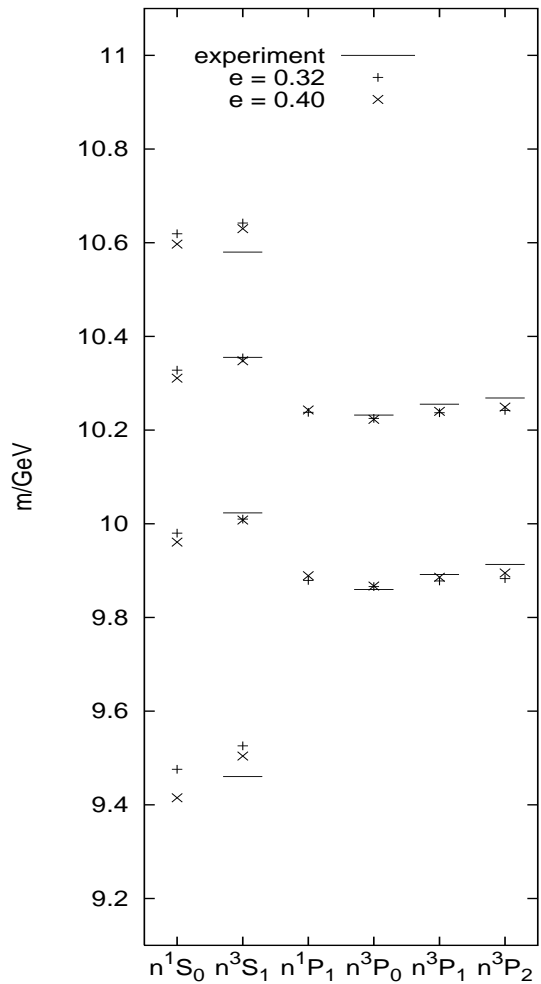}}
\caption{The bottomonium spectrum from the lattice
potentials~\cite{Bali:1997am}.}
\label{fig:botspect}
\end{figure}

The Hamiltonian of Eqs.~(\ref{eq:h0}) -- (\ref{eq:hsd})
originates from the parametrisations, Eqs.~(\ref{v0pa}) -- (\ref{vepa}),
of $V_0(r)$ -- $V_e(r)$ that are in qualitative agreement with
the lattice data. These lattice inspired parametrisations
for intermediate and large distances
can of course be combined with perturbative
short range
expectations~\cite{Pantaleone:1986uf,Schroder:1998vy,Brambilla:1998vm},
for example along the lines of Ref.~\cite{Buchmuller:1981su},
for the purpose of a phenomenologically more accurate description
of bottomonia states. This is certainly worthwhile doing as soon as
lattice results on the transition matrix elements discussed
in Section~\ref{sec:transi} become available and
some of the NRQCD matching coefficients, in particular $c_F$, have been
determined in a non-perturbative way.
Anticipating such future results we can, however, use the available
lattice data to estimate systematic uncertainties that are due
to radiative and relativistic corrections as well as neglecting
sea quarks.

Before discussing such effects, we reproduce
the bottomonium spectrum obtained in Ref.~\cite{Bali:1997am}
from the lattice potentials in Figure~\ref{fig:botspect}.
The displayed spectrum has been obtained from the parameter values,
\begin{eqnarray}
e&=&0.32,\\
h&=&0.065,\\
b&=&1.81\sigma=1.81(1.65-e)\,r_0^{-1},\\
C_0&=&0,\\
\tilde{m}_b&=&4.676\,\mbox{GeV},\\
r_0^{-1}&=&406\,\mbox{MeV},
\end{eqnarray}
with one loop matching coefficients, $c_F$, $d_v$ and $c_D$.
While $e$, $h$ and $b$ have been computed entirely on the lattice,
the quark mass, $\tilde{m}_b$, and the scale, $r_0$, have been determined from
a fit to the experimental spectrum.

The Figure illustrates the precision to which
experiment can at present be reproduced, without
recourse to phenomenological input other than that
required to fix the quark mass
and the scale.
It is not {\em a priori} clear whether the average deviations of almost 20~MeV
are dominantly caused by relativistic and radiative corrections
or due to ultra-soft gluons that have not yet been incorporated
into the potential approach. 
Uncertainties resulting from the statistical errors on the
potentials as well as differences between data
sets obtained at $\beta=6.0$ and $\beta=6.2$ cannot be resolved on the scale
of the plot. In addition to results based on the fit parameters, extracted
from the quenched simulations ($e=0.32$), results for a stronger Coulomb
coupling, $e=0.40$, are displayed. We intend
to model the changes that one might
expect when including sea quarks by this latter choice of $e$.
It is amusing to notice that, when ignoring the mass dependence
of the matching coefficients, $c_i=1$, all
ratios of splittings come out to be consistent~\cite{Bali:1998pi}
with those
determined in direct lattice NRQCD simulations, indicating
that higher order relativistic corrections as well as
effects due to ultra-soft gluons do not play a prominent r\^ole, at least
for the lowest few levels; all differences between
published lattice NRQCD results
and the spectrum of Figure~\ref{fig:botspect}
are entirely due to different prescriptions for assigning a
physical scale to the lattice results and a different choice of the
matching coefficients, $c_i$.
In the potential case,
overall agreement with experiment has been optimised while
in lattice NRQCD usually the most precisely determined $2^3S_1-1^3S_1$
or $1\overline{{}^3P}-1^3S_1$ splittings are taken as the only input.

\subsection{Systematic uncertainties}
\label{sec:uncert}
Having a Hamiltonian representation of the bound state problem at
hand it is straight forward to investigate how the spectrum changes
when the input parameters are varied. For instance, fine structure splittings
are to first approximation proportional to the matching
coefficient, $c_F^2$. Such effects are discussed in detail in
Ref.~\cite{Bali:1998pi}. Here, we briefly summarise the main results
and discuss the uncertainties common to the potential and
the lattice NRQCD approaches. In addition, the effect of neglecting sea quarks
is investigated and finite volume effects for lattice
studies of $\Upsilon$ properties are estimated.

Unfortunately, no precision
results on the corrections to the static potential in QCD with sea quarks
exist. However, the static potential has been determined
accurately~\cite{Glassner:1996xi,Bali:1997ec,Bali:1997bj,Bali:2000vr,Aoki:1998sb}
for $n_f=2$ (cf.~Section~\ref{sec:seaq}), and an increase of the
effective Coulomb coefficient $e$ by 16~\% to 22~\% has been
detected~\cite{Bali:1997bj,Bali:2000vr} for quark masses,
$m_u=m_d>m_s/3$ (cf.~Figure~(\ref{figcoul}).

\begin{figure}[thb]
\centerline{\epsfxsize=10truecm\epsffile{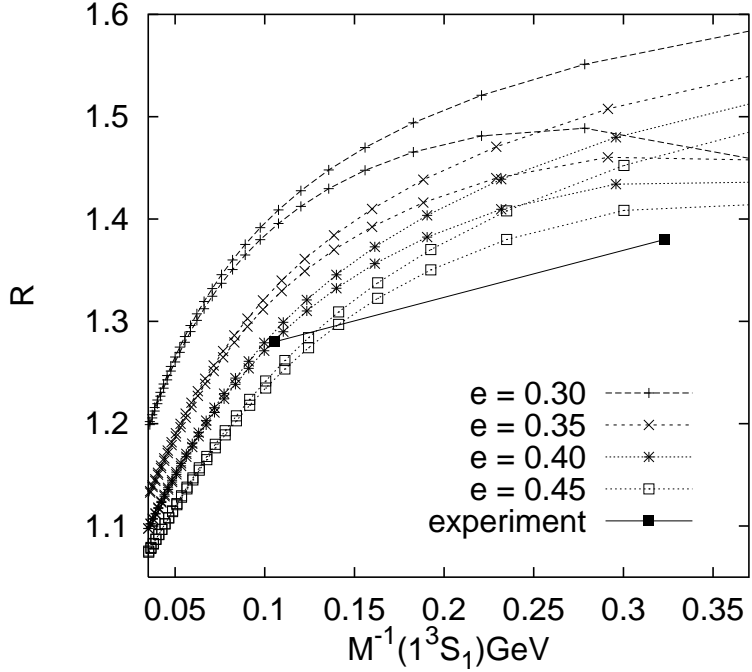}}
\caption{$R=\Delta m_{2S1S}/\Delta m_{1P1S}$ as a function of
  $m_{\Upsilon}^{-1}$ for various values of the Coulomb coupling, $e$.
The upper curves correspond to the lowest order Hamiltonian, the lower
ones incorporate relativistic corrections.}
\label{fig:rdat}
\end{figure}

The ratio,
\begin{equation}
R=\frac{m_{2^3S_1}-m_{1^3S_1}}{m_{1\overline{{}^3P}}-m_{1^3S_1}},
\end{equation}
reacts in a very sensitive way towards quenching. The potentials
yield $R\approx 1.38$ which (while in perfect
agreement with quenched lattice NRQCD~\cite{Davies:1998im})
disagrees with experiment, $R\approx 1.28$. 
The dependence of $R$ on $e$, keeping the parameters
$r_0^{-1}$ and $\tilde{m}$ fixed,
is displayed in Figure~\ref{fig:rdat}. Values, $e\approx 0.4$, appear necessary
for the real world with three active sea quark flavours to reproduce
this ratio. Keeping in mind that two sea quarks resulted in
the effective Coulomb strength to increase by about 20~\%, such an
increase by 30~\% indeed appears to be very reasonable. Our suggested
sizes of quenching effects will be based on this estimate.
From the Figure it is also obvious that while order $v^4$ ($c^{-2}$)
effects
on this ratio are small around the bottom mass, relativistic corrections
explode in an uncontrolled way towards the charm: while
$\langle v^2\rangle_{\Upsilon}\approx 0.1$ for bottomonia,
$\langle v^2\rangle_{J/\psi}\approx 0.4$ is not exactly a small
expansion parameter anymore.

\begin{table}
\caption{Relativistic and radiative corrections to $\Upsilon$ splittings.}
\label{tab:bcorr}

\begin{center}
\begin{tabular}{c|c|c|c}
splitting&${\mathcal O}(v^4)$&${\mathcal O}(v^6)$&radiative\\\hline
$2\overline{S}-1\overline{S}$&13~MeV&2~MeV&5~MeV\\
$2^3S_1-1^3S_1$&17~MeV&2~MeV&6~MeV\\
$1\overline{{}^3P}-1\overline{S}$&24~MeV&1.5~MeV&2~MeV\\
$1\overline{^{3}P}-1^3S_1$&12~MeV&1.5~MeV&5~MeV
\end{tabular}
\end{center}
\end{table}

\begin{table}
\caption{Relativistic and radiative corrections to $J/\psi$ splittings.}
\label{tab:ccorr}

\begin{center}
\begin{tabular}{c|c|c|c}
splitting&${\mathcal O}(v^4)$&${\mathcal O}(v^6)$&radiative\\\hline
$2\overline{S}-1\overline{S}$&65~MeV&25~MeV&15~MeV\\
$1\overline{{}^3P}-1\overline{S}$&35~MeV&15~MeV&15~MeV
\end{tabular}
\end{center}
\end{table}

\begin{table}
\caption{Relativistic and radiative corrections to fine structure splittings.}
\label{tab:finecorr}

\begin{center}
\begin{tabular}{c|c|c|c}
family&${\mathcal O}(v^6)$&radiative&quenching\\\hline
$\Upsilon$&10~\%&25~\%&35~\%\\
$J/\psi$&40~\%&70~\%&30~\%
\end{tabular}
\end{center}
\end{table}

Relativistic ${\mathcal O}(v^4)$ correction terms affect
spin averaged $2\overline{S}-1\overline{S}$ splittings
by about 2.5~\% and 11~\% for bottomonium and charmonium, respectively;
the corresponding numbers for the $1\overline{{}^3P}-1\overline{S}$
splittings are 4~\% and 8~\%. No experimental values
for $\overline{S}$ $\Upsilon$ states are available since
pseudo-scalar $\eta_b$ mesons are not yet discovered. The $\Upsilon$
$2^3S_1-1^3S_1$ and $1\overline{{}^3P}-1^3S_1$ splittings that
are therefore at present of greater interest become reduced
by another 1~\% and 2.5~\% due to spin-spin interactions when
switching on relativistic corrections.
In Table~\ref{tab:bcorr}, we display estimates~\cite{Bali:1998pi}
of the effect of even higher order relativistic correction terms
on various $\Upsilon$ splittings as well as the size of error induced by
ignoring the mass dependence of the matching coefficients between QCD
and NRQCD
when simulating the theory at lattice spacings,
$1.5~$GeV~$\leq a^{-1}\leq 3$~GeV.
In Table~\ref{tab:ccorr}, the corresponding results for $J/\psi$ states are
displayed. To set the scale: $\Delta m_{2S1S}\approx 580$~MeV,
$\Delta m_{1P1S}\approx 430$~MeV.
Note that the order $v^4$ corrections have been calculated while
order $v^6$ and radiative corrections are estimates only.
Within order $v^4$ NRQCD, radiative corrections to the
matching coefficients, that are of size $\alpha_s\log(m/\mu)/m^2$,
dominate over relativistic correction terms that are accompanied
by factors $1/m^{3}$, at least for bottomonia.

In Table~\ref{tab:finecorr} we summarise the estimates of
the uncertainties of the fine structure splittings. Since we
only have results from the lowest order at which the
splittings can occur,
the relative sizes of the relativistic corrections can only
roughly be estimated to be of order $v^2$.
We did not try to
assign quenching errors to individual spin averaged splittings.
Only mass ratios, and not the overall scale, can be determined from the
QCD Lagrangian.
Therefore, assigning a quenching error to
an individual mass is highly subjective since
the result will depend on the experimental input quantity
used to fix the lattice spacing.
One finds different scale determinations to scatter by up to 20~\%
within the quenched approximation which should be interpreted
as the overall systematic uncertainty.
In contradiction to this philosophy, estimates on 
quenching errors are given for the fine structure splittings.
These are explicitly proportional to the Coulomb coupling.
The quenching error estimates have to be interpreted as typical changes
of the size of fine structure splittings with respect to
spin averaged splittings.
Including sea quarks will result in an increase of such ratios.
The effect of radiative corrections goes in the same direction,
this is obvious from the
continuum two loop inspired
estimate of Figure~\ref{fig:cfbottom}.
Besides quenching, the latter uncertainty
again seems to be the dominant source of error. 

Indeed, using tree level matching coefficients,
one underestimates $P$ wave fine structure splittings for $e=0.40$ by
almost a factor two~\cite{Bali:1998pi}, compared to experiment.
However,
for the ratio,
\begin{equation}
R_{FS}=\frac{m_{\chi_{b2}}-m_{\chi_{b1}}}{m_{\chi_{b1}}-m_{\chi_{b0}}},
\end{equation}
from which the dominant radiative correction cancels,
one obtains $R_{FS}\approx 0.56$ which has to be compared to the experimental
value, $R_{FS}\approx 0.66$. By incorporating
running coupling effects into the parametrisation of $V_3$
and calculating
order $v^6$ effects, it should be possible to
further improve the agreement.

\begin{figure}[thb]
\centerline{\epsfxsize=10truecm\epsffile{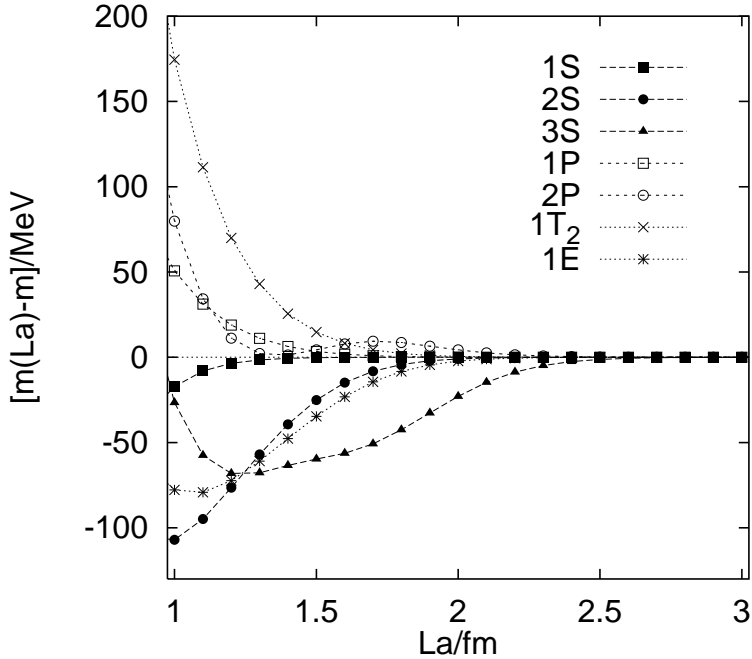}}
\caption{Difference between $\Upsilon$ levels and their infinite volume
values as a function of the lattice extent, $La$.}
\label{fig:fse}
\end{figure}

The potential approach not only offers an
intuitive and transparent representation of quarkonia
bound state properties in a continuum context but it can also
guide lattice simulations. By numerically
solving the Schr\"odinger equation
on a three-dimensional
torus for instance finite size effects can be estimated.
This has been done in Ref.~\cite{Bali:1998pi}, and the main
result is displayed in
Figure~\ref{fig:fse}.
While the approach to the infinite volume limit for $n=1$ states is
monotonous, this is not so for radial excitations. Some states,
in particular the $3S$, show a non-trivial behaviour that results
in infinite volume extrapolations from data obtained at lattices
with $a_{\sigma}L_{\sigma}<2$~fm to become uncontrolled.
The relevant symmetry group
on a torus (as well as on a discrete lattice) is $O_h$, rather than $O(3)$.
The five-dimensional
continuum $O(3)$ $D$ wave representation splits up
into the two-dimensional $O_h$ representation, $E$, and the
three-dimensional representation, $T_2$. It is amusing to see that
rotational symmetry is not only broken for finite lattice spacing but also
at any
finite volume, with $1T_2$ approaching the continuum $1D$ state from above
and $1E$ approaching it from below. If one aims at finite size effects
below 3~MeV, a lattice extent $L_{\sigma}a\approx 1.5$~fm seems to suffice
for $1S$ and $1P$ states while for $2S$, $2P$ and $1D$ states,
$L_{\sigma}a\approx 2$~fm
is required and for $3S$ or $3P$ even 2.5~fm become necessary.
On a 1.5 fm lattice for instance one would underestimate the $3S$ level by
more than 50~MeV.

\section{Conclusions}
QCD contains a rich spectrum of purely gluonic
excitations. Glueballs and torelons that are colour singlet
states can be realised, as well as
glueballinos that transform according to the adjoint
representation of the gauge group.
Chromo electro-magnetic
flux states between static colour sources in the fundamental
or in higher representations of the gauge group
can be constructed that have non-local gauge transformation properties.
Besides mesonic potentials and hybrid
excitations thereof, baryonic three-body potentials and even more
complicated situations can be investigated.

All these excitations can be accessed in lattice simulations with
much more ease
than properties of states containing fermionic constituents.
While the lattice reveals many interesting and non-trivial
aspects of QCD in some cases it is hard
to detect effects that very obviously do exist like the
breaking of the hadronic string.
Lattice results are
extremely useful to test and improve
models of low energy QCD. Moreover, phenomena
like the Casimir scaling found between potentials
between charges in different representations of the gauge group
or the $\Delta$ law of baryonic potentials, provide some insight into
hidden aspects of the dynamics of the theory.
Such results
give reason for optimism
that once out of the chaos there might
arise understanding. On the other hand QCD exhibits
a complex vacuum structure, and
even in the allegedly perturbative short
distance domain non-perturbative effects seem to play a r\^ole
in some cases.

QCD predicts the quark model as a classification
scheme of hadronic states to be incomplete.
Nonetheless, it seems to do quite well;
in particular those gluonic excitations that would make a difference,
come out to be quite heavy.
The quark model can of course be improved
by incorporating the known gluonic excitations. In doing so, one
would expect the lightest quark-gluon spin exotica to be vectors,
$J^{PC}=1^{-+}$. This result can be
systematically derived
for heavy quark bound states.
However, direct lattice simulations show that it also
applies to the light meson sector.

We have demonstrated that in
a non-relativistic situation, it is possible to factorise
gluonic effects from the slower dynamics of the quarks. This
adiabatic approximation is violated when ultra-soft gluons are
radiated, i.e.\ when the nature of the bound state changes during
the interaction time. However, such effects can be incorporated
into the potential formulation by enlarging the basis of states
onto which the Hamiltonian acts. Moreover, the validity of the
adiabatic approximation is tied to that of the non-relativistic
approximation in so far as the transition matrix elements are
suppressed by powers of the velocity, $v$.

Within the adiabatic framework, valence gluons that accompany the quarks
in the form of hybrid excitations of the flux tube and sea gluons,
whose average effect is parameterised in terms of interaction
potentials, can be distinguished from each other. To
lowest order of the relativistic
expansion pure quark model quarkonia and quark-gluon hybrids exist,
which then undergo mixing with each other as higher orders
of the relativistic expansion are incorporated.
Hybrid mesons become a well defined concept in the potential
approach and translation into the variables
used for instance in flux tube models is straight forward.

It has been shown that potential models can be systematically derived from
QCD. An understanding of effective field theory methods
turned out to be essential for this step.
The resulting Hamiltonian representation of the bound state problem
in terms of functions of canonical variables
offers a very intuitive and transparent representation of
quarkonium physics. 
It highlights parallels as well as differences
to well understood atomic physics. 

In view of phenomenological applications,
a non-perturbative determination
of the matching coefficients between QCD and lattice NRQCD is urgent.
As an alternative to lattice NRQCD and the lattice potential approach,
quarkonia properties can also be calculated
from relativistic quarks by introducing an anisotropy,
$a_{\tau}\ll a_{\sigma}$. However, the potential approach is unique
in its capability to access high radial excitations and to
determine wave functions.
From a non-perturbative determination of the matching coefficients,
simulations of heavy-light systems would benefit too.
Another challenge is to generalise the results presented
to heavy-light systems, i.e.\ to achieve a similar factorisation
into sea (gluon and quark) effects and valence (quark and gluon)
effects, within an expansion in terms of the inverse heavy quark mass,
$m^{-1}$.

\section*{Acknowledgements} 
G.B. has been supported by Deutsche Forschungsgemeinschaft
grant Nos.\ Ba 1564/3-1, Ba 1564/3-2 and Ba 1564/3-3
as well as
EU grant HPMF-CT-1999-00353.
The author wishes to express his gratitude in particular
to N.\ Brambilla, J.\ Soto and A.\ Vairo.
He has also benefitted from discussions with
R.\ Faustov, V.\ Galkin, C.\ Morningstar, P.\ Page,
M.\ Polikarpov, Y.\ Simonov and V.\ Zakharov.
S.\ Collins is most gratefully acknowledged for explaining lattice
NRQCD to the author, for motivation and for spotting many errors in the
manuscript. C.\ Davies and in particular N.\ Brambilla, D.\ Ebert
and C.\ Michael
contributed most valuable comments after reading an earlier
version of the manuscript. The author
thanks B.\ Bolder, V.\ Bornyakov, P.\ Boyle, M.\ M\"uller-Preu\ss{}ker,
M.\ Peardon, K.\ Schilling, C.\ Schlichter and A.\ Wachter for collaborating
with him in some of the lattice studies presented. K.\ Juge, J.\ Kuti,
F.\ Knechtli, C.\ Morningstar, M.\ Peardon, O.\ Philipsen and H.\ Wittig
are acknowledged for granting permission to reproduce their Figures and
N.\ Brambilla, A.\ Pineda, J.\ Soto and A.\ Vairo for communicating
their result on $V_{\prime}$ to the author prior to publication.

\newpage
\begin{appendix}
\section{The radial Schr\"odinger equation}
\label{app:schr}

For a rotationally symmetric potential, the standard substitution,
\begin{equation}
\psi_{nll_3}({\mathbf x})=\frac{u_{nl}(r)}{r}Y_{ll_3}(\theta,\phi),
\end{equation}
into the Schr\"odinger equation, Eq.~(\ref{eq:schroe}),
\begin{equation}
\left[\frac{{\mathbf p}^2}{2\mu_R}+V(r)\right]\psi_{nll_3}({\mathbf x})
=E_{nl}\psi_{nll_3}({\mathbf x}),\nonumber
\end{equation}
results in the radial equation,
\begin{equation}
\label{eq:radial}
u''_{nl}(r)+2\mu_R\left[E_{nl}-V(r)-\frac{l(l+1)}{2\mu_R r^2}\right]
u_{nl}(r)=0,
\end{equation}
with $u(0)=0$, $u'(0)=\psi({\mathbf 0})$.
In order to understand the dependence of the spectrum
on the underlying potential, we discuss power law
parametrisations,
\begin{equation}
V(r)=\lambda\, r^{\nu},\quad \nu=-1,1,2,\ldots.
\end{equation}
The virial theorem implies,
\begin{equation}
\label{eq:virial}
\langle T\rangle = E -\langle V\rangle=\frac{\nu}{2+\nu}E,
\end{equation}
where $T=p^2/(2\mu_R)$ denotes the kinetic energy and
$\langle T\rangle=\langle\psi_{nll_3}|T|\psi_{nll_3}\rangle$.
For simplicity we have omitted the quantum numbers from
Eq.~(\ref{eq:virial}).
The average relative velocity of the heavy quarks within the bound state can
easily be determined from Eq.~(\ref{eq:virial}),
\begin{equation}
\langle v^2\rangle=\frac{1}{\mu_R^2}\langle p^2\rangle
=\frac{2}{\mu_R}\frac{\nu}{2+\nu}E,
\end{equation}
whereas
\begin{equation}
\langle r^{\nu}\rangle=\frac{1}{\lambda}\langle V\rangle=
\frac{1}{\lambda}\frac{2}{2+\nu}E.
\end{equation}

Eq.~(\ref{eq:radial}) can be reformulated in terms of
dimensionless variables
by a simple scale transformation,
\begin{eqnarray}
\label{eq:rho}
\rho&=&\left(2\mu_R|\lambda|\right)^{\frac{1}{2+\nu}}r,\\
\label{eq:eps}
\epsilon&=&2\mu_R\left(2\mu_R|\lambda|\right)^{-\frac{2}{2+\nu}}E,\\
w(\rho)&=&u(r).
\end{eqnarray}
As a result, Eq.~(\ref{eq:radial}) reads,
\begin{equation}
\label{eq:radial2}
w_{nl}''+\left[\epsilon_{nl}-\mbox{sign}(\lambda)\rho^{\nu}
-\frac{l(l+1)}{\rho^2}\right]w_{nl}=0.
\end{equation}
The primes now represent derivatives with respect to the argument $\rho$.
The dependence of an energy splitting $\Delta E$ and a length scale
$l$ on the coupling strength $\lambda$ and reduced mass $\mu_R$ is
evident from Eqs.~(\ref{eq:eps}) and (\ref{eq:rho}), respectively,
\begin{equation}
\Delta E\propto |\lambda|^{\frac{2}{2+\nu}}\mu_R^{-\frac{\nu}{2+\nu}},
\quad l\propto\left(|\lambda|\mu_R\right)^{-\frac{1}{2+\nu}}.
\end{equation}
For negative powers $\nu$, level spacings decrease with increasing
quark mass, while for positive exponents, the opposite is the case.

A logarithmic potential, $V(r)=C\,\ln(r/r_0)$,
constitutes the limiting case between positive and negative $\nu$.
Indeed, for such a parametrisation
one obtains a velocity $\langle v^2\rangle=C/\mu_R$
as well as
quark mass independent splittings, $\Delta
E$~\cite{Quigg:1977dd} while $l\propto r_0C^{1/2}/\mu_R^{1/2}$.
For a Coulomb potential ($\nu=-1$) we obtain,
$\Delta E\propto\lambda^2\mu_R$ and $l\propto 1/(\lambda\mu_R)$
while a linear potential ($\nu=1$) yields
$\Delta E\propto\lambda^{2/3}/\mu_R^{1/3}$ and $l\propto 1/(\mu_R\lambda)^{1/3}$.
For a detailed discussion of the connection between spectrum and potential, we
refer the reader to an excellent review article by Quigg and
Rosner~\cite{Quigg:1979vr}.

\section{Euclidean Field Theory}
\label{app:conv}
We summarise the conventions and notations used in this article.
We start by
translating some Euclidean space-time objects
into Minkowski space-time (superscript $M$)
with metric $\eta=\mbox{diag}(1,-1,-1,-1)$
for reference:
\begin{eqnarray}
x_i&=&x^{i,M}=-x_i^M,\quad x_4=ix^{0,M}=ix_0^M,\\
\partial_i&=&\partial_i^M=\partial^{i,M},\quad \partial_4=-i\partial_0^M
=-i\partial^{0,M},\\
\gamma_i&=&i\gamma^{i,M}=-i\gamma_i^M,\quad\gamma_4=\gamma^{0,M}=\gamma_0^M,\\
A_i&=&-A^{i,M}=A_i^M,\quad A_4=-iA^{0,M}=-iA_0^M,\\
F_{i4}&=&iF^{i0,M}=-iF_{i0}^M,\quad F_{ij}=F^{ij,M}=F_{ij}^M,\\
{\mathbf B}&=&{\mathbf B}^M,\quad{\mathbf E}=-i{\mathbf E}^M.
\end{eqnarray}
The above conventions conform to the anti-commutation relations
$\{\gamma_{\mu},\gamma_{\nu}\}=2\delta_{\mu\nu}$ for the Dirac
$\gamma$-matrices. $A_{\mu}$ denotes the electro-magnetic four-potential,
$F_{\mu\nu}$ the Maxwell field strength tensor and ${\mathbf E}$ and
${\mathbf B}$ its components, the electric and magnetic fields.
While in Minkowski notation Lorentz indices assume values $\mu = 0,1,2,3$,
in Euclidean notation, they run from 1 to 4.

We denote the generators of the $SU(N)$ group by $T^a$,
with $a=1,\ldots,N_A$, $N_A=N^2-1$. They fulfil the commutation relations,
\begin{equation}
\label{eq:commu}
[T^a,T^b]=if^{abc}T^c,
\end{equation}
with $f^{abc}$ being real, totally antisymmetric structure constants.
In $SU(2)$ the generators can be represented in terms of Pauli matrices,
$T^i=\sigma^i/2$, in $SU(3)$ by Gell-Mann matrices, $T^a=\lambda^a/2$.
The vector potential lives in the Lie algebra,
\begin{equation}
A_{\mu}(x)=\sum_aA_{\mu}^a(x)T^a,
\end{equation}
while local gauge transformations generate the Lie group,
\begin{equation}
\Omega(x)=\exp[i\omega^a(x)T^a]\in SU(N).
\end{equation}
Under a gauge transformation, $\Omega$, the field $A_{\mu}$ transforms in the
adjoint representation,
\begin{equation}
\label{eq:transa}
A_{\mu}\rightarrow A_{\mu}^{\Omega}=\Omega[A_{\mu}-i\partial_{\mu}]
\Omega^{\dagger}.
\end{equation}
Fields that are in the fundamental representation of
$SU(N)$, e.g.\ Dirac spinors, $q(x)$, transform like,
\begin{equation}
\label{eq:transphi}
q(x)\rightarrow \Omega(x)q(x),\quad \bar{q}(x)\rightarrow
\bar{q}(x)\Omega^{\dagger}(x).
\end{equation}

The Dirac fermionic Lagrangian in Euclidian space reads,
\begin{equation}
\label{eq:Dirac}
{\mathcal L}_f=\bar{q}(\gamma_{\mu}D_{\mu}+m)q,
\end{equation}
with the covariant derivative,
\begin{equation}
D_{\mu}=\partial_{\mu}+iA_{\mu},
\end{equation}
while the Euclidean Yang-Mills Lagrangian,
\begin{equation}
\label{eq:yangmills}
{\mathcal L}_{YM}=\frac{1}{2g^2}\tr(F_{\mu\nu}F_{\mu\nu})
=\frac{1}{2g^2}\tr(F_{\mu\nu}^MF^{\mu\nu,M})=-{\mathcal L_{YM}^M},
\end{equation}
can be constructed from the field strength tensor,
\begin{equation}
F_{\mu\nu}=-i[D_{\mu},D_{\nu}]=\partial_{\mu}A_{\nu}-
\partial_{\nu}A_{\mu}+i[A_{\mu},A_{\nu}].
\end{equation}
The relative minus sign within
Eq.~(\ref{eq:yangmills}) with respect to the Minkowski version implies
that solutions of the classical equations of motion minimise the action,
which is bounded from below. Therefore, quantum fluctuations
are suppressed with respect to classical solutions
by factors, $e^{-\delta S}$,
within the path integral measure.

The phase and normalisation of the field strength tensor above
is chosen such that,
\begin{equation}
\label{eq:ebconv}
gB_i=\frac{1}{2}\epsilon_{ijk}F_{jk},\quad gE_i=F_{i4},
\end{equation}
correspond to the chromo-magnetic and electric fields, respectively.
$g=\sqrt{4\pi\alpha_s}$ denotes the strong coupling ``constant''.
Note that our definition of the electric field,
${\mathbf E}=-i{\mathbf E}^M$, differs by a phase $i$
from some text book conventions, resulting in, $E^a_iE^a_i\geq 0$.
The gauge action expressed in terms of the colour fields reads,
\begin{equation}
S_{YM}=\frac{1}{2}\int\!d^4x\,(E_i^aE_i^a+B_i^aB_i^a).
\end{equation}

\section{The perturbative $\beta$-function}
\label{app:run}
The one, two and three loop coefficients 
of the $\beta$-function, Eq.~(\ref{rge}),
\begin{equation}
\nonumber
\beta(\alpha_s)=\frac{d\alpha_s}{d\ln \mu^2}=
-\beta_0\alpha_s^2-\beta_1\alpha_s^3-\beta_2\alpha_s^4
-\ldots,
\end{equation}
have been calculated
in Refs.~\cite{Gross:1973th,Politzer:1973um},
\cite{Caswell:1974gg,Jones:1974mm,Egorian:1979zx} and
\cite{Tarasov:1980au,Larin:1993tp},
respectively,
in the
modified minimal subtraction $\overline{MS}$
scheme~\cite{'tHooft:1973mm,Bardeen:1978yd,Braaten:1981dv}
of dimensional
regularisation~\cite{'tHooft:1972fi}.
The $n_f$ flavour results for an $SU(N)$ gauge group read,
\begin{eqnarray}
\label{beta0}
\beta_0&=&\left(\frac{11}{3}N-\frac{2}{3}n_f\right)\frac{1}{4\pi},\\
\label{beta1}
\beta_1&=&\left[\frac{34}{3}N^2-
\left(\frac{13}{3}N-\frac{1}{N}\right)n_f\right]\frac{1}{16\pi^2},\\
\beta_2^{\overline{MS}}&=&\left[\frac{2857}{54}N^3-
\left(\frac{1709}{54}N^2-\frac{187}{36}-\frac{1}{4N^2}\right)n_f\right.\\
\nonumber
&+&\left.
\left(\frac{56}{27}N-\frac{11}{18N}\right)n_f^2\right]\frac{1}{64\pi^3},
\end{eqnarray}
while the four-loop coefficient $\beta_3^{\overline{MS}}$ has been calculated
in Ref.~\cite{vanRitbergen:1997va}.
The latter
reference also contains the coefficients for all compact semi-simple Lie
gauge groups.
The conversion between $\overline{MS}$ scheme couplings and
the bare lattice coupling for Wilson gluonic and fermionic
action is know to two
loops~\cite{Hasenfratz:1980kn,Weisz:1981pu,Dashen:1981vm,Kawai:1981ja,Luscher:1995nr,Christou:1998ws}.
The numerical pure gauge result for the $\beta$ function coefficient
$\beta_2^L$
reads~\cite{Alles:1997cy,Luscher:1995np},
\begin{equation}
\label{beta2l}
\beta_2^L\approx\left(-366.2N^3+1433.8N-\frac{2143}{N}\right)\frac{1}{64\pi^3},
\end{equation}
while for $SU(3)$ with $n_f$ flavours
of Wilson fermions one obtains~\cite{Christou:1998ws},
\begin{equation}
\beta_2^L\approx\left(-6299.9-1067n_f+59.89n_f^2\right)\frac{1}{64\pi^3}.
\end{equation}

Translating
between one scheme and another is straight forward: from
\begin{equation}
\alpha'(\mu)=\alpha(\mu)+c_1\alpha^2(\mu)+c_2\alpha^3(\mu)+
c_3\alpha^4(\mu)+\cdots,
\end{equation}
one obtains,
\begin{eqnarray}
\beta_0'&=&\beta_0,\quad\beta_1'=\beta_1,\\
\beta_2'&=&\beta_2-c_1\beta_1+(c_2-c_1^2)\beta_0,\\
\beta_3'&=&\beta_3-2c_1\beta_2+c_1^2\beta_1+2(c_3-3c_1c_2+2c_1^3)\beta_0,\\
\Lambda'&=&\Lambda\, e^{c_1/(2\beta_0)}.
\end{eqnarray}

\section{The centre symmetry}
\label{centres}
On a torus, a global
$Z_N$ symmetry is associated
with each compactified space-time direction, $\mu$,
besides the invariance of the action and path integral measure
under local gauge transformations.
$Z_N$ denotes the set of the $N$ $N$th roots of unity
and $z\in Z_N\subset SU(N)$. This means,
\begin{equation}
[z,U_{x,\mu}]=0.
\end{equation}
Multiplying all links crossing a hypersurface perpendicular to the
$\mu$ direction by a factor $z$,
\begin{equation}
\label{eq:centre}
U_{x,\nu}\rightarrow U_{x,\nu}^z=\left\{
\begin{array}{l}z\,U_{x,\nu}\quad\forall\quad x_{\mu}=0,\quad\nu=\mu\\
U_{x,\nu}\quad\mbox{otherwise}\end{array}\right.,
\end{equation}
leaves traces of closed loops of link variables with trivial winding number
around the boundary in $\mu$-direction invariant:
since every such loop
crosses every hypersurface an even number of times, all factors
$z$ that are collected when crossing in the positive direction
are cancelled by the $z^*$ factors collected from negative crossings.
In particular, this argument applies to
all pure gauge $SU(N)$ actions which are linear combinations of
traces of such loops. The fermionic part of the action, containing
a covariant derivative, however,
explicitly violates this $Z_N$ invariance.

\subsection{The Polyakov line and deconfinement}
The position of the hypersurface of Eq.~(\ref{eq:centre}) can be
moved by means of ordinary gauge transformations. Therefore, in infinite
volume, it can be sent to infinity and the centre symmetry will be in no way
different from an ordinary (large)
gauge transformation. On the torus, however, the surface can still be
moved around but not removed. The simplest object that is sensitive to
the centre symmetry is the Polyakov line,
\begin{equation}
\label{eq:pl}
P({\mathbf x})=\Tr\,\left\{
{\mathcal T}\left[\exp\left(
i\int_0^{aL_{\tau}}\!dx_4\,A_4(x)\right)\right]\right\}=
\Tr\,\left(\prod_{x_4=0}^{aL_{\tau}}U_{x,4}\right),
\end{equation}
a loop encircling the temporal boundary.
${\mathcal T}$ denotes time ordering of the argument.
Obviously, under a centre transformations with respect to the
4-direction, $P^z=zP$.
In the pure gauge case, where the centre symmetry is a symmetry of the
action and the path integral measure,
this means\footnote{Due to translational invariance, the expectation
value of a Polyakov line does not depend on the position, ${\mathbf x}$.},
$\langle P\rangle = 0$ on any finite spatial volume.
In the infinite volume limit there is, however, the possibility
of spontaneously breaking global symmetries and, indeed, similar to Ising and
Potts spin models, at high temperatures, the $Z_N$ symmetry is broken.

The expectation value of the Polyakov loop can be related to
the free energy of an isolated static
colour source~\cite{Kuti:1981gh,McLerran:1981pb},
\begin{equation}
|\langle P\rangle|=
\left|\frac{1}{L_{\sigma}^3}
\sum_{\mathbf x}\langle P({\mathbf x})\rangle\right|
\rightarrow e^{-\beta F_q}\quad (L_{\sigma}\rightarrow\infty),
\end{equation}
where the inverse temperature $\beta=T^{-1}=aL_{\tau}$,
that should not be confused with the
inverse Yang-Mills coupling, is related to the temporal
lattice extent. The vanishing expectation value of the
Polyakov line
observed in low temperature lattice simulations implies
an infinite free energy of an isolated quark and, therefore, confinement.
Vice versa, above a critical temperature $\beta^{-1}\geq T_c$ in
$SU(N)$ gauge theories, the expectation
value will move into the direction of one of the $N$th roots of unity,
implying a finite free energy and the possibility to eventually find
isolated quarks.
The case of QCD with sea quarks is interesting in so far as the centre symmetry
is explicitly broken: an isolated quark comes along
with only a finite free
energy penalty.
However, the energy
required to isolate a quark is still sufficiently high to create a
quark anti-quark pair out of the vacuum. Therefore, despite the
fact that chromo-electric strings between opposite charges can break,
the theory is still {\em effectively} confining colour sources
at zero temperature.

\subsection{Torelons}
It is not only worth considering
Polyakov lines wrapping around the lattice in the
temporal direction but also to discuss their analogue, which we will call the
Wilson line, that encircles a spatial lattice direction $i$,
\begin{equation}
L_i(t)=\frac{1}{L_{\sigma}^3}\sum_{\mathbf x}
\Tr\,\left(\prod_{x_i=0}^{aL_{\sigma}}U_{({\mathbf x},t),i}\right).
\end{equation}
We have already included the projection onto {\em zero} momentum into
the definition by summing over all spatial points.
Note that the above sum over the component
$x_i$ yields $L_{\sigma}$ identical
contributions. In principle, a projection
onto any momentum orthogonal to the direction of the Wilson line is possible.
From the correlation function,
\begin{equation}
\langle \re L_i(t)\re L_i(0)
\rangle\propto e^{-m_Tt}\quad (t\rightarrow\infty),
\end{equation}
the mass $m_T$ of a
torelon~\cite{Michael:1987cj,Michael:1989vh}
can be extracted, an excitation that only exists
on the torus and that corresponds to a colour flux tube wrapping around a
periodic boundary~\cite{'tHooft:1979uj}.

While for small spatial extents, $aL_{\sigma}$, the centre
symmetry of the classical Lagrangian with respect to spatial directions can be
dynamically broken, analogous to the finite temperature case, for sufficiently
large $aL_{\sigma}\gg T_c^{-1}$, the centre symmetry implies for\footnote{
For $SU(2)$, $L_i=L_i^*$ is real.} $N\geq 3$,
\begin{equation}
0=\langle L_i(t)L_i(0)\rangle=\langle \re L_i(t)\re L_i(0)
\rangle - \langle \im L_i(t)\im L_i(0)
\rangle.
\end{equation}
Note that the imaginary part, $i\langle\re L_i(t)\im L_i(0)+
\im L_i(t)\re L_i(0)\rangle$,
of the correlation function vanishes by
charge invariance. From the above equality, it follows that,
\begin{equation}
\langle \re L_i(t)\re L_i(0)
\rangle = \langle \im L_i(t)\im L_i(0)
\rangle=\frac{1}{2}\langle L_i(t)L_i^*(0)\rangle.
\end{equation}
Moreover, centre symmetry yields,
\begin{equation}
\langle \re L_i(t)\re L_j(0)
\rangle = \delta_{ij}
\langle \re L_i(t)\re L_i(0)
\rangle.
\end{equation}
The above two equations imply that all correlation functions
between linear combinations
of imaginary or real parts of Wilson lines are proportional to each other.
Therefore, all torelon states that correspond to one unit of flux are
degenerate.
In addition to torelons
corresponding to one unit of flux, torelons wrapping several times around
different boundary directions can be constructed and labelled according to
${\mathbf n}=(n_1,n_2,n_3)$, $n_1\geq n_2\geq n_3$,
$n_i=0,1,\ldots,n_{\mbox{\scriptsize max}}$,
where $N/2-1<n_{\mbox{\scriptsize max}}\leq N/2$ since the centre symmetry
implies that states with winding numbers $N\pm n$ are indistinguishable
from $n$ wrappings.
As soon as fermions are included into the action, the
$Z_N$ symmetry is
broken and torelons corresponding to different representations of the cubic
group $O_h\otimes C$
will in general assume different masses.

In the limit of large $aL_{\sigma}$ the situation of a closed
flux tube encircling
a periodic boundary becomes indistinguishable from
a flux tube with fixed ends,
created between point-like charge and anti-charge at infinite
separation. Therefore, in pure gauge theories
the energy stored per unit length
will become identical to the string tension, $\sigma$,
the infinite distance slope of
the static potential:
\begin{equation}
m_T\rightarrow\sigma a L_{\sigma}\quad (aL_{\sigma}\rightarrow\infty).
\end{equation}

\section{Matching NRQCD to QCD}
\label{app:match}
In this Appendix, we display results on the matching coefficients
between NRQCD,
Eqs.~(\ref{eq:lagnrqcd}) -- (\ref{eq:lagnrqcdend}),
and QCD, calculated in the $\overline{MS}$
scheme\footnote{Note that the result for $c_D$ derived
in Ref.~\cite{Chen:1995dg} turned out to be
incorrect~\cite{Balzereit:1996yy}.}~\cite{Amoros:1997rx,Balzereit:1996yy,Bauer:1998gs,Chen:1995dg,Pineda:1998kj,Manohar:1997qy},
\begin{eqnarray}
c_F&=&
\left[\frac{\alpha_s(m)}{\alpha_s(\mu)}\right]^{\frac{\gamma_0}{2\beta_0}}
\left\{1+\frac{13}{6\pi}\alpha_s(m)
\right.\nonumber\\\label{eq:cfper}
&+&\left.\frac{\gamma_1\beta_0
-\gamma_0\beta_1}{2\beta_0^2}\left[\alpha_s(m)-\alpha_s(\mu)\right]
\right\},\\\label{eq:cdper}
c_D&=&\left[
\frac{\alpha_s(m)}{\alpha_s(\mu)}\right]^{\frac{\gamma_0}{\beta_0}}
+\frac{308}{117}\left\{1-\left[
\frac{\alpha_s(m)}{\alpha_s(\mu)}\right]^{\frac{13\gamma_0}{12\beta_0}}
\right\},\\\label{eq:d1}
d_{ss}&=&
-\frac{4}{9}\frac{\alpha_s^2(\mu)}{m_1^2-m_2^2}\left[
m_1^2\left(\ln\frac{m_2}{\mu}+\frac{1}{6}\right)
-m_2^2\left(\ln\frac{m_1}{\mu}+\frac{1}{6}\right)\right],\\
d_{sv}&=&\frac{4}{9}\frac{\alpha_s^2(\mu)}{m_1^2-m_2^2}m_1m_2
\ln\frac{m_1}{m_2},\\
d_{vs}&=&
\frac{1}{2}\frac{\alpha_s^2(\mu)}{m_1^2-m_2^2}\left\{
-\frac{5}{3}\left[m_1^2\left(\ln\frac{m_2}{\mu}+\frac{1}{6}\right)
-m_2^2\left(\ln\frac{m_1}{\mu}+\frac{1}{6}\right)\right]\right.\nonumber\\
&+&\left.
\frac{6}{m_1m_2}\left[m_1^4\left(\ln\frac{m_2}{\mu}+\frac{20}{3}\right)
-m_2^4\left(\ln\frac{m_1}{\mu}+\frac{20}{3}\right)\right]\right\},\\
d_{vv}&=&
\frac{1}{2}\frac{\alpha_s^2(\mu)}{m_1^2-m_2^2}\left\{\frac{5}{3}m_1m_2
\ln\frac{m_1}{m_2}\right.\label{eq:d4}\\\nonumber
&+&\left.
3\left[
m_1^2\left(\ln\frac{m_2}{\mu}+\frac{3}{2}\right)
-m_2^2\left(\ln\frac{m_1}{\mu}+\frac{3}{2}\right)\right]\right\},\\
b_1
&=&1-\frac{\alpha_s(\mu)}{6\pi}\left(\ln\frac{m_1}{\mu}+\ln\frac{m_2}{\mu}
\right),\\
b_{2,i}
&=&\frac{\alpha_s(\mu)}{120\pi},\\
b_{3,i}
&=&\frac{13\,\alpha_s(\mu)}{720\pi},
\end{eqnarray}
where $\gamma_0=6/(4\pi)$ and $\gamma_1=
\left(68-52n_f/6\right)/(16\pi^2)$ are the first two
coefficients of the quark mass
anomalous dimension function.
The above values for the $d_i$'s only apply to the non-equal mass case.
In the equal mass case, one encounters additional contributions that are
due to annihilation diagrams\footnote{
We ignore imaginary parts within $d_{ss}^c$ and $d_{vs}^c$.
Such contributions, however, appear in the matching
calculation and are related to the fact that
deep inelastic QCD
cross sections cannot be obtained correctly
within NRQCD.}~\cite{Pineda:1998kj}:
\begin{eqnarray}
d_{ss}^{c,\mbox{\scriptsize a.}}&=&\frac{4}{9}\alpha_s(m)\alpha_s(\mu)
(1-\ln 2),\label{eq:ddd1}\\
d_{sv}^{c,\mbox{\scriptsize a.}}&=&{\mathcal O}(\alpha_s^3),\\
d_{vs}^{c,\mbox{\scriptsize a.}}&=&\frac{5}{6}\alpha_s(m)\alpha_s(\mu)
(1-\ln 2),\\
d_{vs}^{c,\mbox{\scriptsize a.}}&=&-\pi\alpha_s(m)\left\{1+
\frac{31}{6\pi}\alpha_s(\mu)\right.\\\nonumber&\times&\left.
\left[\left(1-\frac{2n_f}{31}\right)\ln\frac{m}{\mu}-\frac{119}{186}+
\left(\frac{5}{3}-2\ln 2\right)\frac{n_f}{31}\right]\right\}.
\end{eqnarray}
In addition there are
the non-annihilation contributions of Eqs.~(\ref{eq:d1}) --
(\ref{eq:d4}) that yield for $m_1=m_2$,
\begin{eqnarray}
d_{ss}^{\mbox{\scriptsize n.a.}}&=&
-\frac{4}{9}\alpha^2_s(\mu)\left(\ln\frac{m}{\mu}-\frac{1}{3}\right),\\
d_{sv}^{\mbox{\scriptsize n.a.}}&=&\frac{2}{9}\alpha_s^2(\mu),\\
d_{vs}^{\mbox{\scriptsize n.a.}}&=&-\frac{13}{6}\alpha_s^2(\mu)\left(
\ln\frac{m}{\mu}-\frac{97}{78}\right),\\
d_{vv}^{\mbox{\scriptsize n.a.}}
&=&\frac{3}{2}\alpha_s^2(\mu)\left(\ln\frac{m}{\mu}+\frac{23}{18}
\right).\label{eq:ddd8}
\end{eqnarray}
Note that a renormalisation group improved result for $d_{vv}$
that agrees with the above equations has also been derived in
Ref.~\cite{Chen:1995dg}.

Within the potentials of Eqs.~(\ref{cepo}),
(\ref{sdpo}), (\ref{eq:hdelta}) and (\ref{eq:hsd}),
the matching coefficients,
\begin{eqnarray}
d_s&=&\frac{1}{4\pi C_F\alpha_s}\left[
d_{ss}(m_1,m_2,\mu)+C_Fd_{vs}(m_1,m_2,\mu)\right]\\
&=&\frac{1}{4\pi\alpha_s}\frac{N_A}{4}
\left[-d_{ss}^c(m_1,m_2,\mu)-3d_{sv}^c(m_1,m_2,\mu)\right],\\
d_v&=&\frac{1}{4\pi C_F\alpha_s}
\left[d_{sv}(m_1,m_2,\mu)+C_Fd_{vv}(m_1,m_2,\mu)\right]\\
&=&\frac{1}{4\pi\alpha_s}\frac{N_A}{4}
\left[-d_{ss}^c(m_1,m_2,\mu)+d_{sv}^c(m_1,m_2,\mu)\right],
\end{eqnarray}
are required. For the equal mass case, we obtain from Eqs.~(\ref{eq:ddd1})
-- (\ref{eq:ddd8}) and (\ref{eq:rela1}) -- (\ref{eq:rela4}),
\begin{eqnarray}
\label{eq:ds2}
d_s&=&-\frac{5}{2}4\pi\alpha_s(\mu)\left[\ln\frac{m}{\mu}-\frac{17}{18}-
\frac{4}{15}\ln 2\right],\\\label{eq:dv}
d_v&=&\frac{3}{8}4\pi\alpha_s(\mu)\left[\ln\frac{m}{\mu}+\frac{17}{18}+
\frac{4}{9}\ln 2\right].
\end{eqnarray}
\end{appendix}

\newpage
\parskip=0pt
\partopsep=0pt
\parsep=0pt
\itemsep=0pt
\small
\bibliography{biblio}        

\begin{thebibliography}{100}
\parskip=0pt
\partopsep=0pt
\parsep=0pt
\itemsep=0pt
\small

\bibitem{Wilson:1974sk}
K.~G. Wilson,
\newblock Phys. Rev. {\bf D10}, 2445 (1974).

\bibitem{Creutz:1980zw}
M.~Creutz,
\newblock Phys. Rev. {\bf D21}, 2308 (1980).

\bibitem{Kuti:1981gh}
J.~Kuti, J.~Polonyi, and K.~Szlachanyi,
\newblock Phys. Lett. {\bf 98B}, 199 (1981).

\bibitem{McLerran:1981pk}
L.~D. McLerran and B.~Svetitsky,
\newblock Phys. Lett. {\bf 98B}, 195 (1981).

\bibitem{McLerran:1981pb}
L.~D. McLerran and B.~Svetitsky,
\newblock Phys. Rev. {\bf D24}, 450 (1981).

\bibitem{Gupta:1986vt}
R.~Gupta, G.~Guralnik, G.~W. Kilcup, A.~Patel, and S.~R. Sharpe,
\newblock Phys. Rev. Lett. {\bf 57}, 2621 (1986).

\bibitem{Gottlieb:1987eg}
S.~Gottlieb, W.~Liu, D.~Toussaint, R.~L. Renken, and R.~L. Sugar,
\newblock Phys. Rev. {\bf D35}, 3972 (1987).

\bibitem{Shifman:1979bx}
M.~A. Shifman, A.~I. Vainshtein, and V.~I. Zakharov,
\newblock Nucl. Phys. {\bf B147}, 385 (1979).

\bibitem{Novikov:1978cn}
V.~A. Novikov {\em et~al.},
\newblock Phys. Rept. {\bf 41}, 1 (1978).

\bibitem{Kang:1975cq}
J.~S. Kang and H.~J. Schnitzer,
\newblock Phys. Rev. {\bf D12}, 841 (1975).

\bibitem{Appelquist:1975ya}
T.~Appelquist and H.~D. Politzer,
\newblock Phys. Rev. {\bf D12}, 1404 (1975).

\bibitem{Eichten:1975af}
E.~Eichten {\em et~al.},
\newblock Phys. Rev. Lett. {\bf 34}, 369 (1975).

\bibitem{Eichten:1977jk}
E.~Eichten and K.~Gottfried,
\newblock Phys. Lett. {\bf 66B}, 286 (1977).

\bibitem{Gromes:1977np}
D.~Gromes,
\newblock Nucl. Phys. {\bf B131}, 80 (1977).

\bibitem{Richardson:1979bt}
J.~L. Richardson,
\newblock Phys. Lett. {\bf 82B}, 272 (1979).

\bibitem{Quigg:1979vr}
C.~Quigg and J.~L. Rosner,
\newblock Phys. Rept. {\bf 56}, 167 (1979).

\bibitem{Eichten:1980ms}
E.~Eichten, K.~Gottfried, T.~Kinoshita, K.~D. Lane, and T.~M. Yan,
\newblock Phys. Rev. {\bf D21}, 203 (1980).

\bibitem{Martin:1980jx}
A.~Martin,
\newblock Phys. Lett. {\bf 93B}, 338 (1980).

\bibitem{Martin:1981rm}
A.~Martin,
\newblock Phys. Lett. {\bf 100B}, 511 (1981).

\bibitem{Buchmuller:1981su}
W.~Buchm{\"u}ller and S.~H.~H. Tye,
\newblock Phys. Rev. {\bf D24}, 132 (1981).

\bibitem{Quigg:1981bj}
C.~Quigg and J.~L. Rosner,
\newblock Phys. Rev. {\bf D23}, 2625 (1981).

\bibitem{Eichten:1979pu}
E.~Eichten and F.~L. Feinberg,
\newblock Phys. Rev. Lett. {\bf 43}, 1205 (1979).

\bibitem{Eichten:1981mw}
E.~Eichten and F.~L. Feinberg,
\newblock Phys. Rev. {\bf D23}, 2724 (1981).

\bibitem{Peskin:1983up}
M.~E. Peskin,
\newblock in Proc. of 11th Int. SLAC Summer Inst. on Particle Physics: Dynamics
  and Spectroscopy at High Energy, Stanford, CA, 1983, ed. P.M. McDonough,
  (SLAC, Stanford, 1984).

\bibitem{Gromes:1984pm}
D.~Gromes,
\newblock Z. Phys. {\bf C22}, 265 (1984).

\bibitem{Barchielli:1988zs}
A.~Barchielli, E.~Montaldi, and G.~M. Prosperi,
\newblock Nucl. Phys. {\bf B296}, 625 (1988),
\newblock erratum, ibid. {\bf B303}, 752.

\bibitem{Barchielli:1990zp}
A.~Barchielli, N.~Brambilla, and G.~M. Prosperi,
\newblock Nuovo Cim. {\bf 103A}, 59 (1990).

\bibitem{Chen:1995dg}
Y.-Q. Chen, Y.-P. Kuang, and R.~J. Oakes,
\newblock Phys. Rev. {\bf D52}, 264 (1995), hep-ph/9406287.

\bibitem{Bali:1997am}
G.~S. Bali, K.~Schilling, and A.~Wachter,
\newblock Phys. Rev. {\bf D56}, 2566 (1997), hep-lat/9703019.

\bibitem{Bali:1998pi}
G.~S. Bali and P.~Boyle,
\newblock Phys. Rev. {\bf D59}, 114504 (1999), hep-lat/9809180.

\bibitem{Brambilla:1999ja}
N.~Brambilla and A.~Vairo,
\newblock (1999), hep-ph/9904330.

\bibitem{Brambilla:2000gk}
N.~Brambilla, A.~Pineda, J.~Soto, and A.~Vairo,
\newblock (2000), hep-ph/0002250.

\bibitem{Caswell:1986ui}
W.~E. Caswell and G.~P. Lepage,
\newblock Phys. Lett. {\bf 167B}, 437 (1986).

\bibitem{Thacker:1991bm}
B.~A. Thacker and G.~P. Lepage,
\newblock Phys. Rev. {\bf D43}, 196 (1991).

\bibitem{Lepage:1992tx}
G.~P. Lepage, L.~Magnea, C.~Nakhleh, U.~Magnea, and K.~Hornbostel,
\newblock Phys. Rev. {\bf D46}, 4052 (1992), hep-lat/9205007.

\bibitem{Chodos:1974je}
A.~Chodos, R.~L. Jaffe, K.~Johnson, C.~B. Thorn, and V.~F. Weisskopf,
\newblock Phys. Rev. {\bf D9}, 3471 (1974).

\bibitem{DeGrand:1975cf}
T.~DeGrand, R.~L. Jaffe, K.~Johnson, and J.~Kiskis,
\newblock Phys. Rev. {\bf D12}, 2060 (1975).

\bibitem{Hasenfratz:1978dt}
P.~Hasenfratz and J.~Kuti,
\newblock Phys. Rept. {\bf 40}, 75 (1978).

\bibitem{Hasenfratz:1980jv}
P.~Hasenfratz, R.~R. Horgan, J.~Kuti, and J.~M. Richard,
\newblock Phys. Lett. {\bf 95B}, 299 (1980).

\bibitem{DeTar:1983rw}
C.~E. DeTar and J.~F. Donoghue,
\newblock Ann. Rev. Nucl. Part. Sci. {\bf 33}, 235 (1983).

\bibitem{Kogut:1976zr}
J.~Kogut, D.~K. Sinclair, and L.~Susskind,
\newblock Nucl. Phys. {\bf B114}, 199 (1976).

\bibitem{Buchmuller:1982fr}
W.~Buchm{\"u}ller,
\newblock Phys. Lett. {\bf 112B}, 479 (1982).

\bibitem{Isgur:1983wj}
N.~Isgur and J.~Paton,
\newblock Phys. Lett. {\bf 124B}, 247 (1983).

\bibitem{Isgur:1985bm}
N.~Isgur and J.~Paton,
\newblock Phys. Rev. {\bf D31}, 2910 (1985).

\bibitem{Goddard:1973qh}
P.~Goddard, J.~Goldstone, C.~Rebbi, and C.~B. Thorn,
\newblock Nucl. Phys. {\bf B56}, 109 (1973).

\bibitem{Chodos:1974gt}
A.~Chodos and C.~B. Thorn,
\newblock Nucl. Phys. {\bf B72}, 509 (1974).

\bibitem{Luscher:1981iy}
M.~L{\"u}scher, G.~M{\"u}nster, and P.~Weisz,
\newblock Nucl. Phys. {\bf B180}, 1 (1981).

\bibitem{Dosch:1987sk}
H.~G. Dosch,
\newblock Phys. Lett. {\bf B190}, 177 (1987).

\bibitem{Dosch:1988ha}
H.~G. Dosch and Y.~A. Simonov,
\newblock Phys. Lett. {\bf 205B}, 339 (1988).

\bibitem{Simonov:1988rn}
Y.~A. Simonov,
\newblock Nucl. Phys. {\bf B307}, 512 (1988).

\bibitem{Baker:1983bt}
M.~Baker, J.~S. Ball, and F.~Zachariasen,
\newblock Nucl. Phys. {\bf B229}, 445 (1983).

\bibitem{Baker:1986mh}
M.~Baker, J.~S. Ball, and F.~Zachariasen,
\newblock Phys. Rev. {\bf D34}, 3894 (1986).

\bibitem{Baker:1995nq}
M.~Baker, J.~S. Ball, and F.~Zachariasen,
\newblock Phys. Rev. {\bf D51}, 1968 (1995).

\bibitem{Baker:1996mk}
M.~Baker, J.~S. Ball, N.~Brambilla, G.~M. Prosperi, and F.~Zachariasen,
\newblock Phys. Rev. {\bf D54}, 2829 (1996), hep-ph/9602419,
\newblock erratum, ibid. {\bf D56}, 2475 (1997).

\bibitem{Maedan:1988yi}
S.~Maedan and T.~Suzuki,
\newblock Prog. Theor. Phys. {\bf 81}, 229 (1989).

\bibitem{Shuryak:1982ff}
E.~V. Shuryak,
\newblock Nucl. Phys. {\bf B203}, 93 (1982).

\bibitem{Diakonov:1986eg}
D.~I. Diakonov and V.~Y. Petrov,
\newblock Nucl. Phys. {\bf B272}, 457 (1986).

\bibitem{Schafer:1998wv}
T.~Sch{\"a}fer and E.~V. Shuryak,
\newblock Rev. Mod. Phys. {\bf 70}, 323 (1998), hep-ph/9610451.

\bibitem{Godfrey:1985xj}
S.~Godfrey and N.~Isgur,
\newblock Phys. Rev. {\bf D32}, 189 (1985).

\bibitem{Baker:1996mu}
M.~Baker, J.~S. Ball, N.~Brambilla, and A.~Vairo,
\newblock Phys. Lett. {\bf B389}, 577 (1996), hep-ph/9609233.

\bibitem{Brambilla:1997aq}
N.~Brambilla and A.~Vairo,
\newblock Phys. Rev. {\bf D55}, 3974 (1997), hep-ph/9606344.

\bibitem{Baker:1997bg}
M.~Baker, J.~S. Ball, and F.~Zachariasen,
\newblock Phys. Rev. {\bf D56}, 4400 (1997), hep-ph/9705207.

\bibitem{Brambilla:1998bt}
N.~Brambilla,
\newblock (1998), hep-ph/9809263.

\bibitem{Regge:1959mz}
T.~Regge,
\newblock Nuovo Cim. {\bf 14}, 951 (1959).

\bibitem{Chew:1961ev}
G.~F. Chew and S.~C. Frautschi,
\newblock Phys. Rev. Lett. {\bf 7}, 394 (1961).

\bibitem{Veneziano:1968yb}
G.~Veneziano,
\newblock Nuovo Cim. {\bf A57}, 190 (1968).

\bibitem{Neveu:1971rx}
A.~Neveu and J.~H. Schwarz,
\newblock Nucl. Phys. {\bf B31}, 86 (1971).

\bibitem{Goto:1971ce}
T.~Goto,
\newblock Prog. Theor. Phys. {\bf 46}, 1560 (1971).

\bibitem{Nambu:1974zg}
Y.~Nambu,
\newblock Phys. Rev. {\bf D10}, 4262 (1974).

\bibitem{Gell-Mann:1964nj}
M.~Gell-Mann,
\newblock Phys. Lett. {\bf 8}, 214 (1964).

\bibitem{Zweig:1964jf}
G.~Zweig,
\newblock (unpublished, 1964) CERN preprint CERN-TH-412.

\bibitem{Greenberg:1964pe}
O.~W. Greenberg,
\newblock Phys. Rev. Lett. {\bf 13}, 598 (1964).

\bibitem{Han:1965pf}
M.~Y. Han and Y.~Nambu,
\newblock Phys. Rev. {\bf 139}, B1006 (1965).

\bibitem{Chew:1962eu}
G.~F. Chew and S.~C. Frautschi,
\newblock Phys. Rev. Lett. {\bf 8}, 41 (1962).

\bibitem{Brink:1973ja}
L.~Brink and H.~B. Nielsen,
\newblock Phys. Lett. {\bf 45B}, 332 (1973).

\bibitem{Bjorken:1969dy}
J.~D. Bj\o{}rken,
\newblock Phys. Rev. {\bf 179}, 1547 (1969).

\bibitem{Feynman:1969ej}
R.~P. Feynman,
\newblock Phys. Rev. Lett. {\bf 23}, 1415 (1969).

\bibitem{Bjorken:1969ja}
J.~D. Bj\o{}rken and E.~A. Paschos,
\newblock Phys. Rev. {\bf 185}, 1975 (1969).

\bibitem{Gross:1969jf}
D.~J. Gross and C.~H.~L. Smith,
\newblock Nucl. Phys. {\bf B14}, 337 (1969).

\bibitem{Fritzsch:1973pi}
H.~Fritzsch, M.~Gell-Mann, and H.~Leutwyler,
\newblock Phys. Lett. {\bf 47B}, 365 (1973).

\bibitem{Weinberg:1973un}
S.~Weinberg,
\newblock Phys. Rev. Lett. {\bf 31}, 494 (1973).

\bibitem{Caso:1998tx}
C.~Caso {\em et~al.},
\newblock Eur. Phys. J. {\bf C3}, 1 (1998).

\bibitem{Quigg:1997fy}
C.~Quigg,
\newblock Phys. Today {\bf 50}, 520 (1997).

\bibitem{Kogut:1975ag}
J.~Kogut and L.~Susskind,
\newblock Phys. Rev. {\bf D11}, 395 (1975).

\bibitem{Creutz:1979dw}
M.~Creutz,
\newblock Phys. Rev. Lett. {\bf 43}, 553 (1979),
\newblock erratum, ibid. {\bf 43}, 890.

\bibitem{Wilson:1979wp}
K.~G. Wilson,
\newblock in Recent Progresses in Gauge Theories, Cargese, France, Aug 26 - Sep
  8, 1979, (Plenum Press, New York, 1980).

\bibitem{Osterwalder:1975tc}
K.~Osterwalder and R.~Schrader,
\newblock Commun. Math. Phys. {\bf 42}, 281 (1975).

\bibitem{Creutz:1984mg}
M.~Creutz,
\newblock {\em Quarks, Gluons and Lattices} (Cambridge University Press,
  Cambridge, UK, 1983).

\bibitem{Itzykson:1989sx}
C.~Itzykson and J.~M. Drouffe,
\newblock {\em Statistical Field Theory, Vol. 1: From Brownian Motion to
  Renormalisation and Lattice Gauge Theory} (Cambridge University Press,
  Cambridge, UK, 1989).

\bibitem{Rothe:1992nt}
H.~J. Rothe,
\newblock {\em Lattice gauge theories: An Introduction} (World Scientific,
  Singapore, 1982).

\bibitem{Montvay:1994cy}
I.~Montvay and G.~M{\"u}nster,
\newblock {\em Quantum fields on a lattice} (Cambridge University Press,
  Cambridge, UK, 1994).

\bibitem{Drouffe:1978dn}
J.~M. Drouffe and C.~Itzykson,
\newblock Phys. Rept. {\bf 38}, 133 (1978).

\bibitem{Kogut:1983ds}
J.~B. Kogut,
\newblock Rev. Mod. Phys. {\bf 55}, 775 (1983).

\bibitem{Hasenfratz:1985pd}
A.~Hasenfratz and P.~Hasenfratz,
\newblock Ann. Rev. Nucl. Part. Sci. {\bf 35}, 559 (1985).

\bibitem{Gupta:1997nd}
R.~Gupta,
\newblock (1998), hep-lat/9807028.

\bibitem{Wittig:1999kb}
H.~Wittig,
\newblock (1999), hep-ph/9911400.

\bibitem{Aoki:1999yr}
CP-PACS, S.~Aoki {\em et~al.},
\newblock Phys. Rev. Lett. {\bf 84}, 238 (2000), hep-lat/9904012.

\bibitem{Kuramashi:1999ij}
CP-PACS, Y.~Kuramashi,
\newblock (1999), hep-lat/9904003.

\bibitem{Davies:1994pz}
C.~T.~H. Davies {\em et~al.},
\newblock Phys. Rev. Lett. {\bf 73}, 2654 (1994), hep-lat/9404012.

\bibitem{Eicker:1997ws}
SESAM, N.~Eicker {\em et~al.},
\newblock Phys. Lett. {\bf B407}, 290 (1997), hep-lat/9704019.

\bibitem{Garden:1999fg}
ALPHA and UKQCD, J.~Garden, J.~Heitger, R.~Sommer, and H.~Wittig,
\newblock (1999), hep-lat/9906013.

\bibitem{AliKhan:1999zp}
CP-PACS, A.~AliKhan {\em et~al.},
\newblock Nucl. Phys. Proc. Suppl. {\bf 83-84}, 176 (2000), hep-lat/9909050.

\bibitem{Capitani:1998mq}
ALPHA, S.~Capitani, M.~L{\"u}scher, R.~Sommer, and H.~Wittig,
\newblock Nucl. Phys. {\bf B544}, 669 (1999), hep-lat/9810063.

\bibitem{Manton:1980ts}
N.~S. Manton,
\newblock Phys. Lett. {\bf 96B}, 328 (1980).

\bibitem{Berg:1984yj}
B.~Berg, A.~Billoire, and K.~Koller,
\newblock Nucl. Phys. {\bf B233}, 50 (1984).

\bibitem{Michael:1988nd}
C.~Michael and M.~Teper,
\newblock Nucl. Phys. {\bf B305}, 453 (1988).

\bibitem{Symanzik:1983dc}
K.~Symanzik,
\newblock Nucl. Phys. {\bf B226}, 187 (1983).

\bibitem{Wilson:1993dy}
K.~G. Wilson,
\newblock Rev. Mod. Phys. {\bf 55}, 583 (1983).

\bibitem{Weisz:1983zw}
P.~Weisz,
\newblock Nucl. Phys. {\bf B212}, 1 (1983).

\bibitem{Luscher:1985xn}
M.~L{\"u}scher and P.~Weisz,
\newblock Commun. Math. Phys. {\bf 97}, 59 (1985),
\newblock erratum, ibid. {\bf 98}, 433.

\bibitem{Luscher:1985zq}
M.~L{\"u}scher and P.~Weisz,
\newblock Phys. Lett. {\bf 158B}, 250 (1985).

\bibitem{Alford:1995hw}
M.~Alford, W.~Dimm, G.~P. Lepage, G.~Hockney, and P.~B. Mackenzie,
\newblock Phys. Lett. {\bf B361}, 87 (1995), hep-lat/9507010.

\bibitem{Pennanen:1997ni}
P.~Pennanen and J.~Peisa,
\newblock Nucl. Phys. Proc. Suppl. {\bf 63}, 919 (1998), hep-lat/9709048.

\bibitem{Morningstar:1998da}
C.~J. Morningstar, K.~J. Juge, and J.~Kuti,
\newblock Nucl. Phys. Proc. Suppl. {\bf 73}, 590 (1999), hep-lat/9809098.

\bibitem{Morningstar:1997ff}
C.~J. Morningstar and M.~Peardon,
\newblock Phys. Rev. {\bf D56}, 4043 (1997), hep-lat/9704011.

\bibitem{Morningstar:1999rf}
C.~J. Morningstar and M.~Peardon,
\newblock Phys. Rev. {\bf D60}, 034509 (1999), hep-lat/9901004.

\bibitem{Shakespeare:1998uu}
N.~H. Shakespeare and H.~D. Trottier,
\newblock Phys. Rev. {\bf D59}, 014502 (1999), hep-lat/9803024.

\bibitem{Beinlich:1997ia}
B.~Beinlich, F.~Karsch, E.~Laermann, and A.~Peikert,
\newblock Eur. Phys. J. {\bf C6}, 133 (1999), hep-lat/9707023.

\bibitem{Parisi:1980pe}
G.~Parisi,
\newblock in Proc. of the 20th Int. Conf. on High Energy Physics, Madison, Jul
  17-23, 1980, eds.\ L.\ Durand and L.G.\ Pondrom, (American Inst.\ of Physics,
  New York, 1981).

\bibitem{Lepage:1993xa}
G.~P. Lepage and P.~B. Mackenzie,
\newblock Phys. Rev. {\bf D48}, 2250 (1993), hep-lat/9209022.

\bibitem{Iwasaki:1997sn}
Y.~Iwasaki, K.~Kanaya, T.~Kaneko, and T.~Yoshie,
\newblock Phys. Rev. {\bf D56}, 151 (1997), hep-lat/9610023.

\bibitem{Iwasaki:1983ck}
Y.~Iwasaki,
\newblock (unpublished, 1983), Tsukuba preprint UTHEP-118.

\bibitem{Iwasaki:1984cj}
Y.~Iwasaki and T.~Yoshie,
\newblock Phys. Lett. {\bf 143B}, 449 (1984).

\bibitem{Hasenfratz:1994sp}
P.~Hasenfratz and F.~Niedermayer,
\newblock Nucl. Phys. {\bf B414}, 785 (1994), hep-lat/9308004.

\bibitem{DeGrand:1995ji}
T.~DeGrand, A.~Hasenfratz, P.~Hasenfratz, and F.~Niedermayer,
\newblock Nucl. Phys. {\bf B454}, 587 (1995), hep-lat/9506030.

\bibitem{Niedermayer:1997eb}
F.~Niedermayer,
\newblock Nucl. Phys. Proc. Suppl. {\bf 53}, 56 (1997), hep-lat/9608097.

\bibitem{Bietenholz:1996cy}
W.~Bietenholz and U.~J. Wiese,
\newblock Nucl. Phys. {\bf B464}, 319 (1996), hep-lat/9510026.

\bibitem{Wilson:1975id}
K.~G. Wilson,
\newblock in Proc. Int. School of Subnuclear Physics, Erice, July 11 - 31,
  1975, ed. A.~Zichichi, (Plenum Press, New York, 1977).

\bibitem{Susskind:1977jm}
L.~Susskind,
\newblock Phys. Rev. {\bf D16}, 3031 (1977).

\bibitem{Sheikholeslami:1985ij}
B.~Sheikholeslami and R.~Wohlert,
\newblock Nucl. Phys. {\bf B259}, 572 (1985).

\bibitem{Luscher:1997ug}
M.~L{\"u}scher, S.~Sint, R.~Sommer, P.~Weisz, and U.~Wolff,
\newblock Nucl. Phys. {\bf B491}, 323 (1997), hep-lat/9609035.

\bibitem{Naik:1989bn}
S.~Naik,
\newblock Nucl. Phys. {\bf B316}, 238 (1989).

\bibitem{Eguchi:1984xr}
T.~Eguchi and N.~Kawamoto,
\newblock Nucl. Phys. {\bf B237}, 609 (1984).

\bibitem{Kaplan:1992bt}
D.~B. Kaplan,
\newblock Phys. Lett. {\bf B288}, 342 (1992), hep-lat/9206013.

\bibitem{Shamir:1993zy}
Y.~Shamir,
\newblock Nucl. Phys. {\bf B406}, 90 (1993), hep-lat/9303005.

\bibitem{Ginsparg:1982bj}
P.~H. Ginsparg and K.~G. Wilson,
\newblock Phys. Rev. {\bf D25}, 2649 (1982).

\bibitem{Luscher:1998du}
M.~L{\"u}scher,
\newblock Nucl. Phys. {\bf B549}, 295 (1999), hep-lat/9811032.

\bibitem{Hasenfratz:1998ri}
P.~Hasenfratz, V.~Laliena, and F.~Niedermayer,
\newblock Phys. Lett. {\bf B427}, 125 (1998), hep-lat/9801021.

\bibitem{Narayanan:1993wx}
R.~Narayanan and H.~Neuberger,
\newblock Phys. Lett. {\bf B302}, 62 (1993), hep-lat/9212019.

\bibitem{Narayanan:1995gw}
R.~Narayanan and H.~Neuberger,
\newblock Nucl. Phys. {\bf B443}, 305 (1995), hep-th/9411108.

\bibitem{Neuberger:1997fp}
H.~Neuberger,
\newblock Phys. Lett. {\bf B417}, 141 (1998), hep-lat/9707022.

\bibitem{Luscher:1994gh}
M.~L{\"u}scher, R.~Sommer, P.~Weisz, and U.~Wolff,
\newblock Nucl. Phys. {\bf B413}, 481 (1994), hep-lat/9309005.

\bibitem{Teper:1987wt}
M.~Teper,
\newblock Phys. Lett. {\bf 183B}, 345 (1987).

\bibitem{Albanese:1987ds}
APE, M.~Albanese {\em et~al.},
\newblock Phys. Lett. {\bf 192B}, 163 (1987).

\bibitem{Gusken:1989ad}
S.~G{\"u}sken {\em et~al.},
\newblock Phys. Lett. {\bf B227}, 266 (1989).

\bibitem{Perantonis:1989uz}
S.~Perantonis, A.~Huntley, and C.~Michael,
\newblock Nucl. Phys. {\bf B326}, 544 (1989).

\bibitem{Bali:1992ab}
G.~S. Bali and K.~Schilling,
\newblock Phys. Rev. {\bf D46}, 2636 (1992).

\bibitem{Gupta:1993vp}
R.~Gupta, D.~Daniel, and J.~Grandy,
\newblock Phys. Rev. {\bf D48}, 3330 (1993), hep-lat/9304009.

\bibitem{Lacock:1995qx}
UKQCD, P.~Lacock, A.~McKerrell, C.~Michael, I.~M. Stopher, and P.~W.
  Stephenson,
\newblock Phys. Rev. {\bf D51}, 6403 (1995), hep-lat/9412079.

\bibitem{Davies:1994mp}
C.~T.~H. Davies {\em et~al.},
\newblock Phys. Rev. {\bf D50}, 6963 (1994), hep-lat/9406017.

\bibitem{Philipsen:1999wf}
O.~Philipsen and H.~Wittig,
\newblock Phys. Lett. {\bf B451}, 146 (1999), hep-lat/9902003.

\bibitem{Eicker:1998vx}
N.~Eicker {\em et~al.},
\newblock Phys. Rev. {\bf D57}, 4080 (1998), hep-lat/9709002.

\bibitem{Spitz:1999tu}
SESAM, A.~Spitz {\em et~al.},
\newblock Phys. Rev. {\bf D60}, 074502 (1999), hep-lat/9906009.

\bibitem{Luscher:1977ms}
M.~L{\"u}scher,
\newblock Commun. Math. Phys. {\bf 54}, 283 (1977).

\bibitem{Osterwalder:1973dx}
K.~Osterwalder and R.~Schrader,
\newblock Commun. Math. Phys. {\bf 31}, 83 (1973).

\bibitem{Osterwalder:1978pc}
K.~Osterwalder and E.~Seiler,
\newblock Ann. Phys. {\bf 110}, 440 (1978).

\bibitem{Luscher:1984is}
M.~L{\"u}scher and P.~Weisz,
\newblock Nucl. Phys. {\bf B240}, 349 (1984).

\bibitem{Butler:1994em}
F.~Butler, H.~Chen, J.~Sexton, A.~Vaccarino, and D.~Weingarten,
\newblock Nucl. Phys. {\bf B430}, 179 (1994), hep-lat/9405003.

\bibitem{Bali:1993fb}
UKQCD, G.~S. Bali {\em et~al.},
\newblock Phys. Lett. {\bf B309}, 378 (1993), hep-lat/9304012.

\bibitem{Sexton:1995kd}
J.~Sexton, A.~Vaccarino, and D.~Weingarten,
\newblock Phys. Rev. Lett. {\bf 75}, 4563 (1995), hep-lat/9510022.

\bibitem{Michael:1988wf}
C.~Michael and M.~Teper,
\newblock Phys. Lett. {\bf 206B}, 299 (1988).

\bibitem{Michael:1989jr}
C.~Michael and M.~Teper,
\newblock Nucl. Phys. {\bf B314}, 347 (1989).

\bibitem{deForcrand:1985rs}
P.~de~Forcrand, G.~Schierholz, H.~Schneider, and M.~Teper,
\newblock Phys. Lett. {\bf 152B}, 107 (1985).

\bibitem{Sommer:1993ce}
R.~Sommer,
\newblock Nucl. Phys. {\bf B411}, 839 (1994), hep-lat/9310022.

\bibitem{Guagnelli:1998ud}
ALPHA, M.~Guagnelli, R.~Sommer, and H.~Wittig,
\newblock Nucl. Phys. {\bf B535}, 389 (1998), hep-lat/9806005.

\bibitem{Schilling:1993bk}
K.~Schilling and G.~S. Bali,
\newblock Int. J. Mod. Phys. {\bf C4}, 1167 (1993), hep-lat/9308014.

\bibitem{Bali:1997bj}
SESAM, G.~S. Bali {\em et~al.},
\newblock Nucl. Phys. Proc. Suppl. {\bf 63}, 209 (1998), hep-lat/9710012.

\bibitem{Edwards:1997xf}
R.~G. Edwards, U.~M. Heller, and T.~R. Klassen,
\newblock Nucl. Phys. {\bf B517}, 377 (1998), hep-lat/9711003.

\bibitem{Bode:1998hd}
ALPHA, A.~Bode, U.~Wolff, and P.~Weisz,
\newblock Nucl. Phys. {\bf B540}, 491 (1999), hep-lat/9809175.

\bibitem{Makeenko:1982bb}
Y.~M. Makeenko and M.~I. Polikarpov,
\newblock Nucl. Phys. {\bf B205}, 386 (1982).

\bibitem{Samuel:1985ub}
S.~Samuel, O.~Martin, and K.~Moriarty,
\newblock Phys. Lett. {\bf 153B}, 87 (1985).

\bibitem{Fingberg:1993ju}
J.~Fingberg, U.~Heller, and F.~Karsch,
\newblock Nucl. Phys. {\bf B392}, 493 (1993), hep-lat/9208012.

\bibitem{Bali:1993ru}
G.~S. Bali and K.~Schilling,
\newblock Phys. Rev. {\bf D47}, 661 (1993), hep-lat/9208028.

\bibitem{Wegner:1984qt}
F.~J. Wegner,
\newblock J. Math. Phys. {\bf 12}, 2259 (1971).

\bibitem{Brown:1979ya}
L.~S. Brown and W.~I. Weisberger,
\newblock Phys. Rev. {\bf D20}, 3239 (1979).

\bibitem{Landau:1987qm}
L.~D. Landau and E.~M. Lifschitz,
\newblock {\em Lehrbuch der theoretischen Physik, Band 3: Quantenmechanik}
  (Akademie Verlag, Berlin, DDR, 1979).

\bibitem{Seiler:1978ur}
E.~Seiler,
\newblock Phys. Rev. {\bf D18}, 482 (1978).

\bibitem{Bachas:1986xs}
C.~Bachas,
\newblock Phys. Rev. {\bf D33}, 2723 (1986).

\bibitem{Simon:1982yv}
B.~Simon and L.~G. Yaffe,
\newblock Phys. Lett. {\bf 115B}, 145 (1982).

\bibitem{Wilson:1976zj}
K.~G. Wilson,
\newblock Phys. Rept. {\bf 23}, 331 (1976).

\bibitem{Balian:1975xw}
R.~Balian, J.~M. Drouffe, and C.~Itzykson,
\newblock Phys. Rev. {\bf D11}, 2104 (1975),
\newblock erratum, ibid. {\bf D19}, 2514 (1979).

\bibitem{Creutz:1978yy}
M.~Creutz,
\newblock Rev. Mod. Phys. {\bf 50}, 561 (1978).

\bibitem{Drouffe:1983fv}
J.~M. Drouffe and J.~B. Zuber,
\newblock Phys. Rept. {\bf 102}, 1 (1983).

\bibitem{Munster:1980vk}
G.~M{\"u}nster and P.~Weisz,
\newblock Phys. Lett. {\bf 96B}, 119 (1980),
\newblock erratum, ibid. {\bf 100B}, 519 (1981).

\bibitem{Munster:1981es}
G.~M{\"u}nster,
\newblock Nucl. Phys. {\bf B190}, 439 (1981),
\newblock errata, ibid. {\bf B200}, 536 (1982) and {\bf B205}, 648 (1982).

\bibitem{Dyson:1952tj}
F.~J. Dyson,
\newblock Phys. Rev. {\bf 85}, 631 (1952).

\bibitem{Zinn-Justin:1981uk}
J.~Zinn-Justin,
\newblock Phys. Rept. {\bf 70}, 109 (1981).

\bibitem{Kogut:1980pm}
J.~B. Kogut and J.~Shigemitsu,
\newblock Phys. Rev. Lett. {\bf 45}, 410 (1980).

\bibitem{Smit:1982fx}
J.~Smit,
\newblock Nucl. Phys. {\bf B206}, 309 (1982).

\bibitem{Luscher:1981ac}
M.~L{\"u}scher,
\newblock Nucl. Phys. {\bf B180}, 317 (1981).

\bibitem{Drouffe:1981dp}
J.~M. Drouffe and J.~B. Zuber,
\newblock Nucl. Phys. {\bf B180}, 264 (1981).

\bibitem{Banks:1977cc}
T.~Banks, R.~Myerson, and J.~Kogut,
\newblock Nucl. Phys. {\bf B129}, 493 (1977).

\bibitem{Guth:1980gz}
A.~H. Guth,
\newblock Phys. Rev. {\bf D21}, 2291 (1980).

\bibitem{Creutz:1979zg}
M.~Creutz, L.~Jacobs, and C.~Rebbi,
\newblock Phys. Rev. {\bf D20}, 1915 (1979).

\bibitem{Lautrup:1980xr}
B.~Lautrup and M.~Nauenberg,
\newblock Phys. Lett. {\bf 95B}, 63 (1980).

\bibitem{Abrikosov:1957sx}
A.~A. Abrikosov,
\newblock Sov. Phys. JETP {\bf 5}, 1174 (1957).

\bibitem{Nielsen:1973ve}
H.~B. Nielsen and P.~Olesen,
\newblock Nucl. Phys. {\bf B61}, 45 (1973).

\bibitem{'tHooft:1974jz}
G.~'t~Hooft,
\newblock Nucl. Phys. {\bf B72}, 461 (1974).

\bibitem{Migdal:1984gj}
A.~A. Migdal,
\newblock Phys. Rept. {\bf 102}, 199 (1983).

\bibitem{Luscher:1980fr}
M.~L{\"u}scher, K.~Symanzik, and P.~Weisz,
\newblock Nucl. Phys. {\bf B173}, 365 (1980).

\bibitem{Arvis:1983fp}
J.~F. Arvis,
\newblock Phys. Lett. {\bf 127B}, 106 (1983).

\bibitem{Caselle:1987ek}
M.~Caselle, R.~Fiore, and F.~Gliozzi,
\newblock Phys. Lett. {\bf B200}, 525 (1988).

\bibitem{Caselle:1997ii}
M.~Caselle, R.~Fiore, F.~Gliozzi, M.~Hasenbusch, and P.~Provero,
\newblock Nucl. Phys. {\bf B486}, 245 (1997), hep-lat/9609041.

\bibitem{Akhmedov:1996mw}
E.~T. Akhmedov, M.~N. Chernodub, M.~I. Polikarpov, and M.~A. Zubkov,
\newblock Phys. Rev. {\bf D53}, 2087 (1996), hep-th/9505070.

\bibitem{Polyakov:1997nc}
A.~M. Polyakov,
\newblock Nucl. Phys. {\bf B486}, 23 (1997), hep-th/9607049.

\bibitem{Chernodub:1998ie}
M.~N. Chernodub and D.~A. Komarov,
\newblock JETP Lett. {\bf 68}, 117 (1998), hep-th/9809183.

\bibitem{Antonov:1998wi}
D.~Antonov and D.~Ebert,
\newblock Phys. Lett. {\bf B444}, 208 (1998), hep-th/9809018.

\bibitem{Baker:1999xn}
M.~Baker and R.~Steinke,
\newblock (1999), hep-ph/9905375.

\bibitem{Polchinski:1991ax}
J.~Polchinski and A.~Strominger,
\newblock Phys. Rev. Lett. {\bf 67}, 1681 (1991).

\bibitem{Ambjorn:1984yu}
J.~Ambj\o{}rn, P.~Olesen, and C.~Peterson,
\newblock Nucl. Phys. {\bf B244}, 262 (1984).

\bibitem{'tHooft:1979uj}
G.~'t~Hooft,
\newblock Nucl. Phys. {\bf B153}, 141 (1979).

\bibitem{Michael:1987cj}
C.~Michael,
\newblock J. Phys. {\bf G13}, 1001 (1987).

\bibitem{Michael:1994ej}
UKQCD, C.~Michael and P.~W. Stephenson,
\newblock Phys. Rev. {\bf D50}, 4634 (1994), hep-lat/9403004.

\bibitem{Teper:1998te}
M.~J. Teper,
\newblock Phys. Rev. {\bf D59}, 014512 (1999), hep-lat/9804008.

\bibitem{deForcrand:1985cz}
P.~de~Forcrand, G.~Schierholz, H.~Schneider, and M.~Teper,
\newblock Phys. Lett. {\bf 160B}, 137 (1985).

\bibitem{Kaczmarek:1999mm}
O.~Kaczmarek, F.~Karsch, E.~Laermann, and M.~L{\"u}tgemeier,
\newblock (1999), hep-lat/9908010.

\bibitem{Bigi:1994em}
I.~I. Bigi, M.~A. Shifman, N.~G. Uraltsev, and A.~I. Vainshtein,
\newblock Phys. Rev. {\bf D50}, 2234 (1994), hep-ph/9402360.

\bibitem{Beneke:1998rk}
M.~Beneke,
\newblock Phys. Lett. {\bf B434}, 115 (1998), hep-ph/9804241.

\bibitem{Appelquist:1977tw}
T.~Appelquist, M.~Dine, and I.~J. Muzinich,
\newblock Phys. Lett. {\bf 69B}, 231 (1977).

\bibitem{Brambilla:1999qa}
N.~Brambilla, A.~Pineda, J.~Soto, and A.~Vairo,
\newblock Phys. Rev. {\bf D60}, 091502 (1999), hep-ph/9903355.

\bibitem{Heller:1985hx}
U.~Heller and F.~Karsch,
\newblock Nucl. Phys. {\bf B251}, 254 (1985).

\bibitem{Martinelli:1998vt}
G.~Martinelli and C.~T. Sachrajda,
\newblock Nucl. Phys. {\bf B559}, 429 (1999), hep-lat/9812001.

\bibitem{Nadkarni:1986cz}
S.~Nadkarni,
\newblock Phys. Rev. {\bf D33}, 3738 (1986).

\bibitem{Quigg:1977dd}
C.~Quigg and J.~L. Rosner,
\newblock Phys. Lett. {\bf 71B}, 153 (1977).

\bibitem{Fischler:1977yf}
W.~Fischler,
\newblock Nucl. Phys. {\bf B129}, 157 (1977).

\bibitem{Kuhn:1988ty}
J.~H. Kuhn and P.~M. Zerwas,
\newblock Phys. Rept. {\bf 167}, 321 (1988).

\bibitem{Lucha:1991vn}
W.~Lucha, F.~F. Sch{\"o}berl, and D.~Gromes,
\newblock Phys. Rept. {\bf 200}, 127 (1991).

\bibitem{Eichten:1994gt}
E.~J. Eichten and C.~Quigg,
\newblock Phys. Rev. {\bf D49}, 5845 (1994), hep-ph/9402210.

\bibitem{Lang:1982tj}
C.~B. Lang and C.~Rebbi,
\newblock Phys. Lett. {\bf 115B}, 137 (1982).

\bibitem{Stack:1983wb}
J.~D. Stack,
\newblock Phys. Rev. {\bf D27}, 412 (1983).

\bibitem{Griffiths:1983ah}
L.~A. Griffiths, C.~Michael, and P.~E.~L. Rakow,
\newblock Phys. Lett. {\bf 129B}, 351 (1983).

\bibitem{Otto:1984qr}
S.~W. Otto and J.~D. Stack,
\newblock Phys. Rev. Lett. {\bf 52}, 2328 (1984).

\bibitem{Hasenfratz:1984gc}
A.~Hasenfratz, P.~Hasenfratz, U.~Heller, and F.~Karsch,
\newblock Z. Phys. {\bf C25}, 191 (1984).

\bibitem{Barkai:1984ca}
D.~Barkai, K.~J.~M. Moriarty, and C.~Rebbi,
\newblock Phys. Rev. {\bf D30}, 1293 (1984).

\bibitem{Sommer:1985du}
R.~Sommer and K.~Schilling,
\newblock Z. Phys. {\bf C29}, 95 (1985).

\bibitem{Huntley:1986ts}
A.~Huntley and C.~Michael,
\newblock Nucl. Phys. {\bf B270}, 123 (1986).

\bibitem{Hoek:1987uy}
J.~Hoek,
\newblock Z. Phys. {\bf C35}, 369 (1987).

\bibitem{Ford:1988ki}
I.~J. Ford, R.~H. Dalitz, and J.~Hoek,
\newblock Phys. Lett. {\bf 208B}, 286 (1988).

\bibitem{Perantonis:1990dy}
S.~Perantonis and C.~Michael,
\newblock Nucl. Phys. {\bf B347}, 854 (1990).

\bibitem{Michael:1992az}
C.~Michael and S.~J. Perantonis,
\newblock J. Phys. {\bf G18}, 1725 (1992).

\bibitem{Booth:1992bm}
UKQCD, S.~P. Booth {\em et~al.},
\newblock Phys. Lett. {\bf B294}, 385 (1992), hep-lat/9209008.

\bibitem{Bali:1995de}
G.~S. Bali, K.~Schilling, and C.~Schlichter,
\newblock Phys. Rev. {\bf D51}, 5165 (1995), hep-lat/9409005.

\bibitem{Itoh:1986gy}
S.~Itoh, Y.~Iwasaki, Y.~Oyanagi, and T.~Yoshie,
\newblock Nucl. Phys. {\bf B274}, 33 (1986).

\bibitem{Born:1994cq}
K.~D. Born, E.~Laermann, R.~Sommer, P.~M. Zerwas, and T.~F. Walsh,
\newblock Phys. Lett. {\bf B329}, 325 (1994).

\bibitem{Heller:1994rz}
U.~M. Heller, K.~M. Bitar, R.~G. Edwards, and A.~D. Kennedy,
\newblock Phys. Lett. {\bf B335}, 71 (1994), hep-lat/9401025.

\bibitem{Glassner:1996xi}
SESAM, U.~Gl{\"a}ssner {\em et~al.},
\newblock Phys. Lett. {\bf B383}, 98 (1996), hep-lat/9604014.

\bibitem{Aoki:1998sb}
CP-PACS, S.~Aoki {\em et~al.},
\newblock Nucl. Phys. Proc. Suppl. {\bf 73}, 216 (1999), hep-lat/9809185.

\bibitem{Bali:2000vr}
SESAM and $T\chi L$, G.~S. Bali {\em et~al.},
\newblock (2000), hep-lat/0003012.

\bibitem{Bernard:2000gd}
C.~Bernard {\em et~al.},
\newblock (2000), hep-lat/0002028.

\bibitem{Teper:1986ek}
M.~Teper,
\newblock Phys. Lett. {\bf B171}, 86 (1986).

\bibitem{Diakonov:1998rk}
D.~Diakonov and V.~Petrov,
\newblock (1998), hep-lat/9810037.

\bibitem{Coste:1985mn}
A.~Coste, A.~Gonzalez-Arroyo, J.~Jurkiewicz, and C.~P. Korthals-Altes,
\newblock Nucl. Phys. {\bf B262}, 67 (1985).

\bibitem{Bali:1999gq}
G.~S. Bali,
\newblock Fizika {\bf B8}, 229 (1999), hep-lat/9901023.

\bibitem{Allton:1998gi}
UKQCD, C.~R. Allton {\em et~al.},
\newblock Phys. Rev. {\bf D60}, 034507 (1999), hep-lat/9808016.

\bibitem{Bali:1997ec}
SESAM, G.~S. Bali {\em et~al.},
\newblock Nucl. Phys. Proc. Suppl. {\bf 53}, 239 (1997), hep-lat/9608096.

\bibitem{Manohar:1999xd}
A.~V. Manohar and I.~W. Stewart,
\newblock (1999), hep-ph/9912226.

\bibitem{Melles:2000dq}
M.~Melles,
\newblock (2000), hep-ph/0001295.

\bibitem{Schroder:1998vy}
Y.~Schr{\"o}der,
\newblock Phys. Lett. {\bf B447}, 321 (1999), hep-ph/9812205.

\bibitem{Peter:1997ig}
M.~Peter,
\newblock Phys. Rev. Lett. {\bf 78}, 602 (1997), hep-ph/9610209.

\bibitem{Peter:1997me}
M.~Peter,
\newblock Nucl. Phys. {\bf B501}, 471 (1997), hep-ph/9702245.

\bibitem{'tHooft:1977am}
G.~'t~Hooft,
\newblock in Proc.\ Int. School of Subnuclear Physics, Erice, Sicily, Jul 23 -
  Aug 10, 1977, ed. A. Zichichi, (Plenum Press, New York, 1979).

\bibitem{Lautrup:1977hs}
B.~Lautrup,
\newblock Phys. Lett. {\bf B69}, 109 (1977).

\bibitem{Shifman:1979cg}
M.~A. Shifman, A.~I. Vainshtein, and V.~I. Zakharov,
\newblock Nucl. Phys. {\bf B147}, 385 (1979).

\bibitem{Mueller:1985vh}
A.~H. Mueller,
\newblock Nucl. Phys. {\bf B250}, 327 (1985).

\bibitem{Beneke:1992mn}
M.~Beneke and V.~I. Zakharov,
\newblock Phys. Rev. Lett. {\bf 69}, 2472 (1992).

\bibitem{Brown:1992ic}
L.~S. Brown and L.~G. Yaffe,
\newblock Phys. Rev. {\bf D45}, 398 (1992).

\bibitem{Zakharov:1992bx}
V.~I. Zakharov,
\newblock Nucl. Phys. {\bf B385}, 452 (1992).

\bibitem{Vainshtein:1994ff}
A.~I. Vainshtein and V.~I. Zakharov,
\newblock Phys. Rev. Lett. {\bf 73}, 1207 (1994), hep-ph/9404248,
\newblock erratum, ibid. {\bf 75}, 3588 (1995).

\bibitem{Baker:1998jw}
M.~Baker, N.~Brambilla, H.~G. Dosch, and A.~Vairo,
\newblock Phys. Rev. {\bf D58}, 034010 (1998), hep-ph/9802273.

\bibitem{Akhoury:1998by}
R.~Akhoury and V.~I. Zakharov,
\newblock Phys. Lett. {\bf B438}, 165 (1998), hep-ph/9710487.

\bibitem{Grunberg:1998ix}
G.~Gr{\"u}nberg,
\newblock JHEP {\bf 11}, 006 (1998), hep-ph/9807494.

\bibitem{Gubarev:1999zq}
F.~V. Gubarev, M.~I. Polikarpov, and V.~I. Zakharov,
\newblock Mod. Phys. Lett. {\bf A14}, 2039 (1999).

\bibitem{Simonov:1999gk}
Y.~A. Simonov,
\newblock Phys. Rept. {\bf 320}, 265 (1999), hep-ph/9902233.

\bibitem{Gubarev:1999ie}
F.~V. Gubarev, M.~I. Polikarpov, and V.~I. Zakharov,
\newblock (1999), hep-ph/9908292.

\bibitem{Bali:1999ai}
G.~S. Bali,
\newblock Phys. Lett. {\bf B460}, 170 (1999), hep-ph/9905387.

\bibitem{Billoire:1980ih}
A.~Billoire,
\newblock Phys. Lett. {\bf 92B}, 343 (1980).

\bibitem{Jezabek:1998wk}
M.~Je{\.z}abek, M.~Peter, and Y.~Sumino,
\newblock Phys. Lett. {\bf B428}, 352 (1998), hep-ph/9803337.

\bibitem{Luscher:1992an}
M.~L{\"u}scher, R.~Narayanan, P.~Weisz, and U.~Wolff,
\newblock Nucl. Phys. {\bf B384}, 168 (1992), hep-lat/9207009.

\bibitem{Drummond:1998ar}
I.~T. Drummond,
\newblock Phys. Lett. {\bf B434}, 92 (1998), hep-lat/9805012.

\bibitem{Drummond:1998eh}
I.~T. Drummond and R.~R. Horgan,
\newblock Phys. Lett. {\bf B447}, 298 (1999), hep-lat/9811016.

\bibitem{Gliozzi:1999wq}
F.~Gliozzi and P.~Provero,
\newblock Nucl. Phys. {\bf B556}, 76 (1999), hep-lat/9903013.

\bibitem{DeTar:1998qa}
C.~DeTar, O.~Kaczmarek, F.~Karsch, and E.~Laermann,
\newblock Phys. Rev. {\bf D59}, 031501 (1999), hep-lat/9808028.

\bibitem{Buerger:1993bq}
W.~Buerger, M.~Faber, H.~Markum, and M.~M{\"u}ller,
\newblock Phys. Rev. {\bf D47}, 3034 (1993).

\bibitem{Faber:1988pi}
M.~E. Faber, H.~Markum, P.~de~Forcrand, M.~Meinhart, and I.~Stamatescu,
\newblock Phys. Lett. {\bf B200}, 348 (1988).

\bibitem{Michael:1999nq}
UKQCD, C.~Michael and P.~Pennanen,
\newblock Phys. Rev. {\bf D60}, 054012 (1999), hep-lat/9901007.

\bibitem{Pennanen:2000yk}
UKQCD, P.~Pennanen and C.~Michael,
\newblock (2000), hep-lat/0001015.

\bibitem{Schilling:1999mv}
K.~Schilling,
\newblock Nucl. Phys. Proc. Suppl. {\bf 83-84}, 140 (2000), hep-lat/9909152.

\bibitem{Michael:1998sg}
UKQCD, C.~Michael and J.~Peisa,
\newblock Phys. Rev. {\bf D58}, 034506 (1998), hep-lat/9802015.

\bibitem{Jorysz:1988qj}
I.~H. Jorysz and C.~Michael,
\newblock Nucl. Phys. {\bf B302}, 448 (1988).

\bibitem{Griffiths:1985ip}
L.~A. Griffiths, C.~Michael, and P.~E.~L. Rakow,
\newblock Phys. Lett. {\bf 150B}, 196 (1985).

\bibitem{Campbell:1986kp}
N.~A. Campbell, I.~H. Jorysz, and C.~Michael,
\newblock Phys. Lett. {\bf 167B}, 91 (1986).

\bibitem{Poulis:1997nn}
G.~I. Poulis and H.~D. Trottier,
\newblock Phys. Lett. {\bf B400}, 358 (1997), hep-lat/9504015.

\bibitem{Knechtli:1999av}
F.~Knechtli,
\newblock Nucl. Phys. Proc. Suppl. {\bf 83-84}, 673 (2000), hep-lat/9909164.

\bibitem{Michael:1992nc}
C.~Michael,
\newblock Nucl. Phys. Proc. Suppl. {\bf 26}, 417 (1992).

\bibitem{Stephenson:1999kh}
P.~W. Stephenson,
\newblock Nucl. Phys. {\bf B550}, 427 (1999), hep-lat/9902002.

\bibitem{deForcrand:1999kr}
P.~de~Forcrand and O.~Philipsen,
\newblock Phys. Lett. {\bf B475}, 280 (2000), hep-lat/9912050.

\bibitem{Kallio:2000jc}
K.~Kallio and H.~D. Trottier,
\newblock (2000), hep-lat/0001020.

\bibitem{Bock:1990kq}
W.~Bock {\em et~al.},
\newblock Z. Phys. {\bf C45}, 597 (1990).

\bibitem{Philipsen:1998de}
O.~Philipsen and H.~Wittig,
\newblock Phys. Rev. Lett. {\bf 81}, 4056 (1998), hep-lat/9807020.

\bibitem{Knechtli:1998gf}
ALPHA, F.~Knechtli and R.~Sommer,
\newblock Phys. Lett. {\bf B440}, 345 (1998), hep-lat/9807022.

\bibitem{Sommer:1996fr}
R.~Sommer,
\newblock Phys. Rept. {\bf 275}, 1 (1996), hep-lat/9401037.

\bibitem{Mandelstam:1974pi}
S.~Mandelstam,
\newblock Phys. Rept. {\bf 23}, 245 (1976).

\bibitem{'tHooft:1975pr}
G.~'t~Hooft,
\newblock in Proc. Int. School of Subnuclear Physics, Erice, Jul 11-31, 1975,
  ed. A. Zichichi, (Plenum Press, New York, 1977).

\bibitem{'tHooft:1981ht}
G.~'t~Hooft,
\newblock Nucl. Phys. {\bf B190}, 455 (1981).

\bibitem{Nielsen:1979xu}
H.~B. Nielsen and P.~Olesen,
\newblock Nucl. Phys. {\bf B160}, 380 (1979).

\bibitem{Ambjorn:1980xi}
J.~Ambj\o{}rn and P.~Olesen,
\newblock Nucl. Phys. {\bf B170}, 60 (1980).

\bibitem{Mack:1980zr}
G.~Mack,
\newblock Phys. Rev. Lett. {\bf 45}, 1378 (1980).

\bibitem{Cornwall:1979hz}
J.~M. Cornwall,
\newblock Nucl. Phys. {\bf B157}, 392 (1979).

\bibitem{Mack:1982gy}
G.~Mack and E.~Pietarinen,
\newblock Nucl. Phys. {\bf B205}, 141 (1982).

\bibitem{DelDebbio:1998uu}
L.~D. Debbio, M.~Faber, J.~Giedt, J.~Greensite, and S.~Olejnik,
\newblock Phys. Rev. {\bf D58}, 094501 (1998), hep-lat/9801027.

\bibitem{Chernodub:1998vk}
M.~N. Chernodub, M.~I. Polikarpov, A.~I. Veselov, and M.~A. Zubkov,
\newblock Nucl. Phys. Proc. Suppl. {\bf 73}, 575 (1999), hep-lat/9809158.

\bibitem{Ezawa:1982bf}
Z.~F. Ezawa and A.~Iwazaki,
\newblock Phys. Rev. {\bf D25}, 2681 (1982).

\bibitem{Bali:1996dm}
G.~S. Bali, V.~Bornyakov, M.~M{\"u}ller-Preu{\ss}ker, and K.~Schilling,
\newblock Phys. Rev. {\bf D54}, 2863 (1996), hep-lat/9603012.

\bibitem{Hioki:1991ai}
S.~Hioki {\em et~al.},
\newblock Phys. Lett. {\bf B272}, 326 (1991),
\newblock erratum, ibid. {\bf B281}, 416 (1992).

\bibitem{Kronfeld:1987vd}
A.~S. Kronfeld, G.~Schierholz, and U.~J. Wiese,
\newblock Nucl. Phys. {\bf B293}, 461 (1987).

\bibitem{Bali:1997cp}
G.~S. Bali, C.~Schlichter, and K.~Schilling,
\newblock Prog. Theor. Phys. Suppl. {\bf 131}, 645 (1998), hep-lat/9802005.

\bibitem{Bali:1998de}
G.~S. Bali,
\newblock (1998), hep-ph/9809351.

\bibitem{Gubarev:1999yp}
F.~V. Gubarev, E.~M. Ilgenfritz, M.~I. Polikarpov, and T.~Suzuki,
\newblock Phys. Lett. {\bf B468}, 134 (1999), hep-lat/9909099.

\bibitem{Lee:1999kv}
W.~Lee and D.~Weingarten,
\newblock Phys. Rev. {\bf D61}, 014015 (2000), hep-lat/9910008.

\bibitem{Jaffe:1976fd}
R.~L. Jaffe and K.~Johnson,
\newblock Phys. Lett. {\bf 60B}, 201 (1976).

\bibitem{Horn:1978rq}
D.~Horn and J.~Mandula,
\newblock Phys. Rev. {\bf D17}, 898 (1978).

\bibitem{Berg:1983kp}
B.~Berg and A.~Billoire,
\newblock Nucl. Phys. {\bf B221}, 109 (1983).

\bibitem{Michael:1990ry}
C.~Michael,
\newblock Acta Phys. Polon. {\bf B21}, 119 (1990).

\bibitem{Lacock:1996vy}
UKQCD, P.~Lacock, C.~Michael, P.~Boyle, and P.~Rowland,
\newblock Phys. Rev. {\bf D54}, 6997 (1996), hep-lat/9605025.

\bibitem{Hamermesh:1962xx}
M.~Hamermesh,
\newblock {\em Group Theory and its Application to Physical Problems}
  (Addison-Wesley, Reading, USA, 1962).

\bibitem{Lacock:1997ny}
UKQCD, P.~Lacock, C.~Michael, P.~Boyle, and P.~Rowland,
\newblock Phys. Lett. {\bf B401}, 308 (1997), hep-lat/9611011.

\bibitem{Bernard:1997ib}
MILC, C.~Bernard {\em et~al.},
\newblock Phys. Rev. {\bf D56}, 7039 (1997), hep-lat/9707008.

\bibitem{Lacock:1998be}
SESAM, P.~Lacock and K.~Schilling,
\newblock Nucl. Phys. Proc. Suppl. {\bf 73}, 261 (1999), hep-lat/9809022.

\bibitem{Campbell:1984fe}
N.~A. Campbell, L.~A. Griffiths, C.~Michael, and P.~E.~L. Rakow,
\newblock Phys. Lett. {\bf 142B}, 291 (1984).

\bibitem{Ford:1989as}
I.~J. Ford,
\newblock J. Phys. {\bf G15}, 1571 (1989).

\bibitem{Collins:1997cb}
UKQCD, S.~Collins, G.~Bali, and C.~Davies,
\newblock Nucl. Phys. Proc. Suppl. {\bf 63}, 335 (1998), hep-lat/9710058.

\bibitem{Juge:1999ie}
K.~J. Juge, J.~Kuti, and C.~J. Morningstar,
\newblock Phys. Rev. Lett. {\bf 82}, 4400 (1999), hep-ph/9902336.

\bibitem{Michael:1998sm}
C.~Michael,
\newblock (1998), hep-ph/9809211.

\bibitem{Foster:1998wu}
UKQCD, M.~Foster and C.~Michael,
\newblock Phys. Rev. {\bf D59}, 094509 (1999), hep-lat/9811010.

\bibitem{Brambilla:1999xf}
N.~Brambilla, A.~Pineda, J.~Soto, and A.~Vairo,
\newblock (1999), hep-ph/9907240.

\bibitem{Jaffe:1986qp}
R.~L. Jaffe, K.~Johnson, and Z.~Ryzak,
\newblock Ann. Phys. {\bf 168}, 344 (1986).

\bibitem{Ambjorn:1984dp}
J.~Ambj\o{}rn, P.~Olesen, and C.~Peterson,
\newblock Nucl. Phys. {\bf B240}, 533 (1984).

\bibitem{Bernard:1982pg}
C.~Bernard,
\newblock Phys. Lett. {\bf 108B}, 431 (1982).

\bibitem{Ambjorn:1984mb}
J.~Ambj\o{}rn, P.~Olesen, and C.~Peterson,
\newblock Nucl. Phys. {\bf B240}, 189 (1984).

\bibitem{Michael:1985ne}
C.~Michael,
\newblock Nucl. Phys. {\bf B259}, 58 (1985).

\bibitem{Trottier:1995fx}
H.~D. Trottier,
\newblock Phys. Lett. {\bf B357}, 193 (1995), hep-lat/9503017.

\bibitem{Deldar:1998ne}
S.~Deldar,
\newblock Nucl. Phys. Proc. Suppl. {\bf 73}, 587 (1999), hep-lat/9809137.

\bibitem{Bali:1999hx}
G.~S. Bali,
\newblock Nucl. Phys. Proc. Suppl. {\bf 83-84}, 422 (2000), hep-lat/9908021.

\bibitem{Deldar:1999vi}
S.~Deldar,
\newblock (1999), hep-lat/9911008.

\bibitem{Bernard:1983my}
C.~Bernard,
\newblock Nucl. Phys. {\bf B219}, 341 (1983).

\bibitem{Ohta:1986pc}
S.~Ohta, M.~Fukugita, and A.~Ukawa,
\newblock Phys. Lett. {\bf B173}, 15 (1986).

\bibitem{Markum:1988na}
H.~Markum and M.~E. Faber,
\newblock Phys. Lett. {\bf B200}, 343 (1988).

\bibitem{Muller:1991xj}
M.~M{\"u}ller, W.~Beirl, M.~Faber, and H.~Markum,
\newblock Nucl. Phys. Proc. Suppl. {\bf 26}, 423 (1992).

\bibitem{Johnson:1976sg}
K.~Johnson and C.~B. Thorn,
\newblock Phys. Rev. {\bf D13}, 1934 (1976).

\bibitem{Wetzorke:1999rt}
I.~Wetzorke, F.~Karsch, and E.~Laermann,
\newblock Nucl. Phys. Proc. Suppl. {\bf 83-84}, 218 (2000), hep-lat/9909037.

\bibitem{Green:1993yw}
A.~M. Green, C.~Michael, and J.~E. Paton,
\newblock Nucl. Phys. {\bf A554}, 701 (1993), hep-lat/9209019.

\bibitem{Green:1995cg}
A.~M. Green, C.~Michael, and M.~E. Sainio,
\newblock Z. Phys. {\bf C67}, 291 (1995), hep-lat/9404004.

\bibitem{Green:1996df}
A.~M. Green, J.~Lukkarinen, P.~Pennanen, and C.~Michael,
\newblock Phys. Rev. {\bf D53}, 261 (1996), hep-lat/9508002.

\bibitem{Green:1998nt}
A.~M. Green and P.~Pennanen,
\newblock Phys. Rev. {\bf C57}, 3384 (1998), hep-lat/9804003.

\bibitem{Pennanen:1998nu}
P.~Pennanen, A.~M. Green, and C.~Michael,
\newblock Phys. Rev. {\bf D59}, 014504 (1999), hep-lat/9804004.

\bibitem{Artru:1975zn}
X.~Artru,
\newblock Nucl. Phys. {\bf B85}, 442 (1975).

\bibitem{Dosch:1976gf}
H.~G. Dosch and V.~F. M{\"u}ller,
\newblock Nucl. Phys. {\bf B116}, 470 (1976).

\bibitem{Brambilla:1994zw}
N.~Brambilla, P.~Consoli, and G.~M. Prosperi,
\newblock Phys. Rev. {\bf D50}, 5878 (1994), hep-th/9401051.

\bibitem{Kalashnikova:1997px}
Y.~S. Kalashnikova and A.~V. Nefediev,
\newblock Phys. Atom. Nucl. {\bf 60}, 1333 (1997), hep-ph/9604411.

\bibitem{Bali:2000ab}
G.~S. Bali, M.~Peardon, and K.~Schilling,
\newblock in preparation.

\bibitem{Cornwall:1977xd}
J.~M. Cornwall,
\newblock Nucl. Phys. {\bf B128}, 75 (1977).

\bibitem{Cornwall:1996xr}
J.~M. Cornwall,
\newblock Phys. Rev. {\bf D54}, 6527 (1996), hep-th/9605116.

\bibitem{Sommer:1986da}
R.~Sommer and J.~Wosiek,
\newblock Nucl. Phys. {\bf B267}, 531 (1986).

\bibitem{Flower:1986ex}
J.~Flower,
\newblock (1986, unpublished) CALT-68-1377.

\bibitem{Thacker:1987aq}
H.~B. Thacker, E.~Eichten, and J.~C. Sexton,
\newblock Nucl. Phys. Proc. Suppl. {\bf 4}, 234 (1988).

\bibitem{Capstick:1986bm}
S.~Capstick and N.~Isgur,
\newblock Phys. Rev. {\bf D34}, 2809 (1986).

\bibitem{Pantaleone:1986uf}
J.~Pantaleone, S.~H.~H. Tye, and Y.~J. Ng,
\newblock Phys. Rev. {\bf D33}, 777 (1986).

\bibitem{Isgur:1989vq}
N.~Isgur and M.~B. Wise,
\newblock Phys. Lett. {\bf B232}, 113 (1989).

\bibitem{Eichten:1990zv}
E.~Eichten and B.~Hill,
\newblock Phys. Lett. {\bf B234}, 511 (1990).

\bibitem{Georgi:1990um}
H.~Georgi,
\newblock Phys. Lett. {\bf B240}, 447 (1990).

\bibitem{Pineda:1997bj}
A.~Pineda and J.~Soto,
\newblock Nucl. Phys. Proc. Suppl. {\bf 64}, 428 (1998), hep-ph/9707481.

\bibitem{Beneke:1999ff}
M.~Beneke, A.~Signer, and V.~A. Smirnov,
\newblock (1999), hep-ph/9906476,
\newblock in Proc., RADCOR 1998, Sept.\ 8 -- 12, 1998, Barcelona, ed.\ J.\ Sola
  (World Scientific, Singapore, 1999).

\bibitem{Bodwin:1998mn}
G.~T. Bodwin and Y.-Q. Chen,
\newblock Phys. Rev. {\bf D60}, 054008 (1999), hep-ph/9807492.

\bibitem{Grinstein:1998xb}
B.~Grinstein,
\newblock (1998), hep-ph/9811264.

\bibitem{Grinstein:1998gv}
B.~Grinstein and I.~Z. Rothstein,
\newblock Phys. Rev. {\bf D57}, 78 (1998), hep-ph/9703298.

\bibitem{Foldy:1950wa}
L.~L. Foldy and S.~A. Wouthuysen,
\newblock Phys. Rev. {\bf 78}, 29 (1950).

\bibitem{Itzykson:1980xx}
C.~Itzykson and J.-B. Zuber,
\newblock {\em Quantum field Theory} (McGraw-Hill, 1980).

\bibitem{Morningstar:1994qe}
C.~J. Morningstar,
\newblock Phys. Rev. {\bf D50}, 5902 (1994), hep-lat/9406002.

\bibitem{Luke:1999kz}
M.~E. Luke, A.~V. Manohar, and I.~Z. Rothstein,
\newblock (1999), hep-ph/9910209.

\bibitem{Pineda:1998kj}
A.~Pineda and J.~Soto,
\newblock Phys. Rev. {\bf D58}, 114011 (1998), hep-ph/9802365.

\bibitem{Manohar:1997qy}
A.~V. Manohar,
\newblock Phys. Rev. {\bf D56}, 230 (1997), hep-ph/9701294.

\bibitem{Luke:1992cs}
M.~Luke and A.~V. Manohar,
\newblock Phys. Lett. {\bf B286}, 348 (1992), hep-ph/9205228.

\bibitem{Chen:1993sx}
Y.-Q. Chen,
\newblock Phys. Lett. {\bf B317}, 421 (1993).

\bibitem{Finkemeier:1997re}
M.~Finkemeier, H.~Georgi, and M.~McIrvin,
\newblock Phys. Rev. {\bf D55}, 6933 (1997), hep-ph/9701243.

\bibitem{Morningstar:1993de}
C.~J. Morningstar,
\newblock Phys. Rev. {\bf D48}, 2265 (1993), hep-lat/9301005.

\bibitem{Davies:1998im}
UKQCD, C.~T.~H. Davies {\em et~al.},
\newblock Phys. Rev. {\bf D58}, 054505 (1998), hep-lat/9802024.

\bibitem{Trottier:1997bn}
H.~D. Trottier and G.~P. Lepage,
\newblock Nucl. Phys. Proc. Suppl. {\bf 63}, 865 (1998), hep-lat/9710015.

\bibitem{Dimm:1995fy}
W.~Dimm, G.~P. Lepage, and P.~B. Mackenzie,
\newblock Nucl. Phys. Proc. Suppl. {\bf 42}, 403 (1995), hep-lat/9412100.

\bibitem{AliKhan:1996ub}
A.~AliKhan, C.~T.~H. Davies, S.~Collins, J.~Sloan, and J.~Shigemitsu,
\newblock Phys. Rev. {\bf D53}, 6433 (1996), hep-lat/9512025.

\bibitem{Collins:1999ff}
S.~Collins {\em et~al.},
\newblock Phys. Rev. {\bf D60}, 074504 (1999), hep-lat/9901001.

\bibitem{Hasenfratz:1980kn}
A.~Hasenfratz and P.~Hasenfratz,
\newblock Phys. Lett. {\bf 93B}, 165 (1980).

\bibitem{Weisz:1981pu}
P.~Weisz,
\newblock Phys. Lett. {\bf 100B}, 331 (1981).

\bibitem{Dashen:1981vm}
R.~Dashen and D.~J. Gross,
\newblock Phys. Rev. {\bf D23}, 2340 (1981).

\bibitem{Kawai:1981ja}
H.~Kawai, R.~Nakayama, and K.~Seo,
\newblock Nucl. Phys. {\bf B189}, 40 (1981).

\bibitem{DiGiacomo:1981wt}
A.~D. Giacomo and G.~C. Rossi,
\newblock Phys. Lett. {\bf B100}, 481 (1981).

\bibitem{DiGiacomo:1982dp}
A.~D. Giacomo and G.~Paffuti,
\newblock Phys. Lett. {\bf 108B}, 327 (1982).

\bibitem{Manke:1997gt}
UKQCD, T.~Manke, I.~T. Drummond, R.~R. Horgan, and H.~P. Shanahan,
\newblock Phys. Lett. {\bf B408}, 308 (1997), hep-lat/9706003.

\bibitem{Brambilla:1998vm}
N.~Brambilla and A.~Vairo,
\newblock Nucl. Phys. Proc. Suppl. {\bf 74}, 201 (1999), hep-ph/9809230.

\bibitem{Gromes:1984ma}
D.~Gromes,
\newblock Z. Phys. {\bf C26}, 401 (1984).

\bibitem{Bali:1995yz}
G.~S. Bali, K.~Schilling, and A.~Wachter,
\newblock (1995), hep-lat/9506017,
\newblock in Proc., Confinement 1995, March 22-24, 1995, Osaka, eds.\ H. Toki
  et al. (World Scientific, Singapore, 1995).

\bibitem{Bali:1997cj}
G.~S. Bali, K.~Schilling, and A.~Wachter,
\newblock Phys. Rev. {\bf D55}, 5309 (1997), hep-lat/9611025.

\bibitem{Michael:1985wf}
C.~Michael and P.~E.~L. Rakow,
\newblock Nucl. Phys. {\bf B256}, 640 (1985).

\bibitem{Ebert:1999xv}
D.~Ebert, R.~N. Faustov, and V.~O. Galkin,
\newblock (1999), hep-ph/9911283.

\bibitem{Brambilla:1998qg}
N.~Brambilla and A.~Vairo,
\newblock Fizika {\bf B7}, 261 (1999), hep-ph/9902360.

\bibitem{Gupta:1982kp}
S.~N. Gupta, S.~F. Radford, and W.~W. Repko,
\newblock Phys. Rev. {\bf D26}, 3305 (1982).

\bibitem{Gromes:1988zx}
D.~Gromes,
\newblock Phys. Lett. {\bf B202}, 262 (1988).

\bibitem{Voloshin:1979hc}
M.~B. Voloshin,
\newblock Nucl. Phys. {\bf B154}, 365 (1979).

\bibitem{Leutwyler:1981tn}
H.~Leutwyler,
\newblock Phys. Lett. {\bf 98B}, 447 (1981).

\bibitem{Balitsky:1985iw}
I.~I. Balitsky,
\newblock Nucl. Phys. {\bf B254}, 166 (1985).

\bibitem{Gromes:1982su}
D.~Gromes,
\newblock Phys. Lett. {\bf B115}, 482 (1982).

\bibitem{Bertlmann:1983pf}
R.~A. Bertlmann and J.~S. Bell,
\newblock Nucl. Phys. {\bf B227}, 435 (1983).

\bibitem{DelDebbio:1994zn}
L.~D. Debbio, A.~D. Giacomo, and Y.~A. Simonov,
\newblock Phys. Lett. {\bf B332}, 111 (1994), hep-lat/9403016.

\bibitem{Bali:1998aj}
G.~S. Bali, N.~Brambilla, and A.~Vairo,
\newblock Phys. Lett. {\bf B421}, 265 (1998), hep-lat/9709079.

\bibitem{Ilgenfritz:1999tg}
E.-M. Ilgenfritz and S.~Thurner,
\newblock (1999), hep-lat/9905012.

\bibitem{Marquard:1987pe}
U.~Marquard and H.~G. Dosch,
\newblock Phys. Rev. {\bf D35}, 2238 (1987).

\bibitem{Labelle:1992hd}
P.~Labelle,
\newblock hep-ph/9209266,
\newblock in Proc., 14th Annual MRST Meeting, May 7 -- 8, 1992, Toronto, ed.\
  P.J.\ O'Donnell (University of Toronto, 1992).

\bibitem{Labelle:1998en}
P.~Labelle,
\newblock Phys. Rev. {\bf D58}, 093013 (1998), hep-ph/9608491.

\bibitem{Pineda:1998kn}
A.~Pineda and J.~Soto,
\newblock Phys. Rev. {\bf D59}, 016005 (1999), hep-ph/9805424.

\bibitem{Michael:1986rh}
C.~Michael,
\newblock Phys. Rev. Lett. {\bf 56}, 1219 (1986).

\bibitem{Huntley:1987de}
A.~Huntley and C.~Michael,
\newblock Nucl. Phys. {\bf B286}, 211 (1987).

\bibitem{deForcrand:1985zc}
P.~de~Forcrand and J.~D. Stack,
\newblock Phys. Rev. Lett. {\bf 55}, 1254 (1985).

\bibitem{Campostrini:1986ki}
M.~Campostrini, K.~Moriarty, and C.~Rebbi,
\newblock Phys. Rev. Lett. {\bf 57}, 44 (1986).

\bibitem{Campostrini:1987hu}
M.~Campostrini, K.~Moriarty, and C.~Rebbi,
\newblock Phys. Rev. {\bf D36}, 3450 (1987).

\bibitem{Koike:1987jh}
Y.~Koike,
\newblock Phys. Rev. Lett. {\bf 59}, 962 (1987).

\bibitem{Koike:1989jf}
Y.~Koike,
\newblock Phys. Lett. {\bf B216}, 184 (1989).

\bibitem{Born:1994cp}
K.~D. Born, E.~Laermann, T.~F. Walsh, and P.~M. Zerwas,
\newblock Phys. Lett. {\bf B329}, 332 (1994).

\bibitem{Gross:1973th}
D.~J. Gross and F.~Wilczek,
\newblock Phys. Rev. Lett. {\bf 30}, 1343 (1973).

\bibitem{Politzer:1973um}
H.~D. Politzer,
\newblock Phys. Rev. Lett. {\bf 30}, 1346 (1973).

\bibitem{Caswell:1974gg}
W.~E. Caswell,
\newblock Phys. Rev. Lett. {\bf 33}, 244 (1974).

\bibitem{Jones:1974mm}
D.~R.~T. Jones,
\newblock Nucl. Phys. {\bf B75}, 531 (1974).

\bibitem{Egorian:1979zx}
E.~Egorian and O.~V. Tarasov,
\newblock Theor. Math. Phys. {\bf 41}, 863 (1979).

\bibitem{Tarasov:1980au}
O.~V. Tarasov, A.~A. Vladimirov, and A.~Y. Zharkov,
\newblock Phys. Lett. {\bf 93B}, 429 (1980).

\bibitem{Larin:1993tp}
S.~A. Larin and J.~A.~M. Vermaseren,
\newblock Phys. Lett. {\bf B303}, 334 (1993), hep-ph/9302208.

\bibitem{'tHooft:1973mm}
G.~'t~Hooft,
\newblock Nucl. Phys. {\bf B61}, 455 (1973).

\bibitem{Bardeen:1978yd}
W.~A. Bardeen, A.~J. Buras, D.~W. Duke, and T.~Muta,
\newblock Phys. Rev. {\bf D18}, 3998 (1978).

\bibitem{Braaten:1981dv}
E.~Braaten and J.~P. Leveille,
\newblock Phys. Rev. {\bf D24}, 1369 (1981).

\bibitem{'tHooft:1972fi}
G.~'t~Hooft and M.~Veltman,
\newblock Nucl. Phys. {\bf B44}, 189 (1972).

\bibitem{vanRitbergen:1997va}
T.~van Ritbergen, J.~A.~M. Vermaseren, and S.~A. Larin,
\newblock Phys. Lett. {\bf B400}, 379 (1997), hep-ph/9701390.

\bibitem{Luscher:1995nr}
M.~L{\"u}scher and P.~Weisz,
\newblock Phys. Lett. {\bf B349}, 165 (1995), hep-lat/9502001.

\bibitem{Christou:1998ws}
C.~Christou, A.~Feo, H.~Panagopoulos, and E.~Vicari,
\newblock Nucl. Phys. {\bf B525}, 387 (1998), hep-lat/9801007.

\bibitem{Alles:1997cy}
B.~All{\'e}s, A.~Feo, and H.~Panagopoulos,
\newblock Nucl. Phys. {\bf B491}, 498 (1997), hep-lat/9609025.

\bibitem{Luscher:1995np}
M.~L{\"u}scher and P.~Weisz,
\newblock Nucl. Phys. {\bf B452}, 234 (1995), hep-lat/9505011.

\bibitem{Michael:1989vh}
C.~Michael,
\newblock Phys. Lett. {\bf B232}, 247 (1989).

\bibitem{Balzereit:1996yy}
C.~Balzereit and T.~Ohl,
\newblock Phys. Lett. {\bf B386}, 335 (1996), hep-ph/9604352.

\bibitem{Amoros:1997rx}
G.~Amor{\'o}s, M.~Beneke, and M.~Neubert,
\newblock Phys. Lett. {\bf B401}, 81 (1997), hep-ph/9701375.

\bibitem{Bauer:1998gs}
C.~Bauer and A.~V. Manohar,
\newblock Phys. Rev. {\bf D57}, 337 (1998), hep-ph/9708306.

\end{thebibliography}
\bibliographystyle{h-physrev3}

\end{document}